\documentclass [12pt]{article}
\usepackage{amsmath,amssymb,cite}

\usepackage{tabularx}
\usepackage[english]{babel}
\usepackage[dvips]{graphicx}
\usepackage{indentfirst}
\usepackage{epsfig}
\numberwithin{equation}{section}

\begin{document}

\title{
Impact of Supersymmetry on Emergent Geometry in Yang-Mills Matrix Models\footnote{An early version of this work titled "Noncommutative Gauge Theory As/Is Matrix Models Around Fuzzy Vacua And Emergent Geometry" was presented to BM Annaba University as a habilitation thesis on 20 january 2011.}}

\author{Badis Ydri\footnote{Emails:ydri@stp.dias.ie, badis.ydri@univ-annaba.org.}\\
Institute of Physics, BM Annaba University,\\
BP 12, 23000, Annaba, Algeria.
}

\maketitle
\abstract{We present a study of $D=4$ supersymmetric Yang-Mills matrix models with $SO(3)$ mass terms based on the cohomological approach and the Monte Carlo method.

In the bosonic models we show the existence of an exotic first/second order transition from a phase with a well defined background geometry (the fuzzy sphere) to a phase with commuting matrices with no geometry in the sense of Connes. At the transition point the sphere expands abruptly to infinite size then  it  evaporates as we increase the temperature (the gauge coupling constant). The transition looks first order due to the discontinuity in the action whereas it looks second order due to the divergent peak in the specific heat.

The fuzzy sphere is stable for the supersymmetric models in the sense that the bosonic phase transition is turned into a very slow crossover transition. The transition point  is found to scale to zero with $N$. We conjecture that the transition from the background sphere to the phase of commuting matrices is associated with spontaneous supersymmetry breaking.

The eigenvalues distribution of any of the bosonic matrices in the matrix phase is found to be given by a non-polynomial law obtained from the fact that the joint probability distribution of the four matrices is uniform inside a solid ball with radius $R$. The eigenvalues of the gauge field on the background geometry are also found to be distributed according to this non-polynomial law.

We also discuss the $D=3$ models and by using cohomological deformation, localization techniques and the saddle-point method we give a derivation of the $D=3$ eigenvalues distribution starting from a particular $D=4$ model.

}
\newpage
\tableofcontents
\newpage
\section{Introduction}

It can be argued using the principles of quantum mechanics and classical general relativity that the picture of spacetime at the very large as being a smooth manifold must necessarily break down at the Planck scale ${\lambda}_p$ \cite{Doplicher:1994tu,Doplicher:1994zv}. At this scale localization looses its operational meaning due to intense gravitational fields and formation of black holes and as a consequence one expects spacetime uncertainty relations which in turn strongly suggest that spacetime has a quantum structure expressed by $[x_{\mu},x_{\nu}]=i{\lambda}_p^2Q_{\mu \nu}$. The geometry of spacetime at the very small is therefore noncommutative. 

Noncommutative geometry \cite{Connes:1994yd}, see also \cite{Madore:2000aq,Landi:1997sh,GraciaBondia:2001tr,Varilly:1997qg,Coquereaux:1992wa}  and \cite{Frohlich:1993es}, allows for the description of the geometry of smooth differentiable manifolds  in terms of the underlying $C^{*}-$algebra of functions defined on these manifolds. Indeed given the following three data $a)$ the algebra ${\cal A}=C^{\infty}(M)$ of complex valued smooth and continous functions on a manifold $M$, $b)$ the Hilbert space ${\cal H}=L^2(M,S)$ of square integrable sections of the irreducible spinor bundle over $M$ and $c)$ the Dirac operator ${\cal D}={\gamma}^{\mu}({\partial}_{\mu}+\frac{1}{2}{\omega}_{\mu ab}{\gamma}^a{\gamma}^b)$ associated with the Levi-Civita connection $\omega$ one can reconstruct completely the differential geometry of the manifold $M$. These three data compose the so-called spectral triple $({\cal A},{\cal H},{\cal D})$ corresponding to the Riemannian manifold $M$. In the absence of spin structure it is sufficient to use the Laplacian ${\Delta}$  instead of the Dirac operator in the spectral triple.

Noncommutative geometry is also more general than ordinary differential geometry in that it also enables us to describe algebraically the geometry of arbitrary spaces (which a priori do not need to consist of points) in terms of  spectral triples. The paradigm of noncommutative geometry adopted so often in physics is to generalize the ordinary commutative space $M$ by replacing the commutative algebra ${\cal A}$ by a noncommutative algebra ${\cal A}_{\theta}$. The result of this deformation is in general a noncommutative space $M_{\theta}$ defined precisely by the spectral triple $({\cal A}_{\theta},{\cal H}_{\theta},{\cal D}_{\theta}/{\Delta}_{\theta})$ where the Hilbert space ${\cal H}_{\theta}$ is the representation space of the noncommutative algebra ${\cal A}_{\theta}$ and ${\cal D}_{\theta}/{\Delta}_{\theta}$ is the deformation of the commutative Dirac operator/Laplacian ${\cal D}/{\Delta}$  \cite{Connes:1994yd,Frohlich:1993es}.

Noncommutative geometry was also proposed  (in fact earlier than renormalization)  as a possible way to eliminate ultraviolet divergences in quantum field theories  \cite{Snyder:1946qz,Yang:1947ud}. The quantum spacetime of \cite{Doplicher:1994tu,Doplicher:1994zv} is Lorentz-covariant based on the commutation relations $[x_{\mu},x_{\nu}]=i{\lambda}_p^2Q_{\mu \nu}$ with $Q_{\mu \nu}$ satisfying $[x_{\lambda},Q_{\mu \nu}]=0$, $Q_{\mu \nu}Q^{\mu \nu}=0$ and $(\frac{1}{2}{\epsilon}_{\mu \nu \lambda \rho}Q^{\mu \nu}Q^{\lambda \rho})^2=1$. As it turns out quantum field theory on this space is ultraviolet finite \cite{Bahns:2003vb} which is a remarkable consequence of spacetime quantization. This result is however not  very surprising. 
Indeed this phenomena of ``regularization by quantization'' already happens in quantum mechanics. For example while classical mechanics fail to explain the blackbody radiation in the ultraviolet (the UV catastroph) quantum mechanics reproduces the correct (finite) answer given by the famous experimentally verified Stefan-Boltzman law.


Noncommutative field theory is by definition a field theory based on a noncommutative spacetime \cite{Douglas:2001ba,Szabo:2001kg}. The most studied examples in the literature are the Moyal-Weyl spaces ${\bf R}^d_{\theta}$ which correspond to the case $Q_{\mu \nu}={\theta}_{\mu \nu}$  where ${\theta}_{\mu \nu}$ are rank $2$ (or $1$) antisymmetric constant tensors, i.e. 
\begin{eqnarray}
[x_{\mu},x_{\nu}]=i{\theta}_{\mu\nu}.
\end{eqnarray}
This clearly breakes Lorentz symmetry. The corresponding quantum field theories are not UV finite \cite{Filk:1996dm} and furthermore they are plagued with the so-called UV-IR mixing phenomena \cite{Minwalla:1999px}. 

For example in the case of  scalar field theory on Moyal-Weyl spaces the UV-IR mixing was shown to destroy the 
perturbative renormalizability of the theory in \cite{Chepelev:1999tt,Chepelev:2000hm}. The UV-IR mixing in this model means in particular that the non-planar contribution to the two-point function which is finite for generic values of the  external momentum behaves as $1/(\theta p)^2$ in the limit $p\longrightarrow 0$ and/or $\theta\longrightarrow 0$. The physics at very large distances is thus altered by the noncommutativity which is supposed to be relevant only at very short distances. Equivalently it was  shown in \cite{Griguolo:2001ez} that although the renormalization group equations are finite in the IR regime their perturbative approximations are IR divergent. In \cite{Grosse:2003aj,Grosse:2004yu,Grosse:2003nw} a modified scalar field theory on noncommutative ${\bf R}^d_{\theta}$ was proposed and shown to be renormalizable to all orders in perturbation theory in $d=2$ and $4$. The modification consists of making  the action invariant under the duality transformation of \cite{Langmann:2002cc} by adding a harmonic oscillator potential to the kinetic term which precisely modifies the free theory as desired. 


For scalar field theory with ${\phi}^4$ interaction the phase diagram consists of three phases instead of the usual two phases found in commutative scalar field theory. We observe a disordered phase, a uniform ordered phase and a novel nonuniform ordered phase which meet at a triple point possibly a Lifshitz point \cite{Gubser:2000cd,Chen:2001an}. The nonuniform phase is the analogue of the matrix phase in pure gauge models (see below) in the sense that in this phase the spacetime metric is modified by quantum fluctuations of the noncommutative field theory since the Laplacian at the transition point is found to be  $({\partial}_{\mu}^2)^2$ and not  ${\partial}_{\mu}^2$  \cite{Chen:2001an}. In the nonuniform ordered phase we have 
spontaneous breakdown of translational invariance \cite{Gubser:2000cd,Castorina:2003zv,Girotti:2002kr}.  The transition from the disordered phase to the nonuniform ordered phase is thought to be first order and the nonuniform ordered phase is a periodically modulated phase which for small values of the coupling constant is dominated by stripes \cite{Gubser:2000cd}. This should hold even in two dimensions.

This behavior was confirmed in Monte Carlo simulations on the noncommutative torus \cite{Ambjorn:2000cs,Ambjorn:1999ts,Ambjorn:2000nb} in \cite{Ambjorn:2002nj,Bietenholz:2004xs} and on the fuzzy sphere \cite{Hoppe:1982,Madore:1991bw} in \cite{Martin:2004un,GarciaFlores:2005xc,Panero:2006bx}. 

It is therefore natural to conjecture that there must exist two fixed points in this theory, the usual Wilson-Fisher fixed point at $\theta=0$ and a novel fixed point at $\theta=\infty$ which is intimately related to the underlying 
matrix model structure of the model \cite{Ydri:2012nw}. The crucial input in a scalar field theory on a noncommutative space is the scalar potential which when written in the matrix base is a random matrix theory which for a ${\phi}^4$ interaction is given by
\begin{eqnarray}
V=N(aTr M^2+bTr M^4).\label{matrix}
\end{eqnarray}
The main property of this matrix model which plays a central role in the phase structure of a noncommutative ${\phi}^4$ interaction is the existence of one-cut/two-cut transition, with the one-cut (disordered) phase corresponding to $a{\geq}a_*$ where \cite{Brezin:1977sv,Shimamune:1981qf}
\begin{eqnarray}
a_*=-2\sqrt{b}.
\end{eqnarray}
 The kinetic term is trying to add a geometry to the dynamics of the matrix $M$ which is at the heart of the rich phase structure we observe. For small $b$ the usual Ising model transition is expected and the $\theta=0$ fixed point should control the physics. We expect on the other hand that the  $\theta=\infty$ fixed point should control most of the phase diagram since generically the  ${\phi}^4$ interaction, i.e. the coupling $b$ is not weak.


Noncommutative gauge theories  attracted a lot of interest in recent years because of their appearance in string theory \cite{Seiberg:1999vs,Connes:1997cr,Schomerus:1999ug}. For example it was discovered that the dynamics  of open strings moving in a flat space in the presence of a non-vanishing Neveu-Schwarz B-field   and with Dp-branes is equivalent to leading order in the string tension to a gauge theory on  a Moyal-Weyl space ${\bf R}^d_{\theta}$. Extension of this result to curved spaces is also possible at least in one particular instance, namely the case of open strings moving in a curved space with  ${\bf S}^3$ metric. The resulting effective gauge theory lives on a noncommutative fuzzy sphere ${\bf S}^2_N$  \cite{Alekseev:1999bs,Alekseev:2000fd,Hikida:2001py}.

Another class of noncommutative spaces, besides Moyal-Weyl spaces, which will be very important for us in this article is fuzzy spaces \cite{O'Connor:2003aj,Balachandran:2002ig}. Fuzzy physics is by definition a field theory based on fuzzy spaces \cite{Ydri:2001pv,Kurkcuoglu:2004gf}. The original idea of discretization by quantization (fuzzification) works well for co-adjoint orbits such as  projective spaces.  A seminal example of fuzzy spaces is the fuzzy two-dimensional sphere ${\bf S}^2_N$  \cite{Hoppe:1982,Madore:1991bw}. The fuzzy sphere is defined by three $N\times N$ matrices $x_i$, $i=1,2,3$ playing the role of coordinates operators  satisfying $\sum_ix_i^2=1$ and the commutation relations
\begin{eqnarray}
[x_i,x_j]=i\theta{\epsilon}_{ijk}x_k~,~\theta=\frac{1}{\sqrt{c_2}}~,~c_2=\frac{N^2-1}{4}.
\end{eqnarray}
The fuzzy sphere which is the simplest among fuzzy projective spaces was actually proposed as a nonperturbative regularization of ordinary quantum field theory in \cite{Grosse:1996mz}. See also  \cite{Grosse:1995ar,Grosse:1998gn,Grosse:1995pr,Klimcik:1999uk}. However we know now that this can not be correct because of the complicated phase structure of noncommutative $\phi^4$ scalar field theories on the fuzzy sphere \cite{O'Connor:2007ea,Steinacker:2005wj}. Instead it is becoming very clear that it is more appropriate to think of the fuzzy sphere as a nonperturbative regularization of noncommutative quantum field theory on Moyal-Weyl plane. For example we can show in a particular double scaling limit that the nonperturbative phase structure of scalar fields on the Moyal-Weyl plane can be  rigorously identified with  the nonperturbative  phase structure of scalar fields on the fuzzy sphere, and in particular the matrix or nonuniform ordered phase on the fuzzy sphere goes precisely to the stripe phase on the Moyal-Weyl plane \cite{Medina:2007nv}.

In $3-$dimensions we have the fuzzy three sphere ${\bf S}^3_N$ \cite{Dolan:2003kq}. In $4-$dimensions we have $3$ fuzzy manifolds which are obtained from co-adjoint orbits. The direct product of fuzzy two spheres ${\bf S}^2_N{\times}{\bf S}^2_N$ \cite{Vaidya:2003ew}, fuzzy ${\bf CP}^2_N$ \cite{Grosse:1999ci,Alexanian:2001qj} and fuzzy ${\bf S}^4_N$ as a  squashed fuzzy ${\bf CP}^3_N$ \cite{Medina:2002pc}. 

The most appealing aspect of discretization by quantization remains its remarkable success  in preserving continuous symmetries including supersymmetry and capturing correctly topological  properties \cite{Ydri:2001pv,Kurkcuoglu:2004gf,Balachandran:2005ew}. Indeed the fuzzy approach does not suffer from fermion doubling \cite{Grosse:1998gn,Balachandran:1999qu,Balachandran:2003ay}, it extends naturally to supersymmetry \cite{Grosse:1995pr,Klimcik:1999uk,Valavane:2000iz,Balachandran:2002jf} and it captures correctly nontrivial field configurations such as monopoles and instantons using the language of projective modules and K-theory \cite{Grosse:1995jt,Baez:1998he,Balachandran:1999hx,CarowWatamura:2004ct}.

The noncommutative Moyal-Weyl spaces should be thought of as infinite dimensional matrix algebras not as continuum manifolds.  Quantum fluctuations of a gauge theory on ${\bf R}_{\theta}^d$ will generically make the vacuum which is here the Moyal-Weyl space itself unstable. To see this effect we formally rewrite gauge theory on ${\bf R}^d_{\theta}$ as a matrix model with $N-$dimensional matrices $\hat{D}_i$ where $N\longrightarrow\infty$. This is given by
\begin{eqnarray}
S=\frac{\sqrt{{\rm det}(\pi \theta B)}}{2g^2}Tr_{\cal H}\bigg(i[\hat{D}_i,\hat{D}_j]-\frac{1}{{\theta}}B^{-1}_{ij}\bigg)^2.\label{action}
\end{eqnarray}
By computing the effective potential in the configuration $\hat{D}_i=-\phi B^{-1}_{ij} \hat{x}_j$ we will verify explicitly that the Moyal-Weyl space itself, i.e the algebra  $[x_{\mu},x_{\nu}]=i{\theta}_{\mu \nu}$, ceases to exist above a certain value $g_*$  of the gauge coupling constant. We find $g_*^2={\pi}^2/N$ in four dimensions (see section $2$). In other words by moving in the phase diagram from strong coupling to weak coupling the geometry of the Moyal-Weyl spaces (including the star product and the representation of operators by fields)  emerges at the critical point $g_*$. As a consequence a natural regularization of Moyal-Weyl spaces must include in a fundamental way matrix degrees of freedom. The regularization which will be employed in this article is given by fuzzy projective spaces \cite{Balachandran:2001dd}. The main reason behind this choice is the fact that the  phenomena of emergent geometry \cite{Bombelli:1987aa,Seiberg:2006wf,Ambjorn:2006hu} which we observed here on  Moyal-Weyl spaces shows up also in all matrix models on fuzzy projective spaces.

Let us note the deep connection which exists between the geometry in transition we observe here on  Moyal-Weyl space  ${\bf R}_{\theta}^4$ and the perturbative UV-IR mixing and beta function of $U(1)$ gauge theory computed in \cite{Martin:2000bk,Krajewski:1999ja}. The structure of the effective potential in $2$ dimensions indicates that there is no transition and no UV-IR mixing in $U(1)$ gauge theory on ${\bf R}_{\theta}^2$ and therefore the  theory is expected to be renormalizable. This was shown numerically to be true in \cite{Bietenholz:2002ch}. We also expect that noncommutative ${\cal N}=1$ supersymmetric $U(1)$ gauge theory in $4$ dimensions is renormalizable.

Let us note also  that fuzzy regularization is different from the usual one which is based on the Eguchi-Kawai model \cite{Eguchi:1982nm} and the noncommutative torus \cite{Ambjorn:2000cs,Ambjorn:1999ts,Ambjorn:2000nb}, i.e the twisted Eguchi-Kawai model.  The twisted Eguchi-Kawai model was employed as a nonperturbative regularization of noncommutative gauge theory  in \cite{Bietenholz:2004wk,Bietenholz:2006cz,Bietenholz:2005iz} where the instability and the phase transition discussed here were also obtained. 

Finally we note  that the strong coupling phase (above $g_*$) of noncommutative gauge theory on ${\bf R}_{\theta}^d$ corresponds to $\theta=\infty$. It is dominated by commuting operators. The limit $\theta\longrightarrow\infty$ is the planar theory (only planar graphs survive) \cite{Filk:1996dm} and it is intimately related to large $N$ limits of hermitian matrix models \cite{DiFrancesco:1993nw,eynard} and   \cite{Gubser:2000cd}. In this phase of commuting operators supersymmetry may be broken.

The phenomena of emergent geometry associated with noncommutative gauge theory is therefore a major motivation behind choosing fuzzy projective spaces as a nonperturbative regularization of Moyal-Weyl spaces. The other motivation comes of course from the novel phase known variously as stripe, nonuniform ordered or matrix phase found in noncommutative scalar field theory which we have only briefly discussed in this introduction. The phenomena  of emergent geometry is also, on the other hand, intimately tied to reduced Yang-Mills models which will be the topic of central interest in the remainder of this introduction and in most of this article.

It is well established that reduced Yang-Mills theories play a central role in the nonperturbative definitions of $M$-theory and superstrings. The BFSS conjecture \cite{Banks:1996vh} relates discrete light-cone quantization (DLCQ) of $M-$theory to the theory of $N$ coincident $D0$ branes which at low energy (small velocities and/or string coupling) is the reduction to $0+1$ dimension  of the $10$ dimensional $U(N)$ supersymmetric Yang-Mills gauge theory \cite{Witten:1995im}. The BFSS model is therefore a Yang-Mills quantum mechanics  which is supposed to be the UV completion of $11$ dimensional supergravity. As it turns out the BFSS action is nothing else but the regularization of the supermembrane action in the light cone gauge \cite{deWit:1988ig}. 

The BMN model \cite{Berenstein:2002jq} is a generalization of the BFSS model to curved backgrounds. It is obtained by adding to the BFSS action a one-parameter mass deformation corresponding to the maximally supersymmetric pp-wave background of $11$ dimensional supergravity. See for example \cite{KowalskiGlikman:1984wv,Blau:2001ne,Blau:2002dy}. We note, in passing, that all maximally supersymmetric pp-wave geometries can arise as Penrose limits of $AdS_p\times S^q$ spaces \cite{penrose}.

The IKKT model \cite{Ishibashi:1996xs} is, on the other hand, a Yang-Mills matrix model obtained by dimensionally reducing $10$ dimensional $U(N)$ supersymmetric Yang-Mills gauge theory to $0+0$ dimensions.  The IKKT model is postulated to provide a constructive definition of type II B superstring theory and for this reason it is also called type IIB matrix model. Supersymmetric analogue of the IKKT model also exists in dimensions $d=3,4$ and $6$ while the partition functions converge only in dimensions $d=4,6$ \cite{Krauth:1998yu,Austing:2001pk}. 

The IKKT Yang-Mills matrix models can be thought of as continuum Eguchi-Kawai reduced models as opposed to the usual lattice Eguchi-Kawai reduced model formulated in \cite{Eguchi:1982nm}. We point out here the similarity between the conjecture that the lattice Eguchi-Kawai reduced model allows us  to recover the full gauge theory in the large $N$ theory and the conjecture that the IKKT matrix model allows us to recover type II B superstring. 

The relation between the BFSS Yang-Mills quantum mechanics and the IKKT Yang-Mills matrix model is discussed at length in the seminal paper \cite{Connes:1997cr} where it is also shown that toroidal compactification of the D-instanton action (the bosonic part of the IKKT action) yields, in a very natural way,  a noncommutative Yang-Mills theory on a dual noncommutative torus \cite{Connes:1987ue}. From the other hand, we can easily check that the ground state of the D-instanton action is given by commuting matrices which can be diagonalized simultaneously with the eigenvalues giving the coordinates of the D-branes. Thus at tree-level an ordinary spacetime emerges from the bosonic truncation of the IKKT action while higher order quantum corrections will define a noncommutative spacetime.

In summary, Yang-Mills matrix models which provide a constructive definition of string theories will naturally lead to emergent geometry \cite{Seiberg:2006wf} and non-commutative gauge theory \cite{Aoki:1999vr,Aoki:1998vn}. Furthermore, non-commutative geometry \cite{Connes:1994yd,GraciaBondia:2001tr} and their non-commutative field theories \cite{Douglas:2001ba,Szabo:2001kg} play an essential role in the non-perturbative dynamics of superstrings and $M$-theory. Thus the connections between non-commutative field theories, emergent geometry and matrix models from one side and string theory from the other side run deep. 

It seems therefore natural that Yang-Mills matrix models provide a non-perturbative framework for emergent spacetime geometry and non-commutative gauge theories. Since non-commutativity  is the only extension which preserves maximal supersymmetry, we also hope that Yang-Mills matrix models will provide a regularization which preserves supersymmetry \cite{Nishimura:2009xm}.

In this article we will explore in particular the possibility of using IKKT Yang-Mills matrix models in dimensions $4$ and $3$ to provide a non-perturbative definition of emergent spacetime geometry, non-commutative gauge theory and supersymmetry in two dimensions. From our perspective in this article, the phase of commuting matrices has no geometry in the sense of Connes and thus we need to modify the models so that a geometry with a well defined spectral triple can also emerge alongside the phase of commuting matrices. 

There are two solutions to this problem. The first solution is given by adding mass deformations which preserve supersymmetry to the flat IKKT Yang-Mills matrix models \cite{Bonelli:2002mb} or alternatively by an Eguchi-Kawai reduction of the mass deformed BFSS Yang-Mills quantum mechanics constructed in \cite{Kim:2006wg,Kim:2002cr,Hyun:2002fk,Hyun:2003se}. The second solution, which we have also considered in this article, is given by deforming the flat Yang-Mills matrix model in $D=4$ using the powerful formalism  of cohomological Yang-Mills theory  \cite{Moore:1998et,Hirano:1997ai,Kazakov:1998ji,Hoppe:1999xg}. 

These mass deformed or cohomologically deformed IKKT Yang-Mills matrix models are the analogue of the BMN model and they  typically include a Myers term \cite{Myers:1999ps} and thus they will sustain the geometry of the fuzzy sphere \cite{Hoppe:1982,Madore:1991bw} as a ground state which at large $N$ will approach the geometry of the ordinary sphere, the ordinary plane or the non-commutative plane depending on  the scaling limit. Thus a non-perturbative formulation of non-commutative gauge theory in two dimensions can be captured rigorously   within these models \cite{CarowWatamura:1998jn,Iso:2001mg,Presnajder:2003ak}. See also \cite{Ishiki:2008vf,Ishiki:2009vr}.

This can in principle be generalized to other fuzzy spaces \cite{Balachandran:2005ew} and higher dimensional non-commutative gauge theories by considering appropriate mass deformations of the flat IKKT Yang-Mills matrix models.

The problem or virtue of this construction, depending on the perspective, is that in these Yang-Mills matrix models the geometry of the fuzzy sphere collapses under quantum fluctuations  into the phase of commuting matrices. Equivalently, it is seen that the geometry of the fuzzy sphere emerges from the dynamics of a random matrix theory \cite{DelgadilloBlando:2007vx,Azuma:2004zq}. Supersymmetry is naturally expected to stabilize the spacetime geometry, and in fact the non stability of the non-supersymmetric vacuum should have come  as no surprise  to us \cite{Witten:2000zk}.  

We should mention here the approach  of \cite{Steinacker:2003sd} in which a noncommutative Yang-Mills gauge theory on the fuzzy sphere emerges also from the dynamics of a random matrix theory. The fuzzy sphere is stable in the sense that the transition to commuting matrices is pushed towards infinite gauge coupling at large $N$ \cite{O'Connor:2006wv}. This was achieved by considering a very special non-supersymmetric mass deformation which is quartic in the bosonic matrices. This construction was extended  to a  noncommutative gauge theory on the fuzzy sphere based on co-adjoint orbits in \cite{Steinacker:2007iq}. 
 
Let us also note here that the instability and the phase transition discussed here were also observed on the non-commutative torus in \cite{Bietenholz:2004wk,Bietenholz:2006cz,Bietenholz:2005iz,Azeyanagi:2008bk,Azeyanagi:2007su} where the twisted Eguchi-Kawai model was employed as a non-perturbative regularization of non-commutative Yang-Mills gauge theory  \cite{Ambjorn:2000cs,Ambjorn:1999ts,Ambjorn:2000nb}.

In this article we will study, using cohomological matrix theory and the Monte Carlo method, the mass deformed Yang-Mills matrix model in $D=4$ as well as a particular truncation to $D=3$. We will derive and study a one-parameter cohomological deformation of the Yang-Mills matrix model which coincides with the mass deformed model in $D=4$ when the parameter is tuned appropriately. We will show that the first/second order phase transition  from the fuzzy sphere to the phase of commuting matrices observed in the bosonic models is converted  in the supersymmetric models into a very slow crossover transition with an arbitrary small transition point in the large $N$ limit. We will determine the eigenvalues distributions for both $D=4$ and $D=3$ throughout the phase diagram. The $D=3$ eigenvalues distribution can be obtained from a particular  $D=4$ model by means of the methods of cohomological Yang-Mills matrix theory, large $N$ saddle point and localization techniques \cite{Witten:1991we,Witten:1992xu}.

This article is organized as follows: In section $2$ we give a brief discussion of the phenomena of emergent geometry on the Moyal-Weyl space using the effective potential. In section $3$ we give a short review on fuzzy projective spaces. In section $4$ we review results on emergent geometry in bosonic $D=3$ Yang-Mills matrix models. In section $5$ we will derive the mass deformed Yang-Mills quantum mechanics  from the requirement of supersymmetry and then reduce it further to obtain Yang-Mills matrix model in $D=4$ dimensions. In section $6$ we will derive a one-parameter family of cohomologically deformed models and then show that the mass deformed model constructed in section $5$ can be obtained for a particular value of the parameter. In section $7$ we report our first Monte Carlo results for the model $D=4$ including the eigenvalues distributions and also comment on the $D=3$ model obtained by simply setting the fourth matrix to $0$. In section $8$ we solve a particular $D=4$ model and show by means of cohomological Yang-Mills matrix theory and localization techniques that it is equivalent to the matrix $D=3$ Chern-Simons theory. By using the saddle point method we derive the  $D=3$ eigenvalues distribution. We conclude in section $9$ with a comprehensive summary of the results and discuss future directions. The detail of the simulations are found in appendices $A$ and $B$ while appendix $C$ contains a detail calculation of the supersymmetric mass deformation and appendix $D$ contains a calculation of the star product on the Moyal-Weyl plane starting from the star product on  the fuzzy sphere.

\section{Gauge Theory on Noncommutative  Moyal-Weyl Spaces}

The
basic noncommutative gauge theory action of interest to us in
this article can be obtained from  a matrix model of the form (see \cite{Douglas:2001ba} and references therein)
\begin{eqnarray}
S=\frac{\sqrt{{\theta}^d{\rm det}(\pi B)}}{2g^2}Tr_{\cal H}\hat{F}_{ij}^2=\frac{\sqrt{{\theta}^d{\rm det}(\pi B)}}{2g^2}Tr_{\cal H}\bigg(i[\hat{D}_i,\hat{D}_j]-\frac{1}{{\theta}}B^{-1}_{ij}\bigg)^2.\label{action}
\end{eqnarray}
Here $i,j=1,...,d$ with $d$ even and ${\theta}$ has dimension of length squared so that the connection operators
$\hat{D}_i$  have dimension of $({\rm length})^{-1}$.  The
coupling constant $g$ is of dimension $(\rm
mass)^{2-\frac{d}{2}}$ and $B^{-1}$ is  an invertible tensor which in
$2$ dimensions is given by $B^{-1}_{ij}={\epsilon}^{-1}_{ij}=-{\epsilon}_{ij}$ while in higher dimensions is given by
\begin{eqnarray}
B^{-1}_{ij}=\left(\begin{array}{ccccccc}
-{\epsilon}_{ij}&&&&&&\\
&.&&&&&\\
&&&&.&&\\
&&&&&&-{\epsilon}_{ij}
\end{array}\right).
\end{eqnarray}
The operators $\hat{A}_i$ belong to an algebra ${\cal A}$. The trace is taken over some infinite
dimensional Hilbert space ${\cal H}$ and hence
$Tr_{\cal H}[\hat{D}_i,\hat{D}_j]$ is ${\neq}0$ in general, i.e. $Tr_{\cal H}[\hat{D}_i,\hat{D}_j]$ is in fact a topological term \cite{Connes:1997cr}.  Furthermore we will assume Euclidean signature throughout.

Minima of the model (\ref{action}) are connection operators $\hat{D}_i=\hat{B}_i$ satisfying 
\begin{eqnarray}
i[\hat{B}_i,\hat{B}_j]=\frac{1}{\theta}B^{-1}_{ij}.\label{eom} 
\end{eqnarray}
We view the algebra ${\cal A}$ as ${\cal A}={\rm Mat}_n({\bf C})\otimes {\cal A}_n$. The trace $Tr_{\cal H}$ takes the form $Tr_{\cal H}=Tr_n Tr_{{\cal H}_n}$ where ${\cal H}_n$ is the Hilbert space associated with the elements of ${\cal A}_n$. The configurations $\hat{D}_i=\hat{B}_i$ which solve equation (\ref{eom}) can be written as 
\begin{eqnarray}
\hat{B}_i=-\frac{1}{\theta}B^{-1}_{ij}\hat{x}_j\otimes {\bf 1}_n.\label{minima}
\end{eqnarray}
The operators $\hat{x}_i$ which are elements of ${\cal A}_n$ can be identified with the coordinate operators on the noncommutative Moyal-Weyl space ${\bf
R}^d_{\theta}$ with the usual commutation relation
\begin{eqnarray}
[\hat{x}_i,\hat{x}_j]=i{\theta}{B}_{ij}.
\end{eqnarray}
Derivations on  ${\bf R}^d_{\theta}$ are defined by
\begin{eqnarray}
\hat{\partial}_i=i\hat{B}_i.
\end{eqnarray}
Indeed we compute
\begin{eqnarray}
[\hat{{\partial}_i},\hat{x}_j]={\delta}_{ij}.
\end{eqnarray}
The sector of this matrix theory
which corresponds to a noncommutative $U(n)$ gauge field on ${\bf
R}^d_{\theta}$ is therefore obtained by expanding $\hat{D}_i$ around $\hat{B}_i\otimes {\bf 1}_n$.  We write the configurations 
\begin{eqnarray}
\hat{D}_i=-\frac{1}{{\theta}}B^{-1}_{ij}\hat{x}_j{\otimes}{\bf 1}_n+\hat{A}_i,~\hat{A}_i^{+}=\hat{A}_i.\label{expansionMW}
\end{eqnarray}
The operators $\hat{A}_i$ are
identified with the components of the dynamical $U(n)$ noncommutative gauge field. The corresponding $U(n)$ gauge transformations which leave the action (\ref{action})
invariant are implemented by unitary operators
$U=\exp(i{\Lambda})~,~UU^{+}=U^{+}U=1~,~{\Lambda}^{+}={\Lambda}$
which act on the Hilbert space ${\cal H}={\cal H}_n\oplus ...\oplus{\cal H}_n$ as $\hat{D}_i{\longrightarrow}U\hat{D}_iU^{+}$, i.e. $\hat{A}_i{\longrightarrow}U\hat{A}_iU^{+}-iU[\hat{\partial}_i,U^{+}]$ and $\hat{F}_{ij}{\longrightarrow}U\hat{F}_{ij}U^{+}$. In other words $U(n)$ in this setting must be identified with $U({\cal H}_n\oplus ...\oplus{\cal H}_n)$. The action (\ref{action}) can be put into the form
\begin{eqnarray}
S=\frac{\sqrt{{\theta}^d{\rm det}(\pi B)}}{4g^2}Tr_{{\cal H}_n}(\hat{F}_{ij}^C)^2.\label{action0MW}
\end{eqnarray}
The curvature $\hat{F}_{ij}^C$ where $C$ is a $U(n)$ index which runs from $1$ to $n^2$ is given by
\begin{eqnarray}
\hat{F}_{ij}^C=[\hat{\partial}_i,\hat{A}_j^C]-[\hat{\partial}_j,\hat{A}_i^C]-\frac{1}{2}f_{ABC}\{\hat{A}_i^A,\hat{A}_j^B\}+\frac{i}{2}d_{ABC}[\hat{A}_i^A,\hat{A}_j^B].
\end{eqnarray}
In calculating $\hat{F}_{ij}^C$ we used $[T_A,T_B]=if_{ABC}T_C$, $\{T_A,T_B\}=d_{ABC}T_c$ and $Tr T_AT_B=\frac{{\delta}_{AB}}{2}$. More explicitely we have defined $T_a=\frac{{\lambda}_a}{2}$ for the $SU(n)$ part and $T_0=\frac{1}{\sqrt{2n}}{\bf 1}_n$ for the $U(1)$ part. The symbols $d_{ABC}$ are defined such that $d_{abc}$ are the usual $SU(n)$ symmetric symbols while $d_{ab0}=d_{a0b}=d_{0ab}=\sqrt{\frac{2}{n}}{\delta}_{ab}$, $d_{a00}=0$ and $d_{000}=\sqrt{\frac{2}{n}}$.

Finally it is not difficult to show using the Weyl map, viz the map between operators and fields,  that the matrix action (\ref{action0MW}) is
precisely the usual noncommutative $U(n)$ gauge action on ${\bf
R}^d_{\theta}$ with a star product $*$ defined by the parameter
${\theta}B_{ij}$ \cite{Douglas:2001ba,Szabo:2001kg}. In particular the trace $Tr_{{\cal H}_n}$ on the Hilbert space ${\cal H}_n$ can be shown to be equal to the integral over spacetime. We get 
\begin{eqnarray}
S=\frac{1}{4g^2}\int d^dx
~({F}_{ij}^C)^2~,~{F}_{ij}^C={\partial}_i{A}_j^C-{\partial}_j{A}_i^C-\frac{1}{2}f_{ABC}\{{A}_i^A,{A}_j^B\}_*+\frac{i}{2}d_{ABC}[{A}_i^A,{A}_j^B]_*.\label{action1MW}\nonumber\\
\end{eqnarray}
Let us note that although the dimensions  ${\rm dim}{\cal H}$ and ${\rm dim}{\cal H}_n$ of the Hilbert spaces ${\cal H}$ and ${\cal H}_n$ are infinite the ratio ${\rm dim}{\cal H}/{\rm dim}{\cal H}_n$ is finite equal $n$. The number of independent unitary transformations which leave the configuration (\ref{minima}) invariant is equal to ${\rm dim}{\cal H}-{\rm dim}{\cal H}_n-n^2$. This is clearly less than ${\rm dim}{\cal H}$ for any $n\geq 2$. In other words from entropy counting the $U(1)$ gauge group (i.e. $n=1$) is more stable than all higher gauge groups.  As we will show the $U(1)$ gauge group is in fact energetically favorable in  most of the  finite $N$ matrix models which are proposed as non-perturbative regularizations of (\ref{action}) in this article. Stabilizing $U(n)$ gauge groups requires adding potential terms to the action. In the rest of this section we will thus consider only the  $U(1)$ case for simplicity.

Quantization of the matrix model (\ref{action}) consists usually
in quantizing the model (\ref{action1MW}).  As we will argue shortly this makes sense only for small values of the coupling constant $g^2$ which are less than a critical value $g^2_*$. Above $g^2_*$ the configuration $\hat{B}_i$ given by (\ref{minima}) ceases to exist, i.e. it ceases to be the true minimum of the theory and as a consequence the expansion (\ref{expansionMW}) does not make sense.

In order to compute this transition we use the one-loop effective action which can be easily obtained in the Feynamn-'t Hooft background field gauge. We find the result 
\begin{eqnarray}
{\Gamma}=S+\frac{1}{2}Tr_dTr_{\rm ad}\ln\bigg({\cal
D}^2{\delta}_{ij}-2i{\cal F}_{ij}\bigg)-Tr_{\rm ad}\ln{\cal
D}^2.\label{effectiveaction}
\end{eqnarray}
The operators ${\cal D}^2={\cal D}_i{\cal
D}_i$ , ${\cal D}_i$ and ${\cal F}_{ij}$ act by commutators, viz ${\cal D}^2(..)=[\hat{D}_i,[\hat{D}_i,..]]$, ${\cal D}_i(..)=[\hat{D}_i,..]$ and ${\cal F}_{ij}(..)=[\hat{F}_{ij},..]$.  Next we compute the effective potential in the configuration $\hat{D}_i=-\phi B^{-1}_{ij}\hat{x}_j$. The curvature $\hat{F}_{ij}$ in this configuration is given by $\theta \hat{F}_{ij}=({\theta}^2{\phi}^2-1)B_{ij}^{-1}$. The trace over the Hilbert space ${\cal H}$ is regularized such that $Tr_{\cal H}{\bf 1}=N$ is a very large but finite natural number. We will also need $\sum_{i,j}B^{-1}_{ij}B^{-1}_{ij}=d$. The effective potential for $d\neq 2$ is given by
\begin{eqnarray}
\frac{V}{(d-2)N^2}=\alpha({\theta}^2{\phi}^2-1)^2+\ln\phi.
\end{eqnarray}
The coupling constant $\alpha$ is given by
\begin{eqnarray}
\alpha=\frac{d}{d-2}\frac{{\pi}^{\frac{d}{2}}}{2}\frac{1}{{\lambda}^2N}~,~\lambda={\theta}^{1-\frac{d}{4}}g.
\end{eqnarray}
We take the limit $N\longrightarrow \infty$ keeping ${\lambda}^2N$ fixed. It is not difficult to show that the minimum of the above potential is then given by
\begin{eqnarray}
(\theta\phi)^2=\frac{1+\sqrt{1-\frac{1}{\alpha}}}{2}.
\end{eqnarray}
The critical values are therefore given by
\begin{eqnarray}
{\alpha}_*=1\Leftrightarrow {\lambda}^2_*N=\frac{d}{d-2}\frac{{\pi}^{\frac{d}{2}}}{2}.
\end{eqnarray}
Thus the configuration $\hat{D}_i=-\phi B^{-1}_{ij}\hat{x}_j$ exists only for values of the coupling constant $\lambda$ which are less than ${\lambda}_*$. Above ${\lambda}_*$ true minima of the model are given by commuting operators,i.e.
\begin{eqnarray}
i[\hat{B}_i,\hat{B}_j]=0. \label{eom1}
\end{eqnarray}
By comparing with (\ref{eom}) we see that this phase corresponds to $\theta=\infty$. The limit $\theta\longrightarrow\infty$ is the planar theory (only planar graphs survive) \cite{Filk:1996dm} which is intimately related to large $N$ limits of hermitian matrix models \cite{Gubser:2000cd}. 

This transition from the noncommutative Moyal-Weyl space (\ref{eom}) to the commuting operators  (\ref{eom1}) is believed to be intimately related to the perturbative UV-IR mixing \cite{Minwalla:1999px}. Indeed this is true in two dimensions using our formalism here.

In two dimensions we can see that the logarithmic correction to the potential is absent and as a consequence the transition to commuting operators will be absent. The  perturbative UV-IR mixing is, on the other hand, absent in two dimensions.  Indeed in two dimensions the first nonzero  correction to the classical action $S$ in the effective action (\ref{effectiveaction}) is given by
\begin{eqnarray}
{\Gamma}&=&S-Tr_{\rm ad}\frac{1}{{\cal D}^2}{\cal F}_{ij}\frac{1}{{\cal D}^2}{\cal F}_{ij}+...\nonumber\\
&=&S+({\theta}{\pi})^2\int_k\frac{1}{k^2}\int_p\frac{1}{p^2}~Tr_{\cal H}F_{ij}[e^{ip\hat{x}},e^{-ik\hat{x}}]~Tr_{\cal H}F_{ij}[e^{ik\hat{x}},e^{-ip\hat{x}}]\nonumber\\
&=&S+2\int_p Tr |\tilde{F}_{ij}(p)|^2\int_k\frac{1}{k^2}\frac{1}{(p-k)^2}(1-\cos {{\theta}_{ij}p_ik_j})|.
\end{eqnarray}
By including a small mass $m^2$ and using Feynman parameters the planar and non-planar contributions are given respectively by 
\begin{eqnarray}
{\Pi}^{\rm P}&=&\int_k\frac{1}{k^2+m^2}\frac{1}{(p-k)^2+m^2}=\frac{({\theta}_{ij}p_i)^2}{4\pi}\int_0^1\frac{dx}{z^2}.
\end{eqnarray}
\begin{eqnarray}
&&{\Pi}^{\rm NP}=\int_k\frac{1}{k^2+m^2}\frac{1}{(p-k)^2+m^2}\cos {{\theta}_{ij}p_ik_j}=\frac{({\theta}_{ij}p_i)^2}{4\pi}\int_0^1\frac{dx}{z^2}zK_1(z).
\end{eqnarray}
In above $z$ is defined by $z^2=({\theta}_{ij}p_i)^2(m^2+x(1-x)p^2)$ and $K_1(z)$ is  the modified Bessel function given by
\begin{eqnarray}
zK_1(z)=\int_0^{\infty}dt~e^{-t}~e^{-\frac{z^2}{4t}}=1+\frac{z^2}{2}\ln\frac{ze^c}{2}+....
\end{eqnarray}
We observe that in two dimensions both the planar and non-planar functions are UV finite, i.e. renormalization of the vacuum polarization is not required. The infrared divergence seen when $m^2\longrightarrow 0$ cancel in the difference ${\Pi}^{\rm P}-{\Pi}^{\rm NP}$.  Furthermore ${\Pi}^{\rm P}-{\Pi}^{\rm NP}$ vanishes identically in the limit $\theta\longrightarrow 0$ or $p\longrightarrow 0$. In other words there is no UV-IR mixing in the vacuum polarization in two dimensions.

The situation in four dimensions is more involved \cite{Martin:2000bk,Krajewski:1999ja}. Explicitly we find  that the planar contribution to the vacuum polarization is UV
divergent as in the commutative theory, i.e. it is logarithmically divergent and thus it requires a
renormalization. It is found that the UV divergences in the $2-$, $3-$ and $4-$point functions at one-loop can be subtracted by a single counter term and hence the theory is
renormalizable at this order. The beta function of the theory at one-loop is identical to the beta function of the ordinary pure $SU(2)$ gauge theory. The non-planar contribution to the vacuum polarization at one-loop is UV finite because of the noncommutativity and only it becomes singular in the limit of vanishing
noncommutativity and/or vanishing external momentum. This also means that the renormalized vacuum polarization diverges in the infrared limit ${p}{\longrightarrow}0$
 and/or $\theta\longrightarrow0$ which is the definition of the UV-IR mixing.

We expect that supersymmetry will make the  Moyal-Weyl geometry and as a consequence the noncommutative gauge theory more stable. In order to see this effect let ${\lambda}_a$, $a=1,...,M$ be $M$ massless Majorana fermions in the adjoint representation of the gauge group $U({\cal H})$. We consider the modification of the action (\ref{action}) given by
\begin{eqnarray}
S\longrightarrow S^{'}=S+\frac{\sqrt{{\theta}^{d}{\rm det}(\pi B)}}{4g^2}\sum_{a=1}^MTr_{\cal H}\bar{\lambda}_a{\gamma}_i[\hat{D}_i,{\lambda}_a].
\end{eqnarray} 
The irreducible representation of the Clifford algebra in $d$ dimensions is $s=2^{\frac{d}{2}}$ dimensional. Let us remark that in the limit $\theta\longrightarrow 0$ the modified action $S^{'}$ has the same limit as the original action $S$. By integrating  over ${\lambda}_a$ in the path integral we obtain the Pfaffian $\big({\rm pf}({\gamma}_i{\cal D}_i)\big)^{M}$. We will assume that ${\rm pf}({\gamma}_i{\cal D}_i)=\big({\rm det}({\gamma}_i{\cal D}_i)\big)^{\frac{1}{2}}$. The modification of the effective action (\ref{effectiveaction}) is given by
\begin{eqnarray}
{\Gamma }\longrightarrow{\Gamma}^{'}=\Gamma-\frac{M}{4}Tr_{s}Tr_{\rm ad}\ln\bigg({\cal D}^2-\frac{i}{2}{\gamma}_i{\gamma}_j{\cal F}_{ij}\bigg).
\end{eqnarray}
It is not very difficult to check that the coefficient of the logarithmic term in the effective potential is positive definite for all $M$ such that $Ms<2d-4$. For $Ms=2d-4$ the logarithmic term vanishes identically and thus the background (\ref{minima}) is completely stable at one-loop order.  In this case the noncommutative gauge theory (i.e. the star product representation) makes sense at least at one-loop order for all values of the gauge coupling constant $g$. The case $Ms=2d-4$ in $d=4$ (i.e. $M=1$) corresponds to noncommutative ${\cal N}=1$ supersymmetric $U(1)$ gauge theory. In this case the effective action is given by
\begin{eqnarray}
{\Gamma}^{'}=S+\frac{1}{2}Tr_{d}Tr_{\rm ad}\ln\bigg({\delta}_{ij}-2i\frac{1}{{\cal D}^2}{\cal F}_{ij}\bigg)-\frac{M}{4}Tr_{s}Tr_{\rm ad}\ln\bigg(1-\frac{i}{2}{\gamma}_i{\gamma}_j\frac{1}{{\cal D}^2}{\cal F}_{ij}\bigg).
\end{eqnarray}
This is manifestly gauge invariant. In $4$ dimensions we use the identity $Tr_s{\gamma}_i{\gamma}_j{\gamma}_k{\gamma}_l=s\big({\delta}_{ij}{\delta}_{kl}-{\delta}_{ik}{\delta}_{jl}+{\delta}_{il}{\delta}_{jk}\big)$ and the first nonzero  correction to the classical action $S$ is given by the equation
\begin{eqnarray}
{\Gamma}^{'}&=&S+\big(\frac{d-2}{8}-1\big)Tr_{\rm ad}\frac{1}{{\cal D}^2}{\cal F}_{ij}\frac{1}{{\cal D}^2}{\cal F}_{ij}+...\nonumber\\
&=&S+2\big(1-\frac{d-2}{8}\big)\int_p Tr |\tilde{F}_{ij}(p)|^2\int_k\frac{1}{k^2}\frac{1}{(p-k)^2}(1-\cos {{\theta}_{ij}p_ik_j})|.\label{effective}
\end{eqnarray}
This correction is the only one-loop contribution  which contains a quadratic term in the gauge field. The planar and non-planar corrections to the vacuum polarization are given in this case by
\begin{eqnarray}
{\Pi}^{\rm P}&=&\int_k\frac{1}{k^2}\frac{1}{(p-k)^2}=\frac{1}{(4\pi)^{\frac{d}{2}}}\int_0^1{dx}\int_0^{\infty}\frac{dt}{t^{\frac{d}{2}-1}}~e^{-x(1-x)p^2t}.
\end{eqnarray}
\begin{eqnarray}
{\Pi}^{\rm NP}&=&\int_k\frac{1}{k^2}\frac{1}{(p-k)^2}\cos {{\theta}_{ij}p_ik_j}=\frac{1}{(4\pi)^{\frac{d}{2}}}\int_0^1{dx}\int_0^{\infty}\frac{dt}{t^{\frac{d}{2}-1}}~e^{-x(1-x)p^2t-\frac{({\theta}_{ij}p_i)^2}{4t}}.\nonumber\\
\end{eqnarray}
The planar correction is UV divergent coming from the limit $t\longrightarrow 0$. Indeed we compute (including also an arbitrary mass scale $\mu$ and defining $\epsilon=2-\frac{d}{2}$)
\begin{eqnarray}
{\Pi}^{\rm P}&=&\frac{1}{(4\pi)^{\frac{d}{2}}}\int_0^1 {dx}~({\mu}^2)^{\frac{d}{2}-2}(x(1-x)\frac{p^2}{{\mu}^2})^{\frac{d}{2}-2}~{\Gamma}(2-\frac{d}{2})\nonumber\\
&=&\frac{({\mu}^2)^{-\epsilon}}{(4\pi)^{\frac{d}{2}}}\bigg[\frac{1}{\epsilon}-\gamma-\int_0^1 dx \ln x(1-x)\frac{p^2}{{\mu}^2}+O(\epsilon)\bigg].
\end{eqnarray}
The singular high energy behaviour is thus logarithmically divergent. The planar correction  needs therefore a renormalization. 
We add the counter term
\begin{eqnarray}
  \delta S=-2(1-\frac{d-2}{8})\frac{({\mu}^2)^{-\epsilon}}{(4\pi)^{\frac{d}{2}}}\frac{1}{\epsilon}\int d^dx F_{ij}^2=-2(1-\frac{d-2}{8})\frac{({\mu}^2)^{-\epsilon}}{(4\pi)^{\frac{d}{2}}}\frac{1}{\epsilon}\int_p |\tilde{F}_{ij}(p)|^2.\label{counter}
\end{eqnarray}
The effective action at one-loop is obtained by adding (\ref{effective}) and the counter term (\ref{counter}). We get
\begin{eqnarray}
{\Gamma}^{'}_{\rm ren}=\int_p \frac{1}{2g^2(\mu)}|\tilde{F}_{ij}(p)|^2.
\end{eqnarray}
\begin{eqnarray}
\frac{1}{2g^2(\mu)}&=&\frac{1}{2g^2}+2(1-\frac{d-2}{8})({\Pi}^{\rm P}-{\Pi}^{\rm NP})-2(1-\frac{d-2}{8})\frac{({\mu}^2)^{-\epsilon}}{(4\pi)^{\frac{d}{2}}}\frac{1}{\epsilon}\nonumber\\
&=&\frac{1}{2g^2}+\frac{3}{2}\frac{1}{(4\pi)^2}\bigg[-\gamma-\int_0^1 dx \ln x(1-x)\frac{p^2}{{\mu}^2}\bigg]-\frac{3}{2}{\Pi}^{\rm NP}.
\end{eqnarray}
This equation means that the gauge coupling constant runs with the renormalization scale. The beta function is non-zero given by
\begin{eqnarray}
\beta(g(\mu))=\mu\frac{dg(\mu)}{d\mu}=-\frac{3}{16{\pi}^2}g^3(\mu).
\end{eqnarray}
The non-planar correction is UV finite. Indeed we compute the closed expression
\begin{eqnarray}
{\Pi}^{\rm NP}=\frac{2}{(4\pi)^2}\int_0^1{dx}K_0(z)~,~z^2=({\theta}_{ij}p_i)^2x(1-x)p^2.
\end{eqnarray}
In the limit $\theta\longrightarrow 0$ and/or $p\longrightarrow 0$ we can use $K_0(z)=-\ln\frac{z}{2}$ and obtain the IR singular behaviour
\begin{eqnarray}
{\Pi}^{\rm NP}=-\frac{1}{(4\pi)^2}\int_0^1{dx}\ln\frac{({\theta}_{ij}p_i)^2x(1-x)p^2}{4}.
\end{eqnarray}
In summary although the Moyal-Weyl geometry is made stable at one-loop order by the introduction of supersymmetry we still have a UV-IR mixing in the quantum gauge theory. The picture that supersymmetry will generally stabilize the geometry will be confirmed nonperturbatively in this article whereas the precise connection to the UV-IR mixing remains unclear.

\section{Fuzzy Spaces}
\subsection{The Fuzzy Sphere}
The ordinary sphere  $S^2$ is defined in global coordinates by the equation 
\begin{eqnarray}
\vec{n}{\in}{\bf R}^3~,~\sum_{a=1}^3n_a^2=1.
\end{eqnarray}
A general function can be expanded in terms of spherical harmonics 
\begin{eqnarray}
f(\vec{n})=\sum_{k=0}^{\infty}f_{km}Y_{km}(\vec{n}).\label{f}
\end{eqnarray} 
Global derivations are given by the generators of rotations in the adjoint representation of the $SU(2)$ group, namely
\begin{eqnarray}
{\cal L}_a=-i{\epsilon}_{abc}n_b{\partial}_c~,~[{\cal L}_a,{\cal L}_b]=i{\epsilon}_{abc}{\cal L}_c.
\end{eqnarray}
The Laplacian is given by
\begin{eqnarray}
{\cal L}^2={\cal L}_a{\cal L}_a~,~{\rm eigenvalues}~l(l+1)~,l=0,...,\infty.
\end{eqnarray}
According to \cite{Frohlich:1993es,Frohlich:1998zm} all the geometry of the sphere is encoded
in the K-cycle or spectral  triple $({\cal A},{\cal H},{\cal
  L}^2)$. ${\cal A}=C^{\infty}(S^2)$ is the algebra of all functions $f$ of the
form (\ref{f})  and ${\cal H}$ is the infinite dimensional Hilbert
space of square integrable functions on which the functions are
represented. In order to encode the geometry of the sphere in the
presence of spin  structure we use instead 
the K-cycle $({\cal A},{\cal H},{\cal D},{\gamma})$ \cite{Connes:1994yd} where $\gamma$ and ${\cal D}$ are the chirality and the Dirac operators on the sphere. Remark in particular that
\begin{eqnarray}
Y_{11}{\propto}~n_1+in_2~,~Y_{10}{\propto}~n_3~,Y_{1-1}{\propto}~n_1-in_2.
\end{eqnarray} 
The fuzzy sphere is a particular deformation of the above triple which is based on the fact that the sphere is the co-adjoint orbit 
$SU(2)/U(1)$, namely
\begin{eqnarray}
g{\sigma}_3g^{-1}=n_a{\sigma}_a~,~g{\in}SU(2)~,~\vec{n}{\in}S^2.
\end{eqnarray}
This also means that $S^2$ is a symplectic manifold and hence it can be quantized in a canonical fashion by simply quantizing the volume form 
\begin{eqnarray}
{\omega}=\sin\theta d{\theta}{\wedge} d{\phi}~=\frac{1}{2}{\epsilon}_{abc}n_adn_b{\wedge}dn_c.
\end{eqnarray}
The result of this quantization is to replace the algebra $C^{\infty}(S^2)$ by the algebra of matrices $Mat_{L+1}$.

$Mat_{L+1}$ is the algebra of $(L+1){\times}(L+1)$ matrices which acts on an $(L+1)-$dimensional Hilbert space ${H}_L$ with inner product $(f,g)=\frac{1}{L+1}Tr(f^{+}g)$ where $f,g{\in}Mat_{L+1} $. The spin $s={L}/{2}$ IRR of $SU(2)$ is found to act naturally on this Hilbert space. The generators are
\begin{eqnarray}
[L_a,L_b]=i{\epsilon}_{abc}L_c~ ,~ \sum_a L_a^2=c_2{\equiv}\frac{L}{2}
(\frac{L}{2}+1).
\end{eqnarray}
Generally the spherical harmonics $Y_{km}(\vec{n})$ become the
canonical $SU(2)$ polarization tensors $\hat{Y}_{km}$ \cite{Varshalovich:1988ye}. 
They are defined by
\begin{eqnarray}
[L_a,[L_a,\hat{Y}_{lm}]]=l(l+1)\hat{Y}_{lm}~,~[L_{\pm},\hat{Y}_{lm}]=\sqrt{(l{\mp}m)(l{\pm}m+1)}\hat{Y}_{lm{\pm}1}~,~[L_3,\hat{Y}_{lm}]=m\hat{Y}_{lm}.
\end{eqnarray}
They also satisfy
\begin{eqnarray}
\hat{Y}_{lm}^{+}=(-1)^m\hat{Y}_{l-m}~,~\frac{1}{L+1}Tr\hat{Y}_{l_1m_1}\hat{Y}_{l_2m_2}=(-1)^{m_1}{\delta}_{l_1l_2}{\delta}_{m_1,-m_2}.
\end{eqnarray}
Matrix coordinates on ${\bf S}^2_L$ are defined by the $k=1$ tensors as in the continuum, namely

\begin{eqnarray}
x_1^2+x_2^2+x_3^2=1~,~
[x_a,x_a]=\frac{i}{\sqrt{c_2}}{\epsilon}_{abc}x_c,~x_a=\frac{L_a}{\sqrt{c_2}}.
\end{eqnarray}
``Fuzzy'' functions on ${\bf S}^2_L$ are linear operators in the
matrix algebra while derivations are inner defined by the
generators of the adjoint action of $SU(2)$, i.e.

\begin{eqnarray}
{\cal L}_a({\phi})\equiv[L_a,{\phi}].
\end{eqnarray}
A natural choice of the Laplacian operator ${\Delta}$ on the fuzzy sphere is therefore given by the following Casimir operator
\begin{eqnarray}
{\Delta}_L={\cal L}_a^2\equiv [L_a,[L_a,..]].\label{24}
\end{eqnarray}
Thus the algebra of matrices $Mat_{L+1}$ decomposes under the
action of the group $SU(2)$ as
\begin{eqnarray}
\frac{L}{2}{\otimes}\frac{L}{2}=0{\oplus}1{\oplus}2{\oplus}..{\oplus}L.
\end{eqnarray}
The first ${L}/{2}$ stands for the left action of the group
while the other ${L}/{2}$ stands for the right action. 
It
is not difficult to convince ourselves that this Laplacain has a
cut-off spectrum of the form $l(l+1)$ where $l=0,1,...,L$. As a
consequence a general function on ${\bf S}^2_N$ can be expanded in terms of polarization tensors as
follows
\begin{eqnarray}
{f}=\sum_{l=0}^{L}\sum_{m=-l}^{l}{f}_{lm}\hat{Y}_{lm}.\label{sum0}
\end{eqnarray}
The fact that the summation over $l$ involves only angular
momenta which are ${\leq}L$ originates of course from the fact
that the spectrum $l(l+1)$ of the Laplacian ${\Delta}$ is cut-off
at $l=L$. 

The commutative continuum limit is given by $L{\longrightarrow}{\infty}$. 
This is the first limit of interest to us in this article.
Therefore the fuzzy sphere is a sequence of the following triples
\begin{eqnarray}
(Mat_{L+1},H_L,{\Delta}_L).
\end{eqnarray}
Again in the presence of spin structure the fuzzy sphere ${\bf S}_L^2$
will be defined instead by the K-cycle $({\bf A},{\bf H},D,{\Gamma})$  where the chirality operator
${\Gamma}$ and the Dirac operator ${D}$ are given for example in \cite{Balachandran:2003ay} and references therein.

\subsection{Fuzzy ${\bf CP}^{2}$}

In this section we will give the K-cycle associated with the classical K\"ahler manifold ${\bf CP}^2$ (which is also the co-adjoint orbit $SU(3)/U(2)$) as a limit of the K-cycle which defines fuzzy (or quantized) ${\bf CP}^2$ when the noncommutativity parameter goes to $0$. We will follow the construction of \cite{Balachandran:2001dd}.

Let $T_a$, $a=1,...,8$ be the generators of $SU(3)$ in the symmetric irreducible
 representation $(n,0)$ of dimension $N=\frac{1}{2}(n+1)(n+2)$. They satisfy
\begin{eqnarray}
[T_a,T_b]=if_{abc}T_c~.\label{comm}
\end{eqnarray}
\begin{eqnarray}
T_a^2=\frac{1}{3}n(n+3)\equiv
|n|^2~,~d_{abc}T_aT_b=\frac{2n+3}{6}T_c.\label{idd}
\end{eqnarray}
Let $t_a={{\lambda}_a}/{2}$ (where ${\lambda}_a$ are the usual Gell-Mann matrices)  be the generators of $SU(3)$ in the fundamental
 representation $(1,0)$ of dimension $N=3$. They satisfy
\begin{eqnarray}
&&2t_at_b=\frac{1}{3}{\delta}_{ab}+(d_{abc}+if_{abc})t_c\nonumber\\
&&tr_3t_at_b=\frac{1}{2}{\delta}_{ab}~,~tr_3t_at_bt_c=\frac{1}{4}(d_{abc}+if_{abc}).
\end{eqnarray}
The $N-$dimensonal generator $T_a$ can be obtained by taking the symmetric product of $n$ copies of the fundamental $3-$dimensional generator $t_a$, viz
\begin{eqnarray}
T_a=(t_a{\otimes}{\bf 1}{\otimes}...{\otimes}{\bf 1}+{\bf 1}{\otimes}t_a{\otimes}...{\otimes}{\bf 1}+...+{\bf 1}{\otimes}{\bf 1}{\otimes}...{\otimes}t_a)_{\rm symmetric}.
\end{eqnarray}
In the continuum ${\bf CP}^2$ is the space of all unit vectors $|\psi>$ in ${\bf C}^3$ modulo the phase. Thus $e^{i\theta}|\psi>$,
for all $\theta {\in}[0,2\pi[$ define the same point on ${\bf CP}^2$. It is obvious that all these vectors $e^{i\theta}|\psi>$ correspond to
 the same projector $P=|\psi><\psi|$. Hence ${\bf CP}^2$ is the space of all projection operators of rank one on ${\bf C}^3$. Let ${\bf H}_N$
  and ${\bf H}_3$ be the Hilbert spaces of the $SU(3)$ representations  $(n,0)$ and $(1,0)$ respectively.
We will define fuzzy ${\bf CP}^2$ through  the canonical $SU(3)$ coherent states as follows. Let $\vec{n}$ be a vector in ${\bf R}^8$, then
 we define the projector
\begin{eqnarray}
P_3=\frac{1}{3}{\bf 1}+n_at_a
\end{eqnarray}
The requirement $P_3^2=P_3$ leads to the condition that $\vec{n}$ is a point on ${\bf CP}^2$ satisfying the equations

\begin{eqnarray}
[n_a,n_b]=0~,~n_a^2=\frac{4}{3}~,~d_{abc}n_an_b=\frac{2}{3}n_c.
\end{eqnarray}
We can write
\begin{eqnarray}
P_3=|\vec{n},3><3,\vec{n}|.
\end{eqnarray}
We think of $|\vec{n},3>$ as the coherent state in ${\bf H}_3$ (level $3\times 3$ matrices) which is localized at
the point $\vec{n}$ of  ${\bf CP}^2$. Therefore the coherent state $|\vec{n},N>$ in ${\bf H}_N$ (level $N\times N$ matrices)
which is localized around the point $\vec{n}$ of  ${\bf CP}^2$ is defined by the projector
\begin{eqnarray}
P_N=|\vec{n},N><N,\vec{n}|=(P_3{\otimes}P_3{\otimes}...{\otimes}P_3)_{\rm symmetric}.
\end{eqnarray}
We compute that
\begin{eqnarray}
tr_3t_aP_3=<\vec{n},3|t_a|\vec{n},3>=\frac{1}{2}n_a~,~
tr_NT_aP_N=<\vec{n},N|T_a|\vec{n},N>=\frac{n}{2}n_a.
\end{eqnarray}
Hence it is natural to identify fuzzy ${\bf CP}^2$ at level $N=\frac{1}{2}(n+1)(n+2)$ (or  ${\bf CP}^2_N$) by the coordinates operators

\begin{eqnarray}
x_a=\frac{2}{n}T_a.
\end{eqnarray}
They satisfy

\begin{eqnarray}
[x_a,x_b]=\frac{2i}{n}f_{abc}x_c~,~x_a^2=\frac{4}{3}(1+\frac{3}{n})~,~d_{abc}x_ax_b=\frac{2}{3}(1+\frac{3}{2n})x_c.
\end{eqnarray}
Therefore in the large $N$ limit we can see that the algebra of $x_a$ reduces to the continuum algebra of $n_a$. Hence $x_a{\longrightarrow}n_a$
 in the continuum limit $N{\longrightarrow}{\infty}$.

The algebra of ~functions on fuzzy  ${\bf CP}^2_N$ is identified with the algebra of $N{\times}N$ matrices $Mat_N$ generated by
 all polynomials in the coordinates operators $x_a$. Recall that $N=\frac{1}{2}(n+1)(n+2)$. The left action of $SU(3)$
  on this algebra is generated by $(n,0)$ whereas the right action is generated by $(0,n)$. Thus the algebra $Mat_N$ decomposes
  under the action of $SU(3)$ as
\begin{eqnarray}
(n,0){\otimes}(0,n)={\otimes}_{p=0}^n(p,p).
\end{eqnarray}
A general function on fuzzy  ${\bf CP}^2_N$ is therefore written as

\begin{eqnarray}
F=\sum_{p=0}^nF_{I^2,I_3,Y}^{(p)}T_{I^2,I_3,Y}^{(p,p)}.
\end{eqnarray}
The $T_{I^2,I_3,Y}^{(p,p)}$ are $SU(3)$ matrix polarization tensors in the irreducible representation $(p,p)$. $I^2,I_3$ and $Y$ are the square of the isospin,
the third component of the isospin and the hypercharge quantum numbers which characterize $SU(3)$ representations.

The derivations on fuzzy  ${\bf CP}^2_N$ are defined by the commutators $[T_a,..]$. The Laplacian is then obviously given by ${\Delta}_N=[T_a,[T_a,...]]$.
Fuzzy ${\bf CP}^2_N$ is completely determined by the spectral triple ${\bf CP}^2_N=(Mat_N,{\Delta}_N,{\bf H}_N)$. Now we can compute

\begin{eqnarray}
tr_NFP_N=<\vec{n},N|F|\vec{n},N>=F_N(\vec{n})=\sum_{p=0}^nF_{I^2,I_3,Y}^{(p)}Y_{I^2,I_3,Y}^{(p,p)}(\vec{n}).
\end{eqnarray}
The $Y_{I^2,I_3,Y}^{(p,p)}(\vec{n})$ are the $SU(3)$ polarization tensors defined by

\begin{eqnarray}
Y_{I^2,I_3,Y}^{(p,p)}(\vec{n})=<\vec{n},N|T_{I^2,I_3,Y}^{(p,p)}|\vec{n},N>.
\end{eqnarray}
Furthermore we can compute

\begin{eqnarray}
tr_N[T_a,F]P_N=<\vec{n},N|[T_a,F]|\vec{n},N>=({\cal L}_aF_N)(\vec{n})~,~{\cal L}_a=-if_{abc}n_b{\partial}_c.
\end{eqnarray}
And
\begin{eqnarray}
tr_NFGP_N=<\vec{n},N|FG|\vec{n},N>=F_N*G_N(\vec{n}).
\end{eqnarray}
The star product on fuzzy ${\bf CP}^2_N$ is found to be given by (see below)
\begin{eqnarray}
&&F_N*G_N(\vec{n})=\sum_{p=0}^n\frac{(n-p)!}{p!n!}K_{a_1b_1}...K_{a_pb_p}{\partial}_{a_1}...{\partial}_{a_p}F_N(\vec{n}){\partial}_{b_1}...
{\partial}_{b_p}G_N(\vec{n})~\nonumber\\
&&~K_{ab}=\frac{2}{3}{\delta}_{ab}-n_an_b+(d_{abc}+if_{abc})n_c.
\end{eqnarray}

\subsection{Star Products on Fuzzy ${\bf CP}^{N-1}$}


The sphere is the complex projective space ${\bf CP}^1$. The
quantization of higher ${\bf CP}^{N-1}$ with $N>2$ is very similar to
the quantization of ${\bf CP}^1$, and yields fuzzy ${\bf
  CP}^{N-1}_L$. In this section we will explain this result by
constructing the coherent states \cite{Man'ko:1996xv,Perelomov:1986tf,klauder:1985}, the Weyl map  \cite{weyl:1931} and star products \cite{Groenewold:1946kp,Moyal:1949sk,Alexanian:2000uz} on all fuzzy ${\bf
  CP}^{N-1}_L$ following \cite{Balachandran:2001dd}. 

We start with classical ${\bf C}{\bf P}^{N-1}$ defined by the
projectors\footnote{In this
  section we  use $N$ to denote the dimension
  of the space instead of the size of the matrix approximation.}

\begin{equation}
P=\frac{1}{N}{\bf 1}+{\alpha}_Nn^at_a~,~ {\alpha}_N=-\sqrt{\frac{2(N-1)}{N}}.\label{projectorcs}
\end{equation}
The requirement $P^2=P$ will lead to the defining equations of  ${\bf C}{\bf P}^{N-1}$ as embedded in ${\bf
R}^{N^2-1}$ given by
\begin{eqnarray}
{n}_a^2=1~,~
d_{abc}n_an_b=\frac{2}{\alpha_N}\frac{N-2}{N}n_c.
\end{eqnarray}
The fundamental representation ${\bf N}$ of $SU(N)$ is generated by
the Lie algebra of Gell-Mann matrices
${t_a}=\frac{{\lambda}_a}{2},a=1,...,N^2-1$. These matrices satisfy
\begin{eqnarray}
&&[t_a,t_b]=if_{abc}t_c\nonumber\\
&&2t_at_b=\frac{1}{N}{\delta}_{ab}{\bf 1}+(d_{abc}+if_{abc})t_c\nonumber\\
&&Trt_at_bt_c=\frac{1}{4}(d_{abc}+if_{abc})~,~Tr
t_at_b=\frac{{\delta}_{ab}}{2}~,~Tr t_a=0.\label{sun1}
\end{eqnarray}
Let us specialize the projector (\ref{projectorcs}) to the
"north" pole of ${\bf C}{\bf P}^{N-1}$ given by the point
$\vec{n}_0=(0,0,...,1)$. We have then the projector $
P_0=\frac{1}{N}{\bf 1}+\alpha_N t_{N^2-1}={\rm diag}(0,0,...,1)$. 
So at the "north" pole  $P$ projects down onto the state $
|{\psi}_0>=(0,0,...,1)$ 
of the Hilbert space ${\bf C}^{N}$ on which the defining
representation of $SU(N)$ is acting .

A general point $\vec{n}{\in}~{\bf C}{\bf P}^{N-1}$ can be
obtained from $\vec{n}_0$ by the action of an element
$g{\in}SU(N)$ as $
\vec{n}=g\vec{n}_0$. 
$P$ will then project down onto the state $
|\psi>=g|{\psi}_0>$
of ${\bf C}^N$ . One can show that
\begin{equation}
P=|\psi><\psi|=g|{\psi}_0><{\psi}_0|g^{+}=gP_0g^{+}.
\end{equation}
\begin{equation}
gt_{N^2-1}g^{+}=n^{a}t_a.
\end{equation}
This last equation is the usual definition of ${\bf C}{\bf
P}^{N-1}$ . Under $g{\longrightarrow}gh$ where $h{\in}U(N-1)$ we
have $ht_{N^2-1}h^{+}=t_{N^2-1}$ , i.e $U(N-1)$ is the stability
group of $t_{N^2-1}$ and hence
\begin{equation}
{\bf C}{\bf P}^{N-1}=SU(N)/U(N-1).
\end{equation}
Thus points $\vec{n}$ of ${\bf C}{\bf P}^{N-1}$ are then equivalent
classes $[g]=[gh],h{\in}U(N-1)$ .

The fuzzy sphere ${\bf S}^2_L$ is the
algebra of operators acting on the Hilbert space
$H_s^{(2)}$ which is the $(2s+1)-$dimensional
irreducible representation of $SU(2)$. This representation can
be obtained from the symmetric product of $L=2s$ fundamental
representations ${\bf 2}$ of $SU(2)$. Indeed given any element
$g{\in}SU(2)$ its $s-$representation matrix $U^{(s)}(g)$ can be
obtained as follows
\begin{equation}
U^{(s)}(g)=U^{(\bf 2)}(g){\otimes}_s...{\otimes}_sU^{(\bf
2)}(g),2s-{\rm times}.
\end{equation}
$U^{(\bf 2)}(g)$ is the spin ${1}/{2}$ representation of
$g{\in}SU(2)$ .

Similarly fuzzy ${\bf CP}^{N-1}_L$ is the algebra of 
operators acting on the Hilbert space $H_s^{(N)}$ which is the $d_s^{(N)}$-dimensional irreducible reprsentation of $SU(N)$.  This dimension is given explicitly by
\begin{eqnarray}
d_s^{(N)}=(\frac{(N-1+2s)!}{(N-1)!(2s)!}.
\end{eqnarray}
Also this $d_s^{(N)}$-dimensional
irreducible representation of $SU(N)$ can be obtained from the
symmetric product of $L=2s$ fundamental representations ${\bf N}$
of $SU(N)$. 

Remark that for $s=1/2$ we have
$d_{1/2}^{(N)}=N$ and therefore
$H_{1/2}^{(N)}={\bf C}^N$ is the fundamental
representation of $SU(N)$. Clearly the states $|{\psi}_0>$ and $|\psi>$ of
$H_{1/2}^{(N)}$  will correspond in $H_s^{(N)}$ to
the two states $|\vec{n}_0,s>$ and $|\vec{n},s>$ respectively so
that $|{\psi}_0>=|\vec{n}_0,\frac{1}{2}>$ and
$|\psi>=|\vec{n},\frac{1}{2}>$. Furthermore the equation  $|\psi>=g |{\psi}_0>$ 
becomes
\begin{equation}
|\vec{n},s>=U^{(s)}(g)|\vec{n}_0,s>.\label{fundamental}
\end{equation}
$U^{(s)}(g)$ is the representation given
by
\begin{equation}
U^{(s)}(g)=U^{(\bf N)}(g){\otimes}_s...{\otimes}_sU^{(\bf
N)}(g),2s-{\rm times}.
\end{equation}
To any operator $\hat{F}$ on $H_s^{(N)}$ (which can be thought
of as a function on fuzzy ${\bf C}{\bf P}^{N-1}_L$) we
associate a "classical" function $F_s(\vec{n})$ on a classical
${\bf C}{\bf P}^{N-1}$  by
\begin{equation}
F_s(\vec{n})=<\vec{n},s|\hat{F}|\vec{n},s>.\label{maps}
\end{equation}
The product of two such operators $\hat{F}$ and
$\hat{G}$ is mapped to the star product of the corresponding two
functions
\begin{equation}
F_s*G_s(\vec{n})=<\vec{n},s|\hat{F}\hat{G}|\vec{n},s>.\label{starproduct1}
\end{equation}
Now we compute this star product in a closed form. First we will use the result that any operator $\hat{F}$ on
the Hilbert space $H_s^{(N)}$ admits the expansion
\begin{equation}
\hat{F}=\int_{SU(N)}d{\mu}(h)\tilde{F}(h)U^{(s)}(h).\label{expansion}
\end{equation}
$U^{(s)}(h)$ are assumed to satisfy the normalization
\begin{equation}
TrU^{(s)}(h)U^{(s)}(h^{'})=d_s^{(N)}{\delta}(h^{-1}-h^{'}).
\end{equation}
Using the above two equations  one can derive the value of the
coefficient $\tilde{F}(h)$ to be
\begin{equation}
\tilde{F}(h)=\frac{1}{d_s^{(N)}}Tr\hat{F}U^{(s)}(h^{-1}).
\end{equation}
Using the expansion (\ref{expansion}) in (\ref{maps}) we get
\begin{eqnarray}
F_s(\vec{n})=\int_{SU(N)}d{\mu}(h)\tilde{F}(h){\omega}^{(s)}(\vec{n},h)~,~
{\omega}^{(s)}(\vec{n},h)&=&<\vec{n},s|U^{(s)}(h)|\vec{n},s>.
\end{eqnarray}
On the other hand using the expansion (\ref{expansion}) in
(\ref{starproduct1}) will give
\begin{equation}
F_s*G_s(\vec{n})=\int
\int_{SU(N)}d{\mu}(h)d{\mu}(h^{'})\tilde{F}(h)\tilde{G}(h^{'}){\omega}^{(s)}(\vec{n},hh^{'}).
\end{equation}
The computation of this star product boils down to the
computation of ${\omega}^{(l)}(\vec{n},hh^{'})$. We have
\begin{eqnarray}
{\omega}^{(s)}(\vec{n},h)&=&<\vec{n},s|U^{(s)}(h)|\vec{n},s>\nonumber\\
&=&\bigg[<\vec{n},\frac{1}{2}|{\otimes}_s...{\otimes}_s<\vec{n},\frac{1}{2}|\bigg]\bigg[U^{(\bf
N)}(h){\otimes}_s...{\otimes}_sU^{(\bf
N)}(h)\bigg]\bigg[|\vec{n},\frac{1}{2}>{\otimes}_s...{\otimes}_s|\vec{n},\frac{1}{2}>\bigg]\nonumber\\
&=&[{\omega}^{(\frac{1}{2})}(\vec{n},h)]^{2s}.
\end{eqnarray}
\begin{eqnarray}
{\omega}^{(\frac{1}{2})}(\vec{n},h)
&=&<\psi|U^{(\bf N)}(h)|\psi>.
\end{eqnarray}
In the fundamental representation ${\bf N}$ of $SU(N)$ we have
$U^{(\bf N)}(h)=\exp(im^at_a)=c(m){\bf 1}+is^a(m)t_a$ and therefore
\begin{eqnarray}
{\omega}^{(\frac{1}{2})}(\vec{n},h)&=&<\psi|c(m){\bf
  1}+is^a(m)t_a|\psi>=c(m)+is^a(m)<\psi|t_a|\psi>.
\end{eqnarray}
\begin{eqnarray}
{\omega}^{(\frac{1}{2})}(\vec{n},hh^{'})&=&<\psi|U^{(\bf N)}(hh^{'})|\psi>\nonumber\\
&=&<\psi|(c(m){\bf 1}+is^a(m)t_a)(c(m^{'}){\bf 1}+is^a(m^{'})t_a)|\psi>\nonumber\\
&=&c(m)c(m^{'})+i[c(m)s^a(m^{'})+c(m^{'})s^a(m)]<\psi|t_a|\psi>\nonumber\\
&-&s^a(m)s^b(m^{'})<\psi|t_at_b|\psi>.
\end{eqnarray}
Now it is not difficult to check that
\begin{eqnarray}
<\psi|t_a|\psi>&=&Trt_aP=\frac{\alpha_N}{2}n^a\nonumber\\
<\psi|t_at_b|\psi>&=&Trt_at_bP=\frac{1}{2N}{\delta}_{ab}+\frac{\alpha_N}{4}(d_{abc}+if_{abc})n^c.\label{nice}
\end{eqnarray}
Hence we obtain
\begin{eqnarray}
{\omega}^{(\frac{1}{2})}(\vec{n},h)&=&c(m)+i\frac{\alpha_N}{2}\vec{s}(m).\vec{n}.
\end{eqnarray}
\begin{eqnarray}
{\omega}^{(\frac{1}{2})}(\vec{n},hh^{'})&=&c(m)c(m^{'})-\frac{1}{2N}\vec{s}(m).\vec{s}(m^{'})+i\frac{\alpha_N}{2}\bigg[c(m)s^a(m^{'})\nonumber\\
&+&c(m^{'})s^a(m)\bigg]n^a-\frac{\alpha_N}{4}(d_{abc}+if_{abc})n^cs^a(m)s^b(m^{'}).
\end{eqnarray}
These two last equations can be combined to get the 
result
\begin{eqnarray}
{\omega}^{(\frac{1}{2})}(\vec{n},hh^{'})-{\omega}^{(\frac{1}{2})}(\vec{n},h){\omega}^{(\frac{1}{2})}(\vec{n},h^{'})&=&
-\frac{1}{2N}\vec{s}(m).\vec{s}(m^{'})-\frac{\alpha_N}{4}(d_{abc}+if_{abc})n^cs^a(m)s^b(m^{'})\nonumber\\
&+&\frac{{\alpha}_N^2}{4}n^an^bs_a(m)s_b(m^{'}).
\end{eqnarray}
We can remark that in this last equation  we have got ride of
all reference to $c$'s. We would like also to get ride of all
reference to $s$'s. This can be achieved by using the formula
\begin{equation}
s_a(m)=\frac{2}{i\alpha_N}\frac{\partial}{{\partial}n^a}{\omega}^{(\frac{1}{2})}(\vec{n},h).
\end{equation}
We get then
\begin{eqnarray}
{\omega}^{(\frac{1}{2})}(\vec{n},hh^{'})-{\omega}^{(\frac{1}{2})}(\vec{n},h){\omega}^{(\frac{1}{2})}(\vec{n},h^{'})&=&
K_{ab}\frac{\partial}{{\partial}n^a}{\omega}^{(\frac{1}{2})}(\vec{n},h)\frac{\partial}{{\partial}n^b}{\omega}^{(\frac{1}{2})}(\vec{n},h^{'}).
\end{eqnarray}
\begin{eqnarray}
K_{ab}&=&\frac{2}{N{\alpha}_N^2}{\delta}_{ab}-n_an_b+\frac{1}{\alpha_N}(d_{abc}+if_{abc})n^c.
\end{eqnarray}
Therefore we obtain
\begin{eqnarray}
F_s*G_s(\vec{n})&=&\sum_{k=0}^{2s}\frac{(2s)!}{k!(2s-k)!}K_{a_1b_1}....K_{a_kb_k}\nonumber\\
&{\times}&\int_{SU(N)}d{\mu}(h)\tilde{F}(h)[{\omega}^{(\frac{1}{2})}(\vec{n},h)]^{2s-k}\frac{\partial}{{\partial}n^{a_1}}{\omega}^{(\frac{1}{2})}(\vec{n},h)...\frac{\partial}{{\partial}n^{a_k}}{\omega}^{(\frac{1}{2})}(\vec{n},h)\nonumber\\
&{\times}&\int_{SU(N)}d{\mu}(h^{'})\tilde{G}(h^{'})[{\omega}^{(\frac{1}{2})}(\vec{n},h^{'})]^{2s-k}\frac{\partial}{{\partial}n^{b_1}}{\omega}^{(\frac{1}{2})}(\vec{n},h^{'})...\frac{\partial}{{\partial}n^{b_k}}{\omega}^{(\frac{1}{2})}(\vec{n},h^{'}).\nonumber\\
\end{eqnarray}
We have  also the formula
\begin{eqnarray}
\frac{(2s-k)!}{(2s)!}\frac{\partial}{{\partial}n^{a_1}}...\frac{\partial}{{\partial}n^{a_k}}F_s(\vec{n})&=&
\int_{SU(N)}d{\mu}(h)\tilde{F}(h)[{\omega}^{(\frac{1}{2})}(\vec{n},h)]^{2s-k}\frac{\partial}{{\partial}n^{a_1}}{\omega}^{(\frac{1}{2})}(\vec{n},h)...\frac{\partial}{{\partial}n^{a_k}}{\omega}^{(\frac{1}{2})}(\vec{n},h).\nonumber\\
\end{eqnarray}
This allows us to obtain the final result \cite{Balachandran:2001dd}
\begin{equation}
F_s*G_s(\vec{n})=\sum_{k=0}^{2s}\frac{(2s-k)!}{k!(2s)!}K_{a_1b_1}....K_{a_kb_k}\frac{\partial}{{\partial}n^{a_1}}...\frac{\partial}{{\partial}n^{a_k}}F_j(\vec{n})\frac{\partial}{{\partial}n^{b_1}}...\frac{\partial}{{\partial}n^{b_k}}G_j(\vec{n}).\label{starproduct}
\end{equation}
Specialization of this result to the sphere is obvious. In the last appendix we will discuss a (seemingly) different star product on the fuzzy sphere which admits a straightforward flattening limit to the star product on the Moyal-Weyl plane.

Let us also do some examples.
Derivations on ${\bf C}{\bf P}^{N-1}$ are generated by the vector
fields ${\cal L}_a=-if_{abc}n_b{\partial}/{{\partial}n_c}$ which satisfy
$
[{\cal L}_a,{\cal L}_b]=if_{abc}{\cal L}_c$. 
The corresponding action on the Hilbert space $H_s^{(N)}$ is
generated by $L_a$ and is given by
\begin{equation}
<\vec{n},s|U^{(s)}(h^{-1})\hat{F}U^{(s)}(h)|\vec{n},s>=<\vec{n}_0,s|U^{(s)}(g^{-1}h^{-1})\hat{F}U^{(s)}(hg)|\vec{n}_0,s>.
\end{equation}
$U^{(s)}(h)$ is given by $U^{(s)}(h)=\exp(i{\eta}_aL_a)$. Now if
we take ${\eta}$ to be small  then one computes
\begin{equation}
<\vec{n},s|U^{(s)}(h)|\vec{n},s>=1+i{\eta}_a<\vec{n},s|L_a|\vec{n},s>.
\end{equation}
On the other hand we know that the representation $U^{(s)}(h)$
is obtained by taking the symmetric product of $2s$ fundamental
representations ${\bf N}$ of $SU(N)$ and hence
\begin{eqnarray}
<\vec{n},s|U^{(s)}(h)|\vec{n},s>=(<\vec{n},\frac{1}{2}|1+i{\eta}_at_a|\vec{n},\frac{1}{2}>)^{2s}=1+i(2s){\eta}_a\frac{\alpha_N}{2}n_a.
\end{eqnarray}
In above we have used the facts 
$L_a=t_a{\otimes}_s....{\otimes}_st_a$,
$|\vec{n},s>=|\vec{n},\frac{1}{2}>{\otimes}_s...{\otimes}_s|\vec{n},\frac{1}{2}>$
and the first equation of (\ref{nice}). Finally we get the
important result
\begin{equation}
<\vec{n},s|L_a|\vec{n},s>=s{\alpha}_Nn_a.
\end{equation}
We define the fuzzy derivative $[{L}_a,\hat{F}]$ by
\begin{eqnarray}
({\cal L}_aF)_s(\vec{n})&{\equiv}&<\vec{n},s|[L_a,\hat{F}]|\vec{n},s>\nonumber\\
&=&s{\alpha}_N\bigg[n_a*F_s(\vec{n})-F_s*n_a(\vec{n})\bigg]\nonumber\\
&=&if_{abc}n^c\frac{\partial}{{\partial}n^b}F_s(\vec{n}).
\end{eqnarray}
Finally we note the identity
\begin{eqnarray}
\frac{1}{d_s^{(N)}}Tr\hat{F}\hat{G}=\int_{{\bf CP}^{N-1}}  F_s*G_s(\vec{n}).
\end{eqnarray}

\section{Review of Bosonic $D=3$ Yang-Mills Matrix Models}

The principal goal of the present article is the construction of a new nonperturbative method for noncommutative gauge theories with and without supersymmetry based on a class of Yang-Mills matrix models in which the classical minima are given by fuzzy projective spaces. These matrix models are directly related to the celebrated IKKT matrix model \cite{Ishibashi:1996xs,Aoki:1999vr,Aoki:1998vn,Aoki:1998bq} in $d=3$ and $d=4$ dimensions with mass deformation which may or may not preserve supersymmetry.

The flat IKKT models, i.e. without mass deformation,  are obtained by dimensionally reducing ${\rm U}(N)$ super Yang-Mills theory in flat $d$ dimensions onto a point, i.e. to zero dimension. The dynamical variables are $d$  matrices of size $N$ with  action
\begin{eqnarray}
S=-\frac{N}{4}Tr[X_{\mu},X_{\nu}]^2+ Tr\bar{\psi}{\Gamma}_{\mu}[X_{\mu},\psi].
\end{eqnarray}
The partition functions of these models are  convergent in dimensions $d=4,6,10$ \cite{Krauth:1998xh,Austing:2001bd,Krauth:1999qw,Krauth:1998yu,Austing:2001pk,Austing:2003kd,Austing:2001ib}. In $d=3$ the partition function may be made finite by adding appropriate mass deformation consisting of a positive quadratic term in the matrices $X_{\mu}$ which damps flat directions. In $d=3,4$ the determinant of the Dirac operator is positive definite \cite{Krauth:1998xh,Ambjorn:2000bf} and thus there is no sign problem. The IKKT model in $d=10$ dimensions is also called the  IIB matrix model. It is postulated to give a constructive definition of type IIB superstring theory.

 Mass deformations such as the Myers term \cite{Myers:1999ps} are essential in order to reproduce non-trivial geometrical backgrounds such as the fuzzy sphere in Yang-Mills matrix models. Supersymmetric mass deformations in Yang-Mills matrix models and Yang-Mills quantum mechanics models are considered for example in \cite{Bonelli:2002mb,Kim:2006wg}. Yang-Mills quantum mechanics models such as the BFSS models \cite{Banks:1996vh} in various dimensions are a non-trivial escalation over the IKKT models since they involve time. The BMN model \cite{Berenstein:2002jq} which is the unique maximally supersymmetric mass deformation  of the BFSS model in $d=10$ admits the fuzzy sphere background as a solution of its equations of motion. The BFSS and BMN models are postualted  to give a constructive definition of M-theory. 

The central motivation behind these proposals of using Yang-Mills matrix models and Yang-Mills quantum mechanics as nonperturbative definitions of M-theory and superstring theory lies in D-brane physics \cite{Polchinski:1995mt,Polchinski:1996na,Taylor:1997dy}. At low energy the theory on the $(p+1)-$dimensional world-volume of $N$ coincident Dp-branes is the reduction to $p+1$ dimensions of $10$ dimensional supersymmetric Yang-Mills \cite{Witten:1995im}. Thus we get a $(p+1)$ dimensional vector field together with $9-p$ normal scalar fields which play the role of position coordinates of the coincident $N$ Dp-branes. The case $p=0$ corresponds to D0-branes. The coordinates become noncommuting matrices.

As we have already said the class of matrix models of interest to us in this article are IKKT matrix models in $d=3$ and $d=4$ dimensions with mass deformations. The main reason behind this interest is that these matrix models suffer generically from an emergent geometry transition and as a consequence they are very suited for studying nonperturbatively gauge theory on Moyal-Weyl spaces. Furthermore the supersymmetric versions of these matrix models provide a natural nonperturbative regularization of supersymmetry which is very interesting in its own right. Also since these matrix models are related to large $N$ Yang-Mills theory they are of paramount importance to the string/gauge duality which would allow us to study nonperturbative aspects of gravity from the gauge side of the duality. These motivations can also be found elsewhere, for example in \cite{Nishimura:2008ta,Hanada:2008gy,Ishiki:2008te,Hanada:2008ez}.

The motivation for studying large $N$ Yang-Mills matrix models with backgrounds given by fuzzy spaces is therefore four-fold:
\begin{itemize}
\item{$1)$} This is the correct way of defining nonperturbatively noncommutative  gauge theories  using random matrix models. 
\item{$2)$} This provides also a nonperturbative definition of supersymmetry. Indeed supersymmetry in this language may prove to be  tractable in Monte Carlo simulation.

\item{$3)$} They provide concrete models for emergent geometry.  Indeed geometry in transition is possible in all these matrix models.

\item{$4)$} These large $N$ Yang-Mills matrix models are also relevant for the string/gauge duality.    
\end{itemize}
Among the tools that can naturally be used in the context of  Yang-Mills matrix models are Monte Carlo simulation \cite{Rothe:2005nw,Montvay:1994cy}, $1/N$ expansion \cite{'tHooft:1973jz,Moshe:2003xn,Brezin:1977sv,Brezin:1979ba}, renormalization group equation \cite{Polchinski:1983gv,Wilson:1973jj} and random matrix theory  \cite{Di Francesco:1993nw,eynard,Mehta}. Monte Carlo simulation using the Hybrid Monte Carlo algorithm was adapted to this type of matrix models in \cite{Ambjorn:2000bf,Ambjorn:2000dx}. However simulations of matrix models remain much harder than simulations of field theory because of the non-local character of matrix interactions. Renormalization group approach to matrix models can be found for example in \cite{Brezin:1992yc} and \cite{Higuchi:1994rv,Higuchi:1993pu,Cicuta:1992cz,Nishigaki:1996ts,Ferretti:1995zn}.

The central result in this review section is as follows. In the case of gauge theory on fuzzy complex projective spaces \cite{Dolan:2006tx} we  will describe how the corresponding matrix models allow for a new transition to and from a new 
high temperature phase known as Yang-Mills or matrix phase with no background geometrical 
structure. The low temperature phase is a geometrical one with gauge fields fluctuating on a
round complex projective space. We discuss the case of the fuzzy sphere in great detail then towards the end we comment on the other known cases.

Noncommutative gauge theory on the fuzzy sphere was introduced in \cite{Iso:2001mg,CarowWatamura:1998jn}. As we have already mentioned it was derived as the low energy dynamics of open strings moving in a background magnetic field with ${\bf S}^3$ metric  in \cite{Alekseev:1999bs,Alekseev:2000fd,Hikida:2001py}. This theory consists of the Yang-Mills term ${\rm YM}$ which can be obtained from the reduction to zero dimensions of ordinary $U(N)$ Yang-Mills theory in $3$ dimensions and a Chern-Simons term ${\rm CS}$ due to Myers effect \cite{Myers:1999ps}. Thus the model contains three $N\times N$ hermitian matrices $X_1$, $X_2$ and $X_3$ with an action given by
\begin{eqnarray}
S={\rm YM}+{\rm CS}=-\frac{1}{4}Tr[X_a,X_b]^2+\frac{2i\alpha}{3}{\epsilon}_{abc}TrX_aX_bX_c.
\end{eqnarray}
This model contains beside the usual two dimensional gauge field a scalar fluctuation normal  to the sphere which can be given by \cite{Karabali:2001te}
\begin{eqnarray}
\Phi=\frac{X_a^2-{\alpha}^2c_2}{2\sqrt{c_2}}.
\end{eqnarray} 
The model was studied perturbatively in \cite{CastroVillarreal:2004vh} and  in \cite{Azuma:2004ie,Imai:2003vr}. In particular in \cite{CastroVillarreal:2004vh} the effective action for a non-zero gauge fluctuation was computed at one-loop and shown to contain a gauge invariant UV-IR mixing in the large $N$ limit. Indeed the effective action in the commutative limit was found to be given by the expression

\begin{eqnarray}
{\Gamma}&=& \frac{1}{4g^2}\int
\frac{d{\Omega}}{4{\pi}}F_{ab}(1+2g^2{\Delta}_3)F_{ab}-\frac{1}{4g^2}{\epsilon}_{abc}\int
\frac{d{\Omega}}{4{\pi}}F_{ab}(1+2g^2{\Delta}_3)A_c+2\sqrt{N^2-1}\int\frac{d{\Omega}}{4{\pi}}\Phi \nonumber\\
&+&{\rm non~local~ quadratic ~terms}.\label{main1}
\end{eqnarray}
The $1$  in $1+2g^2{\Delta}_3$ corresponds to the classical action whereas $2g^2{\Delta}_3$ is the quantum correction. This provides a non-local renormalization of the inverse coupling constant $1/g^2$. The last terms in (\ref{main1}) are new non-local quadratic terms which have no counterpart in the classical action. The eigenvalues of the operator ${\Delta}_3$ are given by
\begin{eqnarray}
{\Delta}_3(p)&=&\sum_{l_1,l_2}\frac{2l_1+1}{l_1(l_1+1)}\frac{2l_2+1}{l_2(l_2+1)}(1-(-1)^{l_1+l_2+p})\left\{\begin{array}{ccc}
        p & l_1 & l_2 \\
    \frac{L}{2} & \frac{L}{2} & \frac{L}{2} \end{array}\right\}^2\frac{l_2(l_2+1)}{p^2(p+1)^2}\nonumber\\
&\times &\big(l_2(l_2+1)-l_1(l_1+1)\big))\longrightarrow  -\frac{h(p)+2}{p(p+1)}~,~h(p)=-2\sum_{l=1}^{p}\frac{1}{l}.
\end{eqnarray}
In above $L+1=N$. The $1$ in $1-(-1)^{l_1+l_2+p}$ corresponds to the planar contribution whereas $(-1)^{l_1+l_2+p}$ corresponds to the non-planar contribution where $p$ is the external momentum. The fact that ${\Delta}_3\neq 0$ in the limit $N\longrightarrow 0$ means that we have a UV-IR mixing problem. 

The model ${\rm YM}+{\rm CS}$ was solved for $N=2$ and $N=3$ in \cite{Tomino:2003hb}. It was studied nonperturbatively in \cite{Azuma:2004zq} where the geometry in transition was first observed. 

\begin{figure}[htbp]
\begin{center}
\includegraphics[width=10.0cm,angle=-90]{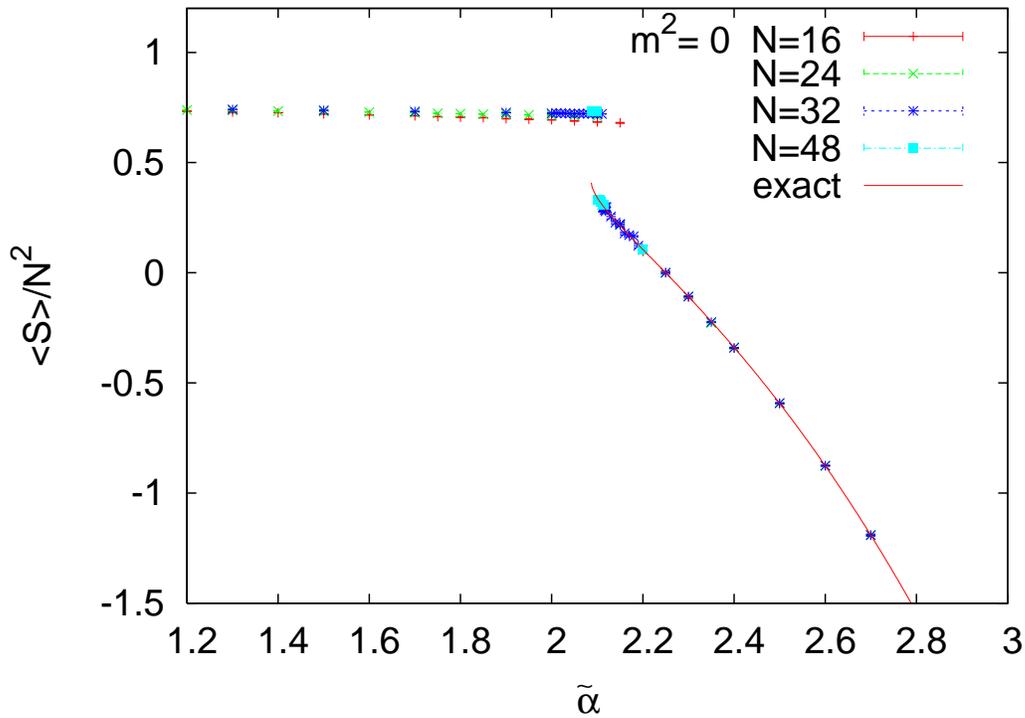}
\caption{The observable $\frac{<S>}{N^2}$ for $m^2=0$ as a function of the coupling constant for different matrix sizes $N$. The solid line corresponds to the theoretical prediction using the local minimum of the effective potential.}\label{obsm0}
\end{center}
\end{figure}

\begin{figure}[htbp]
\begin{center}
\includegraphics[width=10.0cm,angle=-90]{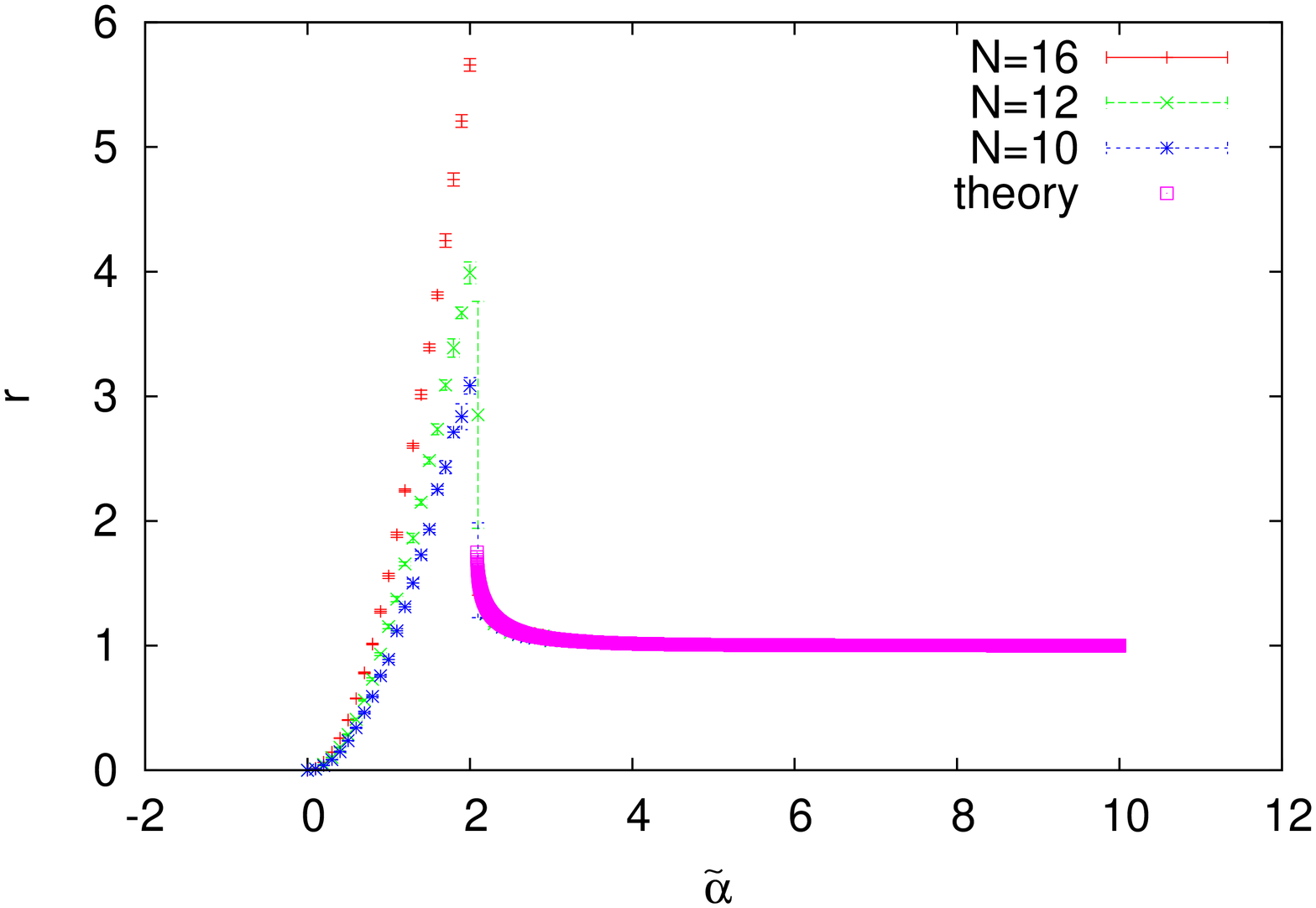}
\includegraphics[width=10.0cm,angle=-90]{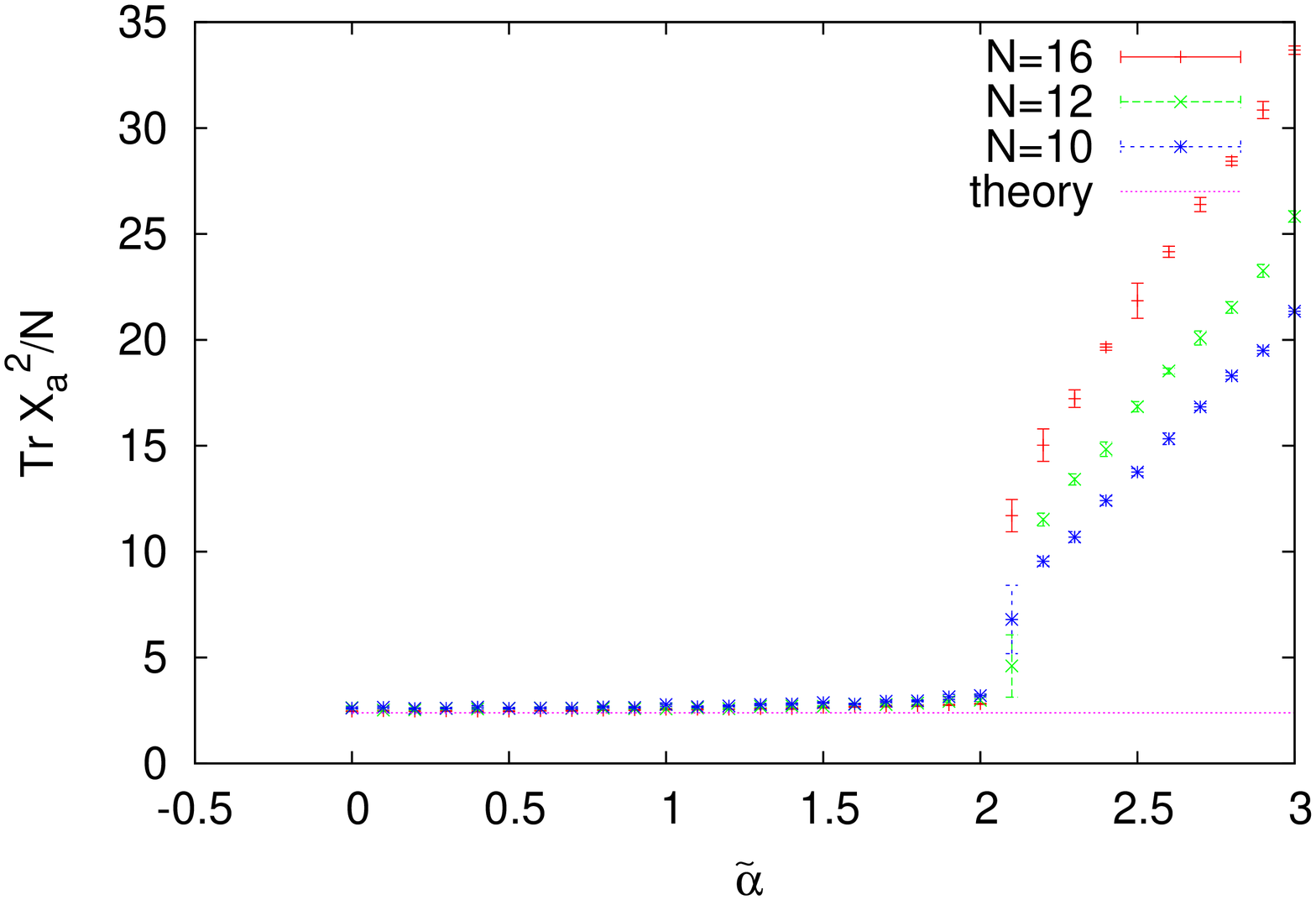}
\caption{The radius for $m^2=0$ as a function of the coupling constant for different matrix sizes $N$. The theoretical prediction is $r=1$ in the fuzzy sphere and $TrX_a^2/N=12/5$ in the matrix phase.}\label{radius}
\end{center}
\end{figure}

\begin{figure}[htbp]
\begin{center}
\includegraphics[width=10.0cm,angle=-90]{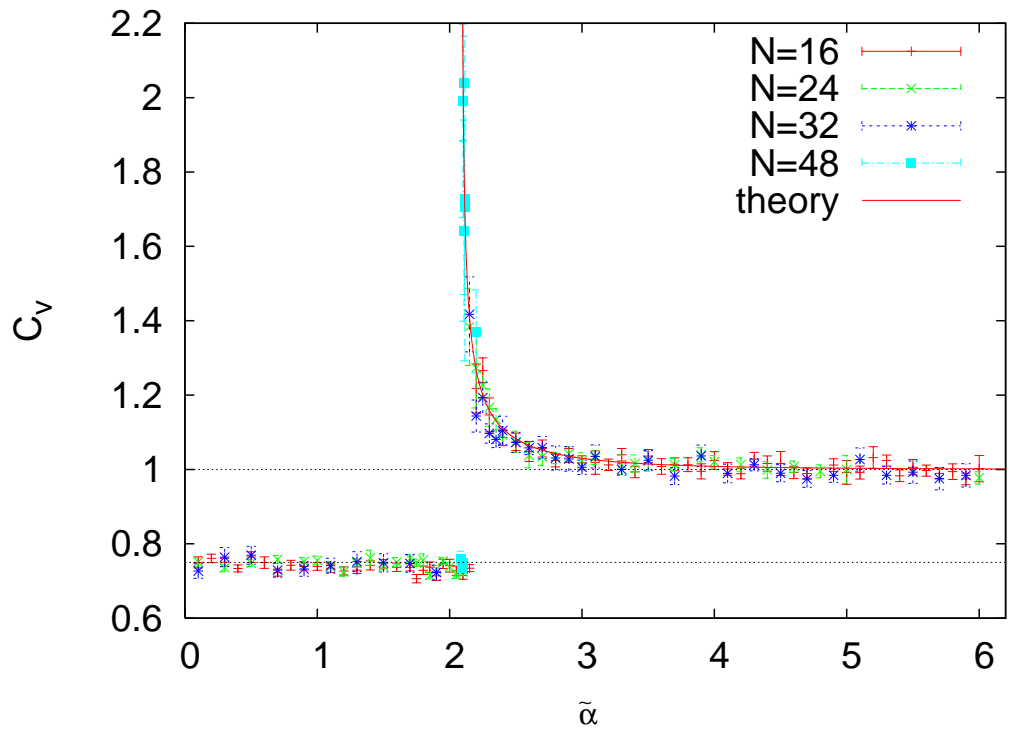}
\caption{The specific heat for $m^2=0$ as a function of the coupling constant for $N=16,24,32$,$48$. The curve corresponds with the theoretical prediction for $m^2=0$.}\label{figcvm0}
\end{center}
\end{figure}

 In \cite{O'Connor:2006wv} a generalized model was proposed and studied in which the normal scalar field was suppressed by giving it a quartic potential $V$ with very large mass. This potential on its own is an $O(3)$ random matrix model given by
\begin{eqnarray}
V&=&N\bigg[\frac{m^2}{2c_2} Tr(X_a^2)^2-{\alpha}^2\mu Tr  (X_a^2)\bigg].\label{O3matrix}
\end{eqnarray}
 The parameter $\mu$ is fixed such that $\mu=m^2$. The model $S+V$ was studied  in \cite{DelgadilloBlando:2008vi} and \cite{DelgadilloBlando:2007vx}  where the instability of the sphere was interpreted along the lines of an emergent geometry phenomena. For vanishing potential $m^2,\mu\longrightarrow 0$ the transition from/to the fuzzy sphere phase was found to have a discontinuity in the internal energy, i.e. a latent heat (figure \ref{obsm0}) and a discontinuity in the order parameter (figure \ref{radius}) indicating that the transition is first order. The order parameter is identified with the radius of the sphere, viz
\begin{eqnarray}
\frac{1}{r}=\frac{1}{Nc_2}Tr D_a^2~,~X_a=\alpha D_a.
\end{eqnarray}
From the other hand the specific heat was found to diverge at the transition point from the sphere side while it remains constant from the matrix side (figure \ref{figcvm0}). This indicates a second order behaviour with critical fluctuations only from one side of the transition. This to our knowledge is quite novel. The scaling of the coupling constant $\alpha$ in the large $N$ limit is found to be given by $\tilde{\alpha}=\alpha \sqrt{N}$. We get the critical value

\begin{eqnarray}
\tilde{\alpha}_s=2.1\pm 0.1.
\end{eqnarray}
The different phases of the model are characterized by 
\begin{center}
\begin{tabular}{|c|c|}
\hline
fuzzy sphere ($\tilde{\alpha}>\tilde{\alpha}_*$ )& matrix phase ($\tilde{\alpha}<\tilde{\alpha}_*$)\\
$r=1$ & $
r=0$\\
$C_v=1$  & $C_v=0.75$  \\
\hline
\end{tabular}
\end{center}
For $m\neq 0$ and/or $\mu\neq 0$ the critical point is replaced by a critical line in the  $\tilde{\beta}-t$ plane where $\tilde{\beta}^4=\tilde{\alpha}^4/(1+m^2)^3$ and $t=\mu(1+m^2)$. In other words for generic values of the parameters the matrix phase persists. The effective potential in these cases was computed in  \cite{CastroVillarreal:2004vh}. We find 
\begin{eqnarray}
V_{\rm eff}=\tilde{\alpha}^4
\big[\frac{1}{4}{\phi}^4-\frac{1}{3}{\phi}^3+\frac{m^2}{4}{\phi}^4-\frac{\mu}{2}{\phi}^2\big]+\ln{\phi}.
\end{eqnarray}
The extrema of
the classical potential occur at
\begin{equation}
\phi=\frac{1}{1+m^2}\left\{0,~  
{\phi}_{\pm}=\frac{1\pm \sqrt{1+4t}}{2}\right\}.
\end{equation} 
For $\mu$ positive the global minimum is ${\phi}_+$. The $0$ is a local maximum and ${\phi}_-$ is a local minimum. In particular for $\mu=m^2$  we obtain the global minimum ${\phi}_+=1$. For $\mu$ negative the global minimum is still ${\phi}_+$ but $0$ becomes a local minimum and ${\phi}_-$ a local maximum. If $\mu$ is sent more negative then 
the global minimum  ${\phi}_+=1$ becomes degenerate with  ${\phi}=0$ 
at $t=-\frac{2}{9}$ and the maximum height of the barrier
is given by $V_-={\tilde\beta^4}/324 $ which occurs at ${\phi}_-=\frac{1}{3}$. The model has a first order transition at $t=-2/9$ where the classical ground states switches from ${\phi}_+$ for $t>-2/9$ to $0$ for $t<2/9$.

\begin{figure}[htbp]
\begin{center}
\includegraphics[width=10.0cm,angle=-90]{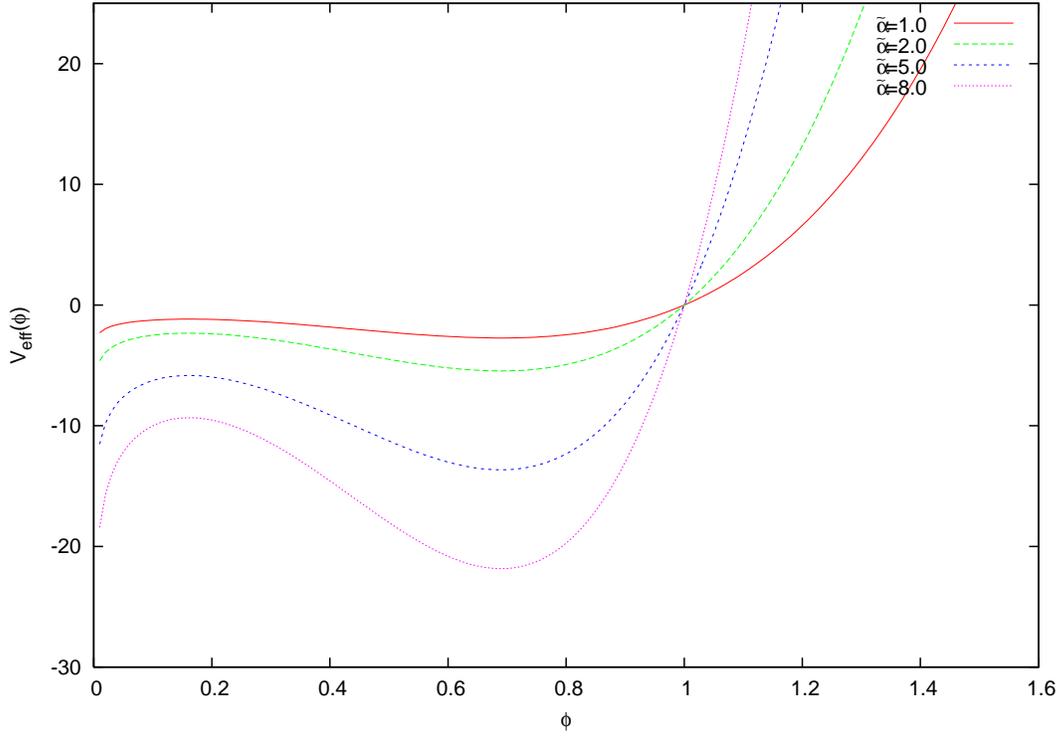}
\caption{The effective potential for $m^2=20$.}\label{V}
\end{center}
\end{figure}

Let us now consider the effect of quantum fluctuations. The condition $V^{'}_{\rm eff}=0$ 
gives us extrema of the model. For large enough $\tilde{\alpha}$  and large enough $m$ and $\mu$ it 
admits two positive solutions. The largest solution can be identified
with the ground state of the system. It will
determine the radius of the sphere. 
The second solution is the local maximum (figure
\ref{V}) of $V_{\rm eff}$ and will determine the height of the 
barrier. As the coupling is decreased these two solutions merge and
the barrier disappears. This is the critical point of the model. For smaller couplings
than the critical value $\tilde\alpha_*$ the fuzzy sphere
solution $D_a= {\phi}L_a$ no longer exists. Therefore the classical transition described above is
significantly affected by quantum fluctuations.

The condition when the barrier disappears is $V_{\rm eff}^{''}=0$. At this point the local minimum  merges with the local maximum (figure \ref{V}). Solving the two equations  $V_{\rm eff}^{'}=V_{\rm eff}^{''}=0$ yield the critical value
\begin{eqnarray}
g_*^2=\frac{1}{\tilde{\alpha}_*^4}=\frac{{\phi}_{*}^2({\phi}_*+2\mu)}{8},
\end{eqnarray}
where
\begin{eqnarray}
{\phi}_{*}=\frac{3}{8(1+m^2)}\bigg[1+\sqrt{1+\frac{32\mu (1+m^2)}{9}}\bigg].
\end{eqnarray}
If we take $\mu$ negative we see that $g_*$ goes to zero at $\mu(1+m^2)=-1/4$ and the critical 
coupling $\tilde{\alpha}_* $ is sent to infinity and therefore
for $\mu(1+m^2)<-\frac{1}{4}$ the model has no fuzzy sphere phase. However in the region $-\frac{1}{4}<\mu(1+m^2) <-\frac{2}{9}$ the action $S+V$ is completely positive. It is therefore
not sufficient to consider only the configuration $D_a=\phi L_a$
but rather all $SU(2)$ representations must be considered. 
Furthermore for large $\tilde{\alpha}$ the ground state will be dominated
by those representations with the smallest Casimir. This means that
there is no fuzzy sphere solution for $\mu(1+m^2)<-\frac{2}{9}$.

The limit of interest is the limit $\mu=m^2{\longrightarrow}\infty$. 
In this case 
\begin{eqnarray}
{\phi}_{*}=\frac{1}{\sqrt{2}}~,~\tilde{\alpha}^4_{*}=\frac{8}{m^2}.\label{pre2}
\end{eqnarray}
This means that the phase transition is located at a smaller value of
the coupling constant $\tilde{\alpha}$ as $m$ is increased.  In other
words the region where the fuzzy sphere is stable is extended to lower
values of the coupling.

\begin{figure}[htbp]
\begin{center}
\includegraphics[width=10.0cm,angle=-90]{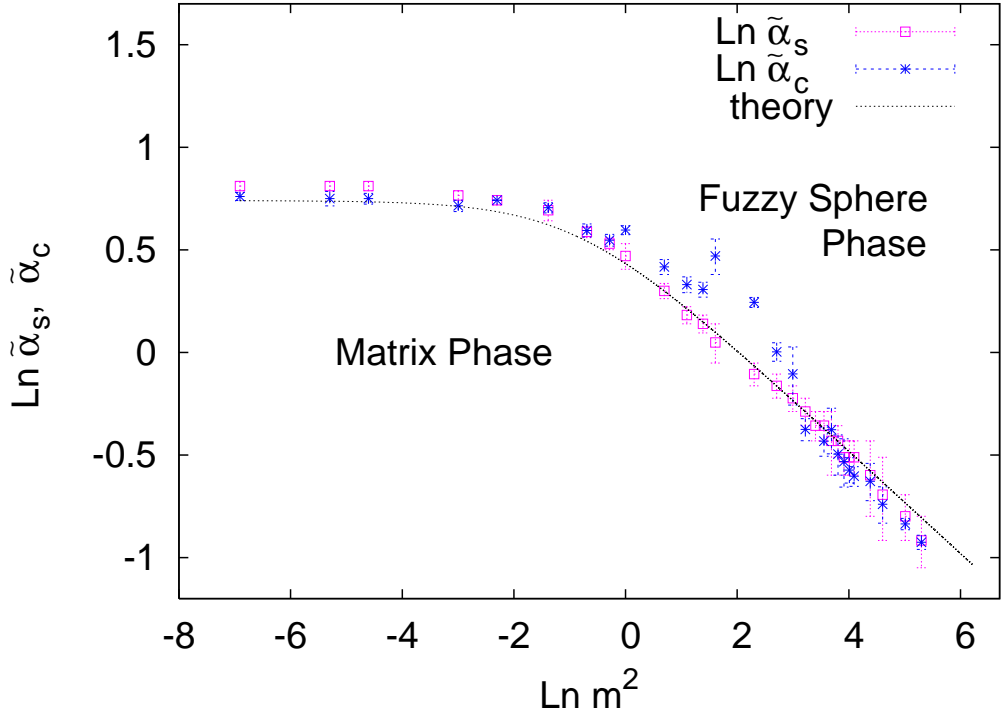}
\caption{The phase diagram.}\label{PD}
\end{center}
\end{figure}

Nonperturbatively the value $\tilde{\alpha}_s$ defined as the value of $\tilde{\alpha}$ 
at which curves of the average value of the action $<S>$ for different $N$ cross
 gives a good estimate of the location of the transition. For large $N$ we observe that the location $\tilde{\alpha}_{\rm max}$ 
of the peak in $C_v$ and the minimum $\tilde{\alpha}_{\rm min}$ coincide and agree well
with $\tilde\alpha_s$. By extrapolating the measured values of $\tilde{\alpha}_{\rm max}$ and $\tilde{\alpha}_{\rm min}$ 
 to $N=\infty$ we obtain the critical value $\tilde{\alpha}_{c}$.  The critical coupling determined either as
$\tilde{\alpha}_{c}$ or as $\tilde{\alpha}_{s}$ gives good agreement with (\ref{pre2}). The phase digaram is given in figure \ref{PD}.

In the case when we include a potential term with $\mu=m^2$ we also found numerical and analytical evidence \cite{Ydri:2007px,Ydri:2006xw} for the existence of another transition on the fuzzy sphere which is of the Gross-Witten type \cite{Gross:1980he}. This is  a field theory transition which occurs within the fuzzy sphere phase before we reach the matrix phase. We also note that a simplified version of our model with $V$ quartic in the matrices, i.e. $m^2=0$ and $\mu\neq 0$ was studied in \cite{Azuma:2005bj,Valtancoli:2002rx}.

In \cite{Steinacker:2003sd} an elegant pure matrix model was shown to be equivalent to a gauge theory on the fuzzy sphere with a very particular form of the potential which in the large $N$ limit leads naturally (at least classically) to a decoupled normal scalar  fluctuation. In \cite{Ydri:2007px,Ydri:2006xw} and \cite{Steinacker:2007iq} an alternative model of gauge theory on the fuzzy sphere was proposed in which field configurations live in the Grassmannian manifold $U(2N)/(U(N+1)\times U(N-1))$.  In \cite{Steinacker:2007iq} this model was shown to possess the same partition function as  commutative gauge theory on the ordinary sphere via the application of the powerful localization techniques \cite{Witten:1991we,Witten:1992xu}.

The matrix phase which is also called the Yang-Mills phase is dominated by commuting matrices. It is found that the eigenvalues of the three matrices $X_1$, $X_2$ and $X_3$ are uniformly distributed  inside a solid ball in $3$ dimensions. This was also observed in higher dimensions in  \cite{Hotta:1998en}. The eigenvalues distribution of a single matrix say $X_3$ can then be derived by assuming that the joint eigenvalues distribution of the the three commuting matrices  $X_1$, $X_2$ and $X_3$ is uniform. We obtain
\begin{eqnarray}
\rho(x)=\frac{3}{4R^3}(R^2-x^2).
\end{eqnarray}
The parameter $R$ is the radius of the solid ball. We find numerically the value $R=2$. A one-loop calculation around the background of commuting matrices gives a value in agreement with this prediction.

These eigenvalues may be interpreted as the positions of D0-branes in spacetime following Witten \cite{Witten:1995im}. In \cite{Berenstein:2008eg} there was an attempt to give this phase a geometrical content along the same lines. However our notion of geometry in this article follows Connes \cite{Connes:1994yd} which requires providing the Dirac or Laplacian operator together with the algebra in order to determine the geometry of the space. Therefore for all practical purposes this phase has no geometry since we can not identify a sensible Laplacian acting on commuting matrices.

We recognize two different scaling limits. In the fuzzy sphere phase the matrices $X_a$  define a round sphere with a radius which scales as $N$ in the commutative limit, i.e. they define a two-dimensional plane  whereas in the matrix phase they define a solid ball in $3$ dimensions. This is the scaling limit of the flat two dimensional plane. In  the second scaling limit the scaled matrices $D_a$  define on the other hand a round sphere with finite radius in the fuzzy sphere phase whereas in the matrix phase they give a single point.

\begin{figure}[htbp]
\begin{center}
\includegraphics[width=10.0cm,angle=-90]{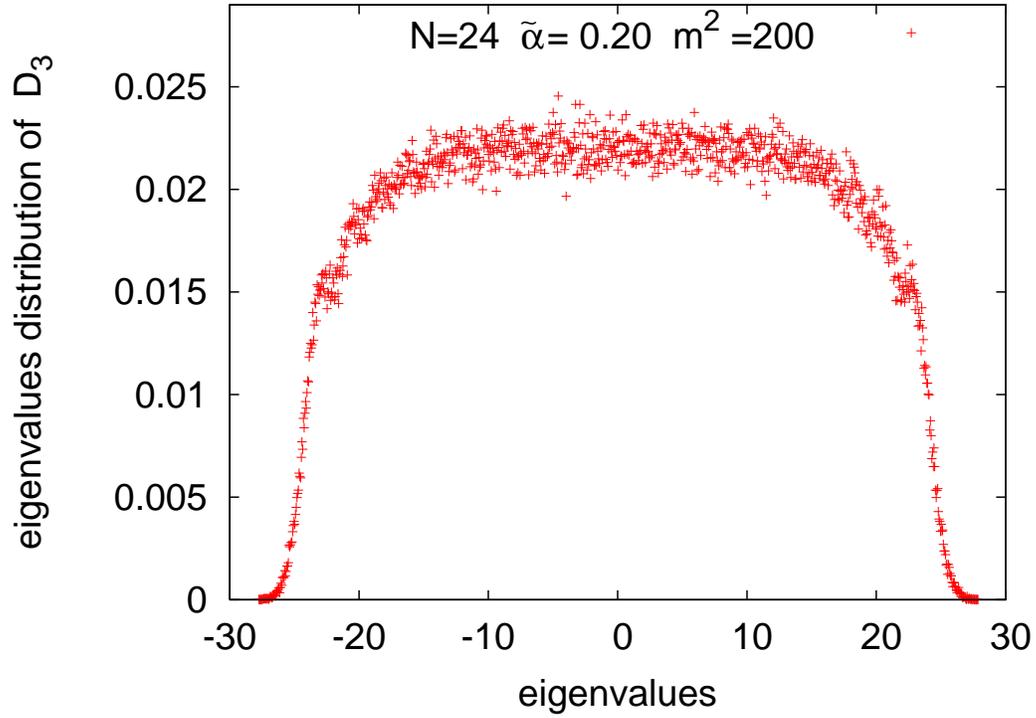}
\includegraphics[width=10.0cm,angle=-90]{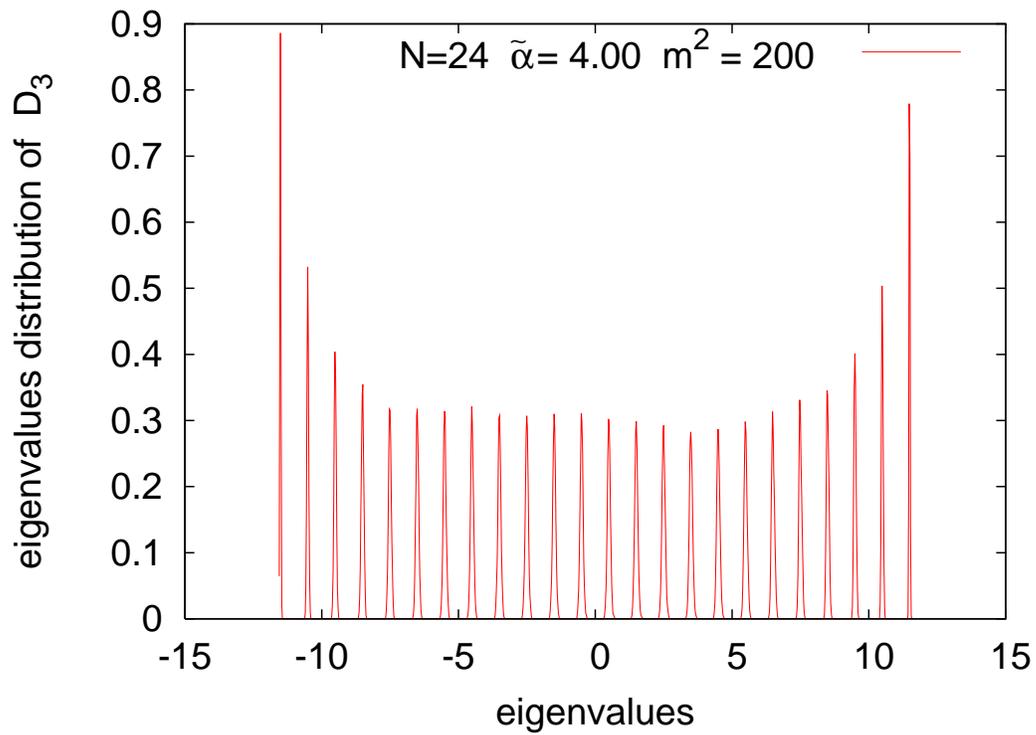}
\caption{The one-cut to N-cut transition. }\label{emerge}
\end{center}
\end{figure}

The essential ingredient in producing this transition is the Chern-Simons term in the action which is due to the Myers effect. This transition is related to the transition found in hermitian quartic matrix models. For example the $O(3)$ matrix model given by the potential $V$ does not have any transition but when the Chern-Simons term is added to it we reproduce the one-cut to the two-cut transition. By adding the Yang-Mills terms, i.e. by considering the full model we should then obtain a  generalization of  the one-cut to the two-cut transition. Indeed the matrix to the fuzzy sphere transition is in fact a one-cut to N-cut transition (figure \ref{emerge}). 

The matrix phase should be identified with the one-cut (disordered) phase of the quartic hermitian matrix model. By formally comparing (\ref{matrix}) and (\ref{O3matrix}) we can make the identification $a=-{\mu}{\alpha}^2$ and $b=m^2/(2c_2)$. The one-cut phase corresponds to the region $a\geq a_*=-2\sqrt{b}$ or equivalently $\tilde{\alpha}^2\leq \tilde{\alpha}_*^2$ where 
\begin{eqnarray}
\tilde{\alpha}_*^4=\frac{8m^2}{{\mu}^2}.
\end{eqnarray}
For $\mu=m^2$ this is precisely the critical point (\ref{pre2}). 

By using the methods of cohomological field theories and topological matrix models  employed in \cite{Kazakov:1998ji,Hoppe:1999xg,Hirano:1997ai}, it might be possible to bring the model into the form of a hermitian matrix model with generalized interaction of the form \cite{ydri_preparation} 
\begin{eqnarray}
S=Na Tr M^2- Tr_{\rm ad}\ln ({\rm ad}M +b).
\end{eqnarray}
In summary we find for pure gauge models with global $SO(3)$ symmetry an exotic line of discontinuous transitions with a jump in the entropy, characteristic of a 1st order
transition, yet with divergent critical fluctuations and a divergent
specific heat with critical exponent $\alpha=1/2$. The low temperature
phase (small values of the gauge coupling constant) is a geometrical one with gauge fields fluctuating on a round sphere. 
As the temperature increased the sphere evaporates 
in a transition to a pure matrix phase with no background geometrical
structure.  These models present an appealing picture of a geometrical phase emerging as the system cools
and suggests a scenario for the emergence of geometry in the early
universe.

Lastly we remark on the effect of fermionic determinants on the transition which is the subject of the remainder of this article. We propose here mass deformed  supersymmetric matrix Yang-Mills which are reduced from mass deformed supersymmetric Yang-Mills quantum  mechanics  as the prime  candidates to study the effect of supersymmetry on emergent geometry and vice versa in noncommutative gauge theory. These mass deformed matrix models or quantum mechanics provide also the prime examples of supersymmetric models which can be put on a computer. It is conjectured that supersymmetry will  remove the transition and stabilizes completely the geometry against the quantum fluctuations of the noncommutative gauge theory or else supersymmetry may be dynamically broken. Indeed our Monte Carlo results reported here (see section $7$) confirms this picture. In \cite{Anagnostopoulos:2005cy} a Monte Carlo simulation of a $4$ dimensional model with a Chern-Simons term was performed. Although the model was not invariant under (flat) supersymmetry transformations the effect of the added Majorana fermions seemed to stabilize the geometry.

Finally we make few remarks on known analogous results in $4$ dimensions. In $4$ dimensions we have $2$ fuzzy projective spaces, fuzzy ${\bf CP}_N^2$ and  fuzzy ${\bf S}_N^2\times {\bf S}_N^2$. Classical gauge theory on fuzzy ${\bf CP}_N^2$ and  fuzzy ${\bf S}_N^2\times {\bf S}_N^2$ are considered in \cite{Grosse:2004wm,Dou:2007in} and \cite{DelgadilloBlando:2006dp,Behr:2005wp} respectively.  A Monte Carlo study of the model ${\rm YM}+{\rm CS}$  on fuzzy ${\bf CP}_N^2$ was conducted in \cite{Azuma:2004qe}. It is observed that in this model both a fuzzy ${\bf CP}_N^2$ phase and a fuzzy sphere phase exist together with the matrix phase. The phase structure is therefore much richer. This was confirmed in  \cite{Dou:2007in} with the calculation of the one-loop effective potential of the model ${\rm YM}+{\rm CS}+V$ on fuzzy ${\bf CP}_N^2$. The one-loop effective potential of the model ${\rm YM}+{\rm CS}+V$ on fuzzy ${\bf S}_N^2\times {\bf S}_N^2$ was computed in \cite{CastroVillarreal:2005uu} and a Monte Carlo study of the same model but with $V=0$ was performed in \cite{Azuma:2005pm} with similar conclusions.  Fuzzy ${\bf S}_N^2\times {\bf S}_N^2$ is also considered in \cite{Imai:2003jb,Imai:2003ja,Kitazawa:2004ef}


\newpage
\section{Mass Deformation of $D=4$ Super Yang-Mills Matrix Model}

\subsection{Dimenional Reduction in $4$D} 

We work with the metric $\eta=(-+++)$. The gamma matrices satisfy $\{{\gamma}_{\mu},{\gamma}_{\nu}\}=2{\eta}_{\mu\nu}$. We consider the representation
\begin{eqnarray}
&&{\gamma}^0=-i\left(
\begin{array}{cc}
0& {\bf 1}_2 \\
{\bf 1}_2&0
\end{array}
\right)~,~{\gamma}^i=-i\left(
\begin{array}{cc}
0&{\sigma}_i \\
-{\sigma}_i&0
\end{array}
\right).
\end{eqnarray}
We have $\beta={\gamma}_4=i{\gamma}^0$, ${\gamma}_5=i{\gamma}_0{\gamma}_1{\gamma}_2{\gamma}_3$. We verify that ${\gamma}_{\mu}^+={\gamma}^0{\gamma}_{\mu}{\gamma}^0$, ${\gamma}_{\mu}^*={\gamma}^2{\gamma}_{\mu}{\gamma}^2$, ${\gamma}_{\mu}^T=C{\gamma}_{\mu}C$. The charge conjugation matrix is defined by
\begin{eqnarray}
C={\gamma}_2\beta=-\epsilon{\gamma}_5~,~C^+=C^{-1}=-C~,~C^T=-C~,~C^2=-1. 
\end{eqnarray}
\begin{eqnarray}
&&\epsilon=\left(
\begin{array}{cc}
i{\sigma}_2& 0 \\
0&i{\sigma}_2
\end{array}
\right)~,~{\gamma}_5=\left(
\begin{array}{cc}
{\bf 1}_2&0 \\
0&-{\bf 1}_2
\end{array}
\right).
\end{eqnarray}
The Majorana condition reads
\begin{eqnarray}
\bar{\psi}=i{\psi}^+{\gamma}^0\equiv -{\psi}^TC.
\end{eqnarray}
The ${\cal N}=1$ supersymmetric Yang-Mills theory is given by the Lagrangian density
\begin{eqnarray}
{\cal L}_0=Tr\bigg(-\frac{1}{4}F_{\mu\nu}F^{\mu\nu}-\frac{1}{2}\bar{\lambda}{\gamma}^{\mu}D_{\mu}\lambda +\frac{1}{2}D^2\bigg).
\end{eqnarray}
\begin{eqnarray}
F_{\mu\nu}={\partial}_{\mu}A_{\nu}-{\partial}_{\nu}A_{\mu}-ig[A_{\mu},A_{\nu}]~,~D_{\mu}\lambda={\partial}_{\mu}\lambda-ig[A_{\mu},\lambda].
\end{eqnarray}
The supersymmetric transformations are explicitly given by
\begin{eqnarray}
&&\delta_0 A_{\mu}=\bar{\epsilon}{\gamma}_{\mu}\lambda\nonumber\\
&&\delta_0 \lambda=\bigg(-\frac{1}{4}[{\gamma}^{\mu},{\gamma}^{\nu}]F_{\mu\nu}+i{\gamma}_5D\bigg)\epsilon\nonumber\\
&&\delta_0 D=i\bar{\epsilon}{\gamma}_5{\gamma}_{\mu}{D}^{\mu}\lambda.
\end{eqnarray}
The reduction of this theory to one dimension is obtained by setting $A_{\mu}=X_{\mu}/g$ where ${\partial}_iX_{\mu}=0$. We also set $\lambda=\psi/g$ and $D=F/g$ where  ${\partial}_i\lambda=0$, ${\partial}_iF=0$. We get the supersymmetric Yang-Mills quantum mechanics given by the Lagrangian density
\begin{eqnarray}
{\cal L}_0=\frac{1}{g^2}Tr\bigg(\frac{1}{2}(D_0X_i)^2+\frac{1}{4}[X_i,X_j]^2-\frac{1}{2}\bar{\psi}{\gamma}^0D_0\psi+\frac{i}{2}\bar{\psi}{\gamma}^i[X_i,\psi] +\frac{1}{2}F^2\bigg).
\end{eqnarray}
\begin{eqnarray}
D_0={\partial}_0-i[X_0,.].
\end{eqnarray}
The supersymmetric transformations become
\begin{eqnarray}
&&\delta_0 X_{0}=\bar{\epsilon}{\gamma}_{0}\psi\nonumber\\
&&\delta_0 X_{i}=\bar{\epsilon}{\gamma}_{i}\psi\nonumber\\
&&\delta_0 \psi=\bigg(-\frac{1}{2}[{\gamma}^{0},{\gamma}^{i}]D_0X_i+\frac{i}{4}[{\gamma}^{i},{\gamma}^{j}][X_i,X_j]+i{\gamma}_5F\bigg)\epsilon\nonumber\\
&&\delta_0 F=-i\bar{\epsilon}{\gamma}_5{\gamma}_{0}{D}_{0}\psi+\bar{\epsilon}{\gamma}_5{\gamma}_{i}[{X}_{i},\psi].
\end{eqnarray}
Note that the variation of the Lagrangian density ${\cal L}_0$ under the supersymmetry transformations is given by (we set $F=0$ for simplicity)
\begin{eqnarray}
\delta\bigg(Tr\frac{i}{2}\bar{\psi}{\gamma}^i[X_i,\psi]-Tr\frac{1}{2}\bar{\psi}{\gamma}^0D_0\psi+\frac{1}{2}(D_0X_i)^2+\frac{1}{4}[X_i,X_j]^2\bigg)
&=&\nonumber\\
-\frac{1}{3}\bigg[(C{\gamma}^{\rho})_{\alpha\beta}(C{\gamma}_{\rho})_{\mu\nu}+(C{\gamma}^{\rho})_{\beta\nu}(C{\gamma}_{\rho})_{\mu\alpha}&+&\nonumber\\
(C{\gamma}^{\rho})_{\nu\alpha}(C{\gamma}_{\rho})_{\mu\beta}\bigg]{\epsilon}_{\mu}Tr{\psi}_{\alpha}\{{\psi}_{\nu},{\psi}_{\beta}\}.
\end{eqnarray}
This vanishes by Fierz identity.
\subsection{Deformed Yang-Mills Quantum Mechanics in $4$D}
Let $\mu$ be a constant mass parameter. A mass deformation of the Lagrangian density ${\cal L}_0$ takes the form
\begin{eqnarray}
{\cal L}_{\mu}={\cal L}_0+\frac{\mu}{g^2}{\cal L}_1+\frac{{\mu}^2}{g^2}{\cal L}_2+...\label{4DL}
\end{eqnarray}
The Lagrangian density ${\cal L}_0$ has mass dimension $4$. The corrections ${\cal L}_1$ and ${\cal L}_2$ must have mass dimension $3$ and $2$ respectively. We recall that the Bosonic matrices $X_0$ and $X_{a}$ have mass dimension $1$ whereas the Fermionic matrices ${\psi}_i$ have mass dimension $\frac{3}{2}$. A typical term in the Lagrangian densities ${\cal L}_1$ and ${\cal L}_2$ will contain $n_f$ Fermion matrices, $n_b$ Boson matrices and $n_t$ covariant time derivatives. Clearly for ${\cal L}_1$ we must have $\frac{3}{2}n_f+n_b+n_t=3$. There are only three solutions $(n_f,n_b,n_t)=(2,0,0),(0,3,0),(0,2,1)$. For ${\cal L}_2$ we must have $\frac{3}{2}n_f+n_b+n_t=2$ and we have only one solution $(n_f,n_b,n_t)=(0,2,0)$. Thus the most general forms of ${\cal L}_1$ and ${\cal L}_2$ are
\begin{eqnarray}
{\cal L}_1=Tr\bigg(\bar{\psi}M{\psi}+\frac{1}{3!}S_{abc}X_aX_bX_c+J_{ab}X_aD_0X_b\bigg).
\end{eqnarray}
\begin{eqnarray}
{\cal L}_2=Tr\bigg(-\frac{1}{2!}S_{ab}X_aX_b\bigg).
\end{eqnarray}
Clearly for ${\cal L}_3$ we must have $\frac{3}{2}n_f+n_b+n_t=1$ which can not be satisfied. Thus the correction ${\cal L}_3$ and all other higher order corrections vanish identically.

We will follow the method of \cite{Kim:2006wg} to determine the exact form of the mass deformation. We start with the fermionic mass term
\begin{eqnarray}
\frac{\mu}{g^2}{\cal L}_{\psi}=\frac{\mu}{g^2} Tr\bar{\psi}M\psi=-\mu Tr\bar{\psi}CM^TC\psi.
\end{eqnarray}
We can verify that only the identity matrix, the gamma five and the cubic terms in the gamma matrices can survive in the expansion of $M$. The cubic terms are either ${\gamma}^0[{\gamma}^i,{\gamma}^j]$ or ${\gamma}^1{\gamma}^2{\gamma}^3$. we have
\begin{eqnarray}
{\cal L}_{\psi}=Tr\bar{\psi}\bigg(ia{\bf 1}_4+b{\gamma}_5+\frac{1}{2}H_{ij}{\gamma}^0[{\gamma}^i,{\gamma}^j]+c{\gamma}^1{\gamma}^2{\gamma}^3\bigg)\psi.
\end{eqnarray}
Under the chiral transformation $\psi\longrightarrow \chi=e^{i\phi {\gamma}_5}\psi$, $\bar{\psi}\longrightarrow \bar{\chi}=\bar{\psi}e^{i\phi{\gamma}_5}$ the Lagrangian density ${\cal L}_0$ remains the same while ${\cal L}_{\psi}$ transforms as
\begin{eqnarray}
{\cal L}_{\psi}=Tr\bar{\chi}\bigg(ia^{'}{\bf 1}_4+b^{'}{\gamma}_5+\frac{1}{2}H_{ij}{\gamma}^0[{\gamma}^i,{\gamma}^j]+c{\gamma}^1{\gamma}^2{\gamma}^3\bigg)\chi,
\end{eqnarray}
where $a^{'}=a\cos 2\phi+b\sin 2\phi$ and $b^{'}=-a\sin 2\phi+b\cos 2\phi $. Thus we can use this symmetry to set $b=0$. We get
\begin{eqnarray}
{\cal L}_{\psi}=Tr\bar{\psi}\bigg(ia{\bf 1}_4+\frac{1}{2}H_{ij}{\gamma}^0[{\gamma}^i,{\gamma}^j]+c{\gamma}^1{\gamma}^2{\gamma}^3\bigg)\psi.
\end{eqnarray}
The numerical coefficients $a$, $H_{ij}$ and $c$ will be constrained further under the requirement of supersymmetry invariance.

Next we consider the bosonic terms. We can choose the coefficients $J_{ab}$ to be antisymmetric without any loss of generality since we have $\int dt ~TrJ_{ab}X^aD_0X^b=\int dt ~Tr\frac{1}{2}(J_{ab}-J_{ba})X^aD_0X^b-\int Tr\frac{1}{2}D_0J_{ab}X^aX^b$ where clearly the last term can be included in ${\cal L}_2$. The Bosonic part of the action reads
\begin{eqnarray}
\frac{1}{g^2}Tr\bigg(\frac{1}{2}(D_0X_a)^2+\frac{1}{4}[X_a,X_b]^2+\frac{\mu}{3!}S_{abc}X_aX_bX_c+\mu J_{ab}X_aD_0X_b -\frac{{\mu}^2}{2!}S_{ab}X_aX_b\bigg).
\end{eqnarray}
We consider the action of a time dependent $SO(3)$ rotation $L$ defined by $Y_a=L_{ab}X_b$, $X_a=L_{ba}Y_b$, $LL^T=L^TL=1$. The action transforms as
\begin{eqnarray}
\frac{1}{g^2}Tr\bigg(\frac{1}{2}(D_0Y_a)^2+\frac{1}{4}[Y_a,Y_b]^2+\frac{\mu}{3!}\hat{S}_{abc}Y_aY_bY_c+\mu \hat{J}_{ab}Y_aD_0Y_b -\frac{{\mu}^2}{2!}\hat{S}_{ab}Y_aY_b\bigg),
\end{eqnarray}
where $\hat{S}_{abc}=S_{a_0b_0c_0}L_{aa_0}L_{bb_0}L_{cc_0}$, $\hat{S}_{ab}=S_{a_0b_0}L_{aa_0}L_{bb_0}-\frac{2}{\mu}J_{a_0b_0}L_{aa_0}{\partial}_0L_{bb_0}-\frac{1}{{\mu}^2}{\partial}_0L_{ac}{\partial}_0L_{bc}$, $\hat{J}_{ab}=J_{a_0b_0}L_{aa_0}L_{bb_0}+\frac{1}{\mu}L_{bc}{\partial}_0L_{ac}$. Thus it is clear that we can choose $L$ such that $\hat{J}_{ab}=0$.

By similar arguments we can show that the coefficients $S_{ab}$ can be chosen to be totally symmetric while the coefficients $S_{abc}$ can be chosen totally antisymmetric. By rotational invariance we must therefore have $S_{ab}=v\delta_{ab}$ and $S_{abc}=6ie\epsilon_{abc}$ for some numerical coefficients $v$ and $e$. In particular the Myers term is
\begin{eqnarray}
\frac{\mu}{g^2}{\cal L}_{\rm myers}=\frac{\mu}{g^2}\bigg(ie{\epsilon}_{ijk}TrX_iX_jX_k\bigg).
\end{eqnarray}

The mass deformed supersymmetric transformations will be taken such that on bosonic fields they will coincide with the non deformed supersymmetric transformations so that the Fierz identity can still be used. The mass deformed supersymmetric transformations on fermionic fields will be different from  the non deformed supersymmetric transformations with a time dependent parameter ${\epsilon}\equiv {\epsilon}(t)$ which satisfies ${\partial}_0{\epsilon}=\mu \Pi{\epsilon}$. We will suppose the supersymmetric transformations
\begin{eqnarray}
&&{\delta}_{\mu} X_{0}={\delta}_0 X_{0}\nonumber\\
&&{\delta}_{\mu} X_{i}={\delta}_0 X_{i}\nonumber\\
&&{\delta}_{\mu} {\psi}={\delta}_0 {\psi}+\mu{\Delta}{\epsilon}.
\end{eqnarray}
By requiring that the  Lagrangian density (\ref{4DL}) is invariant under these transformations we can determine  precisely the form of the  mass deformed Lagrangian density  and the mass deformed supersymmetry transformations. A long calculation yields the mass deformed  Lagrangian density  and mass deformed supersymmetry transformations given respectively by (see Appendix $C$) 
\begin{eqnarray}
{\cal L}_{\mu}&=&{\cal L}_{0}+\frac{\mu}{g^2}Tr\bar{\psi}\big(ia-\frac{3e}{4}{\gamma}^1{\gamma}^2{\gamma}^3\big){\psi}+ie{\epsilon}_{ijk}\frac{\mu}{g^2}TrX_iX_jX_k-\frac{{\mu}^2}{g^2}\frac{1}{2}(e^2-\frac{16}{9}a^2)TrX_i^2.\nonumber\\
\end{eqnarray}
\begin{eqnarray}
&&{\delta}_{\mu} X_{0}=\bar{\epsilon}{\gamma}_{0}\psi\nonumber\\
&&{\delta}_{\mu} X_{i}=\bar{\epsilon}{\gamma}_{i}\psi\nonumber\\
&&{\delta}_{\mu} {\psi}=\bigg(-\frac{1}{2}[{\gamma}^{0},{\gamma}^{i}]D_0X_i+\frac{i}{4}[{\gamma}^{i},{\gamma}^{j}][X_i,X_j]-\frac{4\mu}{3}\big(ia+\frac{3e}{4}{\gamma}^1{\gamma}^2{\gamma}^3\big){\gamma}^iX_i\bigg)\epsilon.
\end{eqnarray}
\begin{eqnarray}
\epsilon\equiv \epsilon(t)=e^{\mu\big(\frac{2ia}{3}{\gamma}^0+\frac{e}{2}{\gamma}^0{\gamma}^1{\gamma}^2{\gamma}^3\big)t}.
\end{eqnarray}
We verify that $({\delta}_{\mu} X_{\mu})^+={\delta}_{\mu} X_{\mu}$ and hence the Hermitian matrices $X_{\mu}$ remains Hermitian under supersymmetry. The corresponding supersymmetric algebra is $su(2|1)$ \cite{Kim:2006wg}.

We prefer to work with the parameters $\mu_1$ and $\mu_2$ defined by $ia\mu={{\mu}_1}/{4}$ and $-\frac{3e\mu}{4}={{\mu}_2}/{4}$. The Lagrangian density and supersymmetric transformations become

\begin{eqnarray}
{\cal L}_{\mu}
&=&{\cal L}_{0}+\frac{1}{4g^2}Tr\bar{\psi}\big({\mu}_1+{\mu}_2{\gamma}^1{\gamma}^2{\gamma}^3\big){\psi}-i{\epsilon}_{ijk}\frac{{\mu}_2}{3g^2}TrX_iX_jX_k-\frac{1}{18g^2}({\mu}_1^2+{\mu}_2^2)TrX_i^2.\nonumber\\
\end{eqnarray}
\begin{eqnarray}
&&{\delta}_{\mu} X_{0}=\bar{\epsilon}{\gamma}_{0}\psi\nonumber\\
&&{\delta}_{\mu} X_{i}=\bar{\epsilon}{\gamma}_{i}\psi\nonumber\\
&&{\delta}_{\mu} {\psi}=\bigg(-\frac{1}{2}[{\gamma}^{0},{\gamma}^{i}]D_0X_i+\frac{i}{4}[{\gamma}^{i},{\gamma}^{j}][X_i,X_j]-\frac{1}{3}\big({\mu}_1-{\mu}_2{\gamma}^1{\gamma}^2{\gamma}^3\big){\gamma}^iX_i\bigg)\epsilon.
\end{eqnarray}
\begin{eqnarray}
\epsilon\equiv \epsilon(t)=e^{\frac{1}{6}\big({\mu}_1{\gamma}^0-{\mu}_2{\gamma}^0{\gamma}^1{\gamma}^2{\gamma}^3\big)t}.
\end{eqnarray}

\subsection{Truncation to Zero Dimension}

We consider now the Lagrangian density (action) given by
\begin{eqnarray}
{\cal L}_{\mu}&=&{\cal L}_{0}+\frac{a}{4g^2}Tr\bar{\psi}\big({\mu}_1+{\mu}_2{\gamma}^1{\gamma}^2{\gamma}^3\big){\psi}-i{\epsilon}_{ijk}\frac{b{\mu}_2}{3g^2}TrX_iX_jX_k-\frac{c}{18g^2}({\mu}_1^2+{\mu}_2^2)TrX_i^2.\nonumber\\
\end{eqnarray}
\begin{eqnarray}
{\cal L}_0=\frac{1}{g^2}Tr\bigg(\frac{1}{4}[X_{\mu},X_{\nu}][X^{\mu},X^{\nu}]+\frac{i}{2}\bar{\psi}{\gamma}^{\mu}[X_{\mu},\psi] \bigg).
\end{eqnarray}
In above we have allowed for the possibility that mass deformations corresponding to the reduction to zero and one dimensions can be different by including different  coefficients $a$, $b$ and $c$ in front of the fermionic mass term, the Myers term and the bosonic mass term respectively. However we will keep the mass deformed supersymmetric transformations unchanged. We have
\begin{eqnarray}
&&{\delta}_{\mu} X_{\mu}=\bar{\epsilon}{\gamma}_{\mu}\psi=-\bar{\psi}{\gamma}_{\mu}\epsilon.
\end{eqnarray}
\begin{eqnarray}
{\delta}_{\mu} {\psi}&=&\bigg(\frac{i}{4}[{\gamma}^{\mu},{\gamma}^{\nu}][X_{\mu},X_{\nu}]-\frac{1}{3}\big({\mu}_1-{\mu}_2{\gamma}^1{\gamma}^2{\gamma}^3\big){\gamma}^iX_i\bigg)\epsilon.
\end{eqnarray}
We compute the supersymmetric variations 
\begin{eqnarray}
g^2{\delta}_{\mu}{\cal L}_0=-\frac{i}{3}Tr\bar{\psi}({\mu}_1+{\mu}_2{\gamma}^1{\gamma}^2{\gamma}^3){\gamma}^0{\gamma}^i[X_0,X_i]\epsilon-\frac{i}{6}Tr\bar{\psi}({\mu}_1-{\mu}_2{\gamma}^1{\gamma}^2{\gamma}^3)[{\gamma}^i,{\gamma}^j][X_i,X_j]\epsilon.\label{var0}\nonumber\\
\end{eqnarray}
\begin{eqnarray}
{\delta}_{\mu}\bigg(\frac{a}{4}Tr\bar{\psi}\big({\mu}_1+{\mu}_2{\gamma}^1{\gamma}^2{\gamma}^3\big){\psi}\bigg)&=&\frac{a}{12}({\mu}_1^2+{\mu}_2^2)Tr{\delta}_{\mu}(X_i^2)\nonumber\\
&+&\frac{ia}{2}Tr\bar{\psi}({\mu}_1+{\mu}_2{\gamma}^1{\gamma}^2{\gamma}^3){\gamma}^0{\gamma}^i[X_0,X_i]\epsilon\nonumber\\
&+&\frac{ia}{8}Tr\bar{\psi}({\mu}_1+{\mu}_2{\gamma}^1{\gamma}^2{\gamma}^3)[{\gamma}^i,{\gamma}^j][X_i,X_j]\epsilon.\label{var1}
\end{eqnarray}
Clearly the first term of (\ref{var0}) must cancel the second term of (\ref{var1}), i.e. 
\begin{eqnarray}
a=\frac{2}{3}~,~c=1.
\end{eqnarray}
We get then
\begin{eqnarray}
{\delta}_{\mu}\bigg(g^2{\cal L}_0+\frac{1}{6}Tr\bar{\psi}\big({\mu}_1+{\mu}_2{\gamma}^1{\gamma}^2{\gamma}^3\big){\psi}-\frac{1}{18}({\mu}_1^2+{\mu}_2^2)TrX_i^2\bigg)&=&\nonumber\\
-\frac{i}{6}Tr\bar{\psi}({\mu}_1-{\mu}_2{\gamma}^1{\gamma}^2{\gamma}^3)[{\gamma}^i,{\gamma}^j][X_i,X_j]\epsilon&+&\nonumber\\
\frac{i}{12}Tr\bar{\psi}({\mu}_1+{\mu}_2{\gamma}^1{\gamma}^2{\gamma}^3)[{\gamma}^i,{\gamma}^j][X_i,X_j]\epsilon.\nonumber\\
\end{eqnarray}
By using the identity $\frac{1}{2}{\gamma}^1{\gamma}^2{\gamma}^3[{\gamma}^i,{\gamma}^j]=-{\epsilon}_{ijk}{\gamma}^k$ we find
\begin{eqnarray}
{\delta}_{\mu}\bigg(-i{\epsilon}_{ijk}\frac{b{\mu}_2}{3}TrX_iX_jX_k\bigg)&=&-\frac{ib{\mu}_2}{4}Tr\bar{\psi}{\gamma}^1{\gamma}^2{\gamma}^3[{\gamma}^i,{\gamma}^j][X_i,X_j]\epsilon.
\end{eqnarray}
We must then have
\begin{eqnarray}
b=1.
\end{eqnarray}
But also we must have
\begin{eqnarray}
{\mu}_1=0.
\end{eqnarray}
Thus we get

\begin{eqnarray}
{\delta}_{\mu}\bigg(g^2{\cal L}_0+\frac{{\mu}_2}{6}Tr\bar{\psi}{\gamma}^1{\gamma}^2{\gamma}^3{\psi}-\frac{{\mu}_2^2}{18}TrX_i^2-i{\epsilon}_{ijk}\frac{{\mu}_2}{3}TrX_iX_jX_k\bigg)&=&0.
\end{eqnarray}
The model of interest is therefore 
\begin{eqnarray}
{\cal L}_{\mu}&=&\frac{1}{g^2}Tr\bigg(\frac{1}{4}[X_{\mu},X_{\nu}][X^{\mu},X^{\nu}]+\frac{i}{2}\bar{\psi}{\gamma}^{\mu}[X_{\mu},\psi] +\frac{{\mu}_2}{6}Tr\bar{\psi}{\gamma}^1{\gamma}^2{\gamma}^3{\psi}-\frac{{\mu}_2^2}{18}TrX_i^2\nonumber\\
&-&i{\epsilon}_{ijk}\frac{{\mu}_2}{3}TrX_iX_jX_k\bigg).
\end{eqnarray}
Since $\psi$ and $\epsilon$ are Majorana spinors we can rewrite them as
\begin{eqnarray}
&&\psi=\left(
\begin{array}{c}
i{\sigma}_2({\theta}^+)^T \\
{\theta}
\end{array}
\right)~,~\epsilon=\left(
\begin{array}{c}
i{\sigma}_2({\omega}^+)^T \\
{\omega}
\end{array}
\right).
\end{eqnarray}
We compute with  $X_0=iX_4$ the action
\begin{eqnarray}
{\cal L}_{\mu}
&=&\frac{1}{g^2}Tr\bigg(\frac{1}{2}[X_{4},X_{i}]^2+\frac{1}{4}\bigg([X_i,X_j]-i\frac{{\mu}_2}{3}{\epsilon}_{ijk}X_k\bigg)^2+{\theta}^+\bigg(i[X_4,..]+{\sigma}_i[X_i,..]+\frac{{\mu}_2}{3}\bigg)\theta\bigg).\nonumber\\
\end{eqnarray}
The supersymmetric transformations are
\begin{eqnarray}
&&{\delta}_{\mu}X_0=i({\omega}^+\theta-{\theta}^+\omega)\nonumber\\
&&{\delta}_{\mu}X_i=i({\theta}^+{\sigma}_i{\omega}-{\omega}^+{\sigma}_i\theta)\nonumber\\
&&{\delta}_{\mu}\theta=\bigg(-i{\sigma}_i[X_0,X_i]-\frac{1}{2}{\epsilon}_{ijk}{\sigma}_k[X_i,X_j]+\frac{i}{3}{\mu}_2{\sigma}_iX_i\bigg)\omega.
\end{eqnarray}

\section{Cohomological Approach}
\subsection{Supersymmetry Transformations}
The ${\cal N}=1$ supersymmetric Yang-Mills theory in four dimensions  is given by the Lagrangian density
\begin{eqnarray}
{\cal L}_0=Tr\bigg(-\frac{1}{4}F_{\mu\nu}F^{\mu\nu}-\frac{1}{2}\bar{\lambda}{\gamma}^{\mu}D_{\mu}\lambda +\frac{1}{2}D^2\bigg).
\end{eqnarray}
\begin{eqnarray}
F_{\mu\nu}={\partial}_{\mu}A_{\nu}-{\partial}_{\nu}A_{\mu}-ig[A_{\mu},A_{\nu}]~,~D_{\mu}\lambda={\partial}_{\mu}\lambda-ig[A_{\mu},\lambda].
\end{eqnarray}
The supersymmetric transformations are explicitly given by
\begin{eqnarray}
&&\delta A_{\mu}=\bar{\epsilon}{\gamma}_{\mu}\lambda\nonumber\\
&&\delta \lambda=\bigg(-\frac{1}{4}[{\gamma}^{\mu},{\gamma}^{\nu}]F_{\mu\nu}+i{\gamma}_5D\bigg)\epsilon\nonumber\\
&&\delta D=i\bar{\epsilon}{\gamma}_5{\gamma}_{\mu}{D}^{\mu}\lambda.
\end{eqnarray}
The Majorana spinors $\lambda$ and $\epsilon$ can be written in terms of two dimensional complex spinors $\theta$ and $\omega$ as
\begin{eqnarray}
&&\lambda=-\frac{1}{g}\left(
\begin{array}{c}
i{\sigma}_2({\theta}^+)^T \\
{\theta}
\end{array}
\right)~,~\epsilon=\left(
\begin{array}{c}
i{\sigma}_2({\omega}^+)^T \\
{\omega}
\end{array}
\right).
\end{eqnarray}
We will also write
\begin{eqnarray}
A_{\mu}=\frac{1}{g}X_{\mu}~,~D=\frac{2i}{g}B.
\end{eqnarray}
We will work with Euclidean metric, i.e $X_0=iX_4$. The reduction of the above theory to zero dimension is obtained by setting $\partial_{\mu}X_4=0$, $\partial_{\mu}X_{a}=0$, $\partial_{\mu}\theta=0$, $\partial_{\mu}\theta^+=0$ and $\partial_{\mu}D=0$. We obtain

\begin{eqnarray}
S
&=&-\frac{1}{g^2}Tr\bigg(\frac{1}{4}[X_{\mu},X_{\nu}]^2+\bar{\theta}\bar{\sigma}^{\mu}[X_{\mu},\theta]-2B^2\bigg).
\end{eqnarray}
We can also trivially check that  (we set $g^2=1$)
\begin{eqnarray}
S
&=&-\frac{1}{4}Tr[X_{\mu},X_{\nu}]^2-Tr{\theta}^+\bigg(i[X_4,..]+{\sigma}_a[X_a,..]\bigg)\theta+2Tr B^2.
\end{eqnarray}
The supersymmetric transformations become
\begin{eqnarray}
&&{\delta}X^{\mu}=i\bar{\omega}\bar{\sigma}^{\mu}\theta-i\bar{\theta}\bar{\sigma}^{\mu}\omega\nonumber\\
&&\delta\theta=i{\sigma}^{\mu\nu}[X^{\mu},X^{\nu}]\omega-2B\omega\nonumber\\
&&\delta\bar{\theta}=-i\bar{\omega}{\sigma}^{\mu\nu}[X^{\mu},X^{\nu}]+2B\bar{\omega}\nonumber\\
&&  \delta B=\frac{1}{2}\bar{\omega}\bar{\sigma}^{\mu}[X^{\mu},\theta]+\frac{1}{2}[X^{\mu},\bar{\theta}]\bar{\sigma}^{\mu}\omega.
\end{eqnarray}
Let us note that since we are in Euclidean signature the transformation law of $X_4$ is antihermitian rather than hermitian.

By using a contour shifting argument for the Gaussian integral over $B$ we can rewrite the auxiliary field $B$ as
\begin{eqnarray}
B=H+\frac{1}{2}[X_1,X_2].
\end{eqnarray}
The matrix $H$ must be taken hermitian. We will also introduce
\begin{eqnarray}
\theta_1=\eta_2+i\eta_1~,~\theta_2=\chi_1+i\chi_2.
\end{eqnarray}
\begin{eqnarray}
\phi=\frac{1}{2}(X_3+iX_4)~,~\bar{\phi}=-\frac{1}{2}(X_3-iX_4).
\end{eqnarray}
Let us now compute
\begin{eqnarray}
S&=&2S_{\rm cohom}.
\end{eqnarray}
\begin{eqnarray}
S_{\rm cohom}
&=&Tr\bigg(H^2+H[X_1,X_2]+[X_i,\phi][X_i,\bar{\phi}]+[\phi,\bar{\phi}]^2-\eta_i[\phi,\eta_i]-\chi_i[\bar{\phi},\chi_i]\nonumber\\
&-&\eta_1\epsilon^{ij}[{\chi}_i,X_j]+\eta_2[\chi_i,X_i]\bigg).
\end{eqnarray}
In above we have used $\epsilon^{12}=1$ and the result $TrF_1[B,F_2]=-Tr\{F_1,F_2\}B$ where $F_i$ are fermionic matrices and $B$ is a bosonic matrix. The indices $i$ and $j$ take the values $1$ and $2$. As we will see $\phi$ and $\bar{\phi}$ must be taken to be independent. Furthermore $\phi$ must be taken antihermitian and $\bar{\phi}$ hermitian.

We have four independent real supersymmetries generated by the four independent grassmannian parameters $\xi_i$, $\rho_i$ defined by the equations $\omega_1=\xi_2+i\xi_1$ and $\omega_2=\rho_1+i\rho_2$. The supercharges $Q_1=Q_{1R}+iQ_{1I}$ and $Q_2=Q_{2R}+iQ_{2I}$ are defined  such that the supersymmetric transformation of any operator ${\cal O}={\cal O}_I+i{\cal O}_R$ is given by
\begin{eqnarray}
\delta {\cal O}&=&[Q^+\omega-\omega^+Q,{\cal O}]=(\delta{\cal O})_R+i(\delta{\cal O})_I,
\end{eqnarray}
where
\begin{eqnarray}
(\delta {\cal O})_R=-2i[{\cal Q},{\cal O}_I]~,~
(\delta {\cal O})_I=2i[{\cal Q},{\cal O}_R].
\end{eqnarray}
The generator ${\cal Q}$ is antihermitian and it can be written as
\begin{eqnarray}
{\cal Q}=\xi_2Q_{1R}+\xi_1Q_{1I}+\rho_1Q_{2R}+\rho_2Q_{2I}.
\end{eqnarray}
For a bosonic field  ${\cal O}$ the imaginary part ${\cal O}_I$ is zero and thus we obatin $(\delta {\cal O})_I=2i[{\cal Q},{\cal O}]=\hat{\delta}{\cal O}=-i\delta{\cal O}$. After a long calculation we obtain
\begin{eqnarray}
\hat{\delta} X_1
&=&2\big(\xi_2\chi_1+\xi_1\chi_2+\rho_1\eta_2+\rho_2\eta_1\big)\nonumber\\
\hat{\delta} X_2
&=&2\big(\xi_2\chi_2-\xi_1\chi_1+\rho_2\eta_2-\rho_1\eta_1\big).
\end{eqnarray}
\begin{eqnarray}
\hat{\delta} \phi
&=&-2\big(\rho_1\chi_1+\rho_2\chi_2\big)\nonumber\\
\hat{\delta} \bar{\phi}&=&-2\big(\xi_1\eta_1+\xi_2\eta_2\big).
\end{eqnarray}
\begin{eqnarray}
\hat{\delta} H&=&2\xi_2[\phi,\eta_1]-2\xi_1[\phi,\eta_2]+2\rho_1[\bar{\phi},\chi_2]-2\rho_2[\bar{\phi},\chi_1]+2[X_1,\rho_1\eta_1-\rho_2\eta_2]\nonumber\\
&+&2[-\rho_1\eta_2-\rho_2\eta_1,X_2].
\end{eqnarray}
For fermion fields we obtain $ (\delta\theta_1)_R=-\hat{\delta}\eta_1$, $(\delta\theta_1)_I=\hat{\delta}\eta_2$, $(\delta\theta_2)_R=-\hat{\delta}{\chi}_2$ and $(\delta\theta_2)_I=\hat{\delta}{\chi}_1$. Another long calculation yields
\begin{eqnarray}
-\hat{\delta} \eta_1&=&2\xi_1[\phi,\bar{\phi}]-2\xi_2H-2i\rho_1[X_1,\bar{\phi}]-2i\rho_2[X_2,\bar{\phi}]\nonumber\\
\hat{\delta} \eta_2&=&-2\xi_2[\phi,\bar{\phi}]-2\xi_1H-2i\rho_2[X_1,\bar{\phi}]+2i\rho_1[X_2,\bar{\phi}].
\end{eqnarray}
\begin{eqnarray}
-\hat{\delta} \chi_2&=&-2\rho_2[\phi,\bar{\phi}]-2\rho_1H-2i\rho_2[X_1,X_2]+2\xi_1[X_1,\phi]+2\xi_2[X_2,\phi]\nonumber\\
\hat{\delta} \chi_1&=&2\rho_1[\phi,\bar{\phi}]-2\rho_2H+2i\rho_1[X_1,X_2]-2\xi_2[X_1,\phi]+2\xi_1[X_2,\phi].
\end{eqnarray}

We look at the supercharge associated with $\xi_2$. We define the exterior derivative $d$ on bosons by $dB=i[Q_{1R},B]$ and on fermions by $dF=i\{Q_{1R},F\}$, i.e $\hat{\delta}B=2\xi_2dB$ and $\hat{\delta}F=2\xi_2dF$. The corresponding supersymmetric transformations are
\begin{eqnarray}
d X_i=\chi_i.
\end{eqnarray}
\begin{eqnarray}
d\phi=0~,~d\bar{\phi}=-\eta_2.
\end{eqnarray}
\begin{eqnarray}
  dH&=&[\phi,\eta_1].
\end{eqnarray}
\begin{eqnarray}
  d \eta_1=H~,~d\eta_2=[\bar{\phi},\phi].
\end{eqnarray}
\begin{eqnarray}
  d\chi_i&=&[\phi,X_i].
\end{eqnarray}
From these transformation laws we can immediately deduce that for any operator ${\cal O}$ we must have
\begin{eqnarray}
d^2{\cal O}=[\phi,{\cal O}]. 
\end{eqnarray}
Thus  $d^2$ is a gauge transformation generated by $\phi$ and as a consequence it is nilpotent on gauge invariant quantities such as the action.

Next we compute
\begin{eqnarray}
\hat{\delta}\bigg(-Tr\chi_i[X_i,\bar{\phi}]\bigg)=2\xi_2Tr\bigg([X_i,\phi][X_i,\bar{\phi}]-\chi_i[\bar{\phi},\chi_i]+\eta_2[\chi_i,X_i]\bigg).
\end{eqnarray}
\begin{eqnarray}
\hat{\delta}\bigg(Tr\eta_1[X_1,X_2]\bigg)=2\xi_2Tr\bigg(H[X_1,X_2]-\eta_1\epsilon_{ij}[\chi_i,X_j]\bigg).
\end{eqnarray}
\begin{eqnarray}
\hat{\delta}\bigg(Tr\eta_1H\bigg)=2\xi_2Tr\bigg(H^2-\eta_1[{\phi},\eta_1]\bigg).
\end{eqnarray}
\begin{eqnarray}
\hat{\delta}\bigg(-Tr\eta_2[\phi,\bar{\phi}]\bigg)=2\xi_2Tr\bigg([{\phi},\bar{\phi}]^2-\eta_2[\phi,\eta_2]\bigg).
\end{eqnarray}
Hence
\begin{eqnarray}
\hat{\delta}TrQ=2\xi_2 S_{\rm cohom}.
\end{eqnarray}
\begin{eqnarray}
Q=-\chi_i[X_i,\bar{\phi}]+\eta_1[X_1,X_2]+\eta_1H-\eta_2[\phi,\bar{\phi}].\label{Q}
\end{eqnarray}
As a consequence
\begin{eqnarray}
dTrQ=S_{\rm cohom}.
\end{eqnarray}
Thus we have
\begin{eqnarray}
d^2TrQ= dS_{\rm cohom}=0.
\end{eqnarray}
In the above equation we have used the result $dS_{\rm cohom}=0$. Thus  $d$ is nilpotent on gauge invariant quantities such as $Q$ which are formed from traces. The term cohomology comes  precisely from the analogy of $d$ with an exterior derivative.
\subsection{Cohomologically Deformed Supersymmetry}
We consider the deformed action and deformed exterior derivative given by 
\begin{eqnarray}
S_{\rm def}=S_{\rm cohom}+\hat{S}=S_{\rm cohom}+\epsilon_1 S_1+\epsilon_2 S_2+...
\end{eqnarray}
\begin{eqnarray}
d_{\rm def}=d+\epsilon T.
\end{eqnarray}
Supersymmetric invariance requires
\begin{eqnarray}
d_{\rm def}S_{\rm def}=0.\label{supersymmetric}
\end{eqnarray}
The fact that $d^2$ is equal $0$ on gauge invariant quantities, i.e. $d^2S_{\rm cohom}=d^2S_i=0$ leads to $d^2S_{\rm def}=0$. We have the identity
\begin{eqnarray}
d_{\rm def}^2S_{\rm def}=0.
\end{eqnarray}
This is equivalent to
\begin{eqnarray}
\{d,T\}S_{\rm cohom}+\epsilon T^2S_{\rm cohom}+\epsilon\{d,T\}\hat{S}+\epsilon T^2\hat{S}=0.
\end{eqnarray}
Thus we must have among other things
\begin{eqnarray}
\{d,T\}S_{\rm cohom}=\{d,T\}\hat{S}=0.
\end{eqnarray}
In other words $\{d,T\}$ generates one of the continuous bosonic symmetries of the action $S_{\rm def}$ which are gauge transformations and the remaining rotations given by the $SO(2)$ subgroup of $SO(4)$. Following \cite{Kazakov:1998ji} we choose $\{d,T\}$ to be the rotation ${U}$ defined by
\begin{eqnarray}
{U}:X_a\longrightarrow i\epsilon_{ab}X_b~,~{\chi}_a\longrightarrow i\epsilon_{ab}{\chi}_b.
\end{eqnarray}
We have then
\begin{eqnarray}
\{d,T\}={ U}.
\end{eqnarray}
The symmetry $T$ must also satisfy
\begin{eqnarray}
T^2=0.
\end{eqnarray}
By following the method of \cite{Austing:2001ib} we can determine precisely the form of the correction $T$ from the two requirements $T^2=0$ and $\{d,T\}=U$ and also from the assumption that $T$ is linear in the fields. A straightforward calculation shows that there are two solutions but we will only consider here the one which  generates mass terms for all the bosonic fields. This is given explicitly by

\begin{eqnarray}
&&TX_i=0~,~T\chi_i=i\epsilon_{ij}X_j~,~T\phi=0\nonumber\\
&&TH=i\gamma \eta_2~,~T\eta_2=0~,~T\eta_1=-i\lambda\phi+i\gamma\bar{\phi}~,~T\bar{\phi}=0.
\end{eqnarray}
The cohomologically deformed supersymmetric transformations are therefore given by
\begin{eqnarray}
d_{\rm def} X_i=\chi_i.
\end{eqnarray}
\begin{eqnarray}
d_{\rm def}\phi=0~,~d_{\rm def}\bar{\phi}=-\eta_2.
\end{eqnarray}
\begin{eqnarray}
  d_{\rm def}H&=&[\phi,\eta_1]+i\epsilon \gamma\eta_2.
\end{eqnarray}
\begin{eqnarray}
  d_{\rm def} \eta_1=H+\epsilon(-i\lambda\phi+i\gamma\bar{\phi})~,~d_{\rm def}\eta_2=[\bar{\phi},\phi].
\end{eqnarray}
\begin{eqnarray}
  d_{\rm def}\chi_i&=&[\phi,X_i]+i\epsilon \epsilon_{ij}X_j.
\end{eqnarray}
Furthermore we have the result
\begin{eqnarray}
d_{\rm def}^2=d^2+\epsilon U.
\end{eqnarray}
On $U-$invariant quantities we have  $d_{\rm def}^2=d^2$. For example $d_{\rm def}^2=d^2$ on quantities independent of $X_a$ and $\chi_a$. Also $d_{\rm def}^2(\chi_aX_a)=d^2(\chi_aX_a)$ and $d_{\rm def}^2(\epsilon^{ab}\chi_aX_b)=d^2(\epsilon^{ab}\chi_aX_b)$. Thus we have on $U-$invariant quantities the identity
\begin{eqnarray}
d_{\rm def}^2(...)=d^2(...)=[\phi,...].
\end{eqnarray}
\subsection{Cohomologically Deformed Action}

Next we need to solve the condition (\ref{supersymmetric}). The deformed action is a trace over some polynomial $P$. In the non-deformed case we have $S=dQ$ where $Q$ is a $U-$invariant expression given by (\ref{Q}). We assume that the deformed action $S_{\rm def}=TrP$ is also $U-$invariant. By using the theorem of Austing \cite{Austing:2001ib} we can conclude that the general solution of the  condition (\ref{supersymmetric}), or equivalently of the equation $d_{\rm def}TrP=0$, is
\begin{eqnarray}
S_{\rm def}=d_{\rm def}TrQ_{\rm def}+TrR_3(\phi).
\end{eqnarray}
For $SU(N)$ gauge group this result holds as long as the degree of $P$ is less than $2N/3$. Clearly when the deformation is sent to zero $d_{\rm def}\longrightarrow d$, $Q_{\rm def}\longrightarrow Q$ and $R\longrightarrow 0$. Thus we take

\begin{eqnarray}
Q_{\rm def}=Q-iR~,~R=\kappa_1R_1+\kappa_2R_2.
\end{eqnarray}
We choose $R_1$ and $R_2$ to be the $U-$invariant quantities given by
\begin{eqnarray}
R_1=\frac{1}{2}\epsilon_{ab}\chi_aX_b~,~R_2=-\eta_1\bar{\phi}.
\end{eqnarray}
We choose $R_3(\phi)$ to be the $U-$invariant quantity given by
 \begin{eqnarray}
R_3(\phi)=-\rho^2\phi^2.
\end{eqnarray}
In order to remove the deformation we must take $\epsilon\longrightarrow 0$ so that $d_{\rm def}\longrightarrow d$ and  $\rho\longrightarrow 0$ so that $S_{\rm def}\longrightarrow d TrQ_{\rm def}$ and $\kappa_i\longrightarrow 0$ so that $Q_{\rm def}\longrightarrow Q$. 

We compute
\begin{eqnarray}
S_{\rm def}&=&dTrQ-idTrR+\epsilon TTrQ-i\epsilon TTrR+TrR_3(\phi)\nonumber\\
&=&S_{\rm cohom}+\hat{S}.
\end{eqnarray}
The first term $S_{\rm cohom}=dTr Q$ is the original action.
By using the result that $d(F_1F_2)=dF_1.F_2-F_1dF_2$ and $T(F_1F_2)=TF_1.F_2-F_1TF_2$ we obtain
\begin{eqnarray}
-idTrR=-iTr\bigg[\kappa_1 \phi[X_1,X_2]-\kappa_1\chi_1\chi_2-\kappa_2H\bar{\phi}-\kappa_2\eta_1\eta_2\bigg].
\end{eqnarray}

\begin{eqnarray}
\epsilon TTrQ=\epsilon Tr\bigg[-i\lambda\phi (H+[X_1,X_2])+i\gamma\bar{\phi}H-i\gamma\eta_1\eta_2+i(\gamma+2)\bar{\phi}[X_1,X_2]\bigg].
\end{eqnarray}

\begin{eqnarray}
-i\epsilon TTrR=\epsilon Tr\bigg(\frac{1}{2}\kappa_1X_i^2+\kappa_2\lambda\phi\bar{\phi}-\kappa_2\gamma\bar{\phi}^2\bigg).
\end{eqnarray}
Also
\begin{eqnarray}
TrR_3(\phi)=Tr(-\rho^2\phi^2).
\end{eqnarray}

We will choose the parameters so that the total action enjoys $SO(3)$ covariance with a Myers (Chern-Simons) term and mass terms for all the bosonic and fermionic matrices.  The relevant fermionic terms are

\begin{eqnarray}
i\kappa_1Tr(\chi_1\chi_2-\eta_1\eta_2)+i(\kappa_1+\kappa_2-\epsilon\gamma)Tr\eta_1\eta_2=i\kappa_1Tr(\chi_1\chi_2-\eta_1\eta_2).
\end{eqnarray}
For $SO(3)$ covariance we have chosen
\begin{eqnarray}
\kappa_1+\kappa_2-\epsilon\gamma=0.\label{cond1}
\end{eqnarray}
Next we remark 
\begin{eqnarray}
Tr\bigg(H^2+H[X_1,X_2]\bigg)+i(\kappa_2+\epsilon\gamma)TrH\bar{\phi}-i\epsilon\lambda Tr H\phi&=&\nonumber\\
Tr\bigg(H+\frac{1}{2}[X_1,X_2]+\frac{i}{2}(\kappa_2+\epsilon\gamma)\bar{\phi}-\frac{i}{2}\epsilon\lambda\phi\bigg)^2-\frac{1}{4}Tr[X_1,X_2]^2&+&\nonumber\\
\frac{1}{4}Tr\bigg((\kappa_2+\epsilon\gamma)\bar{\phi}-\epsilon\lambda\phi\bigg)^2-\frac{i}{2}Tr\bigg((\kappa_2+\epsilon\gamma)\bar{\phi}-\epsilon\lambda\phi\bigg)[X_1,X_2].
\end{eqnarray}
The relevant bosonic mass terms are therefore given by
\begin{eqnarray}
Tr\bigg(\frac{1}{2}\epsilon\kappa_1X_i^2+\epsilon\kappa_2\lambda\phi\bar{\phi}-\epsilon\kappa_2\gamma\bar{\phi}^2-\rho^2\phi^2\bigg)+\frac{1}{4}Tr\bigg((\kappa_2+\epsilon\gamma)\bar{\phi}-\epsilon\lambda\phi\bigg)^2&=&\nonumber\\
\frac{1}{2}\epsilon\kappa_1Tr X_i^2+(\frac{1}{4}\epsilon^2\lambda^2-\rho^2)Tr\phi^2+\frac{1}{4}(\kappa_2-\epsilon\gamma)^2Tr\bar{\phi}^2+\frac{1}{2}\epsilon\lambda(\kappa_2-\epsilon\gamma)Tr\phi\bar{\phi}.
\end{eqnarray}
In order to cancel the cross product we choose
\begin{eqnarray}
\frac{1}{4}\epsilon^2\lambda^2-\rho^2=\frac{1}{4}(\kappa_2-\epsilon\gamma)^2.\label{cond2}
\end{eqnarray}
Then
\begin{eqnarray}
Tr\bigg(\frac{1}{2}\epsilon\kappa_1X_i^2+\epsilon\kappa_2\lambda\phi\bar{\phi}-\epsilon\kappa_2\gamma\bar{\phi}^2-\rho^2\phi^2\bigg)+\frac{1}{4}Tr\bigg((\kappa_2+\epsilon\gamma)\bar{\phi}-\epsilon\lambda\phi\bigg)^2&=&\nonumber\\
\frac{1}{2}\epsilon\kappa_1Tr X_i^2+\frac{1}{8}(\kappa_2-\epsilon\gamma)(\kappa_2-\epsilon\gamma-\epsilon\lambda)TrX_3^2-\frac{1}{8}(\kappa_2-\epsilon\gamma)(\kappa_2-\epsilon\gamma+\epsilon\lambda)TrX_4^2&=&\nonumber\\
\frac{1}{2}\epsilon\kappa_1Tr X_a^2-\frac{1}{8}(\kappa_2-\epsilon\gamma)(\kappa_2-\epsilon\gamma+\epsilon\lambda)TrX_4^2.
\end{eqnarray}
Again for $SO(3)$ covariance we have chosen
\begin{eqnarray}
\frac{1}{2}\epsilon\kappa_1=\frac{1}{8}(\kappa_2-\epsilon\gamma)(\kappa_2-\epsilon\gamma-\epsilon\lambda).\label{cond3}
\end{eqnarray}
The Chern-Simons term can be obtained from the terms
\begin{eqnarray}
Tr\bigg(-i(\kappa_1+\epsilon \lambda)\phi[X_1,X_2]+i\epsilon(\gamma+2)\bar{\phi}[X_1,X_2]\bigg)-\frac{i}{2}Tr\bigg((\kappa_2+\epsilon\gamma)\bar{\phi}-\epsilon\lambda\phi\bigg)[X_1,X_2]&=&\nonumber\\
-\frac{i}{2}(\epsilon\lambda+2\kappa_1)Tr\phi[X_1,X_2]+\frac{i}{2}(-\kappa_2+\epsilon\gamma+4\epsilon)Tr\bar{\phi}[X_1,X_2]&=&\nonumber\\
-\frac{i}{4}(\epsilon\lambda+\epsilon\gamma+2\kappa_1 -\kappa_2+4\epsilon)Tr X_3[X_1,X_2]+\frac{1}{4}(\epsilon\lambda-\epsilon\gamma+2\kappa_1 +\kappa_2-4\epsilon)Tr X_4[X_1,X_2].
\end{eqnarray}
The first term is the Chern-Simons action. We want to impose the condition
\begin{eqnarray}
\epsilon\lambda+2\kappa_1+\kappa_2=\epsilon\gamma+4\epsilon.\label{cond4}
\end{eqnarray}
The solution of equations (\ref{cond1}), (\ref{cond2}), (\ref{cond3}) and (\ref{cond4}) is
\begin{eqnarray}
\gamma=\frac{\kappa_1+\kappa_2}{\epsilon}~,~\lambda=4-\frac{\kappa_1}{\epsilon}~,~\rho^2=2\epsilon(2\epsilon-\kappa_1).
\end{eqnarray}
The deformation action $\Delta{S}_{\rm cohom}$ is therefore

\begin{eqnarray}
\Delta{S}_{\rm cohom}&=&i\kappa_1Tr(\chi_1\chi_2-\eta_1\eta_2)+\frac{1}{2}\epsilon\kappa_1Tr X_a^2-\frac{1}{8}(\kappa_2-\epsilon\gamma)(\kappa_2-\epsilon\gamma+\epsilon\lambda)TrX_4^2\nonumber\\
&-&\frac{i}{4}(\epsilon\lambda+\epsilon\gamma+2\kappa_1 -\kappa_2+4\epsilon)Tr X_3[X_1,X_2]\nonumber\\
&=&\frac{\kappa_1}{2}Tr\theta^+\theta+\frac{1}{2}\epsilon\kappa_1Tr X_a^2+\frac{1}{4}\kappa_1(2\epsilon-\kappa_1)TrX_4^2-\frac{i}{6}(4\epsilon+\kappa_1)\epsilon_{abc}TrX_aX_bX_c.\nonumber\\
\end{eqnarray}
We introduce now 
\begin{eqnarray}
-(4\epsilon+\kappa_1)=\alpha~,~\kappa_1=-\frac{\alpha}{3}+4\zeta_0\alpha.\label{alpha}
\end{eqnarray}
Thus we get
\begin{eqnarray}
\Delta{S}_{\rm cohom}
&=&-\frac{\alpha}{6}(1-12\zeta_0)Tr\theta^+\theta+\frac{\alpha^2}{36}(1+6\zeta_0)(1-12\zeta_0)Tr X_a^2+\frac{\alpha^2}{2}\zeta_0(1-12\zeta_0)TrX_4^2\nonumber\\
&+&\frac{i}{6}\alpha\epsilon_{abc}TrX_aX_bX_c.
\end{eqnarray}
The total action is
\begin{eqnarray}
S_{\rm def}&=&\bigg(S_{\rm cohom}-Tr(H^2+H[X_1,X_2])\bigg)+Tr\bigg(H+\frac{1}{2}[X_1,X_2]+\frac{i}{2}(\kappa_2+\epsilon\gamma)\bar{\phi}-\frac{i}{2}\epsilon\lambda\phi\bigg)^2\nonumber\\
&-&\frac{1}{4}Tr[X_1,X_2]^2+\Delta{S}_{\rm cohom}.
\end{eqnarray}
The first line is effectively equivalent to $S_{\rm cohom}$. Thus the total action is
\begin{eqnarray}
S_{\rm def}&=&S_{\rm cohom}+\Delta{S}_{\rm cohom}.
\end{eqnarray}
Next we perform the scaling
\begin{eqnarray}
X_{\mu}\longrightarrow (2N)^{\frac{1}{4}}X_{\mu}~,~\theta\longrightarrow \sqrt{\frac{2}{N\alpha}}\frac{1}{(2N)^{\frac{1}{8}}}\theta,
\end{eqnarray}
and
\begin{eqnarray}
\alpha\longrightarrow 2(2N)^{\frac{1}{4}}\alpha.\label{scaling}
\end{eqnarray}
We get then the one-parameter family of actions given by (we set $B=0$)

\begin{eqnarray}
S_{\rm def}&=&-\frac{N}{4} Tr[X_{\mu},X_{\nu}]^2+N\frac{2i\alpha}{3}\epsilon_{abc}TrX_aX_bX_c+\frac{2N\alpha^2}{9}(1+6\zeta_0)(1-12\zeta_0)Tr X_a^2\nonumber\\
&+&4N\alpha^2\zeta_0(1-12\zeta_0)TrX_4^2-\frac{1}{N\alpha}Tr{\theta}^+\bigg(i[X_4,..]+{\sigma}_a[{X}_a,..]+\frac{2\alpha}{3}(1-12\zeta_0)\bigg)\theta.\nonumber\\
\end{eqnarray}
For stability the parameter $\zeta_0$ must be in the range
\begin{eqnarray}
0<\zeta_0<1/12.
\end{eqnarray}
This action for $\zeta_0=0$ is precisely the mass deformed action derived in section $2$. The value $\zeta_0=1/12$ will also be of interest to us in this article. This one-parameter family of actions preserves only half of the ${\cal N}=1$ supersymmetry in the sense that we can construct only two mass deformed supercharges \cite{Austing:2001ib}.

\section{Simulation Results for $D=4$ Yang-Mills Matrix Models}
\subsection{Models, Supersymmetry and Fuzzy Sphere}
We are interested in the cohomologically deformed Yang-Mills matrix models
\begin{eqnarray}
S&=&N Tr\bigg[-\frac{1}{4}[X_{\mu},X_{\nu}]^2+\frac{2i\alpha}{3}{\epsilon}_{abc}X_aX_bX_c\bigg]+N{\beta} TrX_a^2+N\beta_4 Tr X_4^2\nonumber\\
&-&\frac{1}{N\alpha}Tr{\theta}^+\bigg(i[X_4,..]+{\sigma}_a[{X}_a,..]+{\zeta}\bigg)\theta.\label{model0}
\end{eqnarray}
The range of the parameters is
 \begin{eqnarray}
\beta=\frac{2}{9}(\alpha+6\xi_0)(\alpha-12\xi_0)~,~\beta_4=4\xi_0(\alpha-12\xi_0)~,~\zeta=\frac{2}{3}(\alpha-12\xi_0).
\end{eqnarray}
\begin{eqnarray}
0\leq \xi_0\leq \frac{\alpha}{12}.
\end{eqnarray}
This action preserves two supercharges compared to the four supercharges of the original non deformed Yang-Mills matrix model \cite{Austing:2001ib}. We will be mainly interested in the "minimally" deformed Yang-Mills matrix model corresponding to the value $\xi_0=\alpha/12$ for which we have
 \begin{eqnarray}
S&=&N Tr\bigg[-\frac{1}{4}[X_{\mu},X_{\nu}]^2+\frac{2i\alpha}{3}{\epsilon}_{abc}X_aX_bX_c\bigg]-\frac{1}{N\alpha}Tr{\theta}^+\bigg(i[X_4,..]+{\sigma}_a[{X}_a,..]\bigg)\theta.\nonumber\\
\end{eqnarray}
The "maximally" deformed Yang-Mills matrix model corresponding to the value $\xi_0=0$ coincides precisely with the mass-deformed model in $D=4$ and as such it has a full ${\cal N}=1$ mass deformed supersymmetry besides the half ${\cal N}=1$  cohomologically deformed supersymmetry.  From this perspective this case is far more important than the previous one. However there is the issue of the convergence of the partition function which we will discuss shortly. In any case the "maximally" deformed Yang-Mills matrix model is given by the action 
\begin{eqnarray}
S&=&N Tr\bigg[-\frac{1}{4}[X_{\mu},X_{\nu}]^2+\frac{2i\alpha}{3}{\epsilon}_{abc}X_aX_bX_c\bigg]+N\frac{2\alpha^2}{9} TrX_a^2\nonumber\\
&-&\frac{1}{N\alpha}Tr{\theta}^+\bigg(i[X_4,..]+{\sigma}_a[{X}_a,..]+\frac{2}{3}{\alpha}\bigg)\theta.
\end{eqnarray}
The above two actions can also be rewritten as (with $X_a=\alpha D_a$ and $\tilde{\alpha}=\alpha\sqrt{N}$)
\begin{eqnarray}
S_{\rm SUSY}&=&NTr\bigg[-\frac{1}{4}[X_a,X_b]^2+\frac{2i\alpha}{3}{\epsilon}_{abc}X_aX_bX_c\bigg]+N\tilde{\beta}\alpha^2TrX_a^2\nonumber\\
&-&\frac{1}{N\alpha}Tr{\theta}^+\bigg(i[X_4,..]+{\sigma}_a[X_a,..]+\alpha\tilde{{\xi}}\bigg)\theta.
\end{eqnarray}
Equivalently
\begin{eqnarray}
S_{\rm SUSY}&=&\frac{\tilde{\alpha}^4}{N} Tr\bigg[-\frac{1}{4}[D_a,D_b]^2+\frac{2i}{3}{\epsilon}_{abc}D_aD_bD_c\bigg]+\frac{\tilde{\beta}\tilde{\alpha}^4}{N}TrD_a^2\nonumber\\
&-&\frac{1}{N}Tr{\theta}^+\bigg(i[D_4,..]+{\sigma}_a[D_a,..]+\tilde{{\xi}}\bigg)\theta.\label{action0}
\end{eqnarray}
\begin{eqnarray}
&&\tilde{\beta}=0~,~\tilde{\xi}=0~~{\rm cohomologically}~{\rm deformed}\nonumber\\
&&\tilde{\beta}=\frac{2}{9}~,~\tilde{\xi}=\frac{2}{3}~~{\rm mass}~{\rm deformed}.
\end{eqnarray}
We remark that the bosonic part of the mass-deformed Yang-Mills matrix action can be rewritten as a complete square, viz
\begin{eqnarray}
S_B&=&N Tr\bigg[-\frac{1}{4}[X_{\mu},X_{\nu}]^2+\frac{2i\alpha}{3}{\epsilon}_{abc}X_aX_bX_c\bigg]+N\frac{2\alpha^2}{9} TrX_a^2\nonumber\\
&=&N Tr\bigg(\frac{i}{2}[X_{\mu},X_{\nu}]+\frac{\alpha}{3}{\epsilon}_{\mu\nu\lambda}X_{\lambda}\bigg)^2.
\end{eqnarray}
Clearly ${\epsilon}_{\mu\nu\lambda}=0$ if any of the indices $\mu$,$\nu$,$\lambda$ takes the value $4$. Generically the bosonic action of interest is given by
\begin{eqnarray}
S_B&=&N Tr\bigg[-\frac{1}{4}[X_{\mu},X_{\nu}]^2+\frac{2i\alpha}{3}{\epsilon}_{abc}X_aX_bX_c\bigg]+N\tilde{\beta}\alpha^2 TrX_a^2.
\end{eqnarray}
Here we allow $\tilde{\beta}$ to take on any value. The variation of the bosonic action for generic values of $\tilde{\beta}$ reads
\begin{eqnarray}
&&{\delta}S_B=NTrJ_4{\delta}X_4+NTrJ_b{\delta}X_b\nonumber\\
&&J_4=[X_a,[X_a,X_4]]~,~J_b=2\tilde{\beta}\alpha^2 X_b+i[F_{ab},X_a]+[X_4,[X_4,X_b]]~,~\nonumber\\
&&F_{ab}=i[X_a,X_b]+\alpha{\epsilon}_{abc}X_c.
\end{eqnarray}
Thus extrema of the model are given by $1)$ reducible representations $J_a$ of $SU(2)$, i.e $X_a=J_a$ and $X_4=0$ and $2)$ commuting matrices, i.e $X_{\mu}$ belong to the Cartan sub-algebra of $SU(N)$. The identity matrix corresponds to an uncoupled mode and thus we have $SU(N)$ instead of $U(N)$. Global minima are given by irreducible representations of $SU(2)$ of dimensions $N$ and $0$. Indeed we find that the configurations $X_a=\phi L_a$, $X_4=0$ solve the equations of motion with $\phi$ satisfying the cubic equation $\phi({\phi}^2-\alpha \phi+\tilde{\beta}\alpha^2)=0$. We get the solutions
\begin{eqnarray}
{\phi}_0=0~,~{\phi}_{\pm}=\alpha\frac{1\pm \sqrt{1-4\tilde{\beta}}}{2}. 
\end{eqnarray}
We can immediately see that we must have $\tilde{\beta}{\leq}{1}/{4}$ which does indeed hold for the values of interest $\tilde{\beta}=0$ and $\tilde{\beta}=2/9$. However the action at ${\phi}_{\pm}$ is given by
\begin{eqnarray}
S_B[{\phi}_{\pm}]=\frac{N^2c_2{\phi}_{\pm}^2}{2}\alpha^2\bigg(\tilde{\beta}-\frac{1}{6}\mp\frac{1}{6}\sqrt{1-4\tilde{\beta}}\bigg).
\end{eqnarray}
We can verify that $S_B[{\phi}_{-}]$ is always positive while $S_B[{\phi}_{+}]$ is  negative for the values of $\tilde{\beta}$ such that  $\tilde{\beta}{\leq}{2}/{9}$. Furthermore we note that $S[{\phi}_0]=0$. In other words for $\tilde{\beta}{\leq}2/9$ the global minimum of the model is the irreducible representation of $SU(2)$ of maximum dimension $N$ whereas for $\tilde{\beta}> 2/9$ the global minimum of the model is the irreducible representation of $SU(2)$ of minimum dimension $0$. 

At $\tilde{\beta}=2/9$ we get ${\phi}_+=2\alpha/3$ and $S[{\phi}_+]=0$. Thus the configuration $X_a=\frac{2\alpha}{3}L_a$ becomes degenerate with the configuration $X_a=0$. However there is an entire $SU(N)$ manifold of configurations $X_a=\frac{2\alpha}{3}UL_aU^+$ which are equivalent to the fuzzy sphere configuration. In other words the fuzzy sphere configuration is  still favored although now due to entropy. Thus there is a first order transition at $\tilde{\beta}=2/9$ when the classical ground state switches from $X_a=\frac{2}{3}L_a$ to $X_a=0$ as we increase $\tilde{\beta}$ through the critical value $\tilde{\beta}=2/9$.  The two values of interest $\tilde{\beta}=0$ and $\tilde{\beta}=2/9$ both lie in the regime where the fuzzy sphere is the stable classical ground state. 

This discussion holds also for the full bosonic model in which we include a mass term for the matrix $X_4$. Quantum correction, i.e. the inclusion of fermions, are expected to alter significantly this picture.



Towards the commutative limit we rewrite the action, the second line of (\ref{action0}), into the form (with $\tilde{\alpha}^4=1/g^2$ and $F_{ab}=i[D_a,D_b]+{\epsilon}_{abc}\phi D_c$)
\begin{eqnarray}
S&=&\frac{1}{4g^2N} TrF_{ab}^2-\frac{3\phi -2}{6g^2N}Tr\left[\epsilon_{abc}F_{ab}D_c+\phi D_a^2\right]+\frac{1}{g^2N}\bigg(\phi(\phi-1)+\tilde{\beta}\bigg)TrD_a^2\nonumber\\
&-&\frac{1}{2g^2N}Tr [D_a,D_4]^2-\frac{1}{N}Tr{\theta}^+ \big(i[D_4,..]+{\sigma}_a[{D}_a,..] + \tilde{\xi} \big){\theta}.
\end{eqnarray}
The $3$rd terms actually cancel for all values of $\beta$. Thus
\begin{eqnarray}
S&=&\frac{1}{4g^2N} TrF_{ab}^2-\frac{3\phi -2}{6g^2N}Tr\left[\epsilon_{abc}F_{ab}D_c+\phi D_a^2\right]-\frac{1}{2g^2N}Tr [D_a,D_4]^2\nonumber\\
&-&\frac{1}{N}Tr{\theta}^+ \big(i[D_4,..]+{\sigma}_a[{D}_a,..] + \tilde{\xi} \big){\theta}.
\end{eqnarray}
The commutative limit $N\longrightarrow \infty$ is then obvious. We write $D_a=\phi (L_a +A_a)$ and we obtain

\begin{eqnarray}
S&=&\frac{1}{4g^2} \int_{S^2} F_{ab}^2-\frac{(3\phi-2)\phi}{4g^2}{\epsilon}_{abc}\int F_{ab}A_c
-\frac{1}{2g^2}\int_{S^2}({\cal L}_aD_4)^2-\int_{S^2} {\psi}^+ \big(\phi {\sigma}_a{\cal L}_a + \tilde{\xi} \big){\psi}.\nonumber\\
\end{eqnarray}
\subsection{Path Integral, Convergence and Observables}

In the quantum theory we will integrate over $N\times N$ bosonic matrices $X_{\mu}$ and $N\times N$ fermionic matrices ${\theta}_{\alpha}^+$ and ${\theta}_{\alpha}$. The trace parts of $X_{\mu}$, ${\theta}_{\alpha}^+$ and ${\theta}_{\alpha}$ will be removed since they correspond to  free degrees of freedom. The  partition function of the model is therefore given by
\begin{eqnarray}
Z&=&\int {\cal D}X_{\mu}~{\cal D} \theta~{\cal D}{\theta}^+~{\delta}\big(Tr X_{\mu}\big)~{\delta}\big(Tr{\theta}_{\alpha}^+\big)~\delta\big(Tr{\theta}_{\alpha}\big)~e^{-S_{\rm SUSY}}\nonumber\\
&=&\int {\cal D}X_{\mu}{\delta}~\big(Tr X_{\mu}\big)~{\rm det}{\cal D}~e^{-S_B}.\label{pathI}
\end{eqnarray}
\begin{eqnarray}
{\rm det}{\cal D}&=&\int d{\theta}d{\theta}^+{\delta}(Tr{\theta}_{\alpha}){\delta}(Tr{\theta}_{\alpha}^+)e^{\frac{1}{N\alpha}Tr {\theta}^+{\cal D}{\theta}}.
\end{eqnarray}
The integration over the fermions yielded  the determinant of the $2(N^2-1)\times 2(N^2-1)$ dimensional matrix ${\cal D}=i[X_4,..]+{\sigma}_a[X_a,..]+\tilde{\xi}\alpha$. This determinant is positive definite since every eigenvalue ${\lambda}$ of ${\cal D}$ is doubly degenerate \cite{Ambjorn:2000bf}. The reason lies in the fact that the Dirac operator ${\cal D}=iX_4-iX_4^R+{\sigma}_aX_a-{\sigma}_aX_a^R+\tilde{\xi}\alpha$ is symmetric under the exchange of left and right operators, viz under $X_a\leftrightarrow -X_a^R$. A much cleaner proof goes as follows. Let $\Psi$ be an eigenstate of ${\cal D}$ with eigenvalue $\lambda$, in other words 
\begin{eqnarray}
i[X_4,\Psi]+{\sigma}_a[X_a,\Psi]+\tilde{\xi}\alpha \psi =\lambda \Psi.
\end{eqnarray}
Taking the hermitian conjugate of this equation we get
\begin{eqnarray}
i[X_4,(\Psi^+)^T]-{\sigma}_a^T[X_a,({\Psi}^+)^T]+\tilde{\xi}\alpha ({\Psi}^+)^T=\lambda ({\Psi}^+)^T.
\end{eqnarray}
In above $({\Psi}^+)^T$ is a column vector with components given by ${\Psi}^+_{1,2}$. Multiplying the above equation by ${\sigma}_2$ and defining the spinor $\tilde{\Psi}={\sigma}_2({\Psi}^+)^T$ we arrive at the equation
\begin{eqnarray}
i[X_4,\tilde{\Psi}]+{\sigma}_a[X_a,\tilde{\Psi}]+\xi\alpha \tilde{\Psi} =\lambda \tilde{\Psi}.
\end{eqnarray}
We have also used the identity ${\sigma}_a=-{\sigma}_2{\sigma}_a^T{\sigma}_2$. We conclude that $\tilde{\Psi}$ is also an eigenstate of ${\cal D}$ with the same eigenvalue $\lambda$. The spinors $\Psi$ and $\tilde{\Psi}$ are charge conjugate to each other. In above we have assumed that ${\lambda}$ is real which follows from the fact that the Dirac operator ${\cal D}=i[X_4,..]+{\sigma}_a[{X}_a,..]+\tilde{\xi}\alpha$ is hermitian. This establishes that the determinant $\det {\cal D}$ is positive definite for any configuration $X_{\mu}$ and hence the model can be accessed directly by Monte Carlo simulation.

Let us also add that the Dirac operator ${\cal D}$ admits an approximate chirality operator and hence there is an approximate chiral symmetry in this model beside exact rotational invariance,  exact gauge invariance and  exact charge conjugation. The existence of chiral symmetry (though approximate) means that there should exist more structure in the low energy fermionic spectrum.

The partition function $Z$ is also invariant under the translation $X_{\mu}\longrightarrow X_{\mu}+\epsilon X_{\mu}$ where $\epsilon$ is a small parameter.  Under this coordinate transformation the  measure $dX_{\mu}$ changes to $(1+4(N^2-1)\epsilon)dX_{\mu}$. The bosonic action $S_B=S_4+S_3+S_2$ changes to $S_B+\epsilon(4S_4+3S_3+2S_2)$ under $X_{\mu}\longrightarrow X_{\mu}+\epsilon X_{\mu}$. The determinant, on the other hand, changes under $X_{\mu}\longrightarrow X_{\mu}+\epsilon X_{\mu}$ as 
\begin{eqnarray}
{\rm det}{\cal D}&\longrightarrow &(1+\epsilon)^{2(N^2-1)}{\rm det}^{'}\bigg({\cal M}^{'}-\epsilon\tilde{\xi}\alpha(1+\gamma)\bigg)\nonumber\\
&\longrightarrow & (1+\epsilon)^{2(N^2-1)}{\rm det}^{'}\bigg(1-\frac{\epsilon}{{\cal M}^{'}}\tilde{\xi}\alpha(1+\gamma)\bigg){\rm det}^{'}{\cal M}^{'}.
\end{eqnarray}
The matrices ${\cal M}^{'}$ and $\gamma$ are given in appendix $B$. We obtain then
\begin{eqnarray}
{\rm det}{\cal D}\longrightarrow \bigg(1+\epsilon\bigg[2(N^2-1)-\tilde{\xi}\alpha Tr^{'}_{\rm ad}\frac{1}{\cal D}-\tilde{\xi}\alpha Tr^{'}_{\rm ad}\frac{1}{\cal D}\gamma\bigg]\bigg) {\rm det}{\cal D}.
\end{eqnarray}
From the invariance of the partition function under the coordinate transformation $X_{\mu}\longrightarrow X_{\mu}+\epsilon X_{\mu}$ we derive therefore the Schwinger-Dyson identity

\begin{eqnarray}
{\rm IDE}&=&4\frac{<{\rm YM}>}{N^2}+4\frac{<{\rm YM}_0>}{N^2}+3\frac{<{\rm CS}>}{N^2}+2\frac{<{\rm RAD}>}{N^2}+\tilde{\xi}\alpha \frac{{\rm COND}}{N^2}+\frac{6}{N^2}\nonumber\\
&\equiv &6.\label{ide}
\end{eqnarray}
This is an exact result. 

The operators ${\rm YM}$, ${\rm YM}_0$ and ${\rm CS}$ are the actions given by
\begin{eqnarray}
&&{\rm YM}= -\frac{N}{4}Tr[X_a,X_b]^2~,~{\rm YM}_0= -\frac{N}{2}Tr[X_0,X_a]^2~,~{\rm CS}=\frac{iN\alpha}{3}{\epsilon}_{abc}Tr[X_a,X_b]X_c.\nonumber\\
\end{eqnarray}
The action ${\rm RAD}$ is related to the radius of the sphere. It is given by
\begin{eqnarray}
{\rm RAD}=N\tilde{\beta}\alpha^2  TrX_a^2.
\end{eqnarray}
We will define the radius $r$ of the sphere through the relation
\begin{eqnarray}
\frac{1}{r}=\frac{1}{N\alpha^2c_2}TrX_a^2.
\end{eqnarray}
The total bosonic action is given by $S={\rm YM}+{\rm YM}_0+{\rm CS}+{\rm RAD}$. The condensation is defined by
\begin{eqnarray}
{\rm COND}&=& \frac{1}{\alpha}\frac{\partial\ln Z}{\partial\tilde{\xi}}=<\frac{1}{N\alpha}Tr\theta^+\theta>.
\end{eqnarray}
We find
\begin{eqnarray}
{\rm COND}
&=&{\rm COND}_0+{\rm COND}_1.
\end{eqnarray}
\begin{eqnarray}
&&{\rm COND}_0= <Tr^{'}_{\rm ad}\frac{1}{i[X_4,..]+{\sigma}_a[{X}_a,..]+\tilde{\xi}\alpha}>,\nonumber\\
&&{\rm COND}_1= <Tr^{'}_{\rm ad}\frac{1}{i[X_4,..]+{\sigma}_a[{X}_a,..]+\tilde{\xi}\alpha}\gamma>.
\end{eqnarray}
These are real quantities. The condensation ${\rm COND}_0$ is obviously real since every eigenvalue of the Dirac operator is doubly degenerate. The expression of the total condensation ${\rm COND}$ suggests that it is real. We have  verified numerically that ${\rm COND}_1$ is real and that the condensation is dominated by ${\rm COND}_0$.

For completeness we will also measure the logarithm of the determinant of the Dirac operator  as an independent observable in Monte Carlo simulation.

Convergence of Yang-Mills path inetgrals such as the one given by (\ref{pathI}) was studied extensively in \cite{Austing:2001ib}  and in \cite{Austing:2001bd,Krauth:1998xh,Krauth:1999qw, Krauth:1998yu, Austing:2001pk,Austing:2003kd}. This question is of paramount importance for analytical analysis as well as for Monte Carlo simulation. The source of the divergence, if any lies in the so-called flat directions, i.e. the set of commuting matrices. 

The path integral (\ref{pathI}) corresponds to a gauge theory with gauge group $SU(N)$ in dimension $D=4$. We will also consider $SU(N)$ gauge theory in dimension $D=3$.  We start the discussion with the model $\alpha=0$, $\tilde{\beta}=0$ and $\tilde{\xi}=0$. It was found in \cite{Austing:2001bd} that the bosonic path integral in $D=3$ is convergent for $N\geq 4$ while the bosonic path integral in $D=4$ is convergent for $N\geq 3$. Since we are interested in large values of $N$ we can safely consider the bosonic path integrals in $D=3,4$ to be convergent for all practical purposes. On the other hand, it is found in \cite{Austing:2001pk}, that the supersymmetric path integral in $D=3$ is not convergent  while the supersymmetric path integral in $D=4$ is convergent for all $N\geq 2$. 

Tuning the parameters  $\alpha$, $\tilde{\beta}$ does not change this picture. For example it was shown in \cite{Austing:2003kd} that adding a Myers term, i.e. considering a non-zero value of $\alpha$, does not change the convergence properties of the $D-$dimensional Yang-Mills matrix path integral. The point is that the Chern-Simons (Myers) term is always small compared to the quartic Yang-Mills term. The same argument should then lead to the conclusion that adding a bosonic mass term, i.e. if we consider a non-zero $\tilde{\beta}$, will not change the above picture.

Tuning the fermion mass term, i.e. considering a non-zero value of the scalar curvature $\tilde{\xi}$, will lead to complications. In this case the Pfaffian, or equivalently the determinant, will be expanded as a polynomial in the scalar curvature $\tilde{\xi}$. The analysis of \cite{Austing:2001pk} should then be repeated for every term in this expansion. We claim that the supersymmetric path integral in $D=4$ is not convergent  for generic values of $\tilde{\xi}$.

We have extensively checked in Monte Carlo simulation the conjecture that Yang-Mills matrix models in dimension $D=4$ does not make sense  for generic values of $\tilde{\xi}$. The major observation is that for $\tilde{\xi}\neq 0$ the fermion determinant for generic values of $\tilde{\alpha}=\sqrt{N}\alpha$  never reaches thermalization \footnote{This happens typically for small values of $\tilde{\alpha}$ far from the fuzzy sphere region but not very close to $\tilde{\alpha}=0$.}. However, we have also observed that for sufficiently small values  of $\tilde{\xi}$ the theory actually makes sense and thus there is some critical value of $\tilde{\xi}$, which we will not determine in this article, above which the path integral is ill defined. The value of interest $\tilde{\xi}=2/3$ corresponding to the mass deformed matrix model lies in this range where the model is actually ill defined.

Therefore in order to access the mass deformed Yang-Mills matrix model by the Monte Carlo method we must regularize the theory in such a way as to make sure that the path integral is absolutely convergent. Unfortunately most regularizations will not maintain neither the full ${\cal N}=1$ mass deformed supersymmetry nor the half ${\cal N}=1$ cohomologically deformed supersymmetry of this model. We adopt here the regularization in which we will simply set $\tilde{\xi}=0$. In other words we make the replacement

\begin{eqnarray}
S_{\rm SUSY}\longrightarrow S_{\rm SUSY}^{'}&=&NTr\bigg[-\frac{1}{4}[X_a,X_b]^2+\frac{2i\alpha}{3}{\epsilon}_{abc}X_aX_bX_c\bigg]+N\tilde{\beta}\alpha^2TrX_a^2\nonumber\\
&-&\frac{1}{N\alpha}Tr{\theta}^+\bigg(i[X_4,..]+{\sigma}_a[X_a,..]\bigg).\label{action1}
\end{eqnarray}
In summary, the value $\tilde{\beta}=2/9$ corresponds to the mass deformed Yang-Mills matrix model with softly broken supersymmetry whereas the value $\tilde{\beta}=0$ is precisely the minimally deformed model which enjoys  half of the ${\cal N}=1$ cohomologically deformed supersymmetry.

The Metropolis algorithm and other algorithms used to study these models numerically  and the detail of the  simulations can be found in appendices $A$ and $B$.

\subsection{Bosonic Theory: Emergent Geometry and Phase Diagram}
\paragraph{Emergent Geometry}
We measure the different observables as a function of the coupling constant $\tilde{\alpha}$ for the two relevant values of $\tilde{\beta}$, i.e. $\tilde{\beta}=0,2/9$. We have verified that the bosonic Schwinger-Dyson equation holds in Monte Carlo simulations. See figure (\ref{ident}). Recall that this identity reads for the bosonic models as follows 
\begin{eqnarray}
{\rm IDE}&=&4\frac{<{\rm YM}>}{N^2}+4\frac{<{\rm YM}_0>}{N^2}+3\frac{<{\rm CS}>}{N^2}+2\frac{<{\rm RAD}>}{N^2}+\frac{4}{N^2}
\equiv 4.\label{ideB}
\end{eqnarray}
The radius which we have defined by the equation $<1/r>=<TrX_a^2>/\tilde{\alpha}^2c_2$ is shown on figure (\ref{radiusB}). For large values of $\tilde{\alpha}$ the result is consistent with the classical prediction
\begin{eqnarray}
<\frac{1}{r}>=<\frac{TrX_a^2}{\tilde{\alpha}^2c_2}>=\phi_+^2~,~\phi_+=\frac{1+\sqrt{1-4\tilde{\beta}}}{2}.
\end{eqnarray}
This means in particular that the system is in the ground state configurations
\begin{eqnarray}
X_4=0~,~X_a=\alpha\phi L_a.
\end{eqnarray}
In other words we have a fuzzy spherical geometry given by the commutation relations
\begin{eqnarray}
[X_4,X_a]=0~,~[X_a,X_b]=i\epsilon_{abc}\alpha\phi X_c.
\end{eqnarray}
We have checked these commutation relations and found them to hold quite well for sufficiently large values of $\tilde{\alpha}$. The coordinates on the sphere are defined by
\begin{eqnarray}
n_a=\frac{X_a}{\sqrt{c_2}\alpha}~,~\sum_an_a^2=\phi^2.
\end{eqnarray}
We observe that as we decrease $\tilde{\alpha}$, the radius $1/r$ jumps abruptly to $0$ then starts to increase again until it becomes infinite at $\tilde{\alpha}=0$. This is the most dramatic effect of the so-called sphere-to-matrix transition in which the sphere suddenly expands  and evaporates at the transition points then it starts shrinking to zero rapidly as we lower the coupling further.  

This is the interpretation advocated in \cite{DelgadilloBlando:2008vi,DelgadilloBlando:2007vx,O'Connor:2006wv} for a similar phenomena observed in the case of three dimensional bosonic models. As far as we know this phenomena was observed in Monte Carlo simulation first in \cite{Azuma:2004zq} and it was found in analytical work on perturbative three dimensional bosonic models in \cite{CastroVillarreal:2004vh} and then in \cite{Azuma:2004ie}.

The transitions for the bosonic mass deformed model with $\tilde{\beta}=2/9$ and the bosonic cohomological model with $\tilde{\beta}=0$ are observed to occur at the following estimated values

\begin{eqnarray}
\tilde{\alpha}_*=4.9\pm 0.1~,~\tilde{\beta}=2/9.
\end{eqnarray}
  \begin{eqnarray}
\tilde{\alpha}_*=2.55\pm 0.1~,~\tilde{\beta}=0.
\end{eqnarray}
The fuzzy sphere phase corresponds to the region $\tilde{\alpha}>\tilde{\alpha}_*$ whereas the matrix phase corresponds to the region  $\tilde{\alpha}<\tilde{\alpha}_*$. In other words the sphere becomes more stable as we make $\tilde{\beta}$ smaller (see below). 

The order of the sphere-to-matrix transition is very difficult to determine. Since the ground state configurations are $X_4=0$ and $X_a=\alpha\phi L_a$, the theoretical analysis based on the effective potential of the three dimensional model done in \cite{DelgadilloBlando:2008vi,DelgadilloBlando:2007vx} should also hold here largely unchanged (see below). As a consequence we will only summarize here the main points omitting much technical details.

The specific heat $C_v=(<S^2>-<S>^2)/(N^2-1)$ shown on figure (\ref{cvB}) diverges from the side of the fuzzy sphere with a critical exponent equal $1/2$. It is equal to $3$ in this phase where $1/2$ is due to the $2$ dimensional $U(1)$ gauge field on the sphere, $1/2$ is due to the normal scalar field on the sphere and $1/2$ is due to the scalar field $X_4$.  This critical behavior is typical of a second order transition. In the matrix phase the specific heat is a constant right up to the transition point and it is equal $1$ where each matrix contributes $1/4$. There is no divergence from this side and the critical exponent is $0$. In other words the behavior above and below the critical coupling are different, which is quite unusual, but still from the specific heat we qualify this transition as second order.  

The expectation values of the Yang-Mills action and the Myers (or Chern-Simons) action are shown on (\ref{YMCSB}). The expectation values of the total bosonic action for the two cases  $\tilde{\beta}=2/9$ and $\tilde{\beta}=0$ are shown on figure (\ref{actionB}). From these observables we observe a discontinuity at the transition point. Thus the sphere-to-matrix transition is associated with  a latent heat equal to $\Delta <S>=<S>_{\rm matrix}-<S>_{\rm sphere}$ which is typical of a first order phase transition. It is straightforward to estimate the value of this latent heat. The latent heat is released by going from the matrix phase to the fuzzy sphere phase  for $\tilde{\beta}=0$ whereas for $\tilde{\beta}=2/9$ the latent heat is released by going in the other direction from the fuzzy sphere phase to the matrix phase.

As we will see from the discussion of the eigenvalues distributions  the matrices $X_{\mu}$ in the matrix phase are commuting matrices centered around $0$.
\paragraph{Phase Diagram}
The last point we would like to address  within the context of the bosonic theory is the construction of the phase diagram in the plane $\tilde{\alpha}-\tilde{\beta}$. We have already measured two points of this phase diagram which correspond to the two values  $\tilde{\beta}=2/9$ and $\tilde{\beta}=0$.  In order to map the phase boundary between the fuzzy sphere and the matrix phase we choose other values of $\tilde{\beta}$ and  measure for each one of them the critical value of $\tilde{\alpha}$ from the discontinuity in the radius $1/r$. The result is shown on figure (\ref{phaseB}).

The effective potential of the $4$ dimensional bosonic Yang-Mills matrix model in the Feynman-'t Hooft background field gauge in the  ground state configurations $X_4=0$ and $X_a=\alpha\phi L_a$ can be calculated using the method of \cite{CastroVillarreal:2004vh}. We find
  \begin{eqnarray}
\frac{V_{\rm eff}}{2c_2}=\tilde{\alpha}^4\bigg[\frac{\phi^4}{4}-\frac{\phi^3}{3}+\tilde{\beta}\frac{\phi^2}{2}\bigg]+2\log\phi^2.
\end{eqnarray}
The difference with the three dimensional bosonic Yang-Mills matrix model lies in the factor of $2$ multiplying the logarithm. The critical line can then be obtained following the method of  \cite{DelgadilloBlando:2008vi}. We get
 \begin{eqnarray}
\phi_*=\frac{3}{8}(1+\sqrt{1-\frac{32\tilde{\beta}}{9}})~,~\tilde{\alpha}_*^4=\frac{16}{\phi_*^2(\phi_*^2-2\tilde{\beta})}.
\end{eqnarray}
This is the fit used on  figure (\ref{phaseB}) with very reasonable agreement with the Monte Carlo data.
\subsection{Dynamical Fermions: Impact of Supersymmetry}

In this section we will discuss the effect of the fermion determinant. First we note that simulations with fermions are much more harder than pure bosonic simulations. The main source of complication is the evaluation of the determinant which is highly non trivial. Thermalization is very difficult and as a consequence taking the limit of large $N$ is not so easy even with the use of the Hybrid Monte Carlo algorithm. In the bosonic case we could go as large as $N=100$ with very decent number of statistics although in this article we have only reported data with $N$ up to $N=16$. In the fermionic case we will report data with $N$ up to $N=10$. 

The first thing we have checked is the Schwinger-Dyson identity (\ref{ide}). The Monte Carlo data are shown on figure (\ref{ident}). The fermionic data agrees well with the prediction $6$ whereas the bosonic data agrees well with the prediction $4$. Note that $6=4+2$ where $4$ is the number of bosonic matrices and $2$ is the number of fermionic matrices.

The most important order parameter with direct significance to the underlying geometry is the radius $1/r$. See figures (\ref{radiusF1}),  (\ref{radiusF2}) and (\ref{radiusF3}). We observe that the transition sphere-to-matrix observed in the bosonic theory disappeared completely. Again it seems here that there is no major difference between the two models with $\tilde{\beta}=2/9$ and $\tilde{\beta}=0$. It is clear from the structure of the action that the theory with $\tilde{\alpha}=0$ can not support the fuzzy sphere geometry and thus a phase with commuting matrices must still exist. However, the transition to the phase of commuting matrices starting from the fuzzy sphere phase seems to be a crossover transition not the second/first order behavior observer in the bosonic model. This seems to be confirmed by the behavior of the specific heat, the Yang-Mills and Myers actions and the total action shown on figures  (\ref{cvF}), (\ref{YMCSF1}-\ref{YMCSF2}) and (\ref{actionF}) respectively. The jump and critical behavior in the specific heat and the discontinuity in the various actions disappeared.

We have to note here that the observable $<Tr X_a^2>/N$ diverges in the supersymmetric theory with $\tilde{\alpha}=\tilde{\beta}=0$ \cite{Krauth:1999qw}. For the mass deformed theory we have $\tilde{\beta}=2/9$ and thus the observable $<Tr X_a^2>/N$ always exists. We observe on the second graph of figure (\ref{radiusF1}) that $<Tr X_a^2>/N$ increases as we decrease $\tilde{\alpha}$ towards $0$ which is consistent with the fact that it will diverge in the limit $\tilde{\alpha}\longrightarrow 0$. Qualitatively the same phenomena is observed for $\tilde{\beta}=0$ on the second graph of figure (\ref{radiusF2}) with more erratic behavior as we decrease  $\tilde{\alpha}$ towards $0$. However in this case we can not infer that $<Tr X_a^2>/N$ exists for all $\tilde{\alpha}$ since $\tilde{\beta}=0$ although it looks that it does from the data. From this perspective the mass deformed model is better than the cohomologically deformed model.

We have not succeeded in determining precisely the value at which the crossover transition occurs  but it seems that it depends on $N$ in such a way that it is pushed to smaller values of $\tilde{\alpha}$ as we increase $N$. From figure (\ref{radiusF3}) we can see that the crossover transition for  $\tilde{\beta}=2/9$ occurs at $\tilde{\alpha}=3.13,2.63,2.38$ and $2.13$ for $N=4,6,8$ and $N=10$ respectively. A simple fit yields the result
\begin{eqnarray}
\tilde{\alpha}_*^4=\frac{61.13}{N^{2.38}}~,~\tilde{\beta}=2/9.
\end{eqnarray}
The conjecture that the crossover transition occurs at arbitrarily small values of $\tilde{\alpha}$ in the large $N$ limit is one of the main results of this article. In this way the fuzzy sphere is truly stable in the supersymmetric theory and does not decay. In any case we are certain that the fuzzy sphere in the supersymmetric theory is more stable compared to the bosonic theory and the crossover transition to the matrix phase is much slower. This conclusion is similar to that of \cite{Anagnostopoulos:2005cy}.

\subsection{Eigenvalues Distributions} 
\paragraph{Bosonic Theory}
A powerful set of order parameters is given by the eigenvalues distributions of the two matrices $X_3$ and $X_4$. See (\ref{distrB}). The eigenvalues distribution of the matrix $X_4$ is qualitatively the same for all values of the coupling constants $\tilde{\alpha}$.  However, the eigenvalues distribution of the matrix $X_3$ suffers a major change as we go across the transition point. In the fuzzy sphere region the eigenvalues distribution of $X_3$ is given by an $N-$cut distribution corresponding to the $N$ eigenvalues $-(N-1)/2,...,(N-1)/2$ whereas in the matrix phase  the  eigenvalues distribution of $X_3$ is identical to the  eigenvalues distribution of $X_4$.

The eigenvalues distribution $\rho_4(x_4)$ of the matrix $X_4$ is always centered around $0$. In the fuzzy sphere phase $\rho_4(x_4)$ depends on the coupling constant  $\tilde{\alpha}$. In the matrix phase below the critical value the eigenvalues distribution $\rho_4(x_4)$ does not depend on $\tilde{\alpha}$ and coincides with the eigenvalues distribution of the non deformed model with  $\tilde{\alpha}=0$. In this region the eigenvalues distributions of the matrices $X_3$ and $X_4$ are identical.

 Motivated by the work \cite{Berenstein:2008eg,Hotta:1998en} it was conjectured \cite{denjoeprivate} that the joint eigenvalues distribution of $d$ matrices $X_1$, $X_2$,...$X_d$ with dynamics given by a reduced Yang-Mills action should be uniform inside a solid ball of some radius $R$. We have already checked that this conjecture works in three dimensions \cite{Ydri:2010kg}. We will check now that this conjecture holds also true in four dimensions. Let $\rho(x_1,x_2,x_3,x_4)$ be the joint eigenvalues distribution of the $4$ matrices $X_1$, $X_2$, $X_3$ and $X_4$. We assume that $\rho(x_1,x_2,x_3,x_4)$ is uniform inside a four dimensional ball of radius $R$. The normalization condition gives $\rho(x_1,x_2,x_3,x_4)=1/V_4=2/\pi^2 R^4$. We want to compute the eigenvalues distribution of a single matrix, say $X_4$, which is induced by integrating out the other three matrices. We compute 

 \begin{eqnarray}
\bigg[\int_{-R}^R dx_4\int_{-R}^R dx_3\int_{-R}^R dx_2\int_{-R}^R dx_1\bigg]_{x_1^2+x_2^2+x_3^2+x_4^2\leq R^2} &=&\nonumber\\
\int_{-R}^R dx_4\bigg[\int_{-R}^R dx_3\int_{-R}^R dx_2\int_{-R}^R dx_1\bigg]_{x_1^2+x_2^2+x_3^2\leq R^2-x_4^2}&=&\nonumber\\
 \int_{-R}^R dx_4\bigg[\int r^2dr d\Omega_3\bigg]_{r^2\leq R^2-x_4^2}&=&\nonumber\\
 \frac{4\pi}{3}\int_{-R}^R dx_4(R^2-x_4^2)^{\frac{3}{2}}.
\end{eqnarray}
We obtain therefore the eigenvalues distribution
\begin{eqnarray}
\rho_4(x_4)=\frac{8}{3\pi R^4}(R^2-x_4^2)^{\frac{3}{2}}.\label{4ball}
\end{eqnarray}
This is precisely the fit shown on figure  (\ref{distrB}) with a measured value of $R$ for $\tilde{\beta}=0$ and $\tilde{\alpha}=0$ given by
\begin{eqnarray}
R_0=1.826\pm 0.004.
\end{eqnarray}
The above eigenvalues distribution works better for the theory with $\tilde{\beta}=2/9$ as shown on figure (\ref{distrF2}) with a similar measured value of $R$ given for $\tilde{\alpha}=0.25$ by
\begin{eqnarray}
R_{2/9}=1.815\pm 0.008.
\end{eqnarray}
We have found that these two measured values are almost the same throughout the matrix phase.

We emphasize that $\rho_4(x_4)$ is the eigenvalues distribution of the matrix $X_4$ not only in the matrix phase but also in the fuzzy sphere phase with a value of $R$ which depends on the coupling constant $\tilde{\alpha}$. We also emphasize that $\rho_4(x_3)$ is the eigenvalues distribution of the matrix $X_3$ in the matrix phase for the bosonic theory for both values $\tilde{\beta}=0$ and $\tilde{\beta}=2/9$.

Another non trivial check for this important conjecture is the theoretical prediction of the radius
  \begin{eqnarray}
<\frac{1}{N}Tr X_a^2>&=&3<\frac{1}{N}Tr X_3^2>\nonumber\\
&=&3\int_{-R}^{R}dx_4 \rho_4(x_4) x_4^2\nonumber\\
&=&\frac{R^2}{2}.
\end{eqnarray}
This means that in the matrix phase the order parameter $<\frac{1}{N}Tr X_a^2>$, which is related to the radius, is  constant. Indeed, this is what we see on figure (\ref{radiusB}) with a good agreement between  the Monte Carlo measurement and the theoretical prediction. The observed value of $<\frac{1}{N}Tr X_a^2>$ is slightly below the theoretical  prediction for $\tilde{\beta}=2/9$ throughout.  This is not the case for $\tilde{\beta}=0$ where the Monte Carlo measurement starts slightly below $R^2/2$ and then rises above it as we approach the transition to the fuzzy sphere. 

This can  potentially be a serious difference between the tow cases $\tilde{\beta}=2/9$ and $\tilde{\beta}=0$. The transition to the sphere in the case of $\tilde{\beta}=2/9$ is in the form of an abrupt jump but in the case of $\tilde{\beta}=0$ there is a slow rise in the matrix phase as we increase $\tilde{\alpha}$  before the actual jump.

The main conclusion of these successful measurements is the fact that the matrices $X_{\mu}$ in the matrix phase are commuting and thus they are diagonalizable with a joint eigenvalues distribution which is uniform inside a ball of dimension $R$.

\paragraph{Supersymmetric Theory} In the supersymmetric case we found it much easier to compute eigenvalues distributions with the value $\tilde{\beta}=2/9$ and thus we will only consider here the mass deformed model. A sample of the eigenvalues distributions of the mass deformed model is shown on figures (\ref{distrF1}) and (\ref{distrF2}). 

Again it is observed that the eigenvalues distribution of $X_3$ in the fuzzy sphere phase is given by an $N-$cut distribution corresponding to the $N$ eigenvalues $-(N-1)/2,...,(N-1)/2$ whereas in the matrix phase  the  eigenvalues distributions of $X_3$ is given by $\rho_4(x_3)$ with a much larger value of $R$ given for $\tilde{\alpha}=0.25$ by 
\begin{eqnarray}
R=2.851\pm 0.009. 
\end{eqnarray}
The eigenvalues distribution of $X_4$ is always centered around $0$ given by $\rho_4(x_4)$ with a value of $R$ which depends on the coupling constant $\tilde{\alpha}$. This  eigenvalues distribution coincides with the eigenvalues distribution of $X_3$ deep inside the matrix phase below $\tilde{\alpha}=0.25$.

The eigenvalues distribution $\rho_4$ is therefore universal in the sense that it describes the behavior of the eigenvalues of the matrix $X_4$ for all values of $\tilde{\alpha}$ and all values of $\tilde{\beta}$ and the behavior of the eigenvalues of the matrix $X_3$ in the matrix phase for all values of $\tilde{\beta}$. We note here the difference between this eigenvalues distribution $\rho_4$  and the eigenvalues distribution of the supersymmetric model with $\tilde{\alpha}=\tilde{\beta}=0$ \cite{Krauth:1999qw}. In the latter case the distribution extends from $-\infty $ to $+\infty $ and goes as $1/x^3$ for large eigenvalues. It is not clear to us at this stage how the two distributions relate to each other.

The point at which the eigenvalues distributions of $X_3$ and $X_4$ coincide may be taken as the measure for the crossover transition point and thus for $N=10$ this occurs at a point between $\tilde{\alpha}=1$ and $\tilde{\alpha}=0.25$.
\subsection{Remarks: $D=3$ Yang-Mills Matrix Models and Scalar Fluctuations}
The $D=3$ Yang-Mills matrix models we can immediately consider here can be obtained from the above $D=4$ models by simply setting the fourth matrix $X_4$ to zero.  This is different from the IKKT supersymmetric Yang-Mills matrix model in $D=3$ by the fact that it involves a determinant instead of a Pffafian and  as a consequence the path integrals of these theories are convergent. 

The physics of these $D=3$ models is identical to the physics of the $D=4$ models in the sense that there is a first/second order phase transition from a background geometry (the fuzzy sphere) to commuting matrices which in the presence of dynamical fermions is turned into a slow crossover transition. The most important difference is the natural expectation that the eigenvalues distributions of the matrices $X_a$ in the matrix phase  must be distributed according to the formula

\begin{eqnarray}
\rho_3(x_3)=\frac{3}{4R^3}(R^2-x_3^2).\label{3ball}
\end{eqnarray}
By analogy with the $D=4$ formula (\ref{4ball}) this distribution can be derived from the assumption that the joint eigenvalues distribution of the $3$ matrices $X_1$, $X_2$, $X_3$ is uniform inside a three dimensional ball of radius $R$. In the next section we will also derive this distribution for the $D=4$ Yang-Mills matrix model with a particular choice of the parameters with $\beta_4\neq 0$ (see (\ref{model0})). 

Monte Carlo measurment of the radius gives the value

\begin{eqnarray}
R\simeq 2.
\end{eqnarray}
As shown on the first graph of the figure (\ref{D3D4}) a sample of the data for the $N=10$ three-dimensional bosonic Yang-Mills matrix model with $\tilde{\beta}=2/9$ and $\tilde{\alpha}=0.5$ is shown. Clearly it can be fit nicely to (\ref{3ball}) with $R=2$. In performing the fitting in three dimensions we have to cut the tails in order to get a sensible answer.  As it turns out the same three dimensional data can also be fit to the four dimensional  prediction (\ref{4ball}).  We also show the data for the $N=10$ four-dimensional bosonic Yang-Mills matrix model with $\tilde{\beta}=2/9$ and $\tilde{\alpha}=0.5$ for comparison.

In the Monte Carlo data of the $D=3$ Yang-Mills matrix models reported in this article it was not possible to resolve the ambiguity between the two fits (\ref{4ball}) and (\ref{3ball}). However high precision runs performed in \cite{rodrigo} seems to indicate that indeed the three dimensional prediction  (\ref{3ball}) is the correct behavior for the eigenvalues distributions of the $D=3$ Yang-Mills matrix models.

The second remark we would like to discuss in this section concerns the dependence on $N$ and $\tilde{\alpha}$ of the eigenvalues distributions  $\rho_4$ and $\rho_3$ given in (\ref{4ball}) and (\ref{3ball}) respectively. As shown on the second graph of the figure (\ref{D3D4}) the distributions $\rho_4$ and $\rho_3$ are independent of $\tilde{\alpha}$. Similarly we can show that these distributions are independent of $N$.

The third remark concerns the eigenvalues distributions of the normal scalar field in the fuzzy sphere phase which is define by $\phi=(X_a^2-\phi^2c_2)/(2\phi\sqrt{c_2})$ \cite{CastroVillarreal:2004vh}. This is shown on figure (\ref{scalar}) for both the $D=3$ and $D=4$ models. The behavior in both cases is the same although we have to note that the effective values $c_{2,\rm eff}$ in $D=3$ and $D=4$ are slightly different. This is quite natural as the three dimensional model is more stable in the sense that it has a lower critical value $\tilde{\alpha}_*$. The central observation here is that we can nicely fit these eigenvalues distributions to the eigenvalues distribution  $\rho_4$ given in (\ref{4ball}) as shown on figure (\ref{scalar}).

\section{Elements of the Solution of the $D=3$ Theory}
\subsection{Localization: Reduction to $D=3$ Chern-Simons Theory}
We start by summarizing our main results of section $3$. The action of the theory of interest consists of two terms $S_{\rm cohom}$ and $\hat{S}$, viz $S_{\rm def}=S_{\rm cohom}+\hat{S}$ \footnote{In fact the action that should be considered is $2S_{\rm def}/g^2$.}. The first piece $S_{\rm cohom}$ is the standard ${\cal N}=1$ Yang-Mills matrix theory in $D=4$ which has $4$ supercharges whereas the second piece $\hat{S}$ is the cohomological deformation which preserves two of the supercharges. The parameters of the deformation were fixed so that the action (after the integration of the BRST field $H$) corresponds to a generalization of the mass deformed action considered in section $2$ with  mass terms for the bosonic and fermionic matrices and a Myers term. Effectively the action   $S_{\rm def}$ will be the sum of two actions $S_{\rm cohom}$ and $\Delta{S}_{\rm cohom}$  given by

\begin{eqnarray}
S_{\rm cohom}
&=&-\frac{1}{8}Tr[X_{\mu},X_{\nu}]^2-\frac{1}{2}Tr{\theta}^+\bigg(i[X_4,..]+{\sigma}_a[X_a,..]\bigg)\theta+Tr B^2.\label{loc1}
\end{eqnarray}
\begin{eqnarray}
\Delta{S}_{\rm cohom}
&=&\frac{\kappa_1}{2}Tr\theta^+\theta+\frac{1}{2}\epsilon\kappa_1Tr X_a^2+\frac{1}{4}\kappa_1(2\epsilon-\kappa_1)TrX_4^2-\frac{i}{6}(4\epsilon+\kappa_1)\epsilon_{abc}TrX_aX_bX_c.\label{loc2}\nonumber\\
\end{eqnarray}
The first supercharge of this cohomologically deformed theory corresponds to the supersymmetry transformations
\begin{eqnarray}
d_{\rm def} X_i=\chi_i.
\end{eqnarray}
\begin{eqnarray}
d_{\rm def}\phi=0~,~d_{\rm def}\bar{\phi}=-\eta_2.
\end{eqnarray}
\begin{eqnarray}
  d_{\rm def}H&=&[\phi,\eta_1]+i(\kappa_2+\kappa_1)\eta_2.
\end{eqnarray}
\begin{eqnarray}
  d_{\rm def} \eta_1=H+(-i(4\epsilon-\kappa_1)\phi+i(\kappa_2+\kappa_1)\bar{\phi})~,~d_{\rm def}\eta_2=[\bar{\phi},\phi].
\end{eqnarray}
\begin{eqnarray}
  d_{\rm def}\chi_i&=&[\phi,X_i]+i\epsilon \epsilon_{ij}X_j.
\end{eqnarray}
The second supercharge can be obtained by an appropriate permutation of the spinors $\eta_i$ and $\chi_i$ \cite{Austing:2001ib}. 

We observe that we have three parameters $\kappa_1$, $\kappa_2$ and $\epsilon$ where $\kappa_2$ does not appear in the action. Thus the path integral does not depend on $\kappa_2$. We note that  in  \cite{Austing:2001ib} a different cohomological deformation was considered for which it was pssible to show explicitly that the path integral is independent of $\kappa_2$. In summary we have then two deformation parameters $\kappa_1$ and $\epsilon$ or equivalently $\alpha$ and $\zeta_0$ defined in equation (\ref{alpha}). For our purposes $\alpha$ will play the role of the gauge coupling constant while $\zeta_0$ is the actual parameter of the cohomological/mass deformation.

The theory with $\kappa_2=0$ will yield the same action (\ref{loc1})+(\ref{loc2}) with the supersymmetry transformations
\begin{eqnarray}
d_{\rm def} X_i=\chi_i.\label{tra1}
\end{eqnarray}
\begin{eqnarray}
d_{\rm def}\phi=0~,~d_{\rm def}\bar{\phi}=-\eta_2.\label{tra2}
\end{eqnarray}
\begin{eqnarray}
  d_{\rm def}H&=&[\phi,\eta_1]+i\kappa_1\eta_2.\label{tra3}
\end{eqnarray}
\begin{eqnarray}
  d_{\rm def} \eta_1=H+(-i(4\epsilon-\kappa_1)\phi+i\kappa_1\bar{\phi})~,~d_{\rm def}\eta_2=[\bar{\phi},\phi].\label{tra4}
\end{eqnarray}
\begin{eqnarray}
  d_{\rm def}\chi_i&=&[\phi,X_i]+i\epsilon \epsilon_{ij}X_j.\label{tra4}
\end{eqnarray}
We note that the action for the value $\kappa_1=2\epsilon$ corresponds precisely to the mass deformed action considered in section $2$ and as a consequence the above supersymmetry transformations (\ref{tra1})-(\ref{tra4}) will correspond to one of the mass deformed supercharges.  

The path integral we wish to evaluate is 
\begin{eqnarray}
Z&=&\int dX_{\mu}~dB~d{\theta}^+~d\theta~\exp(-S_{\rm cohom}-\Delta{S}_{\rm cohom}).
\end{eqnarray}
This can be obtained from the path integral (with $S_{\rm def}[\kappa_2=0]=S_{\rm cohom}+\hat{S}[\kappa_2=0]$)
\begin{eqnarray}
Z
&=&\int dX_1~dX_2~d\bar{\phi}~d\phi~dH~d\chi_1~d\chi_2~d\eta_1~d\eta_2~\exp(-S_{\rm def}[\kappa_2=0]).\label{path}
\end{eqnarray}
In this equation the actions  $S_{\rm cohom}$ and $\hat{S}$ must be expressed in terms of the BRST fields $H$, $\phi$, $\bar{\phi}$, $X_i$, $\eta_i$ and $\chi_i$, viz
\begin{eqnarray}
S_{\rm cohom}
&=&Tr\big(H^2+H[X_1,X_2]+[X_i,\phi][X_i,\bar{\phi}]+[\phi,\bar{\phi}]^2-\eta_i[\phi,\eta_i]-\chi_i[\bar{\phi},\chi_i]\nonumber\\
&-&\eta_1\epsilon^{ij}[{\chi}_i,X_j]+\eta_2[\chi_i,X_i]\big).
\end{eqnarray}
\begin{eqnarray}
\hat{S}[\kappa_2=0]
&=&i\kappa_1Tr(\chi_1\chi_2-\eta_1\eta_2)-4i\epsilon Tr\phi[X_1,X_2]+i(\kappa_1+2\epsilon)Tr\bar{\phi}[X_1,X_2]\nonumber\\
&+&i\kappa_1Tr\bar{\phi}H-i(4\epsilon-\kappa_1)Tr\phi H-2\epsilon(2\epsilon-\kappa_1)Tr\phi^2+\frac{1}{2}\epsilon\kappa_1TrX_i^2.
\end{eqnarray}
To perform the above path integral (\ref{path}) we will now add the terms proportional to  $\kappa_2$ and since the resulting action remains independent of $\kappa_2$ we will take the limit $\kappa_2\longrightarrow\infty$ in which the path integral localizes. We have
\begin{eqnarray}
Z&=&\int dX_{\mu}~dB~d{\theta}^+~d\theta~\exp(-S_{\rm cohom}-\Delta{S}_{\rm cohom})\nonumber\\
&=&\int dX_1~dX_2~d\bar{\phi}~d\phi~dH~d\chi_1~d\chi_2~d\eta_1~d\eta_2~\exp(-S_{\rm def}[\kappa_2=0]-\kappa_2\Delta\hat{S}).\label{path1}
\end{eqnarray}
\begin{eqnarray}
\Delta\hat{S}&=&iTr\eta_1\eta_2+iTr\bar{\phi}H+(4\epsilon-\kappa_1)Tr\bar{\phi}\phi-\kappa_1Tr\bar{\phi}^2.
\end{eqnarray}
In the limit $\kappa_2\longrightarrow\infty$ we are able to integrate the BRST quartet $\eta_1$, $\eta_2$, $H$ and $\bar{\phi}$. The fermionic part of the action which depends on $\kappa_2$ is
\begin{eqnarray}
i\kappa_2Tr\eta_1\eta_2.
\end{eqnarray}
Thus by using the saddle point method we obtain
\begin{eqnarray}
\int d\eta_1~d\eta_2~ f(\eta_1,\eta_2)~e^{-i\kappa_2Tr\eta_1\eta_2}=\kappa_2^{N^2-1}i^{N^2-1}Nf(0,0).
\end{eqnarray}
The actions $S_{\rm cohom}$ and $\hat{S}[\kappa_2=0]$ reduce to
\begin{eqnarray}
S_{\rm cohom}
&=&Tr\bigg(H^2+H[X_1,X_2]+[X_i,\phi][X_i,\bar{\phi}]+[\phi,\bar{\phi}]^2-\chi_i[\bar{\phi},\chi_i]\bigg).
\end{eqnarray} 
\begin{eqnarray}
\hat{S}[\kappa_2=0]
&=&i\kappa_1Tr\chi_1\chi_2-4i\epsilon Tr\phi[X_1,X_2]+i(\kappa_1+2\epsilon)Tr\bar{\phi}[X_1,X_2]\nonumber\\
&+&i\kappa_1Tr\bar{\phi}H-i(4\epsilon-\kappa_1)Tr\phi H-2\epsilon(2\epsilon-\kappa_1)Tr\phi^2+\frac{1}{2}\epsilon\kappa_1TrX_i^2.
\end{eqnarray}
Next we do the path integral over $H$ then $X_3$ in that order using again the saddle point method. The relevant terms are the bosonic contributions which are proportional to $\kappa_2$. They are given by
\begin{eqnarray}
(4\epsilon-\kappa_1)\kappa_2Tr\bar{\phi}{\phi}-\kappa_1\kappa_2Tr\bar{\phi}^2+i\kappa_2Tr\bar{\phi}H.
\end{eqnarray}
For $\epsilon<0$ we can verify that the integral over $X_3$ is exponentially damped and therefore we can shift $X_3$ appropriately. The resulting integral over $H$ turns out also to be damped exponentially. Explicitly we have
\begin{eqnarray}
(4\epsilon-\kappa_1)\kappa_2Tr\bar{\phi}{\phi}-\kappa_1\kappa_2Tr\bar{\phi}^2+i\kappa_2Tr\bar{\phi}H&=&
-\epsilon\kappa_2Tr\big(X_3+\frac{i}{4\epsilon}(H-\kappa_1X_4)\big)^2\nonumber\\
&-&\frac{\kappa_2}{16\epsilon}Tr(H+(4\epsilon-\kappa_1)X_4)^2+\frac{\kappa_2}{2}(4\epsilon-\kappa_1)TrX_4^2.\nonumber\\
\end{eqnarray}
In the limit $\kappa_2\longrightarrow \infty$ the hermitian matrix $H$ will be localized around $-(4\epsilon-\kappa_1)X_4=i(4\epsilon-\kappa_1)(\phi+\bar{\phi})$.  The above equation reduces to
\begin{eqnarray}
(4\epsilon-\kappa_1)\kappa_2Tr\bar{\phi}{\phi}-\kappa_1\kappa_2Tr\bar{\phi}^2+i\kappa_2Tr\bar{\phi}H&=&
-\epsilon\kappa_2Tr(X_3-iX_4)^2+\frac{\kappa_2}{2}(4\epsilon-\kappa_1)TrX_4^2.\nonumber\\\label{5.25}
\end{eqnarray}
Equivalently
\begin{eqnarray}
(4\epsilon-\kappa_1)\kappa_2Tr\bar{\phi}{\phi}-\kappa_1\kappa_2Tr\bar{\phi}^2+i\kappa_2Tr\bar{\phi}H&=&
-4\epsilon\kappa_2Tr\bar{\phi}^2-\frac{\kappa_2}{2}(4\epsilon-\kappa_1)Tr(\phi+\bar{\phi})^2.\nonumber\\
\end{eqnarray}
We can now shift the integral over $X_3$ as $X_3\longrightarrow \bar{\phi}=-\frac{1}{2}(X_3-iX_4)$. Clearly we can now assume that   $\bar{\phi}$ is hermitian. For consistency we will also shift the integral over $X_4$ as $X_4\longrightarrow \phi=iX_4-\bar{\phi}$. In the limit $\kappa_2\longrightarrow \infty$ the hermitian matrix $\bar{\phi}$ is localized around $0$. The matrix $\phi$ is then seen to be antihermitian identified with $iX_4$. 

We get
thus by using the saddle point method the result
\begin{eqnarray}
\int d\bar{\phi}~dH~ f(\bar{\phi},H)~e^{(4\epsilon-\kappa_1)\kappa_2Tr\bar{\phi}{\phi}-\kappa_1\kappa_2Tr\bar{\phi}^2+i\kappa_2Tr\bar{\phi}H}&=&\nonumber\\
\kappa_2^{1-N^2}(2\pi)^{N^2-1}\frac{1}{N}f(0,i(4\epsilon-\kappa_1)\phi)e^{\frac{\kappa_2}{2}(4\epsilon-\kappa_1)Tr\phi^2}.
\end{eqnarray}
 The $\kappa_2$ dependence cancels completely if we choose $4\epsilon=\kappa_1$. The actions $S_{\rm cohom}$ and $\hat{S}[\kappa_2=0]$ reduce to
\begin{eqnarray}
S_{\rm cohom}
&=&Tr\bigg(-(4\epsilon-\kappa_1)^2\phi^2+i(4\epsilon-\kappa_1)\phi[X_1,X_2]\bigg).
\end{eqnarray} 
\begin{eqnarray}
\hat{S}[\kappa_2=0] &=&i\kappa_1Tr\chi_1\chi_2-4i\epsilon Tr\phi[X_1,X_2]+(4\epsilon-\kappa_1)^2Tr\phi^2+2\epsilon(\kappa_1-2\epsilon)Tr\phi^2\nonumber\\
&+&\frac{1}{2}\epsilon\kappa_1TrX_i^2.
\end{eqnarray}
The integration over the last fermionic degrees of freedom $\chi_1$ and $\chi_2$ is now trivial since they are free degrees of freedom decoupled from everything else. We end up with the model
\begin{eqnarray}
Z&=&\int d{\phi}~dX_1~dX_2 \exp\bigg(i\kappa_1Tr\phi[X_1,X_2]-2\epsilon(\kappa_1-2\epsilon)Tr\phi^2-\frac{1}{2}\epsilon\kappa_1TrX_i^2\nonumber\\
&+&\frac{\kappa_2}{2}(4\epsilon-\kappa_1)Tr\phi^2\bigg).
\end{eqnarray}
This is essentially the path integral of two-dimensional gauge theory on the fuzzy sphere studied in \cite{Ishiki:2008vf}. As was shown in \cite{Ishiki:2009vr} it can also be derived from the reduction to zero dimension of Chern-Simons theory on ${\bf S}^3$ . 

We can therefore conclude that the mass deformation considered here reduces the $4$-dimensional supersymmetric theory to a bosonic $D=3$ matrix theory which when expanded about the ground state yields a noncommutative gauge theory on the fuzzy sphere.

 
\subsection{Exact Integration: Emergence of the Fuzzy Sphere}
In the previous section we have found that $\bar{\phi}=-(X_3-iX_4)/2$ is localized around $0$ which means that the most important configurations satisfy $X_3-iX_4=0$. Since $\phi=iX_4-\bar{\phi}$ we can think of $\phi$ as $i X_4$, i.e. $\phi$ is antihermitian, and as a consequence the cubic term is not real and the mass term of $\phi$ is of the wrong sign.

Formally we can also think of $\phi$ as a hermitian matrix which can be identified with $X_3$. The cubic term in this case is real which is precisely a Chern-Simon term and the mass term of $\phi$ is of the correct sign. This can be seen by rewriting in the limit $\kappa_2\longrightarrow\infty$ the path integration over $X_3$ and $X_4$ as (see equation (\ref{5.25}))
\begin{eqnarray}
\int dX_3\int dX_4 \exp\bigg(\epsilon\kappa_2Tr(X_3-iX_4)^2-\frac{\kappa_2}{2}(4\epsilon-\kappa_1)TrX_4^2\bigg)(...)&\sim &\nonumber\\
\int dX_3\int dX_4 \delta(X_3-iX_4)\exp\bigg(-\frac{\kappa_2}{2}(4\epsilon-\kappa_1)TrX_4^2\bigg)(...).
\end{eqnarray}
Instead of doing the integral over $X_3$ first then over $X_4$ by performing the shift $X_3\longrightarrow \bar{\phi}=-(X_3-iX_4)/2$, $X_4\longrightarrow \phi=iX_4-\bar{\phi}$ we will do the integral over $X_4$ first then over $X_3$ by performing the shift  $X_4\longrightarrow \bar{\phi}=-(X_3-iX_4)/2$, $X_3\longrightarrow \phi=X_3+\bar{\phi}$.

In the following we will therefore assume that $\phi$ is a hermitian matrix identified with $X_3$. Furthermore recall that $\epsilon <0$ and as a consequence we will take $\kappa_1<0$ and $\kappa_1-2\epsilon<0$ in order to have  mass terms for $\phi$ and $X_i$. Since $\epsilon=-\alpha(1+6\zeta_0)/6$ and $\kappa_1-2\epsilon=6\zeta_0\alpha$ we conclude that we must have $-1/6\leq \zeta_0<0$. We have already found that we must have $0\leq\zeta_0\leq 1/12$ in order to have a stable classical theory in the direction $X_4$ and that the value $\zeta_0=0$ corresponds precisely to the mass deformed theory of interest which is constructed in section $2$. 

In the current case we have to take the value $4\epsilon=\kappa_1$, or equivalently $\zeta_0=-1/24$, which is the value at which the dependence on $\kappa_2$ drops completely. For this value the field $X_4$ will have a negative mass but the original theory remains stable in the $X_4$ as one might easily check in the fuzzy sphere region.

We introduce
\begin{eqnarray}
t=2\epsilon^3(\kappa_1-2\epsilon)=-\frac{\alpha^4}{18}\zeta_0(1+6\zeta_0)^3.
\end{eqnarray}
Furthermore since $\bar{\phi}$ is localized around $0$ we can identify $\phi$ with $X_3$. In order to make $SO(3)$ covariance manifest we perform the rescaling
\begin{eqnarray}
X_{i}\longrightarrow -2\epsilon\sqrt{\frac{\kappa_1-2\epsilon}{\kappa_1}}X_{i}~,~\phi\longrightarrow -\epsilon \phi.
\end{eqnarray}
The path integral becomes (with  $4\epsilon=\kappa_1$)
\begin{eqnarray}
Z=\int d{\phi}~dX_1~dX_2 \exp t\bigg(-2iTr\phi[X_1,X_2]-Tr\phi^2-TrX_i^2\bigg).
\end{eqnarray}
We perform now the scaling (\ref{scaling}). We get then
\begin{eqnarray}
Z=\int d{\phi}~dX_1~dX_2 \exp N t\bigg(-2iTr\phi[X_1,X_2]-Tr\phi^2-TrX_i^2\bigg).
\end{eqnarray}
\begin{eqnarray}
t
&=&-\frac{16\alpha^4}{9}\zeta_0(1+6\zeta_0)^3.
\end{eqnarray}
For $4\epsilon=\kappa_1$ we have explicitly 
\begin{eqnarray}
t
&=&\frac{\alpha^4}{32}.
\end{eqnarray}
The classical equations of motion are
\begin{eqnarray}
[X_1,X_2]=i\phi~,~[X_2,\phi]=iX_1~,~[\phi,X_1]=iX_2.
\end{eqnarray}
The solutions are given by $SU(2)$ irreducible representations, namely 

\begin{eqnarray}
X_i=L_i~,~\phi=L_3.
\end{eqnarray}
We perform now the integrals over $X_1$ and $X_2$. By integrating over $X_2$ we get
\begin{eqnarray}
Z
&=&\int d\phi~dX_1 \exp Nt\bigg(-Tr[\phi,X_1]^2-Tr\phi^2- TrX_1^2\bigg).
\end{eqnarray}
Next we diagonalize the hermitian matrix $\phi$. We have
\begin{eqnarray}
d\phi=\prod_{i=1}^Nd\phi_i \prod_{i<j}(\phi_i-\phi_j)^2.
\end{eqnarray}
We get then
\begin{eqnarray}
Z&=&\int \prod_{i=1}^Nd\phi_i 
\prod_{i<j}(\phi_i-\phi_j)^2\int dX_1 \exp\bigg(-\sum_{i,j}(X_1)_{ij}(X_1)^*_{ij}\bigg[-Nt(\phi_i-\phi_j)^2+Nt\bigg]-Nt\sum_{i}\phi_i^2\bigg).\nonumber\\
\end{eqnarray}
Integrating over $X_1$ we get

\begin{eqnarray}
Z
&=&\int \prod_{i=1}^Nd\phi_i 
\prod_{i<j}(\phi_i-\phi_j)^2 \det\bigg(-Nt(\phi_i-\phi_j)^2+Nt\bigg)^{-1}\exp\bigg(-Nt\sum_{i}\phi_i^2\bigg).\nonumber\\
\end{eqnarray}
 Let us compute the determinant for $N=3$. Clearly $t \longrightarrow 0$ when $\alpha\longrightarrow 0$. Thus we have in the limit $\alpha\longrightarrow 0$ (with $t_{ij}=-Nt(\phi_i-\phi_j)^2+Nt$)
\begin{eqnarray}
\det\bigg(-Nt(\phi_i-\phi_j)^2+Nt\bigg)=2t_{12}t_{13}t_{23}+t_{11}^3-t_{11}(t_{12}^2+t_{13}^2+t_{23}^2).
\end{eqnarray}
But $t_{11}$ is proportional to $\alpha$. In other words $t_{11}$ vanishes if $\alpha\longrightarrow 0$. In this case the detreminant is given by 
\begin{eqnarray}
\det\bigg(-Nt(\phi_i-\phi_j)^2+Nt\bigg)^{-1}\propto \prod_{i<j}\bigg(-Nt(\phi_i-\phi_j)^2+Nt\bigg)^{-1}.
\end{eqnarray}
The generalization of this result to higher $N$ is straightforward. This result is actually exact and it does not require taking the limit  $\alpha\longrightarrow 0$ \cite{Eynard:1998fn}. The path integral becomes
\begin{eqnarray}
Z
&=&\int \prod_{i=1}^Nd\phi_i 
\prod_{i<j}(\phi_i-\phi_j)^2 \prod_{i<j}\bigg(-Nt(\phi_i-\phi_j)^2+Nt\bigg)^{-1} \exp\bigg(-Nt\sum_{i}\phi_i^2\bigg).\label{patheff}\nonumber\\
\end{eqnarray}
This can also put in the form (with $t=1/g^2$)
\begin{eqnarray}
Z
&=&\int \prod_{i=1}^Nd\phi_i \prod_{i\neq j}\frac{\phi_i-\phi_j}{\phi_i-\phi_j+1}\exp\bigg(-\frac{N}{g^2} \sum_{i}\phi_i^2\bigg).
\end{eqnarray}
We will use the Cauchy formula 
\begin{eqnarray}
\sum_{\sigma\in{\cal S}_N}(-1)^{\sigma}\prod_{i=1}^N\bigg(\frac{f(\phi_i)}{\phi_i-\phi_{\sigma(i)}+1}\bigg)&=&\prod_{i=1}^Nf(\phi_i) \prod_{i\neq j}\frac{\phi_i-\phi_j}{\phi_i-\phi_j+1}.
\end{eqnarray}
It is quite illuminating to illustrate this identity for smaller values of $N$. For example we consider $N=3$. We have
\begin{eqnarray}
\sum_{\sigma\in{\cal S}_3}(-1)^{\sigma}\prod_{i=1}^3\bigg(\frac{f(\phi_i)}{\phi_i-\phi_{\sigma(i)}+1}\bigg)
&=&\bigg[\frac{1}{\phi_1-\phi_{1}+1}\frac{1}{\phi_2-\phi_{2}+1}\frac{1}{\phi_3-\phi_{3}+1}\nonumber\\
&+&\frac{1}{\phi_1-\phi_{3}+1}\frac{1}{\phi_2-\phi_{1}+1}\frac{1}{\phi_3-\phi_{2}+1}\nonumber\\
&+&\frac{1}{\phi_1-\phi_{2}+1}\frac{1}{\phi_2-\phi_{3}+1}\frac{1}{\phi_3-\phi_{1}+1}\nonumber\\
&-&\frac{1}{\phi_1-\phi_{1}+1}\frac{1}{\phi_2-\phi_{3}+1}\frac{1}{\phi_3-\phi_{2}+1}\nonumber\\
&-&\frac{1}{\phi_1-\phi_{3}+1}\frac{1}{\phi_2-\phi_{2}+1}\frac{1}{\phi_3-\phi_{1}+1}\nonumber\\
&-&\frac{1}{\phi_1-\phi_{2}+1}\frac{1}{\phi_2-\phi_{1}+1}\frac{1}{\phi_3-\phi_{3}+1}\bigg]f(\phi_1)f(\phi_2)f(\phi_3).\label{165}\nonumber\\
\end{eqnarray}
We introduce the notation $T_{ij}=\phi_i-\phi_j$. Then we have
\begin{eqnarray}
\sum_{\sigma\in{\cal S}_3}(-1)^{\sigma}\prod_{i=1}^3\bigg(\frac{f(\phi_i)}{\phi_i-\phi_{\sigma(i)}+1}\bigg)
&=&\frac{f(\phi_1)f(\phi_2)f(\phi_3)}{(T_{12}+1)(T_{21}+1)(T_{13}+1)(T_{31}+1)(T_{23}+1)(T_{32}+1)}\bigg[\nonumber\\
&-&(T_{12}^2-1)(T_{13}^2-1)(T_{23}^2-1)+(-T_{13}+1)(T_{12}+1)(T_{23}+1)\nonumber\\
&+&(-T_{12}+1)(-T_ {23}+1)(T_{13}+1)-(T_{12}^2-1)(T_{13}^2-1)\nonumber\\
&-&(T_{12}^2-1)(T_{23}^2-1)-(T_{13}^2-1)(T_{23}^2-1)\bigg]\nonumber\\
&=&\frac{f(\phi_1)f(\phi_2)f(\phi_3)}{(T_{12}+1)(T_{21}+1)(T_{13}+1)(T_{31}+1)(T_{23}+1)(T_{32}+1)}\bigg[-T_{12}^2T_{13}^2T_{23}^2\nonumber\\
&+&(T_{12}-T_{13}+T_{23})^2\bigg]\nonumber\\
&=&f(\phi_1)f(\phi_2)f(\phi_3)\prod_{i\ne j}\frac{T_{ij}}{T_{ij}+1}.
\end{eqnarray}
By using the Cauchy formula the partition function becomes
\begin{eqnarray}
Z
&=&\sum_{\sigma\in{\cal S}_N}(-1)^{\sigma}\int\prod_{i=1}^N\bigg( d\phi_i \frac{e^{-\frac{N}{g^2} \phi_i^2}}{\phi_i-\phi_{\sigma(i)}+1}\bigg).
\end{eqnarray}
From our earlier considerations it is natural to understand the integrals as contour integrals. The poles are on the real line so we regularize this partition function as follows
\begin{eqnarray}
Z_{\beta}
&=&\sum_{\sigma\in{\cal S}_N}(-1)^{\sigma}\oint\prod_{i=1}^N\bigg( d\phi_i \frac{e^{-\frac{N}{g^2} \phi_i^2+i\beta\phi_i}}{\phi_i-\phi_{\sigma(i)}+1+i\beta}\bigg)~,~\beta >0.
\end{eqnarray}
Thus we close the contours in the upper half-plane. The reason behind this way of regularization is twofold. Firstly among the two poles which may appear for each variable only one will be counted. Secondly the result obtained here will be consistent with the result for Yang-Mills quantum mechanics obtained in \cite{Hoppe:1999xg}.

 Since the eigenvalues $\phi_i$ are defined up to a permutation we must also incorporate a combinatorial factor equal $1/N!$. Furthermore the tracelessness condition $Tr\phi=0$ must be included in the form  $N\delta(\phi_1+...+\phi_N)$. In summary we get the partition function
 \begin{eqnarray}
Z_{\beta}
&=&\frac{1}{(N-1)!}\sum_{\sigma\in{\cal S}_N}(-1)^{\sigma}\oint \delta(\phi_1+...+\phi_N)\prod_{i=1}^N\bigg( d\phi_i \frac{e^{-\frac{N}{g^2} \phi_i^2+i\beta\phi_i}}{\phi_i-\phi_{\sigma(i)}+1+i\beta}\bigg).
\end{eqnarray}
 Among the $N!$ integrals there are $(N-1)!$ which are non zero. Let us check this in the case of $N=3$. The contour integrals over $\phi_1$, $\phi_2$ and $\phi_3$ of the last three terms respectively in (\ref{165})  vanish because there are no poles in these variables. The first term  in (\ref{165}) vanishes because the contour integrals over $\phi_1$, $\phi_2$ and $\phi_3$ vanish separately. The second and third term  in (\ref{165}) lead to identical contributions given by
\begin{eqnarray}
Z_{\beta}
&=&(-1)^{2}\frac{2!}{2!}\oint \delta(\phi_1+\phi_2+\phi_3) \prod_{i=1}^3\bigg( d\phi_i \frac{e^{-\frac{3}{g^2} \phi_i^2+i\beta\phi_i}}{\phi_i-\phi_{i-1}+1+i\beta}\bigg)~,~\phi_0=\phi_3.
\end{eqnarray}
In general we get
 \begin{eqnarray}
Z_{\beta}
&=&(-1)^{N-1}\frac{(N-1)!}{(N-1)!}\oint \delta(\phi_1+...+\phi_N)\prod_{i=1}^N\bigg( d\phi_i \frac{e^{-\frac{N}{g^2} \phi_i^2+i\beta\phi_i}}{\phi_i-\phi_{i-1}+1+i\beta}\bigg)~,~\phi_0=\phi_{N}.\nonumber\\
\end{eqnarray}
Next we will perform the integrals over the variables $\phi_2$,...,$\phi_{N-1}$ using the residue theorem. The two remaining integrals over $\phi_1$ and $\phi_N$ will be constrained such that $\phi_1+\phi_N=0$. To see this explicitly we try one more time a small value of $N$ say $N=4$. We have with the notation $i\gamma=1+i\beta$ the partition function
 \begin{eqnarray}
Z_{\beta}
&=&(-1)^{4-1}\oint \delta(\phi_1+\phi_2+\phi_3+\phi_4)\prod_{i=1}^4\bigg( d\phi_i \frac{e^{-\frac{4}{g^2} \phi_i^2+i\beta\phi_i}}{\phi_i-\phi_{i-1}+i\gamma}\bigg).
\end{eqnarray}
The pole in the $\phi_2-$plane is at $\phi_2=\phi_3+i\gamma$ and the pole in the $\phi_3-$plane is at $\phi_3=\phi_4+i\gamma$. By using the residue theorem we get
  \begin{eqnarray}
Z_{\beta}
&=&(-1)^{4-1}\bigg[-(-2\pi i)^2\oint d\phi_1d \phi_4\delta(\phi_1+3\phi_4+3i\gamma)~e^{-\frac{4}{g^2} \big(\phi_1^2+(\phi_4+2i\gamma)^2+(\phi_4+i\gamma)^2+\phi_4^2\big)}\nonumber\\
&\times &e^{i\beta \big(\phi_1+(\phi_4+2i\gamma)+(\phi_4+i\gamma)+\phi_4\big)} \frac{1}{\phi_1-\phi_4+i\gamma}\frac{1}{\phi_1-\phi_4-3i\gamma}\bigg].
\end{eqnarray}
In general there will be a pole in the $\phi_2-$plane  at $\phi_2=\phi_3+i\gamma$, a pole in the $\phi_3-$plane at $\phi_3=\phi_4+i\gamma$, a pole in the $\phi_4-$plane at $\phi_4=\phi_5+i\gamma$...and a pole in the $\phi_{N-1}-$plane  at $\phi_{N-1}=\phi_N+i\gamma$. By using the residue theorem we get
 \begin{eqnarray}
Z_{\beta}
&=&(-1)^{N-1}\bigg[-(-2\pi i)^{N-2}\oint d\phi_1d \phi_N\delta(\phi_1+(N-1)\phi_N+\frac{1}{2}(N-1)(N-2)i\gamma)\nonumber\\
&\times &e^{-\frac{N}{g^2} \big(\phi_1^2+(\phi_N+(N-2)i\gamma)^2+...+(\phi_N+i\gamma)^2+\phi_N^2\big)}e^{i\beta \big(\phi_1+(\phi_N+(N-2)i\gamma)+...+(\phi_N+i\gamma)+\phi_N\big)}\nonumber\\
&\times & \frac{1}{\phi_1-\phi_N+i\gamma}\frac{1}{\phi_1-\phi_N-(N-1)i\gamma}\bigg].
\end{eqnarray}
The integration over $\phi_1$ will be dominated by the pole at $\phi_1=\phi_N+(N-1)i\gamma$. The delta function becomes
 \begin{eqnarray}
\delta(\phi_1+(N-1)\phi_N+\frac{1}{2}(N-1)(N-2)i\gamma)&=&\delta\bigg(N(\phi_N+\frac{1}{2}(N-1)i\gamma)\bigg)\nonumber\\
&=&\delta\bigg(\frac{N}{2}(\phi_1+\phi_N)\bigg).
\end{eqnarray}
In other words we must have (we take the limit $\beta\longrightarrow 0$)
 \begin{eqnarray}
Z_{\beta}
&=&(-1)^{N-1}\bigg[-(-2\pi i)^{N-2}\oint d\phi_1d \phi_N\delta\bigg(\frac{N}{2}(\phi_1+\phi_N)\bigg)e^{-\frac{N}{g^2} \big(\phi_1^2+(\phi_N+(N-2))^2+...+(\phi_N+1)^2+\phi_N^2\big)}\nonumber\\
&\times & \frac{1}{\phi_1-\phi_N+1}\frac{1}{\phi_1-\phi_N-(N-1)}\bigg]\nonumber\\
&=&(-1)^{N-1}\bigg[-(-2\pi i)^{N-1}\oint d \phi_N \frac{1}{N}\delta\bigg(\phi_N+\frac{N-1}{2}\bigg)e^{-\frac{N}{g^2} \big(\phi_N^2+(\phi_N+(N-2))^2+...+(\phi_N+1)^2+\phi_N^2\big)} \frac{1}{N}\bigg]\nonumber\\
&=&(-1)^N(2\pi i)^{N-1}\frac{1}{N^2}\exp\bigg(-\frac{N}{g^2}\sum_{m=-\frac{N-1}{2}}^{m=\frac{N-1}{2}}m^2\bigg)\nonumber\\
&=&(-1)^N(2\pi i)^{N-1}\frac{1}{N^2}\exp\bigg(-\frac{N^2}{3g^2}s(s+1)\bigg)~,~s=\frac{N-1}{2}.
\end{eqnarray}
The smallest eigenvalue is $\phi_N=-(N-1)/2$ and the largest eigenvalue is $\phi_1=(N-1)/2$. We observe that $\phi_1=\phi_N+N-1$. We have in total $N=2s+1$ eigenvalues between $\phi_N$ and $\phi_1$ with a step equal $1$, viz $m=(N-1)/2,(N-3)/2,...,-(N-3)/2,-(N-1)/2$. This vacuum configuration corresponds precisely to the $SU(2)$ irreducible representation $s=(N-1)/2$. The vacuum energy defined as the logarithm of the partition function is clearly proportional to the quadratic Casimir in the irreducible representation $s=(N-1)/2$.  In summary we have found that the partition function is dominated by the integration in the vicinity of the poles $\phi_i-\phi_j+1=0$ which corresponds to the irreducible representation of $SU(2)$ of size $N$.

\subsection{Saddle-Point Method: The Matrix Phase}
\paragraph{Hermitian Case:}
We will keep assuming that the matrix $\phi$ is hermitian. The effective potential derived from the path integral (\ref{patheff}) is given by
\begin{eqnarray}
-V_{\rm eff}(\phi_i)
&=&-Nt\sum_{i}\phi_i^2+\frac{1}{2}\sum_{i\neq j}\ln (\phi_i-\phi_j)^2-\frac{1}{2}\sum_{i\neq j}\ln\bigg(-t(\phi_i-\phi_j)^2+t\bigg).\nonumber\\
\end{eqnarray}
The saddle point associated with this potential is essentially the inverted oscillator problem which is the analytic continuation of the supersymmetric model considered in   \cite{Moore:1998et}. More importantly the observation that the saddle point associated with this potential actually also corresponds to the Baxter's three-colorings problem \cite{baxter}.

As before we will  define $t=1/g^2$. The saddle-point equation reads explicitly
\begin{eqnarray}
\frac{2}{g^2}\phi_k
&=&\frac{2}{N}\sum_{i\neq k}\frac{1}{\phi_k-\phi_i}-\frac{1}{N}\sum_{j\neq k}\bigg[\frac{1}{1+\phi_k+\phi_j}-\frac{1}{1-\phi_k-\phi_j}\bigg].
\end{eqnarray}
We introduce the density of eigenvalues and the resolvent given by
\begin{eqnarray}
\rho(x)=\frac{1}{N}\sum_i\delta(x-\phi_i)~,~W(z)=\int dx \frac{\rho(x)}{z-x}.
\end{eqnarray}
The saddle-point equation becomes
\begin{eqnarray}
\frac{2}{g^2}z=2\int dy\frac{\rho(y)}{z-y}-\int dy\bigg[\frac{\rho(y)}{1+z-y}-\frac{\rho(y)}{1-z+y}\bigg].
\end{eqnarray}
In terms of the resolvent this reads
\begin{eqnarray}
\frac{2}{g^2}z=W(z+i\epsilon)+W(z-i\epsilon)+W(-1-z)+W(1-z).\label{sad2}
\end{eqnarray}
We observe that the term $\sum_j1/(1-\phi_k-\phi_j)$ is singular if one of the eigenvalues approaches $1/2$ whereas the term  $\sum_j1/(1+\phi_k+\phi_j)$ is singular if one of the eigenvalues approaches $-1/2$. We are interested in the case where the eigenvalues live on a single interval $[a,b]\subset[-1/2,1/2]$. Thus $W(z)$ is analytic everywhere in the complex plane except along the cut $[a,b]$. The above equation can be rewritten as
\begin{eqnarray}
W(z+i\epsilon)=\frac{2}{g^2}z-W(z-i\epsilon)-W(-1-z)-W(1-z)~,~z\in[a,b].
\end{eqnarray}
Thus as the complex number $z$ crosses the cut $[a,b]$ from the first sheet into the second sheet we see that the resolvent $W(z)$ becomes a linear combination of $W(z)$, $W(1-z)$ and $W(-1-z)$. In other words we have three cuts $[a,b]$, $[1-b,1-a]$ and $[-1-b,-1-a]$ in the second sheet. By crossing the cuts $[a,b]$, $[1-b,1-a]$ and $[-1-b,-1-a]$ into the third sheet we generate more cuts. Hence the domain of definition of $W(z)$ is a Riemann surface of infinite genus with an infinite number of cuts in each sheet.

By comparison the one-matrix model (we set $2z/g^2$ to $V^{'}(z)/g^2$ and $W(-1-z)+W(1-z)$ to $0$) yields a Riemann surface with one cut in each of the two possible sheets whereas the $O(n)$-matrix model (we set $2z/g^2$ to $V^{'}(z)/g^2$ and $W(-1-z)+W(1-z)$ to $n W(-z)$) yields a Riemann surface with two cuts in each of the two possible sheets.



We introduce now the notation $z_k=(-1)^k(z-k)$. The saddle-point equation (\ref{sad2}) is equivalent to the loop equation \cite{Eynard:1998fn}
\begin{eqnarray}
S(z)=0.\label{sad3}
\end{eqnarray}
\begin{eqnarray}
S(z)
&=&\sum_{k=1}^{+\infty}(f(z_k)+f((1-z)_k)).
\end{eqnarray}
\begin{eqnarray}
f(z)&=&W^2(z)+\frac{1}{2}(W(1-z)+W(-1-z))W(z)-\frac{1}{g^2}(2zW(z)-R(z)).
\end{eqnarray}
The function $S(z)$ has no cut throughout the complex plane and it satisfies $S(z+2)=S(z)$, $S(1-z)=S(z)$. In order to guarantee convergence of the sum we have subtracted the polynomial part $R(z_k)$ of $2z_kW(z_k)$. 

Next we can easily verify that the resolvent can be rewritten in terms of the  moments $t_n$ of the matrix $\phi$ as
\begin{eqnarray}
W(z)
&=&\sum_{n=0}^{\infty}\frac{t_n}{z^{n+1}}~,~t_n=\int dx x^n\rho(x)=\frac{1}{N}Tr \phi^n.
\end{eqnarray}
After a long calculation we can rewrite the saddle-point equation (\ref{sad3}) as
\begin{eqnarray}
T_{q}=\sum_{p=0}^{q-1}T_{q-p-1}T_{p}+2g^2\sum_{p=0}^{\infty}\sum_{m=0}^{\infty}g^{2(p+m)}T_{p}T_{m+q}C^{2(p+m)+1}_{2p}~,~T_q=\frac{t_{2q}}{2g^{2q}}.
\end{eqnarray}
It is not difficult to show now that the free theory $g^2=0$ corresponds to a density of eigenvalues given by the Wigner's semicircle law
\begin{eqnarray}
\rho(x)=\frac{2}{\pi a^2}\sqrt{a^2-x^2}~,~a^2=2g^2.
\end{eqnarray}
A perturbative solution around this free solution can be constructed following the method of \cite{Eynard:1998fn}. However the exact solution as we will see shortly is quite different from the prediction of perturbation theory. This result is rather expected since in the limit of interest $\alpha\longrightarrow 0$ we have $t\longrightarrow 0$ and as a consequence $g^2\longrightarrow \infty$. This conclusion is actually also confirmed By Monte Carlo simulations.

\paragraph{Eigenvalues Distribution:}
As we have already said the analytic continuation of our model is essentially the model studied in  \cite{Kazakov:1998ji}. However the exact solution we would like to construct here is entirely based on the eigenvalues distribution of the matrix $\phi$ from which we can calculate all observables of interest to us. This explicit solution is rather different from the sophisticated implicit solution found in  \cite{Kazakov:1998ji}. The eigenvalues distribution we will derive shortly is also different from the one presented in  \cite{Berenstein:2008eg} and it agrees well with the numerical Monte Carlo simulations.

Let us first recall our notation
\begin{eqnarray}
t=\frac{1}{g^2}=\frac{\alpha^4}{32}.
\end{eqnarray}
Let us next consider the saddle point equation once again. This equation can be rewritten in the form 
\begin{eqnarray}
\frac{z}{g^2}=\int dy \frac{\rho(y)}{(z-y)(1-(z-y)^2)}~,~\int dy \rho(y)=1.\label{sad5}
\end{eqnarray}
We have
\begin{eqnarray}
\frac{z}{g^2}&=&\int_{z-L}^{z+L} dx \frac{\rho(z-x)}{x(1-x^2)}\nonumber\\
&=&\rho(z)\int_{z-L}^{z+L} dx \frac{1}{x(1-x^2)}-\rho^{'}(z)\int_{z-L}^{z+L} dx \frac{1}{1-x^2}+\rho^{''}(z)\int_{z-L}^{z+L} dx \frac{x}{2(1-x^2)}+...\nonumber\\
\end{eqnarray}
We have
\begin{eqnarray}
\int \frac{dx}{x(1-x^2)}
&=&-\frac{1}{2}\bigg(-\frac{1}{x^2}-\frac{1}{2x^4}+... \bigg).
\end{eqnarray}

\begin{eqnarray}
\int dx \frac{1}{1-x^2}
&=&\frac{1}{x}+\frac{1}{3x^3}+...
\end{eqnarray}
\begin{eqnarray}
\int dx \frac{x}{2(1-x^2)}&=&-\frac{1}{4}\ln(x^2-1).
\end{eqnarray}
We  will assume $2$ things:
\begin{itemize}
\item{}A quadratic distribution.
\item{}An (infinite) support $[-L,L]$.
\end{itemize}
We get then to leading contribution in $1/L$ the expansion
\begin{eqnarray}
\frac{z}{g^2}
&=&\rho(z)(-\frac{2z}{L^3}+...)-\rho^{'}(z)(\frac{2}{L}+...)+\rho^{''}(z)(-\frac{z}{L}+...)+...
\end{eqnarray}
We assume a quadratic distribution which is also symmetric given by
\begin{eqnarray}
\rho(z)=a+b z^2.
\end{eqnarray}
Normalization gives
\begin{eqnarray}
a=\frac{1}{2L}-\frac{bL^2}{3}.
\end{eqnarray}
The saddle-point equation  to leading order in $1/L$  becomes
\begin{eqnarray}
\frac{z}{g^2}
&=&-\frac{6bz}{L}+...
\end{eqnarray}
In other words
\begin{eqnarray}
b=-\frac{1}{g^2}\frac{L}{6}.
\end{eqnarray}
The distribution becomes
\begin{eqnarray}
\rho(z)&=&\frac{1}{2L}+\frac{L^3}{18g^2}-\frac{L}{6g^2}z^2\nonumber\\
&=&\frac{L}{6g^2}(L^2-z^2)\nonumber\\
&=&\frac{3}{4L^3}(L^2-z^2).
\end{eqnarray}
We imposed the condition
\begin{eqnarray}
\frac{1}{2L}+\frac{L^3}{18g^2}=\frac{L^3}{6g^2}~\Leftrightarrow L=(\frac{9}{2})^{\frac{1}{4}}\sqrt{g}.
\end{eqnarray}
By performing the rescaling $z\longrightarrow z_0=(L_0/L)z$ we can bring the above distribution to the form
\begin{eqnarray}
\rho_0(z_0)
&=&\frac{3}{4L_0^3}(L_0^2-z_0^2).
\end{eqnarray}
This can be derived from the same saddle-point equation (\ref{sad5}) and therefore the same path integral (\ref{patheff}) with a coupling constant $t_0=1/g_0^2$ where $g_0$ is defined by the equation
 \begin{eqnarray}
L_0=(\frac{9}{2})^{\frac{1}{4}}\sqrt{g_0}.
\end{eqnarray}
The original normalization of the matrices corresponds to $L_0/L=\alpha\phi$. Thus we have
 \begin{eqnarray}
L_0&=&\alpha\phi L\nonumber\\
&=&2\sqrt{3}\phi\nonumber\\
&=&\sqrt{3}(1+\sqrt{1-4\tilde{\beta}}).
\end{eqnarray}
This is independent of $N$ and $\alpha$ which is precisely what we observe in Monte Carlo simulations. For the value of $\tilde{\beta}$ at hand which for $\tilde{\zeta}_0=-1/24$ is $\tilde{\beta}=1/4$ we get the prediction  $L_0=\sqrt{3}$. If we simply apply this formula for $\tilde{\beta}=2/9$ we obtain the prediction $L_0=4/\sqrt{3}=2.31$ which should be compared with the measured value $L_0=2.85$.

Since the above distribution will work only for large $L$ it must only be valid for large $g^2$ which  is equivalent to small $\alpha$. This is the regime of the matrix phase. We must have then
\begin{eqnarray}
L>>1\Leftrightarrow  g^2>>\frac{2}{9}.
\end{eqnarray}
In other words we must have the following lower estimate of the critical value
\begin{eqnarray}
\alpha<<\alpha_*=2\sqrt{3}.
\end{eqnarray}

\paragraph{Antihermitian Case:}
The above distribution is the same distribution which was found for large values of $g^2$ for the antihermitian model in \cite{Berenstein:2008eg}. However the crucial difference is the functional dependence of $L$ on $g$ which we will now briefly discuss. 

To get the model of \cite{Kazakov:1998ji} and \cite{Berenstein:2008eg} we must assume that $\phi$ is antihermitian and take $\kappa_1-2\epsilon>0$ and $\kappa_1<0$, i.e. $\zeta_0>0$. We get the path integral
\begin{eqnarray}
Z=\int d{\phi}~dX_1~dX_2 \exp N t\bigg(-2iTr\phi[X_1,X_2]+Tr\phi^2-TrX_i^2\bigg).
\end{eqnarray}
In this case
\begin{eqnarray}
t
&=&-\frac{\alpha^4}{32}.
\end{eqnarray}
Integration over $X_1$ and $X_2$ yields
\begin{eqnarray}
Z
&=&\int \prod_{i=1}^Nd\phi_i 
\prod_{i<j}(\phi_i-\phi_j)^2 \prod_{i<j}\bigg(-Nt(\phi_i-\phi_j)^2+Nt\bigg)^{-1} \exp\bigg(Nt\sum_{i}\phi_i^2\bigg).\nonumber\\
\end{eqnarray}
We get the effective potential (shifting the eigenvalues as $\phi_i\longrightarrow i\phi_i$)
\begin{eqnarray}
-V_{\rm eff}(\phi_i)
&=&-Nt\sum_{i}\phi_i^2+\frac{1}{2}\sum_{i\neq j}\ln (\phi_i-\phi_j)^2-\frac{1}{2}\sum_{i\neq j}\ln\bigg(t(\phi_i-\phi_j)^2+t\bigg).
\end{eqnarray}
The saddle-point equation is
\begin{eqnarray}
\frac{1}{g^2}\phi_k=\frac{1}{N}\sum_{i\neq k}\frac{1}{(\phi_k-\phi_i)\big(1+(\phi_k-\phi_i)^2\big)}.\label{sad1}
\end{eqnarray}
Therefore in this case the saddle-point equation is given by
\begin{eqnarray}
\frac{z}{g^2}=\int dy \frac{\rho(y)}{(z-y)(1+(z-y)^2)}.
\end{eqnarray}
We go now through the same steps. We have
\begin{eqnarray}
\frac{z}{g^2}&=&\int_{z-L}^{z+L} dx \frac{\rho(z-x)}{x(1+x^2)}\nonumber\\
&=&\rho(z)\int_{z-L}^{z+L} dx \frac{1}{x(1+x^2)}-\rho^{'}(z)\int_{z-L}^{z+L} dx \frac{1}{1+x^2}+\rho^{"}(z)\int_{z-L}^{z+L} dx \frac{x}{2(1+x^2)}+...\nonumber\\
&=&-\frac{1}{2}\rho(z)\ln(1+\frac{1}{x^2})|_{z-L}^{z+L}-\rho^{'}(z)\arctan x|_{z-L}^{z+L}+\frac{1}{4}\rho^{"}(z)\ln(1+x^2)|_{z-L}^{z+L}+...\nonumber\\
\end{eqnarray}
Again we assume a symmetric quadratic distribution. As before only the above three terms will contribute. The first term is still of order $1/L^3$, the third term is still of order $1/L$ whereas the second term becomes of order $1$. 
The saddle-point equation  to leading order in $1/L$  becomes
\begin{eqnarray}
\frac{z}{g^2}
&=&-\pi\rho^{'}(z)+...
\end{eqnarray}
We immediately obtain
\begin{eqnarray}
\rho(z)=-\frac{z^2}{2\pi g^2}+a.
\end{eqnarray}
Normalization gives
\begin{eqnarray}
a=\frac{1}{2L}+\frac{L^2}{6\pi g^2}.
\end{eqnarray}
We obtain the distribution
\begin{eqnarray}
\rho(z)&=&\frac{1}{2\pi g^2}(2\pi g^2 a-z^2)\nonumber\\
&=&\frac{3}{4L^3}(L^2 -z^2).
\end{eqnarray}
We imposed the condition
\begin{eqnarray}
L^2=2\pi g^2 a\Leftrightarrow L=(\frac{3\pi}{2})^{\frac{1}{3}}g^{\frac{2}{3}}.
\end{eqnarray}
By comparing with the numerical results we will find that the hermitian prediction is more accurate than this antihermitian calculation.

\section{Summary and Future Directions}
In this article we employed  supersymmetry, cohomological deformation, localization and the saddle-point method as well as the Monte Carlo method to study nonperturbatively Yang-Mills matrix models in $D=4$ with mass terms. We can summarize the main results, findings and conjectures of this work as follows:
\begin{itemize}
\item{}By imposing the requirement of supersymmetry and $SO(3)$ covariance we have shown that there exists a single mass deformed Yang-Mills quantum mechanics in $D=4$ which preserves all four real supersymmetries of the original theory although in a deformed form. This is the $4$ dimensional analogue of the $10$ dimensional BMN model. Full reduction yields a unique mass deformed $D=4$ Yang-Mills matrix model. This latter $4$ dimensional model is the analogue of the $10$ dimensional IKKT model.
\item{} By using cohomological deformation of supersymmetry we constructed a one-parameter ($\zeta_0$) family of cohomologically  deformed  $D=4$ Yang-Mills matrix models which preserve two supercharges. The mass deformed model is one limit ($\zeta_0\longrightarrow 0$) of this 
one-parameter family of cohomologically  deformed  Yang-Mills models.
\item{}We studied the models with the values $\tilde{\beta}=0$ and $\tilde{\beta}=2/9$ where $\tilde{\beta}$ is the mass parameter of the bosonic matrices $X_a$. The second model is special in the sense that classically the configurations $X_a\sim L_a,X_4=0$ is degenerate with the configuration $X_a=0,X_4=0$.
\item{}The Monte Carlo simulation of the bosonic  $D=4$ Yang-Mills matrix model with mass terms shows the existence of an exotic first/second order transition from a phase with a well defined background geometry given by the famous fuzzy sphere to a phase with commuting matrices with no geometry in the sense of Connes. The transition looks first order due to the jump in the action whereas it looks second order due to the divergent peak in the specific heat.
\item{}The fuzzy sphere is less stable as we increase the mass term of the bosonic matrices $X_a$, i.e. as we increase $\tilde{\beta}$. For $\tilde{\beta}=2/9$ we find the critical value $\tilde{\alpha}_*=4.9$ whereas for $\tilde{\beta}=0$ we find the critical value $\tilde{\beta}=2.55$. 
\item{}The measured critical line in the plane $\tilde{\alpha}-\tilde{\beta}$ agrees well with the theoretical prediction coming from the effective potential calculation.  
\item{}The order parameter of the transition is given by the inverse radius of the sphere defined by $1/r=Tr X_a^2/(\tilde{\alpha}^2c_2)$. The radius is equal to $1/\phi^2$ (where $\phi$ is the classical configuration) in the fuzzy sphere phase. At the transition point the sphere expands abruptly to infinite size. Then as we decrease the inverse temperature (the inverse gauge coupling constant) $\tilde{\alpha}$, the size of the sphere shrinks fast to $0$, i.e. the sphere evaporates.

\item{}The fermion determinant is positive definite for all gauge configurations in $D=4$. We have conjectured that the path integral is convergent as long as the scalar curvature (the mass of the fermionic matrices) is zero.  
\item{}We have simulated the two models $\tilde{\beta}=0$ and $\tilde{\beta}=2/9$ with dynamical fermions. The model with  $\tilde{\beta}=0$ has two supercharges while the model  $\tilde{\beta}=2/9$ has a softly broken supersymmetry since in this case we needed to set by hand the scalar curvature to zero in order to regularize the path integral.

Thus $\tilde{\beta}=0$ is amongst the very few models (which we known of) with exact supersymmetry which can be probed and accessed with the Monte Carlo method.

\item{}The fuzzy sphere is stable for the supersymmetric $D=4$ Yang-Mills matrix model with mass terms in the sense that the bosonic phase transition is turned into a very slow crossover transition. The transition point $\tilde{\alpha}$ is found to scale to zero with $N$. There is no jump in the action nor a peak in the specific heat.

\item{}The fuzzy sphere is stable also in the sense that the radius is equal $1/\phi^2$ over a much larger region then it starts to decrease slowly as we decrease   the inverse temperature $\tilde{\alpha}$ until it reaches $0$ at $\tilde{\alpha}=0$. We claim that the value where the radius starts decreasing becomes smaller as we increase $N$. 

The model at $\tilde{\alpha}=0$ can never sustain the geometry of the fuzzy sphere since it is the non deformed model so in some sense the transition to commuting matrices always occurs and in the limit $N\longrightarrow \infty$ it will occur at  $\tilde{\alpha}_*\longrightarrow 0$. 
 
\item{}We have spent a lot of time in trying to determine the eigenvalues distributions of the matrices $X_{\mu}$ in both the bosonic and supersymmetric theories. A universal behavior seems to emerge with many subtleties. These can be summarized as follows:
\begin{itemize}
\item{}In the fuzzy sphere the matrices $X_a$ are given by the $SU(2)$ irreducible representations $L_a$. For example diagonalizing the matrix $X_3$ gives $N$ eigenvalues between $(N-1)/2$ and $-(N-1)/2$ with a step equal $1$, viz $m=(N-1)/2,(N-3)/2,...,-(N-3)/2,-(N-1)/2$.

\item{}In the matrix phase the matrices $X_{\mu}$ become commuting. More explicitly the eigenvalues distribution of any of the matrices $X_{\mu}$ in the matrix phase is given by the non-polynomial law 
\begin{eqnarray}
\rho_4(x)=\frac{8}{3\pi R^4}(R^2-x^2)^{\frac{3}{2}}.
\end{eqnarray}
This can be obtained from the conjecture that the joint probability distribution of the four matrices $X_{\mu}$ is uniform inside a solid ball with radius $R$. 
\item{}In the matrix phase the eigenvalues distribution of any of the $X_a$, say $X_3$,  is given by the above  non-polynomial law with a radius $R$ independent of $\tilde{\alpha}$ and $N$.

\item{}This  is also confirmed by computing the radius in this distribution and comparing to the Monte Carlo data. 

\item{}A very precise measurement of the transition point can be made by observing the point at which the eigenvalues distribution of $X_3$ undergoes the transition from the $N$-cut distribution to the above  non-polynomial law.

\item{}The eigenvalues distribution of $X_4$  is always given by the above  non-polynomial law, i.e. for all values of $\tilde{\alpha}$, with a radius $R$ which depends on $\tilde{\alpha}$ and $N$.

\item{}Another signal that the matrix phase is fully reached is when the eigenvalues distribution of $X_4$ coincides with that of $X_3$. From this point downward the eigenvalues distribution of $X_4$ ceases to depend on $\tilde{\alpha}$ and $N$. 

\item{}Monte Carlo measurements seems to indicate that $R=1.8$ for bosonic models and $R=2.8$ for supersymmetric models. The distribution becomes wider in the supersymmetric case.
\item{}We have also observed that the eigenvalues of the normal scalar field $X_a^2-c_2$ in the fuzzy sphere are also distributed according to the above non-polynomial law. This led us to the conjecture that the eigenvalues of the gauge field on the background geometry are also distributed according to the above non-polynomial law. Recall that the normal scalar field is the normal component of the gauge field to the background geometry which is the sphere here.
\end{itemize}
\item{}In the $D=3$ Yang-Mills matrix model with mass terms the eigenvalues distribution becomes polyonomial (parabolic) given by
 \begin{eqnarray}
\rho_3(x)=\frac{3}{4R^3}(R^2-x^2).
\end{eqnarray}
It was difficult for us in this article to differentiate with certainty between the two distributions $\rho_4$ and $\rho_3$ in the three dimensional setting.
\item{}We have also attempted to compute the above eigenvalues distributions analytically. Using localization techniques we were able to find a special set of parameters for which the $D=4$ Yang-Mills matrix model with mass terms can be reduced to the three dimensional Chern-Simons (CS) matrix model. The saddle-point method leads then immediately to the eigenvalues distributions $\rho_3$. We believe that our theoretical prediction for the value of $R$ is reasonable compared to the Monte Carlo value.  
\item{}We have also made a preliminary comparison between the dependence of $R$ on $\alpha$ in the hermitian and antihermitian CS matrix models. The hermitian case seems more appropriate for the description of the eigenvalues of $X_3$ whereas the antihermitian case may be relevant to the description of the eigenvalues of $X_4$.
\item{}Finally, we conjecture that the transition from a background geometry to the phase of commuting matrices is associated with spontaneous supersymmetry breaking. Indeed mass deformed supersymmetry preserves the fuzzy sphere configuration but not diagonal matrices.
\end{itemize}

Among the future directions that can be considered we will simply mention the following four points:
\begin{itemize} 
\item{}Higher precision Monte Carlo simulations of the models studied in this article is the first obvious direction for future investigation. The most urgent question (in our view) is the precise determination of the behavior of the eigenvalues distributions in $D=4$ and $D=3$. An analytical derivation of $\rho_3$ and especially $\rho_4$ is an outstanding problem.
\item{}Finding matrix models with emergent $4$ dimensional background geometry is also an outstanding problem. 
\item{}Models for emergent time, and to a lesser extent emergent gravity, and as a consequence emergent cosmology are very rare.
\item{}Monte Carlo simulation of supersymmetry based on matrix models seems to be a very promising goal.
\end{itemize}

\appendix

\section{Algorithms and Simulations}
We will use the Metropolis algorithm to compute the probability distribution
\begin{eqnarray}
d{\cal P}(X_{\mu})=\frac{1}{Z}{\delta}~\big(TrX_{\mu}\big)~{\rm det}{\cal D}~e^{-S_B}.
\end{eqnarray}
In problems involving fermions and/or matrices, the Hybrid Monte Carlo algorithm would have been a better choice as it involves global updates as opposed to local updates involved in the Metropolis and as a consequence it is more efficient for non local theories. The  disadvantage in using the Hybrid Monte Carlo algorithm is the fact that the molecular dynamics part of this algorithm contains two parameters (the time step size and the number of molecular dynamics steps) which require optimization while the Metropolis algorithm is free from extraneous parameters. Since we will be reporting data with the fermion determinant for small values of $N$ up to $N=10$  we prefer here the Metropolis algorithm over the  Hybrid Monte Carlo algorithm. For the bosonic theory the Metropolis algorithm is quite efficient and we can reach up to $N=100$ as in  \cite{DelgadilloBlando:2008vi,DelgadilloBlando:2007vx}. The data with the fermion determinant for large values  of $N$  using the Hybrid Monte Carlo algorithm will be reported elsewhere \cite{ydri_preparation}.

We have used two random numbers in these simulations: RAN2 and RANLUX. We have found no discernible or measurable differences between these two generators. For consistency, the data reported here is all with RAN2. The errors were estimated using the jackknife method which is supposed to take into account and suppress statistical correlations. The determinant of the Dirac operator and the eigenvalues are computed using LAPACK. 

A simulation consists of $T_{\rm th}+T_{\rm mc}$ sweeps where $T_{\rm th}$ is the number of thermalization sweeps and $T_{\rm mc}$ is the number of Monte Carlo sweeps. For all simulations with the fermion determinant we choose   $T_{\rm th}=T_{\rm mc}=2^{13}$. For the bosonic simulations we choose  $T_{\rm th}=T_{\rm mc}=2^{13}-2^{18}$. We have checked that there is no difference between starting from a cold start and a hot start in bosonic simulations with $\tilde{\beta}=0$. The same holds for bosonic simulations with  $\tilde{\beta}\neq 0$ although thermalization becomes somehow difficult. For the supersymmetric models we always start from the  matrices $X_4=0$, $X_a=\phi L_a$ since thermalization is very costly in this case.

In a Monte Carlo sweep we update every single entry of the four matrices $X_{\mu}$ once and thus a Monte Carlo sweep contains $2(N^2+N)$ Metropolis steps. This consists one unit of time or a step of the Monte Carlo evolution.

In the actual code there was no symmetry between the matrix $X_4$ and the matrices $X_a$ in the sense that we chose different intervals  from which we pick the variations of $X_4$ and $X_a$. These intervals are adjusted at each Monte Carlo unit of time (sweep) so that the acceptance rates for $X_4$ and $X_a$  are fixed separately  between $25$ and $30$ per cent.

In actual numerical simulations it is of vital importance to remove the free degrees of freedom $Tr X_{\mu}$, $Tr\theta_{\alpha}$ and $Tr\theta_{\alpha}^+$ in order to achieve a very efficient code.   After updating the diagonal matrix element $i$ of a given matrix $X_{\mu}$ as: $(X_{\mu})_{ii}\longrightarrow (X_{\mu})_{ii}=(X_{\mu})_{ii} + d+d^*$ the trace changes as: $Tr X_{\mu}=0\longrightarrow Tr X_{\mu}=d+d^*$. Thus to restore a traceless matrix   $X_{\mu}$ we subtract from it the matrix $-(d+d^*){\bf 1}_N/N$ and as a consequence we change the diagonal entries of $X_{\mu}$ as 

\begin{eqnarray}
&&(X_{\mu})_{ii}\longrightarrow (X_{\mu})_{ii}= (X_{\mu})_{ii} + d+d^*-\frac{d+d^*}{N}\nonumber\\
&&(X_{\mu})_{jj}\longrightarrow (X_{\mu})_{jj}= (X_{\mu})_{jj} -\frac{d+d^*}{N}~,~j\neq i.
\end{eqnarray}
We found that this is the only prescription of imposing tracelessness that works in the matrix phase. The effect of the tracelessness condition on the determinant of the Dirac operator is discussed in the next appendix.  
\section{The Dirac Operator in Simulations}

The aim is to simulate the action
\begin{eqnarray}
S&=&S_B
-\frac{1}{N\alpha}Tr{\theta}^+\bigg(i[X_4,..]+{\sigma}_a[{X}_a,..]+\alpha\tilde{\xi}\bigg)\theta.
\end{eqnarray}
\begin{eqnarray}
S_B&=&N Tr\bigg[-\frac{1}{4}[X_{\mu},X_{\nu}]^2+\frac{2i\alpha}{3}{\epsilon}_{abc}X_aX_bX_c\bigg]+N\tilde{\beta}\alpha^2 TrX_a^2.
\end{eqnarray}

We work with the effective action
\begin{eqnarray}
S_{\rm eff}&=&S_B-Tr_{\rm ad}\log{\cal D}.
\end{eqnarray}
The Dirac operator is given by
\begin{eqnarray}
{\cal D}=iX_4-iX_4^R+{\sigma}_aX_a-{\sigma}_aX_a^R+\alpha \tilde{\xi}.
\end{eqnarray}
It is possible to rewrite the Dirac action in the following form (with $X_{34}=X_3+iX_4$)
\begin{eqnarray}
Tr{\theta}^+{\cal D}\theta&=&Tr\bigg[{\theta}_1^+(X_{34} +\alpha\tilde{\xi}){\theta}_1+{\theta}_1^+X_{-}{\theta}_2+{\theta}_2^+X_{+}{\theta}_1+{\theta}_2^+(-X_{34}^++\alpha\tilde{\xi}){\theta}_2\bigg]\nonumber\\
&-&Tr\bigg[X_{34}{\theta}_1^+{\theta}_1+X_-{\theta}_1^+{\theta}_2+X_+{\theta}_2^+{\theta}_1-X_{34}^+{\theta}_2^+{\theta}_2\bigg].
\end{eqnarray}
Following \cite{Ambjorn:2000bf, Anagnostopoulos:2005cy}, we expand the $N\times N$ matrices ${\theta}_1,{\theta}_2$ and ${\theta}_1^+,{\theta}_2^+$ as ${\theta}_i=\sum_{A=1}^{N^2}{\theta}_i^AT^A$ and ${\theta}_i^+=\sum_{A=1}^{N^2}({\theta}_i^A)^{*}(T^A)^+$ where the $N\times N$ matrices $T^A$ are defined by $(T^A)_{ij}={\delta}_{ii_A}{\delta}_{jj_A}$ with $i_A$ and $j_A$ are such that $A=N(i_A-1)+j_A$ whereas the matrices $(T^A)^+$ are defined by $(T^A)^+_{ij}={\delta}_{ii_A}{\delta}_{jj_A}$ with $i_A$ and $j_A$ such that $A=N(j_A-1)+i_A$ . Then we find that
\begin{eqnarray}
Tr{\theta}^+{\cal D}\theta = {\chi}_1^{*}{\cal M}_{11}{\chi}_1+ {\chi}_1^{*}{\cal M}_{12}{\chi}_2+ {\chi}_2^{*}{\cal M}_{21}{\chi}_2+ {\chi}_2^{*}{\cal M}_{22}{\chi}_2.
\end{eqnarray} 
The matrices ${\cal M}_{ij}$ are $N^2\times N^2$ defined by
\begin{eqnarray}
&&({\cal M}_{11})^{AB}=Tr(T^A)^+(X_{34} +\alpha\tilde{\xi})T^B-TrX_{34}(T^A)^+T^B\nonumber\\
&&({\cal M}_{12})^{AB}=Tr(T^A)^+X_{-}T^B-TrX_- (T^A)^+T^B\nonumber\\
&&({\cal M}_{21})^{AB}=Tr (T^A)^+X_+T^B-TrX_+(T^A)^+T^B\nonumber\\
&&({\cal M}_{22})^{AB}=Tr (T^A)^+(-X_{34}^+ +\alpha\tilde{\xi})T^B+TrX_{34}^+(T^A)^+T^B.
\end{eqnarray}
We remark that
\begin{eqnarray}
&&Tr(T^A)^+XT^B-TrX(T^A)^+T^B=X_{j_Aj_B}{\delta}_{i_Bi_A}-X_{i_Bi_A}{\delta}_{j_Aj_B}\nonumber\\
&&Tr(T^A)^+T^B={\delta}_{j_Aj_B}{\delta}_{i_Bi_A}\nonumber\\
&&A=N(j_A-1)+i_A~,~B=N(j_B-1)+i_B.
\end{eqnarray}
The $N^2-$dimensional vectors ${\chi}_1$, ${\chi}_2$ and ${\chi}_1^+$, ${\chi}_2^+$ are defined by $({\chi}_i)_A={\theta}_i^A$ and  $({\chi}_i^+)_A=({\theta}_i^A)^{*}$. Thus in terms of  the $2N^2-$dimensional vectors ${\chi}$ and ${\chi}^+$ the above Dirac term becomes
\begin{eqnarray}
Tr{\theta}^+{\cal D}\theta = {\chi}^+{\cal M}{\chi}.
\end{eqnarray} 
We observe that the trace parts of the matrices $X_{\mu}$ drop from the partition function since they correspond to   Gaussian distributions. Thus the measure should read $\int dX_{\mu}{\delta}(TrX_{\mu})$ instead of simply $\int dX_{\mu}$.  Similarly we observe that if we write  $\theta={\theta}_0+\eta {\bf 1}$ then the trace part $\eta$ will decouple from the rest, viz

\begin{eqnarray}
Tr{\theta}^+\bigg(i[X_4,..]+{\sigma}_a[{X}_a,..]+\alpha\tilde{\xi}\bigg)\theta=Tr{\theta}_0^+\bigg(i[X_4,..]+{\sigma}_a[{X}_a,..]+\alpha\tilde{\xi}\bigg){\theta}_0+\alpha\tilde{\xi} {\eta}^+\eta.
\end{eqnarray}
Hence the constant fermion modes ${\eta}_{\alpha}$ can also be integrated out from the partition function and thus we should consider the measure $\int d{\theta}d{\theta}^+{\delta}(Tr{\theta}_{\alpha}){\delta}(Tr{\theta}_{\alpha}^+)$ instead of  $\int d{\theta}d{\theta}^+$.  We are thus led to consider the partition function
\begin{eqnarray}
&&Z=\int dX_a~{\delta}(TrX_a)~ \det {\cal D} ~e^{-S_B}.
\end{eqnarray}
The determinant is given by 
\begin{eqnarray}
{\rm det}{\cal D}&=&\int d{\theta}d{\theta}^+{\delta}(Tr{\theta}_{\alpha}){\delta}(Tr{\theta}_{\alpha}^+)e^{\frac{1}{N\alpha}Tr {\theta}^+{\cal D}{\theta}}\nonumber\\
&=&\int d{\chi}d{\chi}^+{\delta}\bigg(\sum_{A=1}^{N^2}({\chi}_{\alpha})_A{\delta}_{i_Aj_A}\bigg){\delta}\bigg(\sum_{A=1}^{N^2}({\chi}_{\alpha}^{+})_A{\delta}_{i_Aj_A}\bigg)e^{ \frac{1}{N\alpha}{\chi}^+{{\cal M}}{\chi}}\nonumber\\
&=&\int d{\chi}^{'}d{\chi}^{'+}e^{\frac{1}{N\alpha}Tr^{'} {\chi}^{'+}{{\cal M}^{'}}{\chi}^{'}}.
\end{eqnarray}
The vectors ${\chi}_1^{'}$, ${\chi}_2^{'}$ are $(N^2-1)-$dimensional. The matrix ${\cal M}^{'}$ is  $2(N^2-1)\times 2(N^2-1)$ dimensional and it is given by
\begin{eqnarray}
{\cal M}_{\alpha \beta}^{'A^{'}B^{'}}={\cal M}_{\alpha \beta}^{A^{'}B^{'}}-{\cal M}_{\alpha \beta}^{N^2B^{'}}{\delta}_{i_{A^{'}}j_{A^{'}}}-{\cal M}_{\alpha \beta}^{A^{'}N^2}{\delta}_{i_{B^{'}}j_{B^{'}}}+{\cal M}_{\alpha \beta}^{N^2N^2}{\delta}_{i_{A^{'}}j_{A^{'}}}{\delta}_{i_{B^{'}}j_{B^{'}}}.
\end{eqnarray}
We remark that
\begin{eqnarray}
{\cal M}_{\alpha \beta}^{N^2N^2}=\alpha\tilde{\xi} {\delta}_{\alpha \beta}.
\end{eqnarray}
Thus we must have
\begin{eqnarray}
\det {\cal D}=\det^{'} {{\cal M}^{'}}.
\end{eqnarray}
We prefer to write ${\cal D}={\cal D}_0+\alpha\tilde{\xi}$. The massless Dirac operator ${\cal D}_0$ will lead to the $2(N^2-1)\times 2(N^2-1)$ dimensional matrix ${\cal M}_0^{'}$ defined by
\begin{eqnarray}
{\cal M}_{0,\alpha \beta}^{'A^{'}B^{'}}={\cal M}_{0,\alpha \beta}^{A^{'}B^{'}}-{\cal M}_{0,\alpha \beta}^{N^2B^{'}}{\delta}_{i_{A^{'}}j_{A^{'}}}-{\cal M}_{0,\alpha \beta}^{A^{'}N^2}{\delta}_{i_{B^{'}}j_{B^{'}}}.
\end{eqnarray}
\begin{eqnarray}
&&({\cal M}_{0,11})^{AB}=Tr(T^A)^+X_{34}T^B-TrX_{34}(T^A)^+T^B\nonumber\\
&&({\cal M}_{0,12})^{AB}=Tr(T^A)^+X_{-}T^B-TrX_- (T^A)^+T^B\nonumber\\
&&({\cal M}_{0,21})^{AB}=Tr (T^A)^+X_+T^B-TrX_+(T^A)^+T^B\nonumber\\
&&({\cal M}_{0,22})^{AB}=-Tr (T^A)^+X_{34}^+T^B+TrX_{34}^+(T^A)^+T^B.
\end{eqnarray}
The $2N^2\times 2N^2$ dimensional identity matrix $\alpha\tilde{\xi}{\bf 1}_{2N^2}$ will lead to the $2(N^2-1)\times 2(N^2-1)$ dimensional matrix $\alpha\tilde{\xi}(1_{2(N^2-1)}+{\gamma})$. This can be calculated as follows. We have
\begin{eqnarray}
\alpha\tilde{\xi}Tr\theta^+\theta=\alpha\tilde{\xi}\sum_{A=1}^{N^2}(\theta_{\alpha}^A)^*\theta_{\alpha}^A.
\end{eqnarray}
We use $\theta_{\alpha}^{N^2}=-\sum_{A=1}^{N^2-1}\theta_{\alpha}^A\delta_{i_Aj_A}$ where $A=(i_A-1)N+j_A$. We get
\begin{eqnarray}
\alpha\tilde{\xi}Tr\theta^+\theta&=&\alpha\tilde{\xi}\sum_{A=1}^{N^2-1}\sum_{B=1}^{N^2-1}(\theta_{\alpha}^B)^*\bigg(\delta_{AB}+\delta_{i_Aj_A}\delta_{i_Bj_B}\bigg)\theta_{\alpha}^A\nonumber\\
&=&\alpha\tilde{\xi}\sum_{A=1}^{N^2-1}\sum_{B=1}^{N^2-1}(\theta_{\alpha}^B)^*\bigg(\delta_{i_Ai_B}\delta_{j_Aj_B}+\delta_{i_Aj_A}\delta_{i_Bj_B}\bigg)\theta_{\alpha}^A.
\end{eqnarray}
The first term corresponds precisely to the  $2(N^2-1)\times 2(N^2-1)$ dimensional identity matrix whereas the second term defines the matrix $\gamma$. Hence we can write
\begin{eqnarray}
\det {\cal D}=\det^{'} ({\cal M}_0^{'}+\alpha\tilde{\xi}+\alpha\tilde{\xi}\gamma).
\end{eqnarray}
\section{Calculation of the  Mass Deformation}

The mass deformed Lagrangian density is
\begin{eqnarray}
{\cal L}_{\mu}={\cal L}_0+\frac{\mu}{g^2}{\cal L}_1+\frac{{\mu}^2}{g^2}{\cal L}_2.
\end{eqnarray}
\begin{eqnarray}
{\cal L}_1={\cal L}_{\psi}+{\cal L}_{\rm myers}.
\end{eqnarray}
\begin{eqnarray}
{\cal L}_{\psi}=Tr\bar{\psi}\bigg(ia{\bf 1}_4+\frac{1}{2}H_{ij}{\gamma}^0[{\gamma}^i,{\gamma}^j]+c{\gamma}^1{\gamma}^2{\gamma}^3\bigg)\psi.
\end{eqnarray}
\begin{eqnarray}
{\cal L}_{\rm myers}=ie{\epsilon}_{ijk}TrX_iX_jX_k.
\end{eqnarray}
\begin{eqnarray}
{\cal L}_2=Tr\bigg(-\frac{1}{2!}S_{ab}X_aX_b\bigg).
\end{eqnarray}

We will suppose  mass deformed supersymmetric transformations, with a time dependent parameter ${\epsilon}\equiv {\epsilon}(t)$ which satisfies ${\partial}_0{\epsilon}=\mu \Pi{\epsilon}$, given by
\begin{eqnarray}
&&{\delta}_{\mu} X_{0}={\delta}_0 X_{0}\nonumber\\
&&{\delta}_{\mu} X_{i}={\delta}_0 X_{i}\nonumber\\
&&{\delta}_{\mu} {\psi}={\delta}_0 {\psi}+\mu{\Delta}{\epsilon}.
\end{eqnarray}
The variation of ${\cal L}_0$, the Myers term and the fermionic mass term under the new supersymmetric transformations is
\begin{eqnarray}
{\delta}_{\mu}{\cal L}_0&=&\frac{\mu}{g^2}Tr\bar{\psi}{\gamma}^0\bigg(\frac{1}{2}[{\gamma}^0,{\gamma}^i]D_0X_i\Pi-\frac{i}{4}[{\gamma}^i,{\gamma}^j][X_i,X_j]\Pi-i{\gamma}^0{\gamma}^i[X_i,\Delta]-D_0\Delta\nonumber\\
&-&\mu\Delta\Pi\bigg){\epsilon}.
\end{eqnarray}
\begin{eqnarray}
\frac{\mu}{g^2}{\delta}_{\mu}({\cal L}_{\rm myers})
&=&\frac{\mu}{g^2}Tr\bar{\psi}\bigg(\frac{3e}{4}[{\gamma}^i,{\gamma}^j][X_i,X_j]{\gamma}^0{\gamma}_5\bigg)\epsilon.
\end{eqnarray}
\begin{eqnarray}
\frac{\mu}{g^2}{\delta}_{\mu}(Tr\bar{\psi}M{\psi})=\frac{\mu}{g^2} Tr\bar{\psi}M\bigg(-[{\gamma}^{0},{\gamma}^{i}]D_0X_i+\frac{i}{2}[{\gamma}^{i},{\gamma}^{j}][X_i,X_j]+2\mu{\Delta}\bigg){\epsilon}.
\end{eqnarray}
The terms containing the covariant time derivative take the form
\begin{eqnarray}
\frac{\mu}{g^2}Tr\bar{\psi}{\gamma}^0D_0\bigg({\gamma}^0X\Pi-\Delta-2\hat{M}X\bigg)\epsilon+\frac{\mu}{g^2}Tr\bar{\psi}XD_0\Pi.\epsilon+2\frac{\mu}{g^2}Tr\bar{\psi}{\gamma}^0D_0\hat{M}.X\epsilon.
\end{eqnarray}
We have defined $\hat{M}=-{\gamma}^0M{\gamma}^0=ia{\bf 1}_4+\frac{1}{2}H_{ij}{\gamma}^0[{\gamma}^i,{\gamma}^j]-c{\gamma}^1{\gamma}^2{\gamma}^3$. Note that $\Pi$, $M$ and $\hat{M}$ are Clifford algebra valued.  Requiring the second and third terms to vanish means that $\Pi$, $M$ and $\hat{M}$ must be time independent. Requiring the first term to vanish we obtain the constraint
\begin{eqnarray}
\Delta={\gamma}^0X\Pi-2\hat{M}X.
\end{eqnarray}
The remainder
\begin{eqnarray}
{\delta}_{\mu}\bigg({\cal L}_0+\frac{\mu}{g^2}Tr\bar{\psi}M{\psi}+ie{\epsilon}_{ijk}\frac{\mu}{g^2}TrX_iX_jX_k\bigg)
&=&\nonumber\\
\frac{\mu}{g^2}Tr\bar{\psi}\bigg(-\frac{i}{4}[{\gamma}^i,{\gamma}^j][X_i,X_j](3{\gamma}^0\Pi+2\hat{M}-4c{\gamma}^1{\gamma}^2{\gamma}^3+3ie{\gamma}^0{\gamma}_5)&-&\nonumber\\
\mu {\gamma}^0\Delta\Pi +2\mu M{\Delta}-i\{{\gamma}^i,[\hat{M},{\gamma}^j]\}[X_i,X_j]\bigg){\epsilon}.
\end{eqnarray}
We choose 
\begin{eqnarray}
\Pi=\frac{2}{3}{\gamma}^0\hat{M}-\frac{1}{3}(4c+3e){\gamma}^0{\gamma}^1{\gamma}^2{\gamma}^3=\frac{2}{3}{\gamma}^0(ia{\bf 1}_4+{\gamma}^0H)-\frac{1}{3}(6c+3e){\gamma}^0{\gamma}^1{\gamma}^2{\gamma}^3.
\end{eqnarray}
We get
\begin{eqnarray}
{\delta}_{\mu}\bigg({\cal L}_0+\frac{\mu}{g^2}Tr\bar{\psi}M{\psi}+ie{\epsilon}_{ijk}\frac{\mu}{g^2}TrX_iX_jX_k\bigg)&=&\frac{\mu}{g^2}Tr\bar{\psi}\bigg(-\mu {\gamma}^0\Delta\Pi +2\mu M{\Delta}\nonumber\\
&-&i\{{\gamma}^i,[\hat{M},{\gamma}^j]\}[X_i,X_j]\bigg){\epsilon}.
\end{eqnarray}
We can verify that $\{{\gamma}^i,[\hat{M},{\gamma}^j]\}=0$. In other words

\begin{eqnarray}
{\delta}_{\mu}\bigg({\cal L}_0+\frac{\mu}{g^2}Tr\bar{\psi}M{\psi}+ie{\epsilon}_{ijk}\frac{\mu}{g^2}TrX_iX_jX_k\bigg)&=&\frac{\mu}{g^2}Tr\bar{\psi}\bigg(-\mu {\gamma}^0\Delta\Pi +2\mu M{\Delta}\bigg){\epsilon}.\nonumber\\
\end{eqnarray}
This must be identified with
\begin{eqnarray}
{\delta}_{\mu}\bigg({\cal L}_0+\frac{\mu}{g^2}Tr\bar{\psi}M{\psi}+ie{\epsilon}_{ijk}\frac{\mu}{g^2}TrX_iX_jX_k\bigg)
&=&-\frac{{\mu}^2}{g^2} Tr\bar{\psi}\bigg(S_{ij}{\gamma}^iX_j\bigg){\epsilon}.
\end{eqnarray}
In other words we must have
\begin{eqnarray}
S_{ij}{\gamma}^iX_j={\gamma}^0\Delta \Pi-2 M\Delta .
\end{eqnarray}
The final result is

\begin{eqnarray}
{\delta}_{\mu}\bigg({\cal L}_{0}+\frac{\mu}{g^2}Tr\bar{\psi}M{\psi}+ie{\epsilon}_{ijk}\frac{\mu}{g^2}TrX_iX_jX_k-\frac{{\mu}^2}{g^2}\frac{1}{2!}S_{ij}TrX_iX_j\bigg)&=&0.
\end{eqnarray}
We compute
\begin{eqnarray}
\Delta=-\frac{2}{3}{\gamma}^0(XH+3HX)-\frac{4}{3}(ia+\frac{3e}{4}{\gamma}^1{\gamma}^2{\gamma}^3)X.
\end{eqnarray}
We get after some algebra
\begin{eqnarray}
S_{ij}{\gamma}^iX_j&=&\frac{16}{9}{\gamma}^0(ia+\frac{3e}{4}{\gamma}^1{\gamma}^2{\gamma}^3)(XH+3HX)-\frac{8}{3}HXH-4H^2X-\frac{4}{9}XH^2\nonumber\\
&+&\frac{16}{9}\bigg(-a^2+\frac{9}{16}e^2+3ia(c+\frac{3e}{4}){\gamma}^1{\gamma}^2{\gamma}^3\bigg)X.
\end{eqnarray}
We have $H=\frac{1}{2}H_{ij}[{\gamma}^i,{\gamma}^j]=-iH_{ij}{\epsilon}_{ijk}{\gamma}^0{\gamma}_5{\gamma}^k$, $\frac{i}{2}[{\gamma}^i,{\gamma}^j]={\epsilon}_{ijk}{\gamma}^0{\gamma}_5{\gamma}_k$ and ${\gamma}^0{\gamma}_5=i{\gamma}^1{\gamma}^2{\gamma}^3$. Thus we can compute that $HXH$, $H^2X$ and $XH^2$ can only be linear in ${\gamma}^i$, proportional to the cubic product ${\gamma}^1{\gamma}^2{\gamma}^3$ or proportional to the identity. We can also compute that $HX+3HX=4H_{ij}{\gamma}^iX_j+4H_{ij}{\epsilon}_{ijk}X_k{\gamma}^1{\gamma}^2{\gamma}^3$. The term linear in ${\gamma}^0$ reads $-\frac{16}{3}{\gamma}^0eH_{ij}{\epsilon}_{ijk}X_k$. This must vanish. Hence either $H_{ij}=0$ or $e=0$. But since we are interested in  a non-zero Myers term we take $H_{ij}=0$. We get then
\begin{eqnarray}
S_{ij}{\gamma}^iX_j&=&\frac{16}{9}\bigg(-a^2+\frac{9}{16}e^2+3ia(c+\frac{3e}{4}){\gamma}^1{\gamma}^2{\gamma}^3\bigg)X.
\end{eqnarray}
The cubic product must also vanish thus we get 
\begin{eqnarray}
c=-\frac{3e}{4}. 
\end{eqnarray}
In other words we must have
\begin{eqnarray}
S_{ij}=(e^2-\frac{16}{9}a^2){\delta}_{ij}.
\end{eqnarray}
\section{Star Products on ${\bf S}^2_N$ And ${\bf R}^2_{\theta}$ and The Flattening Limit }
In this appendix we will show how does the star product on the fuzzy sphere approach the star product on the Moyal-Weyl plane in the flattening limit. We will follow \cite{Alexanian:2000uz}.

The commutation relations on the sphere read 
\begin{eqnarray}
[x_a,x_b]=\frac{iR}{\sqrt{c_2}}{\epsilon}_{abc}x_c.
\end{eqnarray}
We define the stereographic projections $A$ and $A^{+}$ in terms of the operators $x_a$ as follows
\begin{eqnarray}
A=\frac{1}{2}(x_1-ix_2){B}~,~A^{+}=\frac{1}{2}{B}(x_1+ix_2)~,~{B}=\frac{2}{R-x_3}.\label{because1}
\end{eqnarray}
The commutation relation $[x_1,x_2]=iRx_3/\sqrt{c_2}$  takes now the simpler form 
\begin{eqnarray}
[A,A^{+}]=F(|A|^2)~,~F(|A|^2)=\alpha {B}\bigg[ 1+|A|^2-\frac{\alpha
 B}{4}|A|^2-\frac{R{B}}{2}\bigg]~,~\alpha=\frac{{\theta}^2}{R}~,~|A|^2=AA^{+}.\nonumber\\
\end{eqnarray}
The constraint $\sum_a x_a^2 =R^2$ reads in terms of the new variables 
\begin{eqnarray}
\frac{\alpha}{4}{\beta}{B}^2-({\beta}+\frac{\alpha}{2}){B}+1+|A|^2=0~,~\beta
=R +\alpha |A|^2. 
\end{eqnarray}
This quadratic equation can be solved and one finds the solution
\begin{eqnarray}
B{\equiv}B(|A|^2)=\frac{2}{\alpha}+\frac{1}{R+\alpha |A|^2}\bigg[1-\sqrt{1+\frac{4R^2}{{\alpha}^2}+\frac{4R}{\alpha}|A|^2}\bigg].
\end{eqnarray}
For large $N$  we have $\frac{1}{\beta}=\frac{1}{R}[1-\alpha \frac{|A|^2}{R}]+O({\alpha}^2)$ and hence $\frac{\alpha}{2}{B}=\frac{\alpha}{2R}(1+|A|^2)$ or equivalently $
[A,A^{+}]=\frac{1}{2\sqrt{c_2}}(1+|A|^2)^2+O({\alpha}^2)$. From the formula $|A|^2=L_{-}(\sqrt{c_2}-L_3)^{-2}L_{+}$ it is easy to
find the spectrum of the operator $F(|A|^2)$. This  is given by 
\begin{eqnarray}
F(|A|^2)|s,m>=F({\lambda}_{
s,m})|s,m>~,~s=\frac{N-1}{2}. 
\end{eqnarray}   
\begin{eqnarray}
{\lambda}_{s,m}=\frac{c_2-m(m+1)}{(\sqrt{c_2}-m-1)^2}=\frac{n(N-n)}{(\sqrt{c_2}+s-n)^2}={\lambda}_{n-1}~,~m=-s,...,+s~,~n=s+m+1=1,...,N.\nonumber\\
\end{eqnarray}
Now we introduce ordinary creation and annihilation operators ${a}$
and ${a}^{+}$ which are defined as usual by $[a,a^{+}]=1$ with the
canonical basis $|n>$ of the number operator ${\cal N}={a}^{+}{a}$, i.e.
${\cal N}|n>=n|n>$ , $n=0,1,...$ , and which also satisfy
$a|n>=\sqrt{n}|n-1>$ and $a^{+}|n>=\sqrt{n+1}|n+1>$. Next we embed
the $N-$dimensional Hilbert space $H_N$ generated by the eigenstates
$|s,m>$ into the infinite dimensional Hilbert space generated by the
eigenstates $|n>$ and then define the maps on $H_N$ given by 
\begin{eqnarray}
A=f_N({\cal N}+1){a}~,~A^{+}={a}^{+}f_N({\cal N}+1). 
\end{eqnarray}
It is an identity easy to check that $({\cal N}+1)f_N^2({\cal
  N}+1)=|A|^2$ and hence $({\cal N}+1)f_N^2({\cal N}+1)|n-1>=nf_N^2(n)|n-1>$ and $|A|^2|s,m>={\lambda}_{n-1}|s,m>$. In other words we can identify the first $N$ states $|n>$ in the infinite dimensional Hilbert space of the harmonic oscillator  with the states of $H_N$ via
\begin{eqnarray}
|s,m>{\leftrightarrow}|n-1=s+m>.
\end{eqnarray}
As a consequence we have the result
$f_N(n)=\sqrt{\frac{{\lambda}_{n-1}}{n}}$ which clealry indicates that
the above map is well define (as it should be) only for states
$n{\leq}N$. For example $A|0>=f_L(N+1)a|0>=0$ because $a|0>=0$ but
also because
$A|0>=\frac{1}{2}(x_1-ix_2)B(|A|^2)|s,-s>=B({\lambda}_{s,-s})\frac{R}{2\sqrt{c_2}}L_{-}|s,-s>=0$
. The above map also vanishes identically on $|s,s>=|N-1>$ as one
might check. The relation between $F$ and $f_N$ is easily found to be
given by 
\begin{eqnarray}
F({\lambda}_n)=(1+n)f_N^2(n+1)-nf_N^2(n). 
\end{eqnarray}
Now we define the corresponding coherent states by
\begin{eqnarray}
&&|z;N>=\frac{1}{\sqrt{M_N(x)}}\sum_{n=0}^{N-1}\frac{z^n}{\sqrt{n!}[f_N(n)]!}|n>~,~M_N(x)=\sum_{n=0}^{N-1}\frac{(x)^n}{n!([f_N(n)]!)^2}~,~x=|z|^2\nonumber\\
&&[f_N(n)]!=f_N(0)f_N(1)...f_N(n-1)f_N(n).
\end{eqnarray}
By construction these states are normalized and they are such that 
\begin{eqnarray}
A|z;N>=z|z;N>-\frac{1}{\sqrt{M_N(|z|^2)}}\frac{z^{N}}{\sqrt{(N-1)!}[f_N(N-1)]!}|N-1>.
\end{eqnarray}
In the large $N$ limit we can check that $M_N(x){\longrightarrow}(N-1)(1+x)^{N-2}$ and $\sqrt{(N-1)!}[f_N(N-1)]!{\longrightarrow}\sqrt{{\pi}(N-1)}$ and hence $A|z;N>{\longrightarrow}z|z;N>$ which means that $|z;N>$ becomes exactly an $A-$eigenstate.

As it is the case with standard coherent states the above states $|z;N>$ are not orthonormal since  $<z_1;N|z_2;N>=M_N(|z_1|^2)^{-\frac{1}{2}}M_N(|z_2|^2)^{-\frac{1}{2}}M_N(\bar{z}_1z_2)$. Using this result as well as the completeness relation $\int d{\mu}_N(z,\bar{z})|z;N><z;N|=1$ where $d{\mu}_N(z,\bar{z})$ is the corresponding measure we can deduce the identity
\begin{eqnarray}
M_N(1)=\int d{\mu}_N(z,\bar{z}) \frac{M_N(z)M_N(\bar{z})}{M_N(|z|^2)}.
\end{eqnarray}
This last equation allows us to determine that the measure
$d{\mu}(z,\bar{z})$ is given by 
\begin{eqnarray}
d{\mu}_N(z,\bar{z})=iM_N(|z|^2)X_N(|z|^2)dz{\wedge}d\bar{z}=2M_N(|z|^2)X_N(|z|^2){\rho}d{\rho}d{\theta}.
\end{eqnarray}
The function $X_N$ is defined by
\begin{eqnarray}
\int_{0}^{\infty} dx~ x^{s-1}X_N(x)=\frac{{\Gamma}(s)([f_N(s-1)]!)^2}{2{\pi}}.
\end{eqnarray} 
The solution of this equation was found in \cite{Alexanian:2000uz} and it is
given by 
\begin{eqnarray}
2{\pi}X_N(x)=F_1({\gamma}+N,{\gamma}+N;N+1;-x)~,~
{\gamma}=\sqrt{c_2}-\frac{N-1}{2}. 
\end{eqnarray}
For large $N$ where $|z|^2<<N$ we can find that the behaviour of the measure $d{\mu}_N(z,\bar{z})$ coincides with the ordinary measure on ${\bf S}^2$, viz 
\begin{eqnarray}
d{\mu}_N(z,\bar{z}){\simeq}\frac{N-1}{2{\pi}}\frac{idz{\wedge}d\bar{z}}{(1+|z|^2)^2}.
\end{eqnarray}
With the help of this coherent state $|z;N>$ we can therefore
associate to every operator $O$ a function ${O}_N(z,\bar{z})$ by
setting $<z;N|O|z;N>=O_N(z,\bar{z})$.  It is therefore clear that the
trace of the operator $O$ is mapped to the inetgral of the function
$O_N$ against the measure $d{\mu}_N(z,\bar{z})$, i.e 
\begin{eqnarray}
TrO=\int d{\mu}_N(z,\bar{z})O_N(z,\bar{z}).
\end{eqnarray}
  Given now two such operators $O$ and $P$ their product is associated to the star product of their corresponding functions, namely      
\begin{eqnarray}
&&O_N*P_N(z,\bar{z}){\equiv}<z;N|OP|z;N>=\int
d{\mu}(\eta,\bar{\eta})O_N(\eta,\bar{z})\frac{M_N(\bar{z}{\eta})M_N(\bar{\eta}z)}{M_N(|z|^2)M_N(|\eta|^2)}P_N(z,\bar{\eta})\nonumber\\
&&O_N(\eta,\bar{z})=(<z;N|\eta;N>)^{-1}<z;L|O|\eta;L>~,~P_N(z,\bar{\eta})=(<\eta;N|z;N>)^{-1}<\eta;N|P|z;N>. \nonumber\\
\end{eqnarray}
The large $N$ limit of this star product is given by the Berezin star product on the sphere \cite{Berezin:1974du}, namely
\begin{eqnarray}
O_N*P_N(z,\bar{z})=\frac{N-1}{2{\pi}}\int \frac{id{\eta}{\wedge}d\bar{\eta}}{(1+|\eta|^2)^2}O_N(\eta,\bar{z})\bigg[\frac{(1+\bar{z}\eta)(1+\bar{\eta}z)}{(1+|z|^2)(1+|\eta|^2)}\bigg]^{N-2}P_N(z,\bar{\eta}).
\end{eqnarray}
Finally we comment on the planar (flattening) limit of the above star product which
in fact is the central point of our discussion here. In this limit we
have $x_3=-R $ where the minus sign is due to our definition of the
stereographic coordinate $B$ in (\ref{because1}). The stereographic coordiantes $B$, $A$ and $A^{+}$ are scaled in this limit as 
\begin{eqnarray}
B=\frac{1}{R}~,~A=\frac{1}{2R}\hat{A}~,~\hat{A}=\hat{x}_1-i\hat{x}_2~,~{A}^{+}=\frac{1}{2R}\hat{A}^{+}~,~\hat{A}^+=\hat{x}_1+i\hat{x}_2~,~\hat{x}_i=x_i.\label{this}
\end{eqnarray}
This scaling means in particular that the coordinates $z$ and
$\bar{z}$ must scale as $z=\hat{z}/2R$ and
$\bar{z}=\bar{\hat{z}}/2R$. Since we already know that
$[A,A^{+}]=\frac{1}{2\sqrt{c_2}}(1+|A|^2)^2$ in the large $N$ limit we
can immediately conclude that $[\hat{A},\hat{A}^{+}]=2{\theta}^2$ in
this limit or equivalently $[\hat{x}_1,\hat{x}_2]=-i{\theta}^2$. Next
from the result $M_N(x){\longrightarrow}(N-1)(1+x)^{N-2}$ when
$N{\longrightarrow}{\infty}$ we can conlude that in this large $N$
limit we must have 
\begin{eqnarray}
M_N(|{z}|^2){\longrightarrow}N~e^{\frac{1}{2{\theta}^2}|\hat{z}|^2}.
\end{eqnarray} 
The measure $d{\mu}_N(z,\bar{z})$ behaves as 
\begin{eqnarray}
d{\mu}_N(z,\bar{z})=\frac{i}{4{\pi}{\theta}^2}d\hat{z}{\wedge}d\bar{\hat{z}}. 
\end{eqnarray}
Putting all these results together we obtain the Berezin star product
on the plane \cite{Berezin:1974du}, namely
\begin{eqnarray}
O*P(\hat{z},\bar{\hat{z}})=\frac{i}{4{\pi}{\theta}^2}\int d\hat{\eta}{\wedge}d\bar{\hat{\eta}} O(\hat{\eta},\bar{\hat{z}})~e^{-\frac{1}{2{\theta}^2}(\hat{z}-\hat{\eta})(\bar{\hat{z}}-\bar{\hat{\eta}})}P(\hat{z},\bar{\hat{\eta}}).
\end{eqnarray}
Lastly it is easy to check that the trace behaves in the limit as
follows 
\begin{eqnarray}
\frac{R^2}{N}Tr{\longrightarrow}\frac{i}{8{\pi}}\int d\hat{z}{\wedge}\bar{\hat{z}}.
\end{eqnarray}

\paragraph{Acknowledgment:} I would like to thank Denjoe O'Connor for extensive discussions at various  stages of this project. I would like also to thank  R.Delgadillo-Blando and Adel Bouchareb for their collaboration. The numerical simulations reported in this article were conducted on the clusters of the Dublin Institute for Advanced Studies.

\newpage

\begin{figure}[htbp]
\begin{center}
\includegraphics[width=10.0cm,angle=-90]{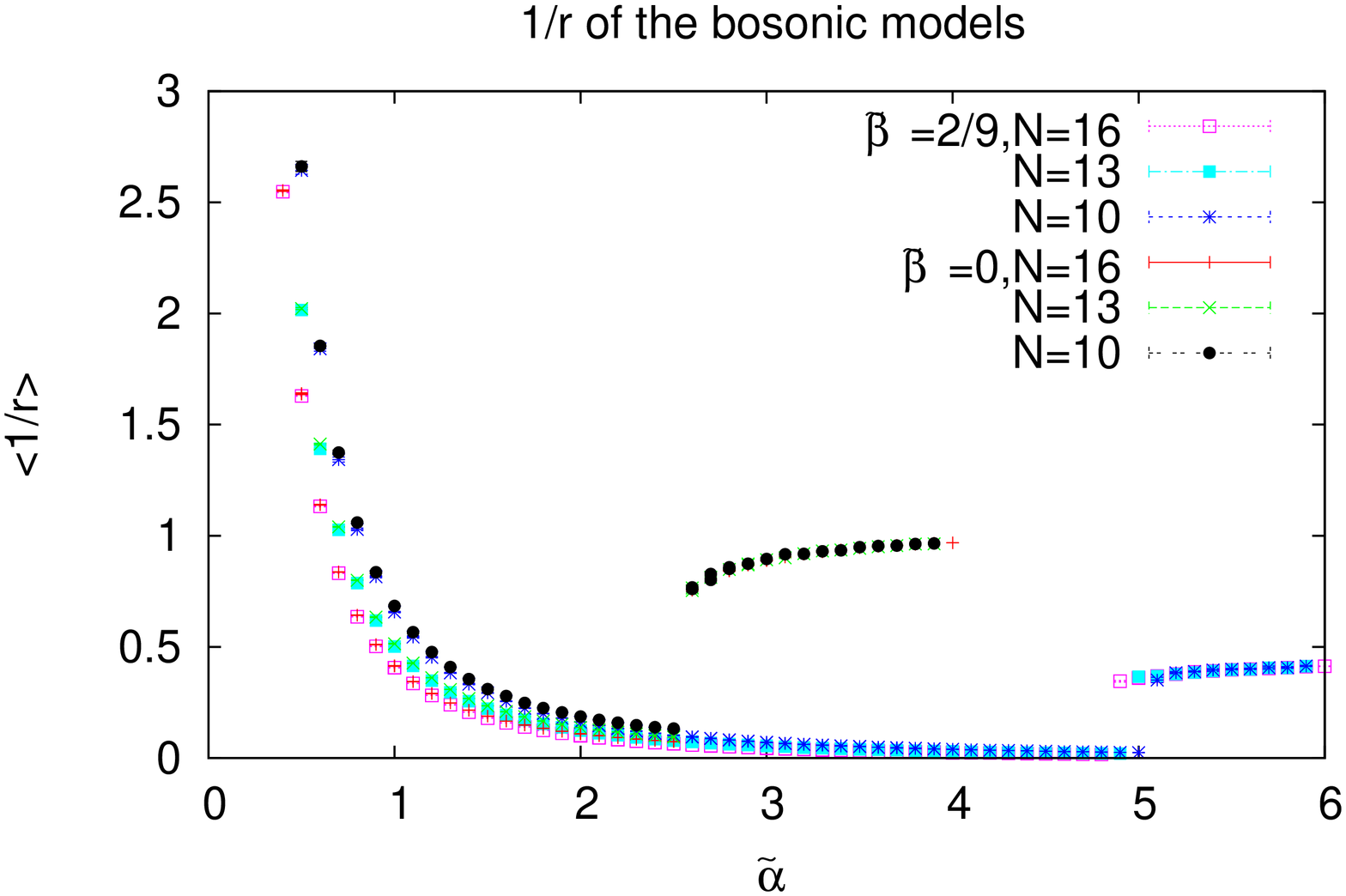}
\includegraphics[width=10.0cm,angle=-90]{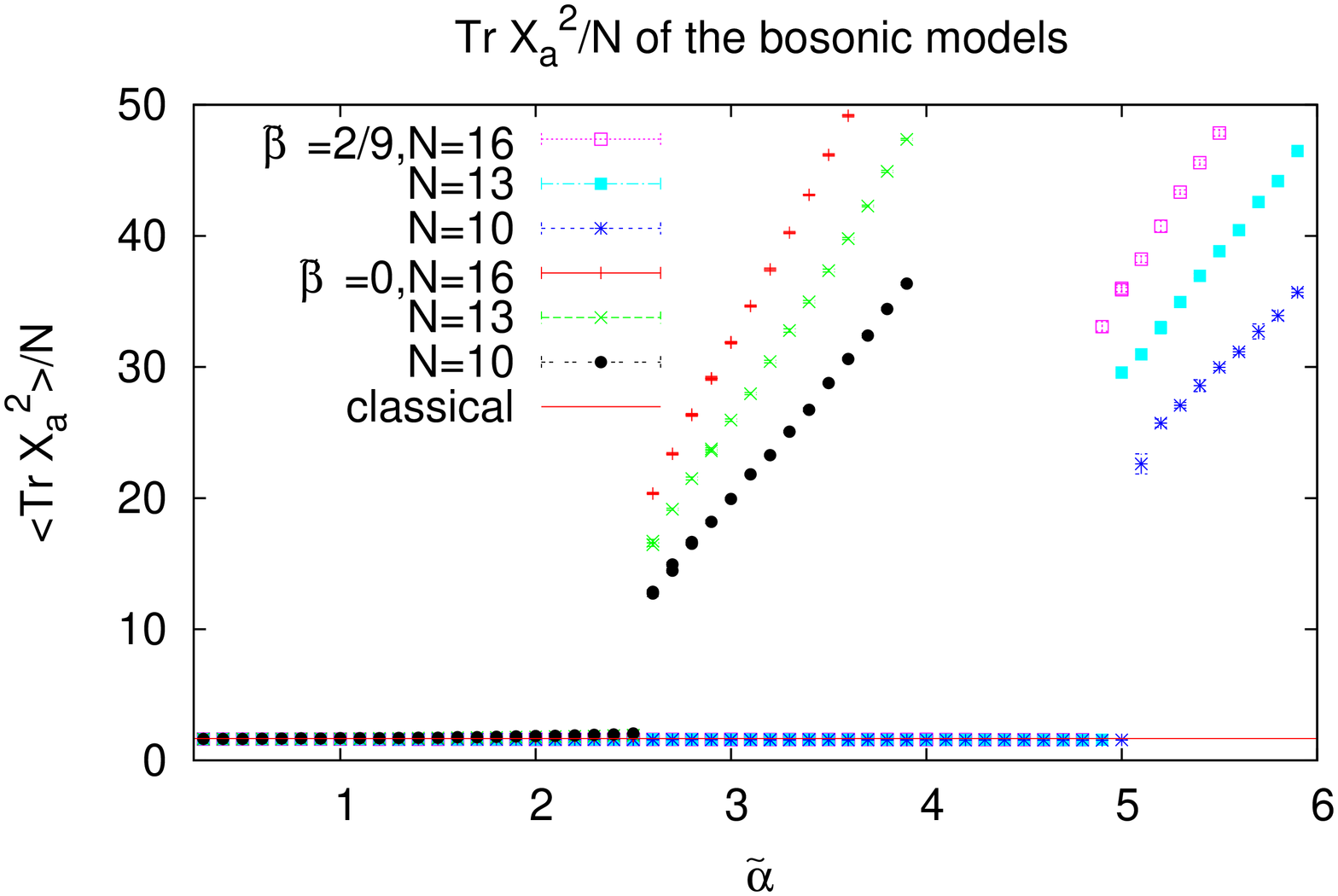}
\caption{The radius of the bosonic models. }\label{radiusB}
\end{center}
\end{figure}

\begin{figure}[htbp]
\begin{center}
\includegraphics[width=10.0cm,angle=-90]{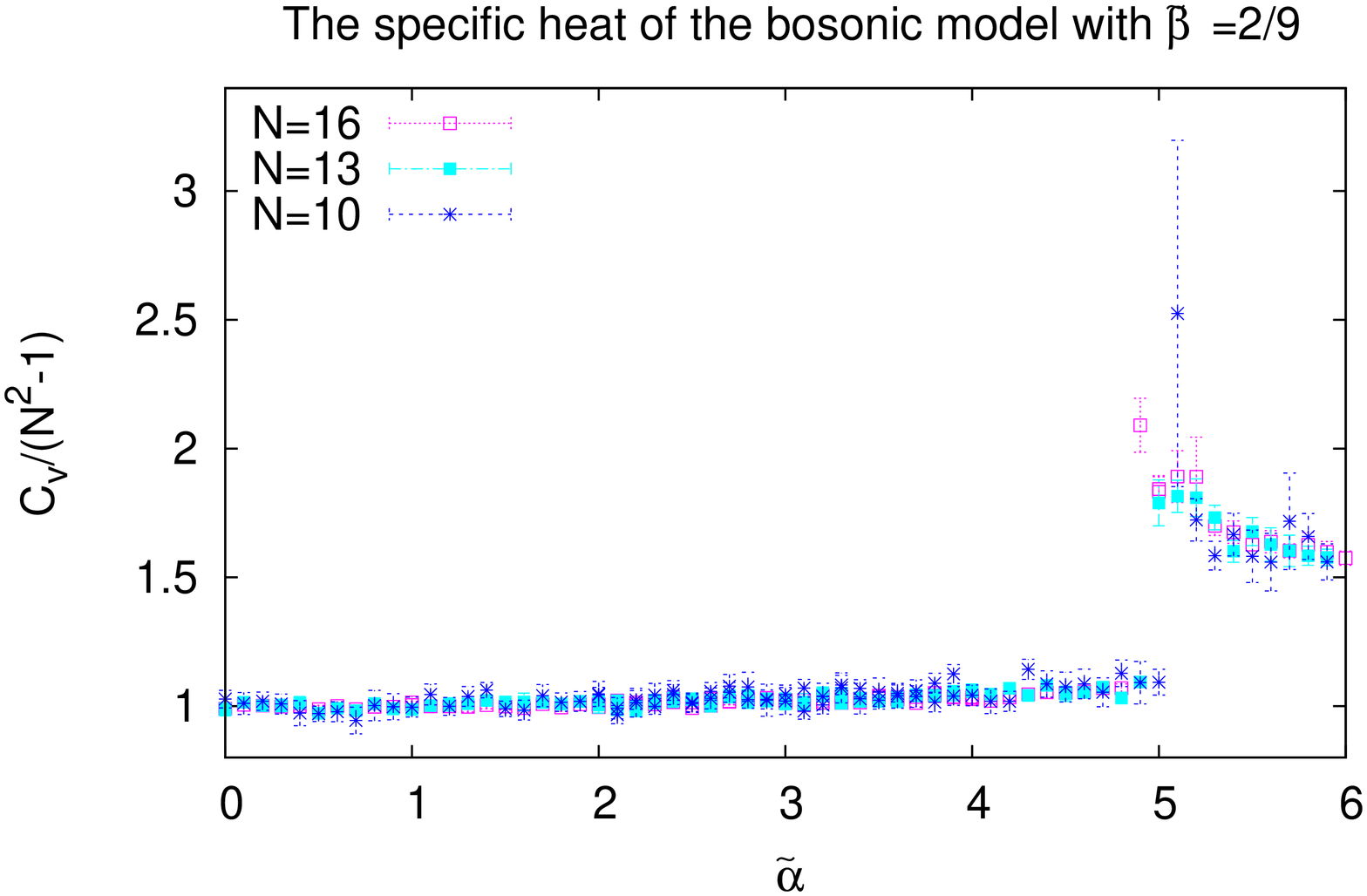}
\includegraphics[width=10.0cm,angle=-90]{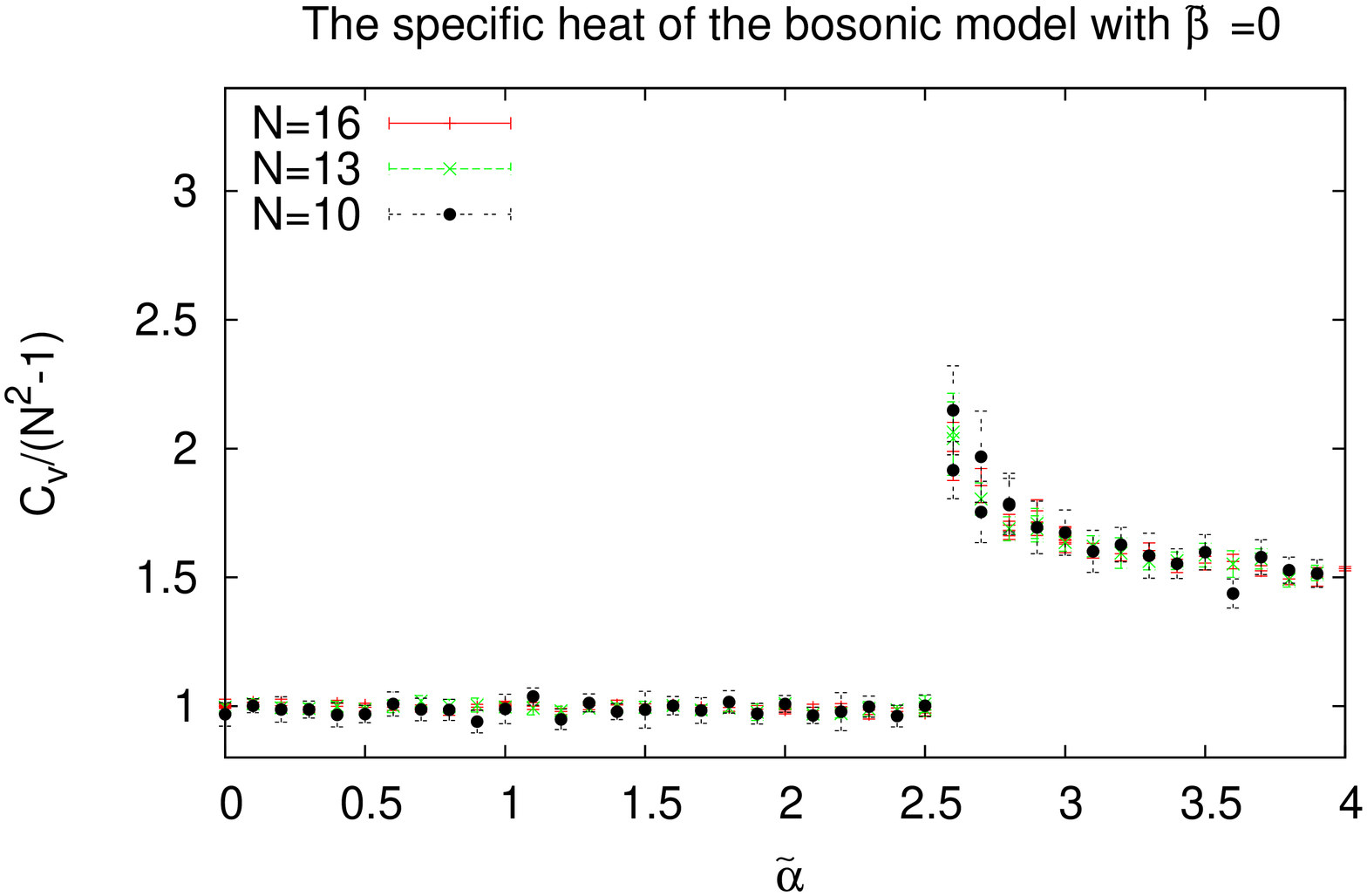}
\caption{The specific heat of the bosonic  models. }\label{cvB}
\end{center}
\end{figure}

\begin{figure}[htbp]
\begin{center}
\includegraphics[width=10.0cm,angle=-90]{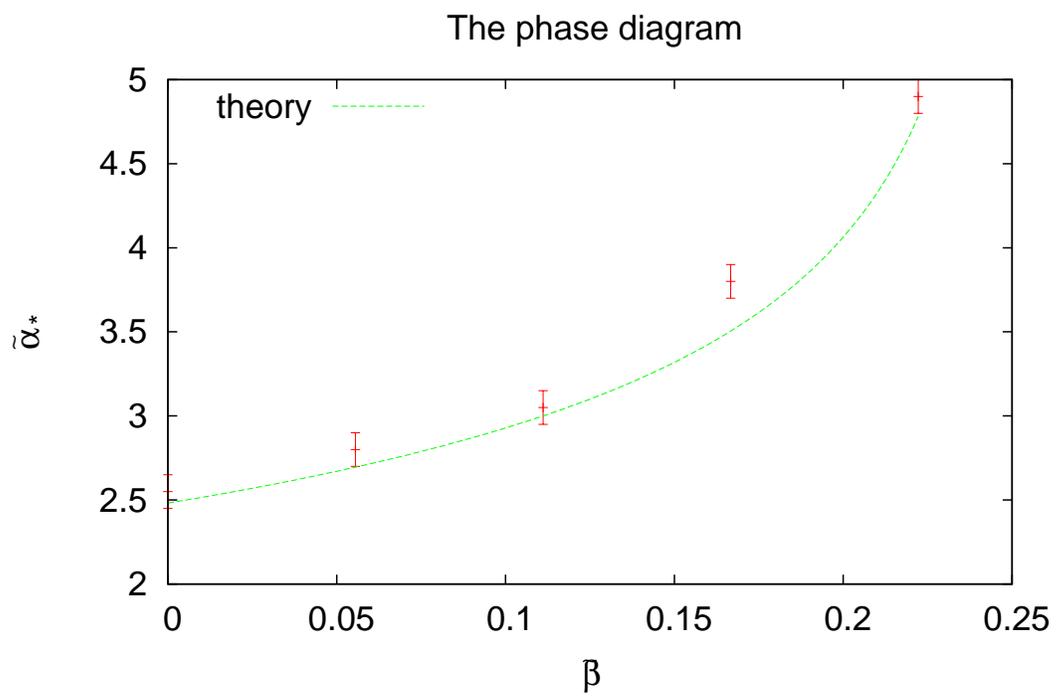}
\caption{The phase diagram of the bosonic model with generic value of $\tilde{\beta}$. }\label{phaseB}
\end{center}
\end{figure}

\begin{figure}[htbp]
\begin{center}
\includegraphics[width=10.0cm,angle=-90]{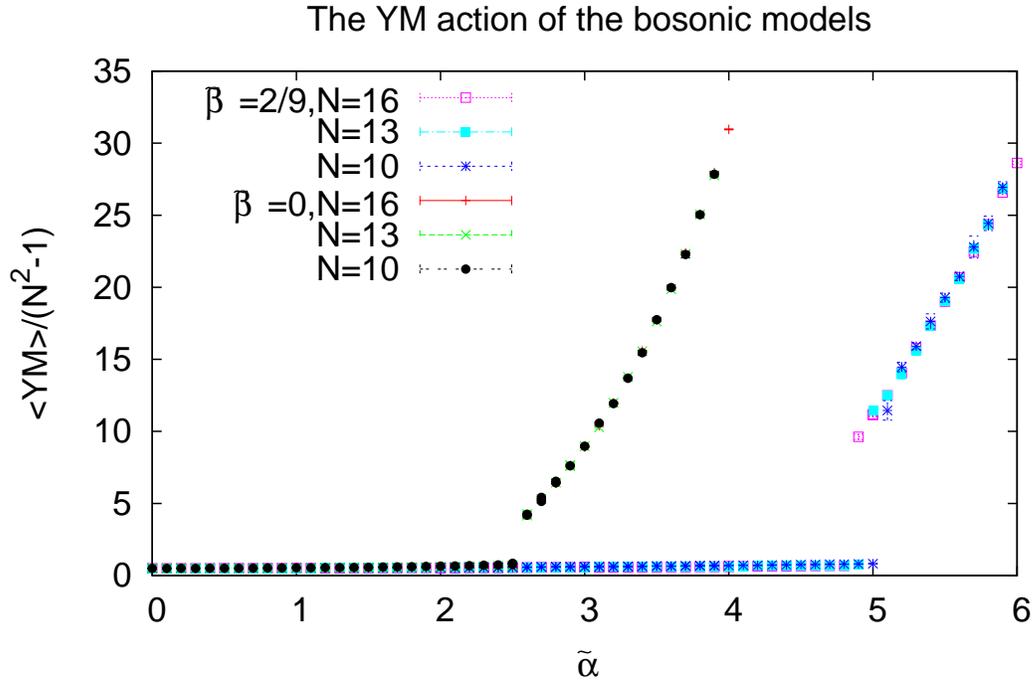}
\includegraphics[width=10.0cm,angle=-90]{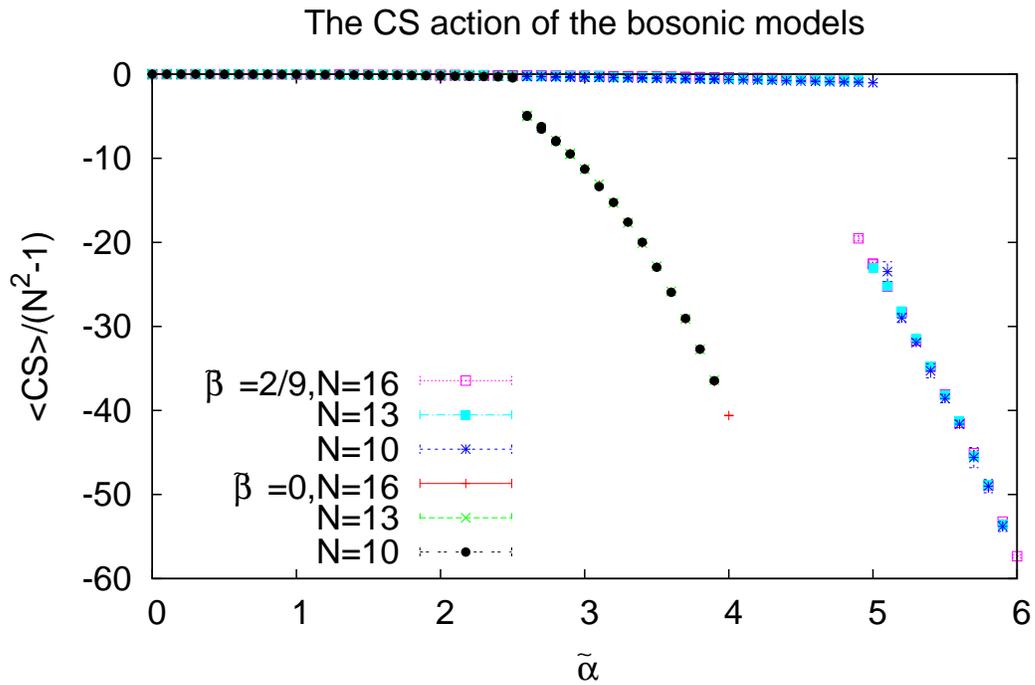}
\caption{The average of the Yang-Mills and Myers actions of the bosonic models. }\label{YMCSB}
\end{center}
\end{figure}

\begin{figure}[htbp]
\begin{center}
\includegraphics[width=10.0cm,angle=-90]{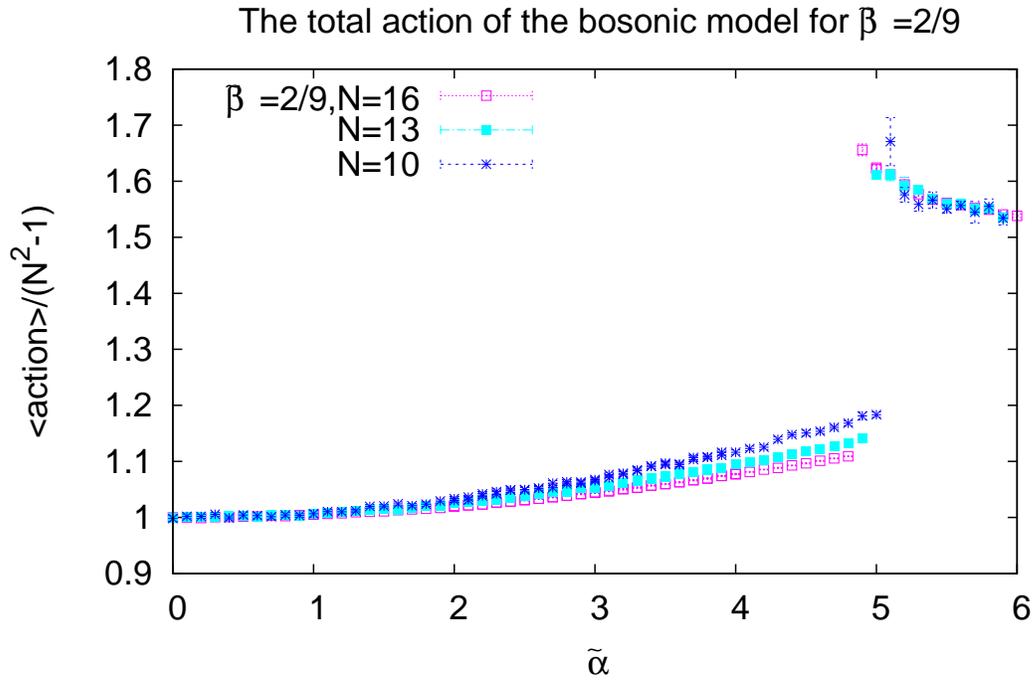}
\includegraphics[width=10.0cm,angle=-90]{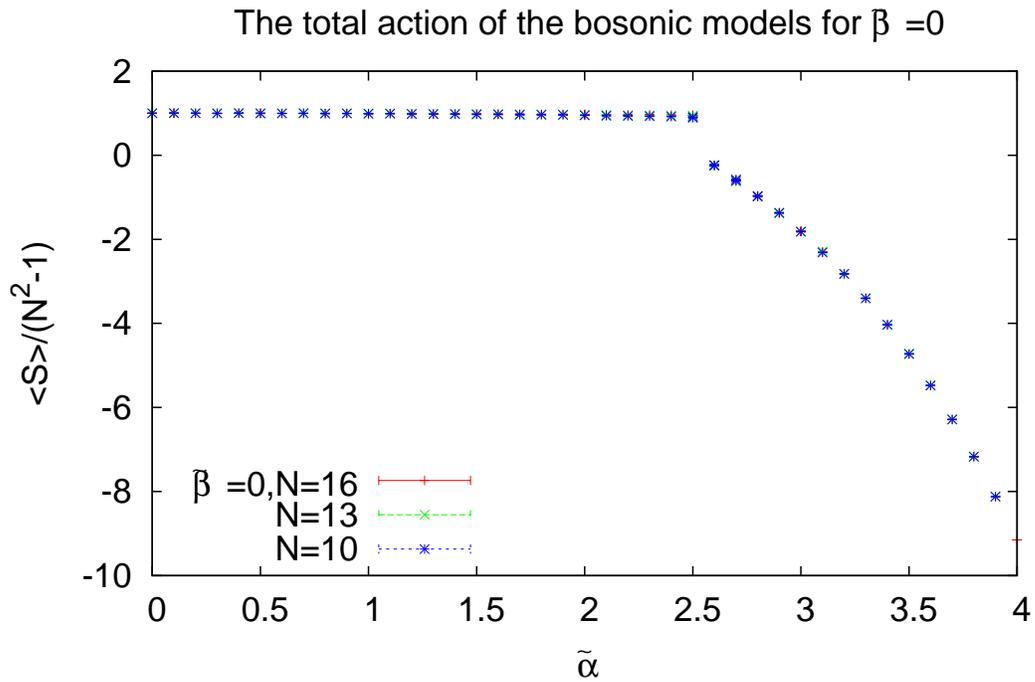}
\caption{The average of the total action of the bosonic models. }\label{actionB}
\end{center}
\end{figure}

\begin{figure}[htbp]
\begin{center}
\includegraphics[width=10.0cm,angle=-90]{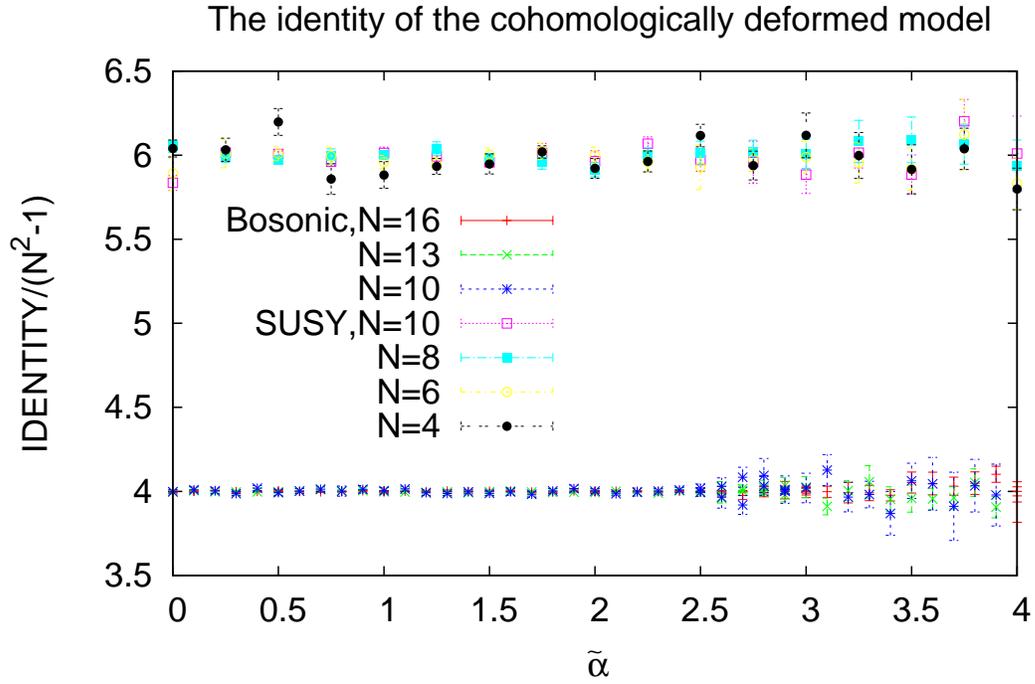}
\includegraphics[width=10.0cm,angle=-90]{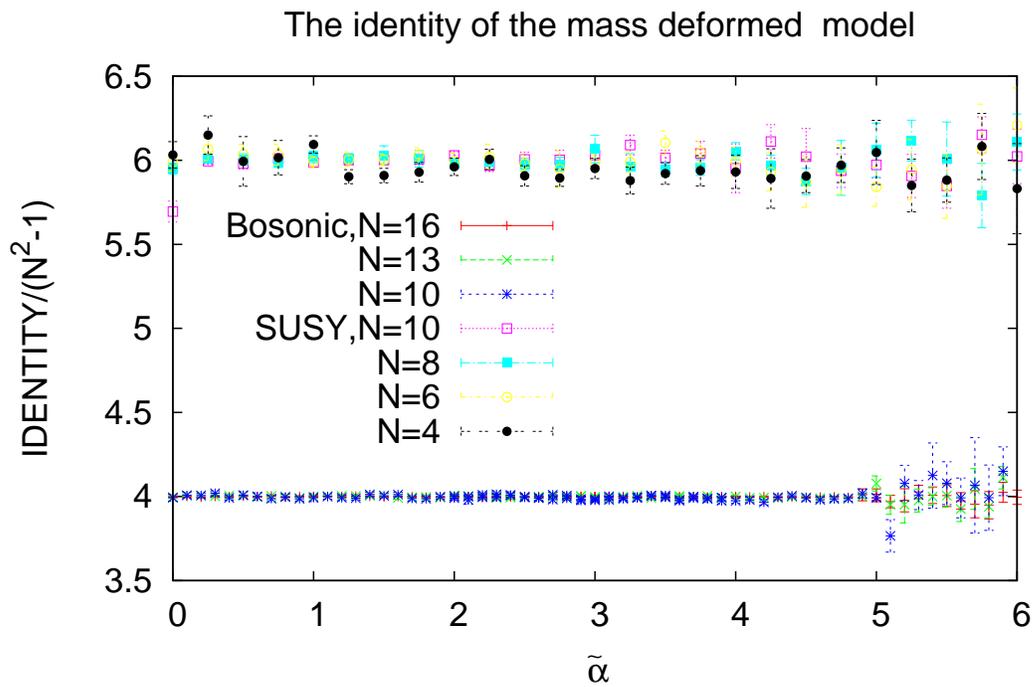}
\caption{The Schwinger-Dyson identity of the bosonic and supersymmetric models. }\label{ident}
\end{center}
\end{figure}

\begin{figure}[htbp]
\begin{center}
\includegraphics[width=10.0cm,angle=-90]{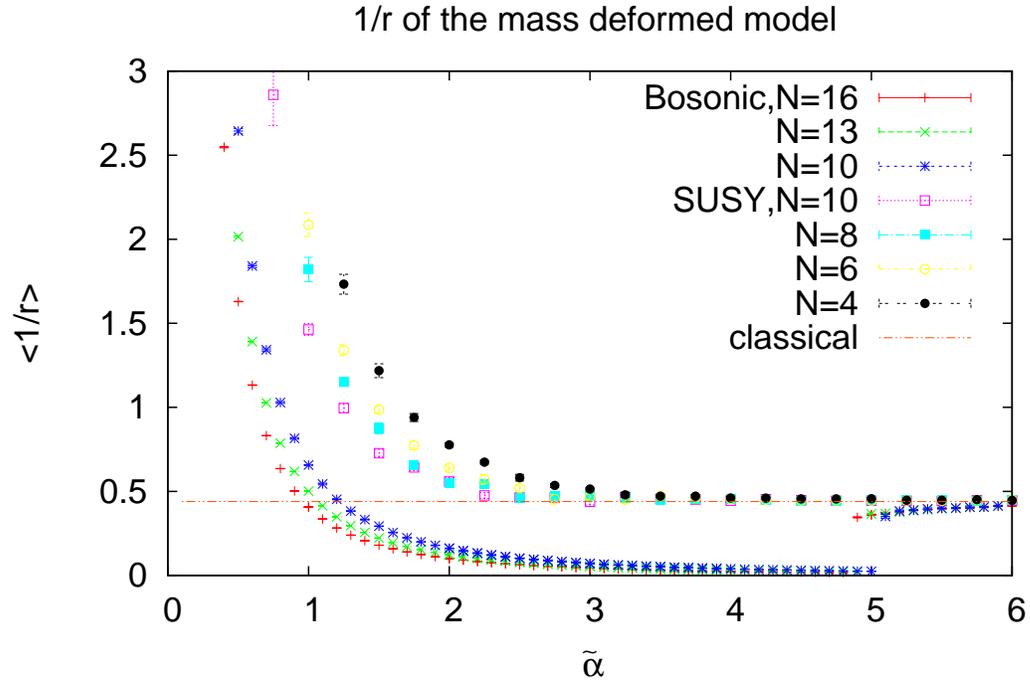}
\includegraphics[width=10.0cm,angle=-90]{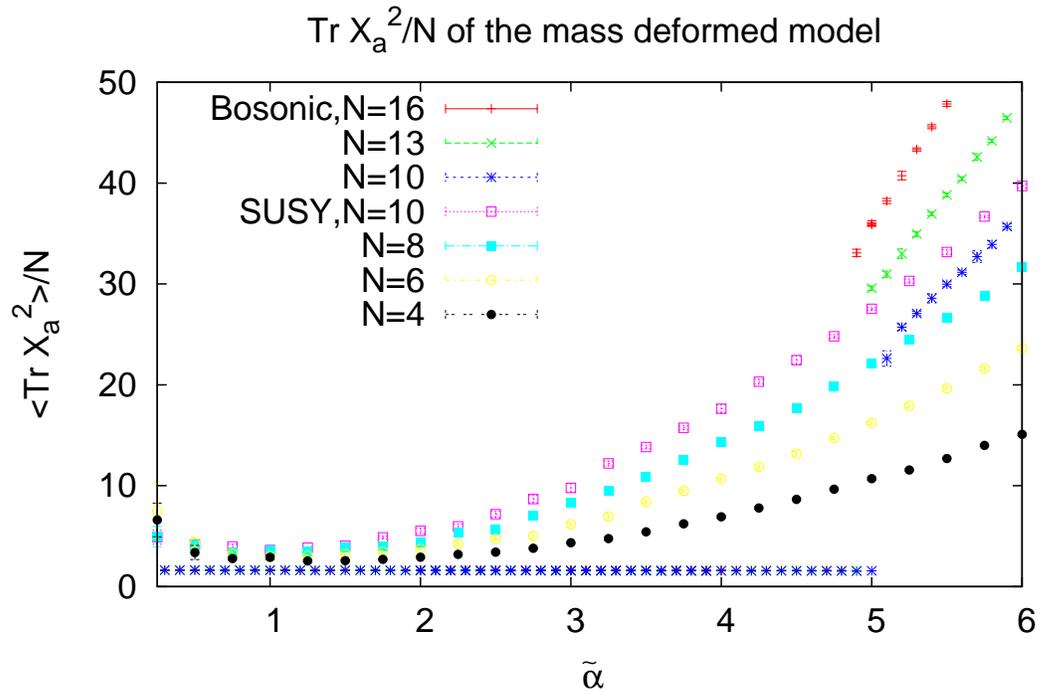}
\caption{The radius of the mass deformed model. }\label{radiusF1}
\end{center}
\end{figure}

\begin{figure}[htbp]
\begin{center}
\includegraphics[width=10.0cm,angle=-90]{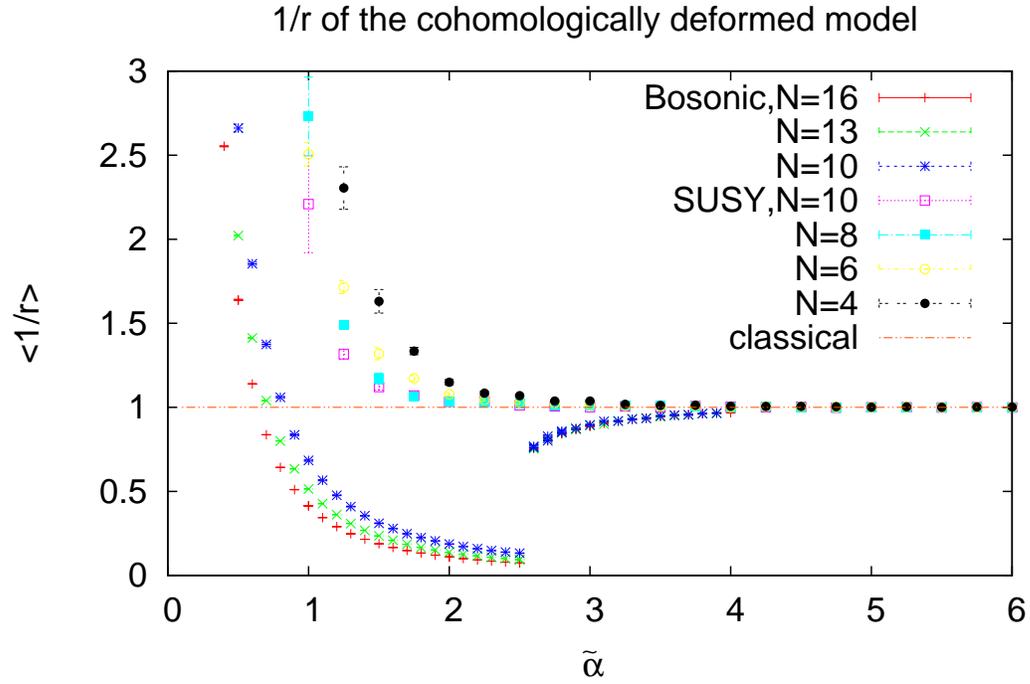}
\includegraphics[width=10.0cm,angle=-90]{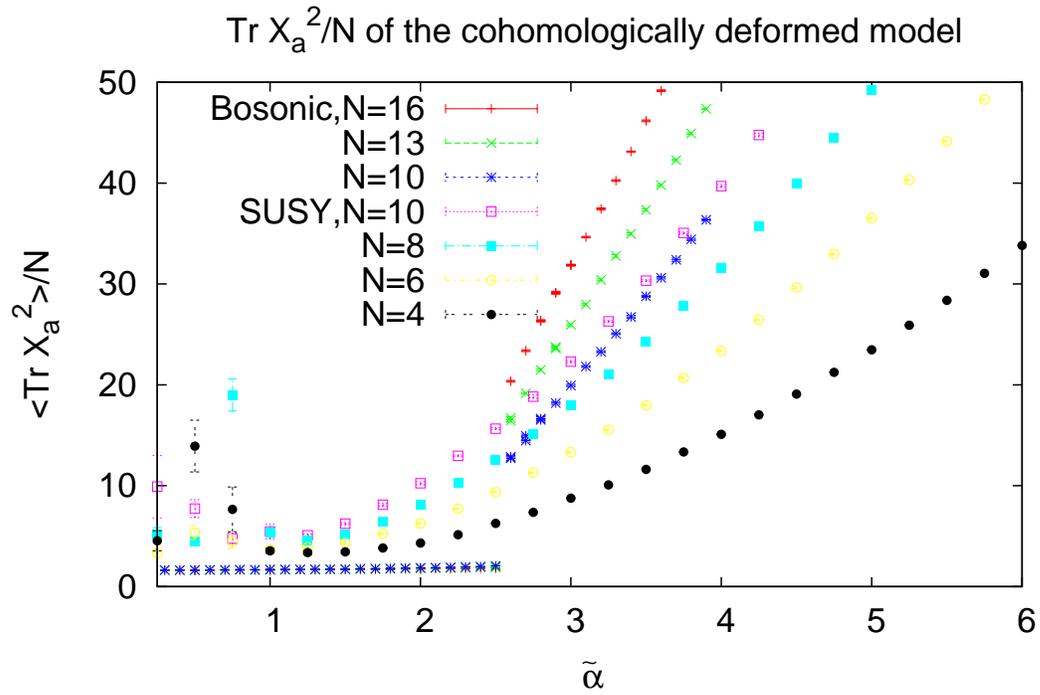}
\caption{The radius of the cohomologically deformed model. }\label{radiusF2}
\end{center}
\end{figure}

\begin{figure}[htbp]
\begin{center}
\includegraphics[width=10.0cm,angle=-90]{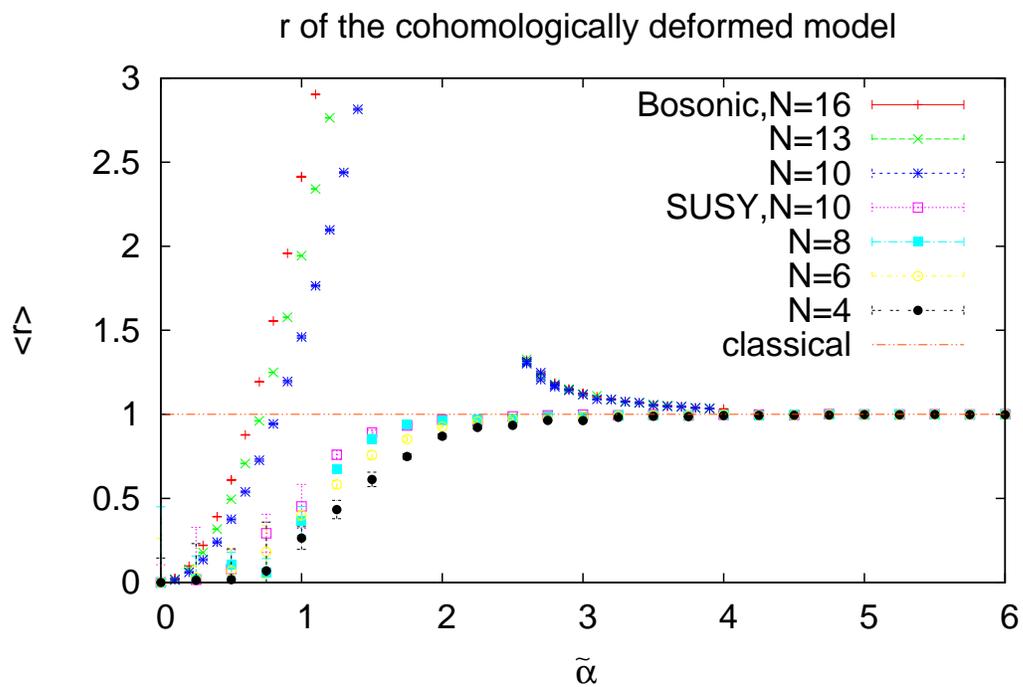}
\includegraphics[width=10.0cm,angle=-90]{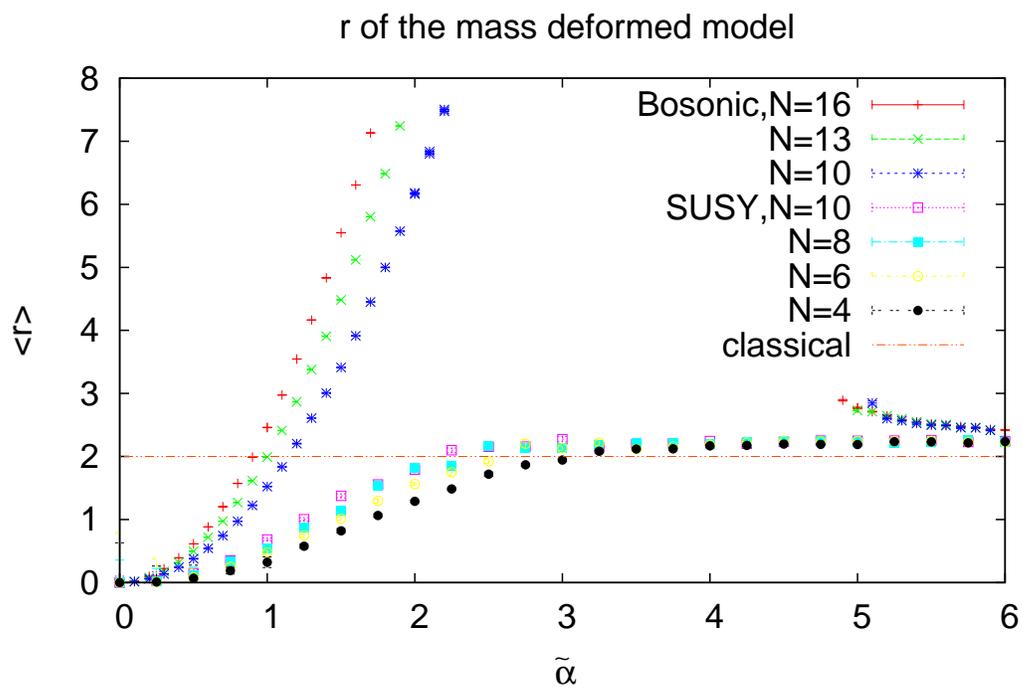}
\caption{The radius of the cohomologically deformed and mass deformed models. }\label{radiusF3}
\end{center}
\end{figure}

\begin{figure}[htbp]
\begin{center}
\includegraphics[width=10.0cm,angle=-90]{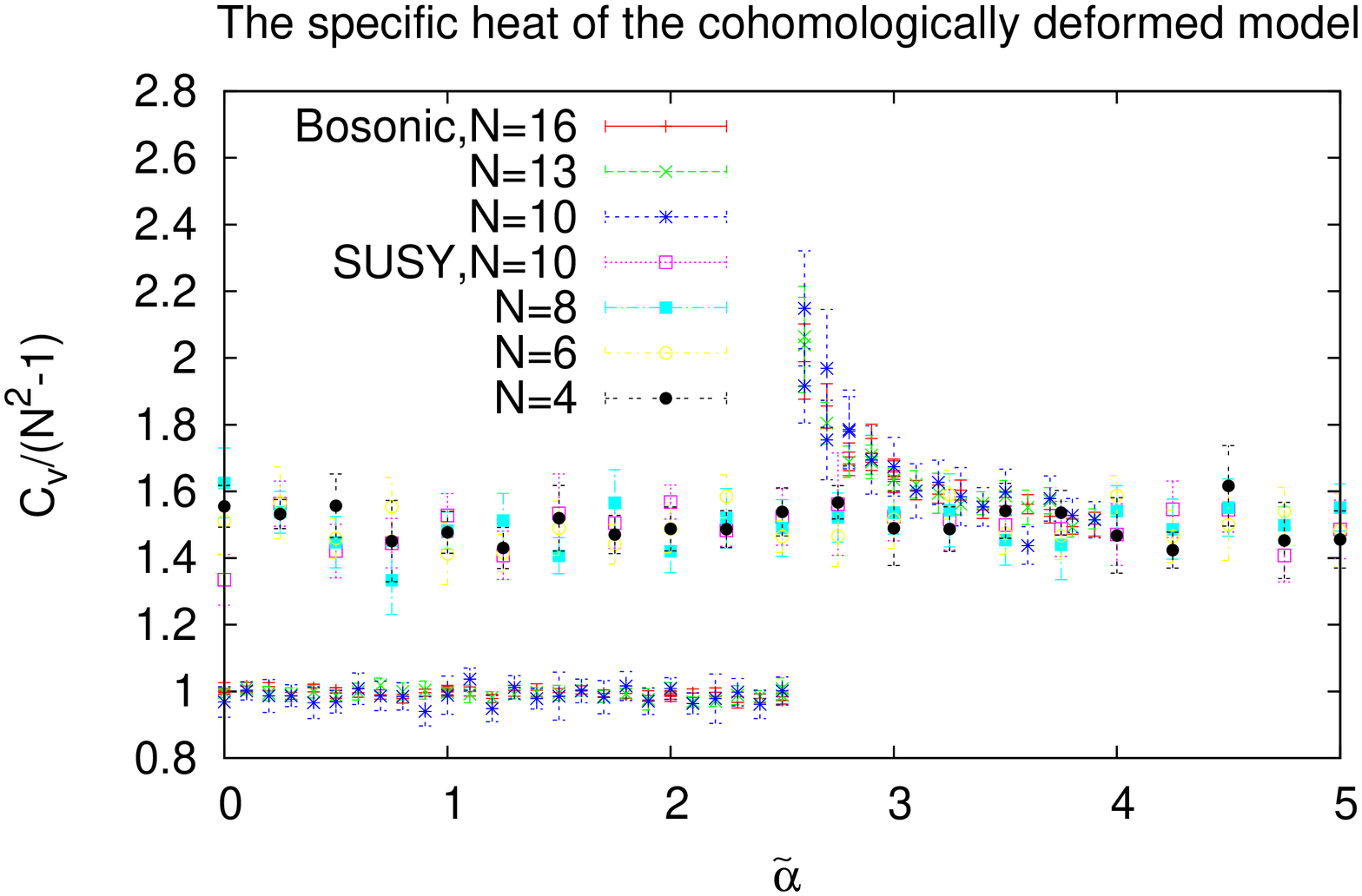}
\includegraphics[width=10.0cm,angle=-90]{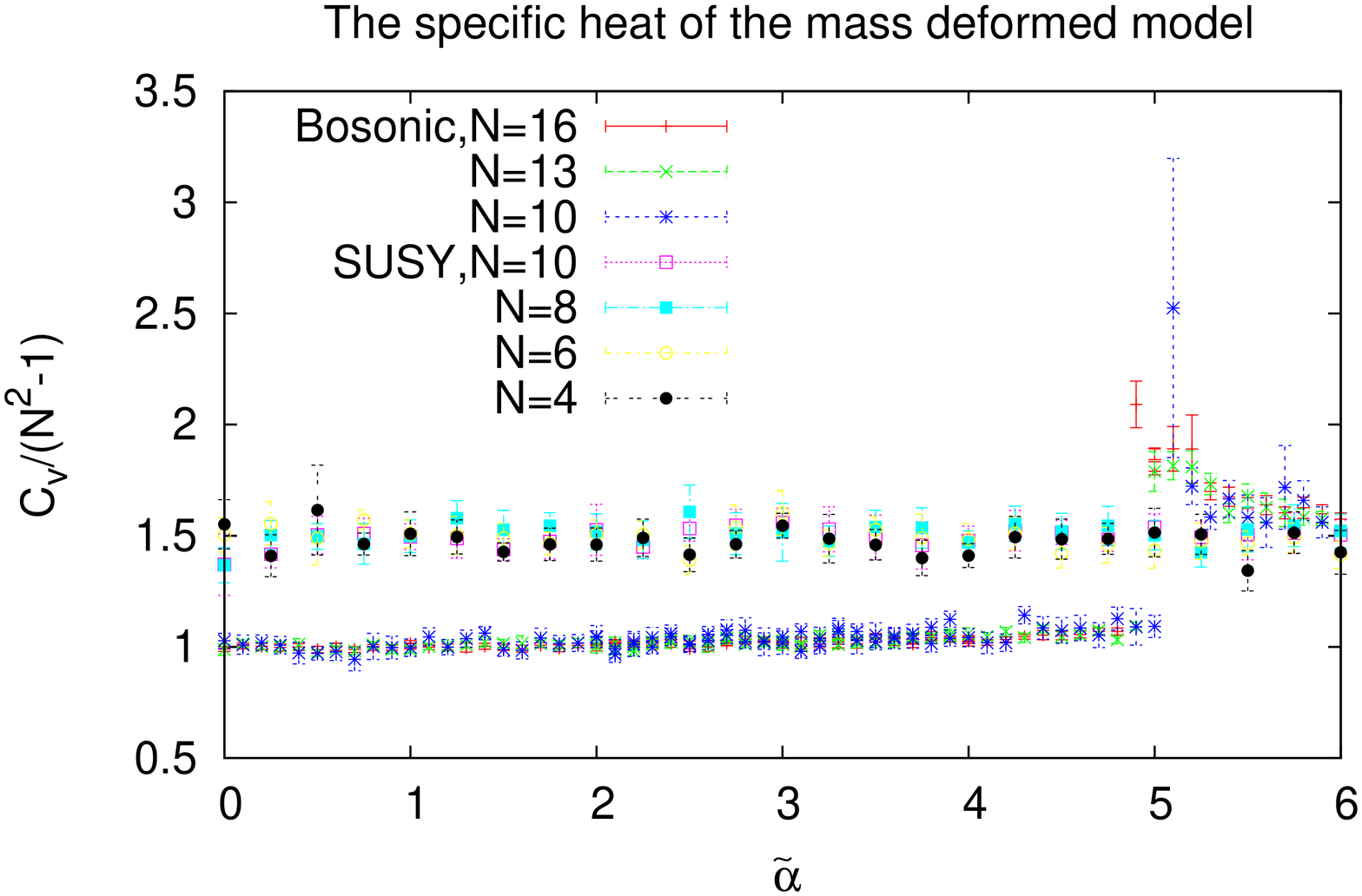}
\caption{The specific heat of the supersymmetric models. }\label{cvF}
\end{center}
\end{figure}

\begin{figure}[htbp]
\begin{center}
\includegraphics[width=10.0cm,angle=-90]{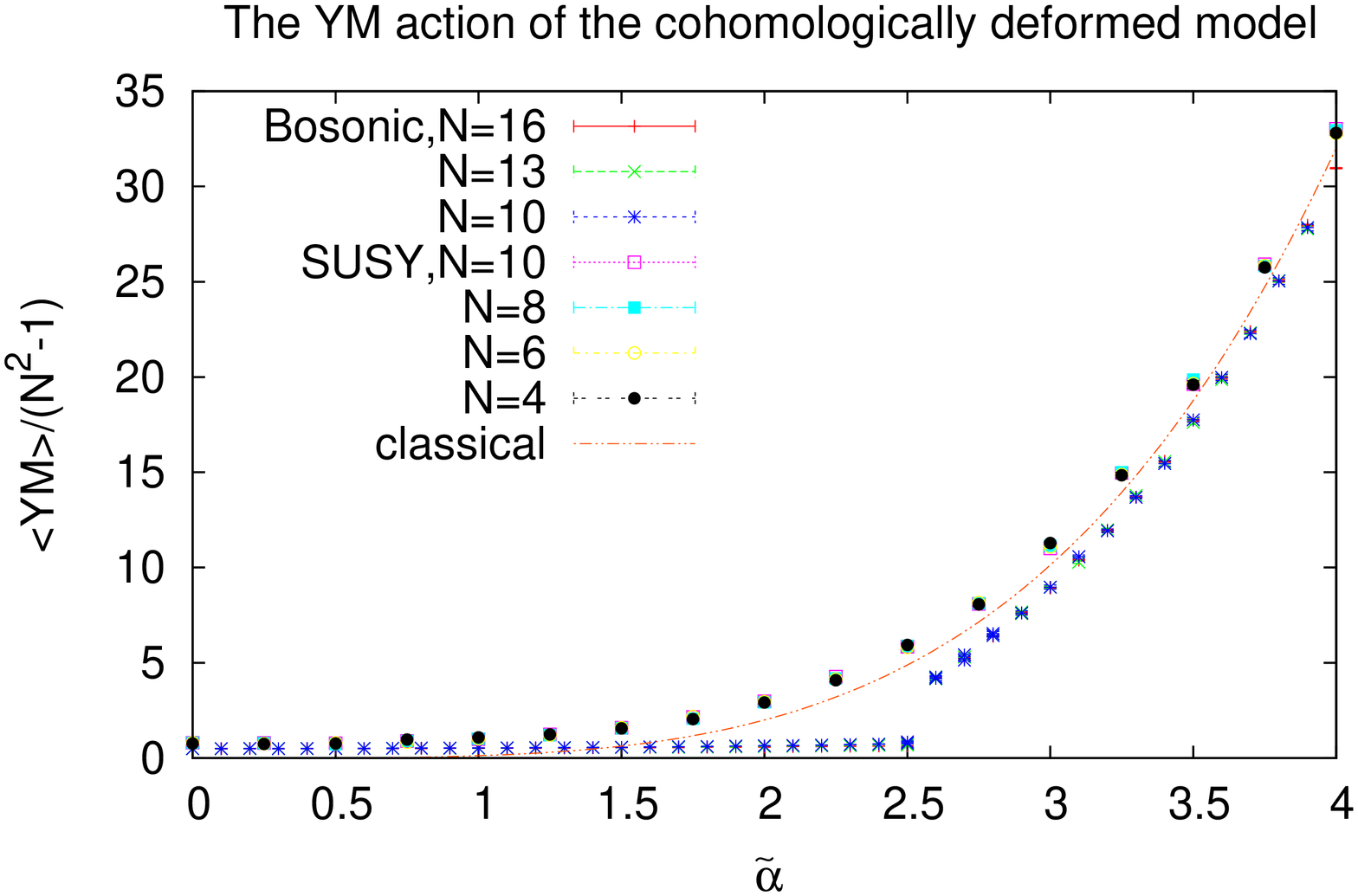}
\includegraphics[width=10.0cm,angle=-90]{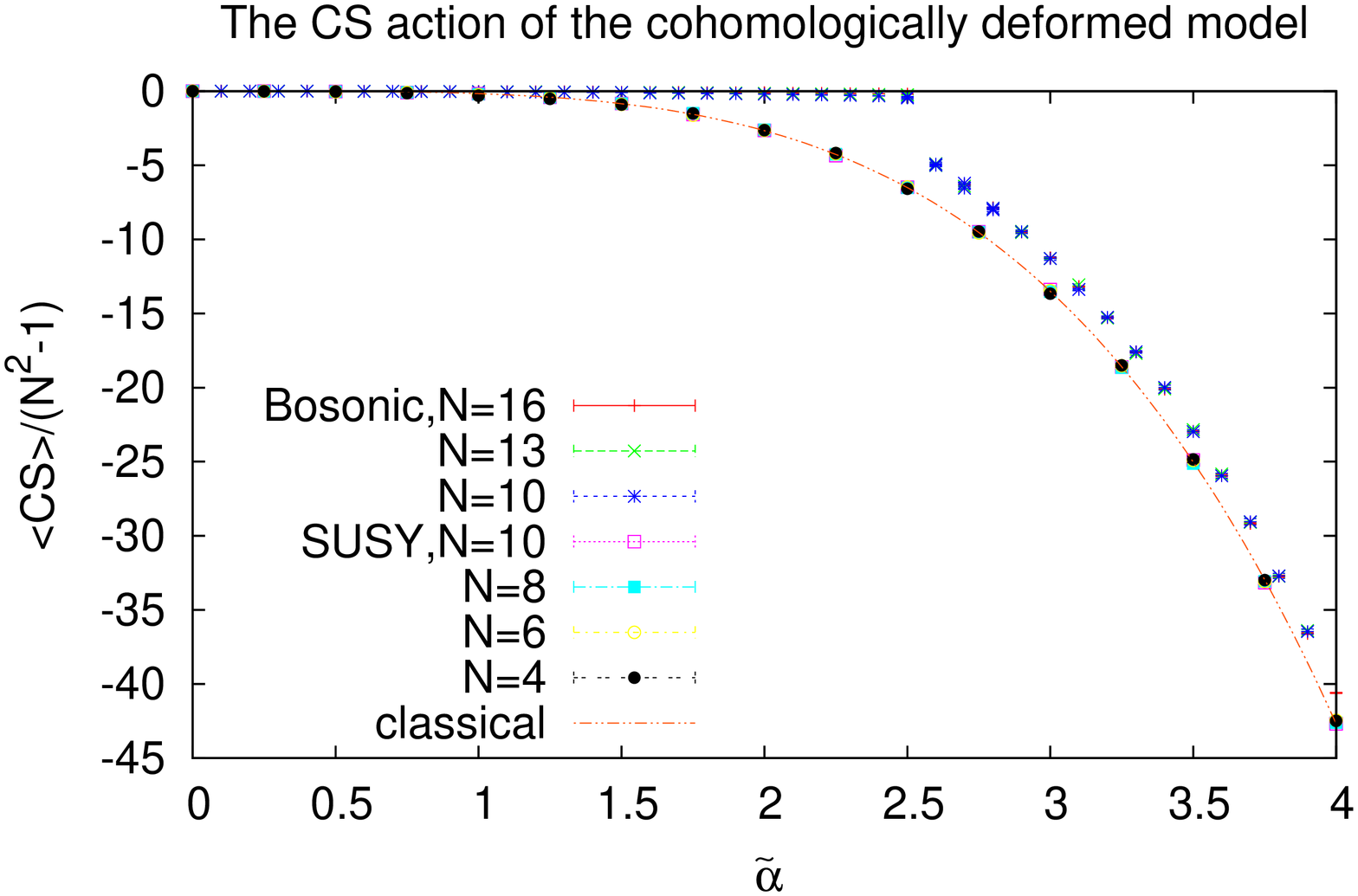}
\caption{The average of the Yang-Mills and Myers actions of the cohomologically deformed  model. }\label{YMCSF1}
\end{center}
\end{figure}

\begin{figure}[htbp]
\begin{center}
\includegraphics[width=10.0cm,angle=-90]{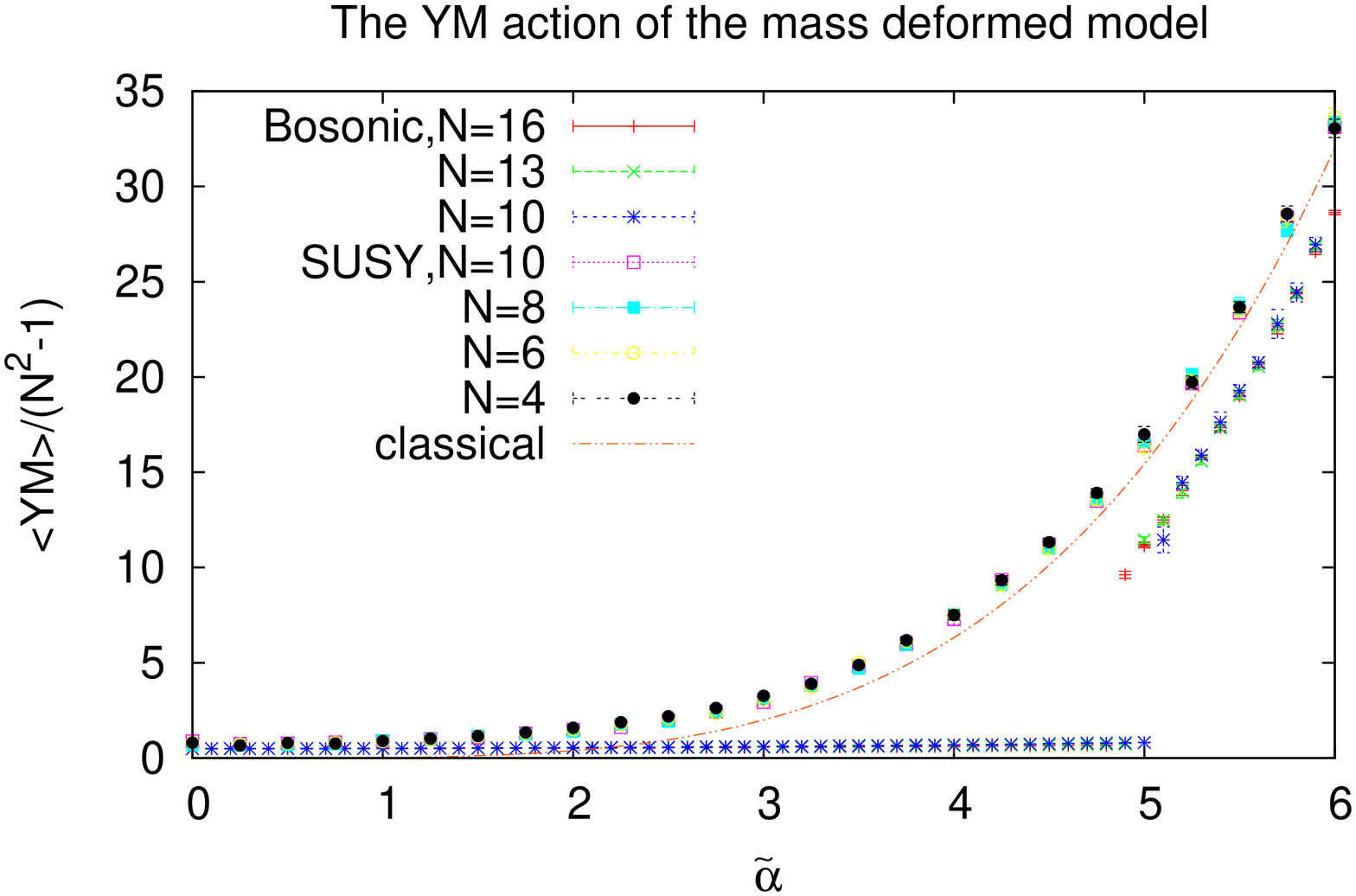}
\includegraphics[width=10.0cm,angle=-90]{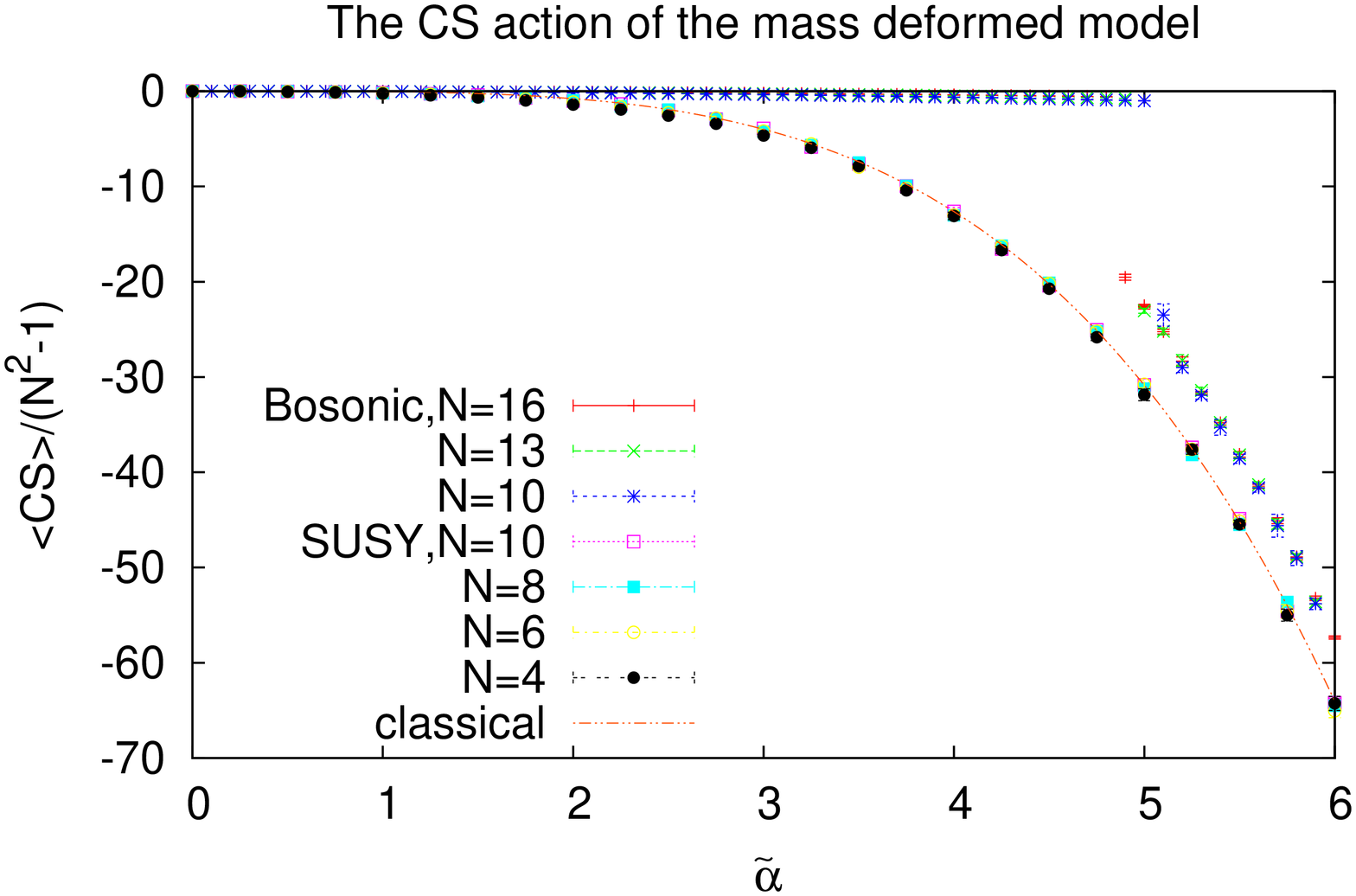}
\caption{The average of the Yang-Mills and Myers actions of the  mass deformed model. }\label{YMCSF2}
\end{center}
\end{figure}

\begin{figure}[htbp]
\begin{center}
\includegraphics[width=10.0cm,angle=-90]{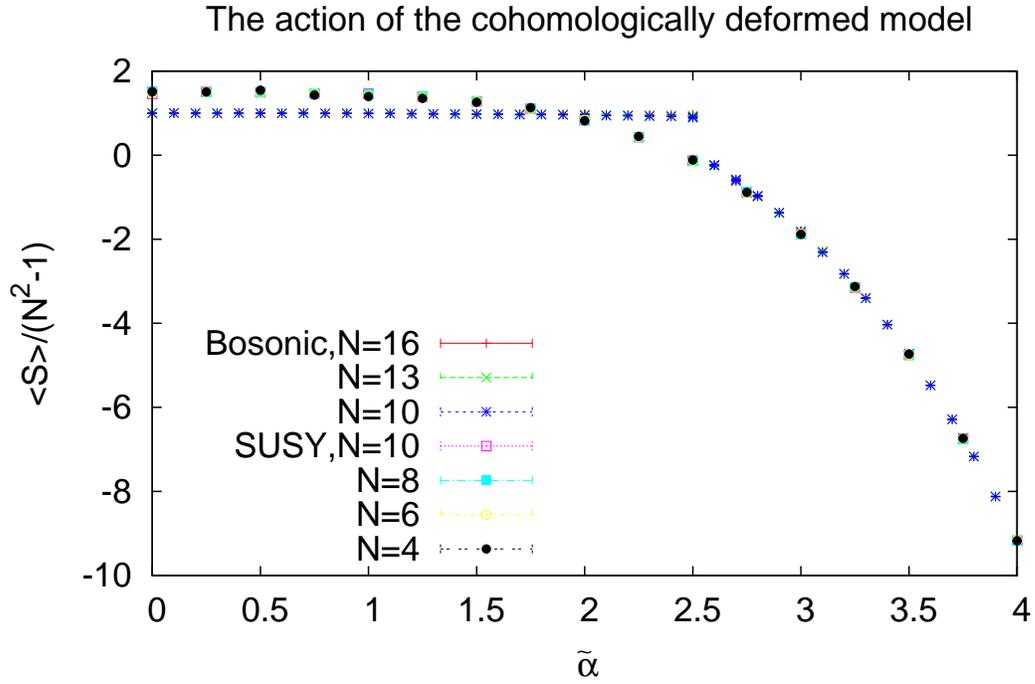}
\includegraphics[width=10.0cm,angle=-90]{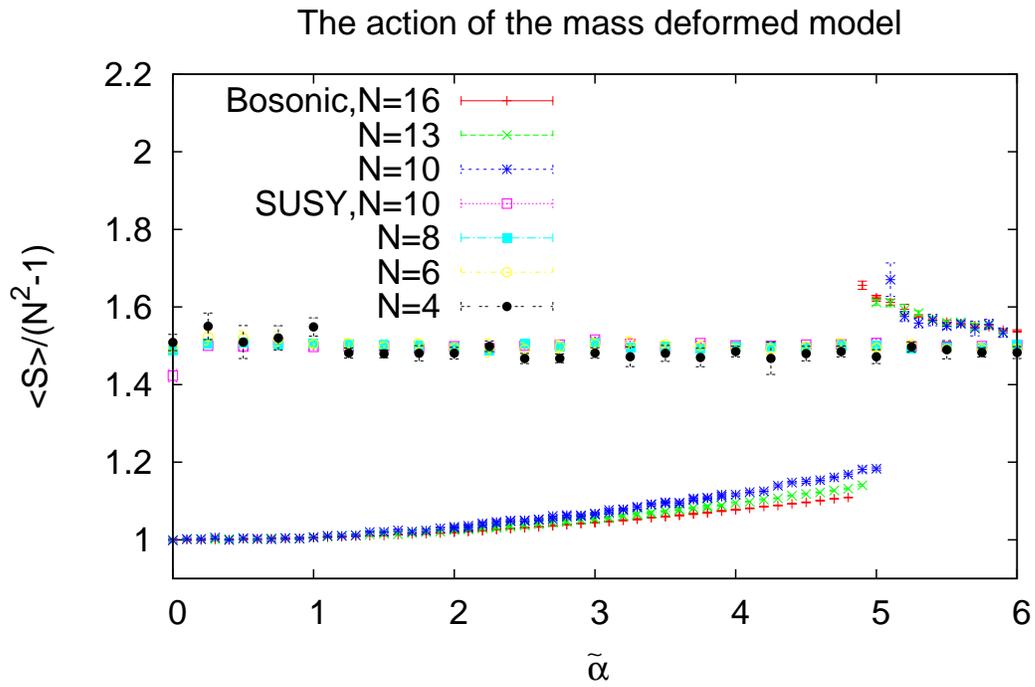}
\caption{The average of the total action of the supersymmetric models. }\label{actionF}
\end{center}
\end{figure}

\begin{figure}[htbp]
\begin{center}
\includegraphics[width=10.0cm,angle=-90]{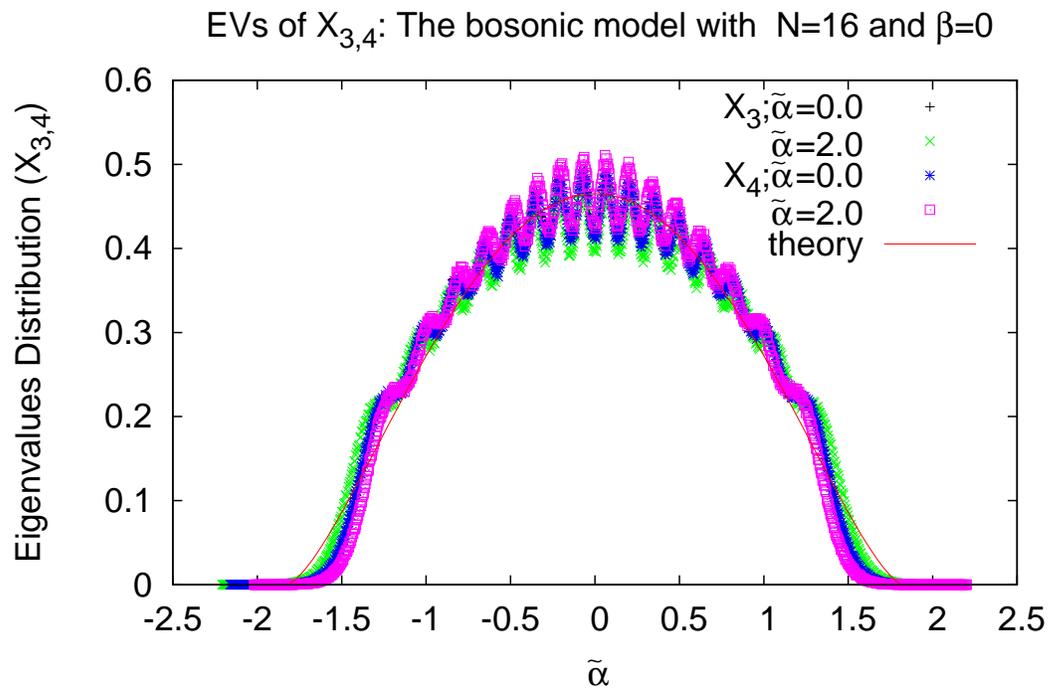}
\includegraphics[width=10.0cm,angle=-90]{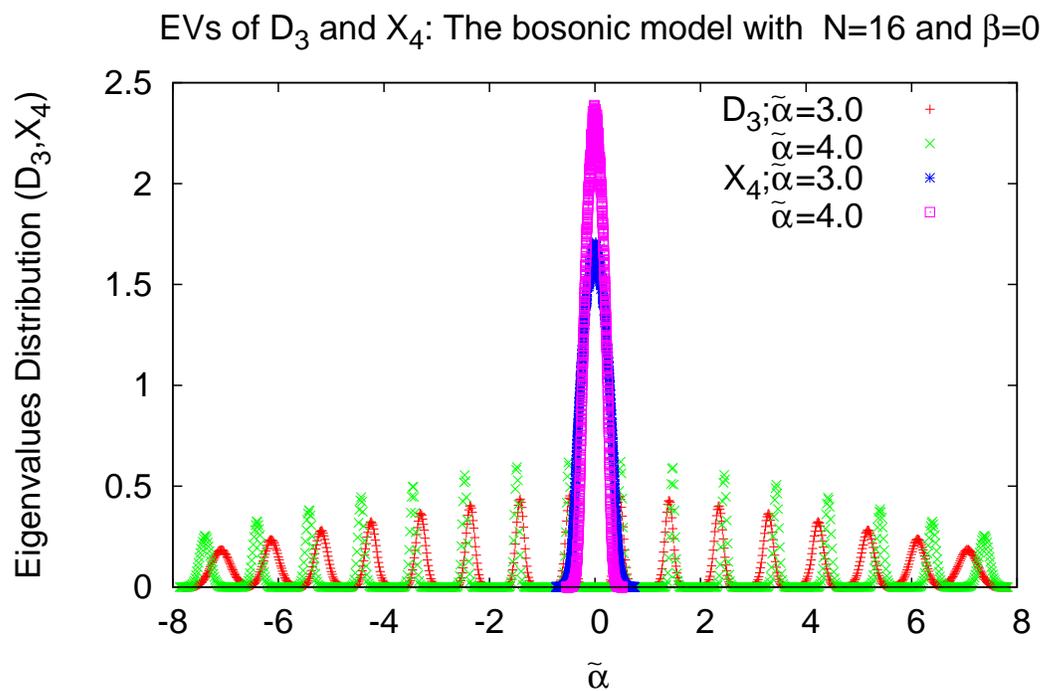}
\caption{The eigenvalues distributions of the bosonic model with $\tilde{\beta}=0$. }\label{distrB}
\end{center}
\end{figure}

\begin{figure}[htbp]
\begin{center}
\includegraphics[width=10.0cm,angle=-90]{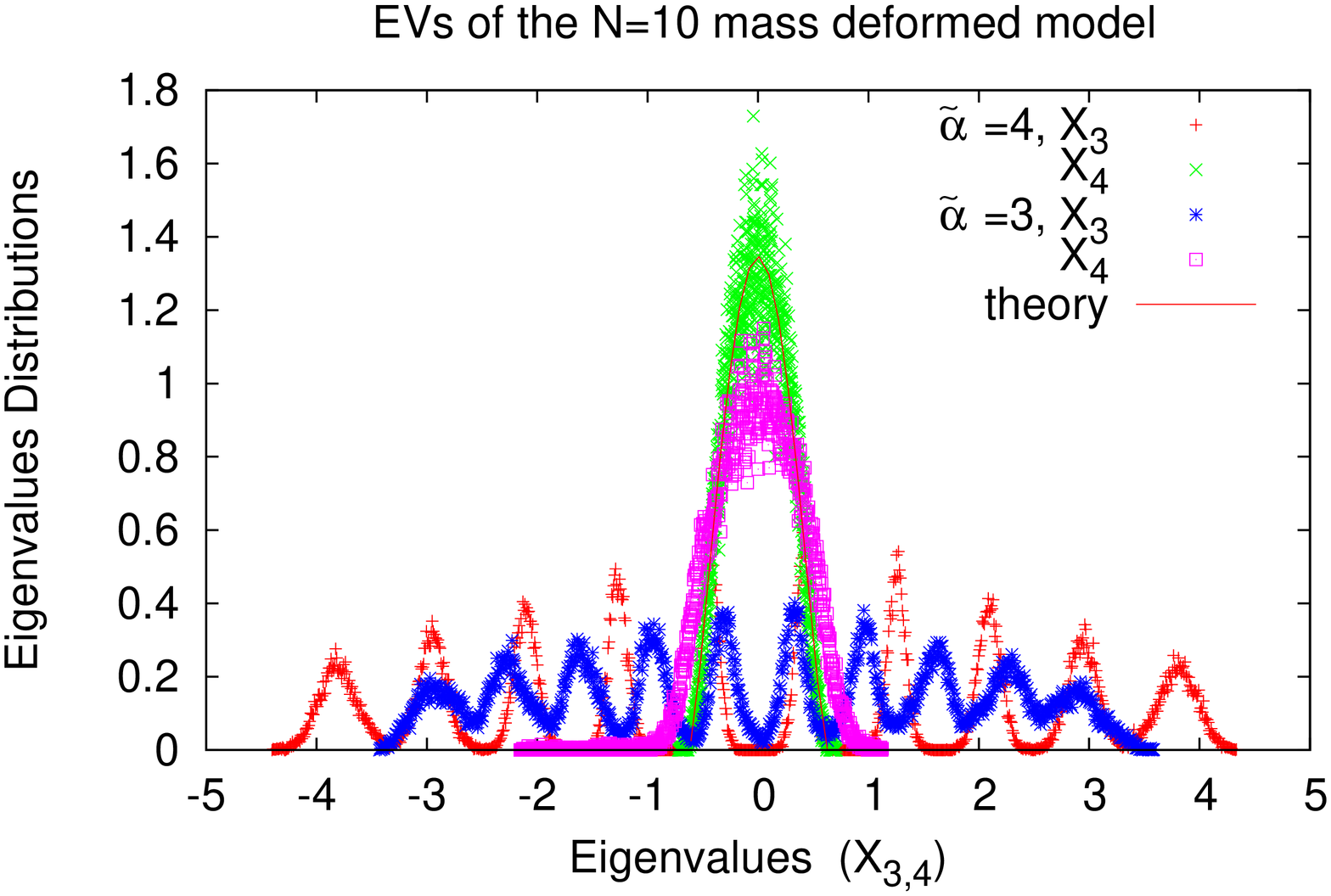}
\includegraphics[width=10.0cm,angle=-90]{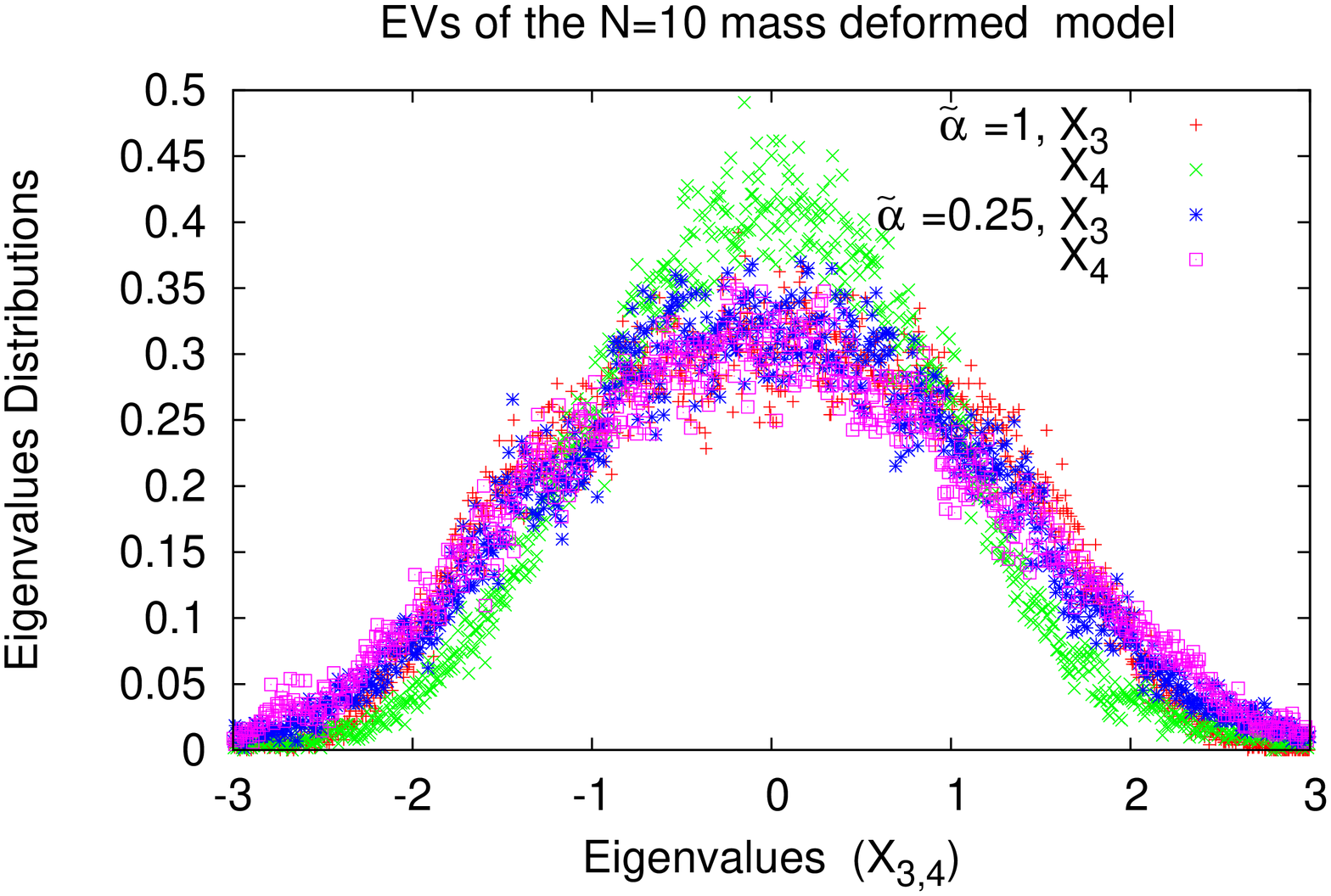}
\caption{The eigenvalues distributions of the mass deformed model. }\label{distrF1}
\end{center}
\end{figure}

\begin{figure}[htbp]
\begin{center}
\includegraphics[width=10.0cm,angle=-90]{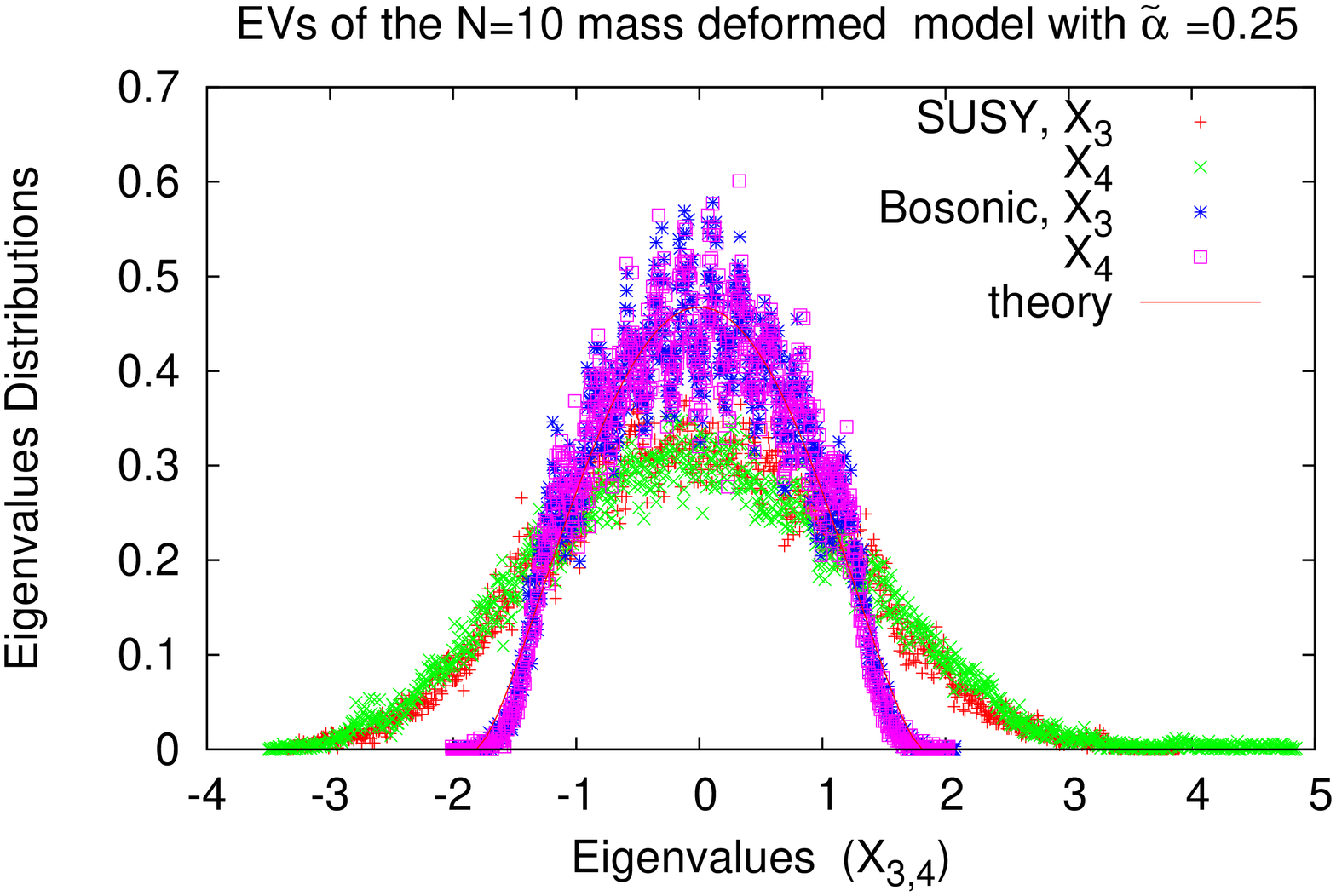}
\includegraphics[width=10.0cm,angle=-90]{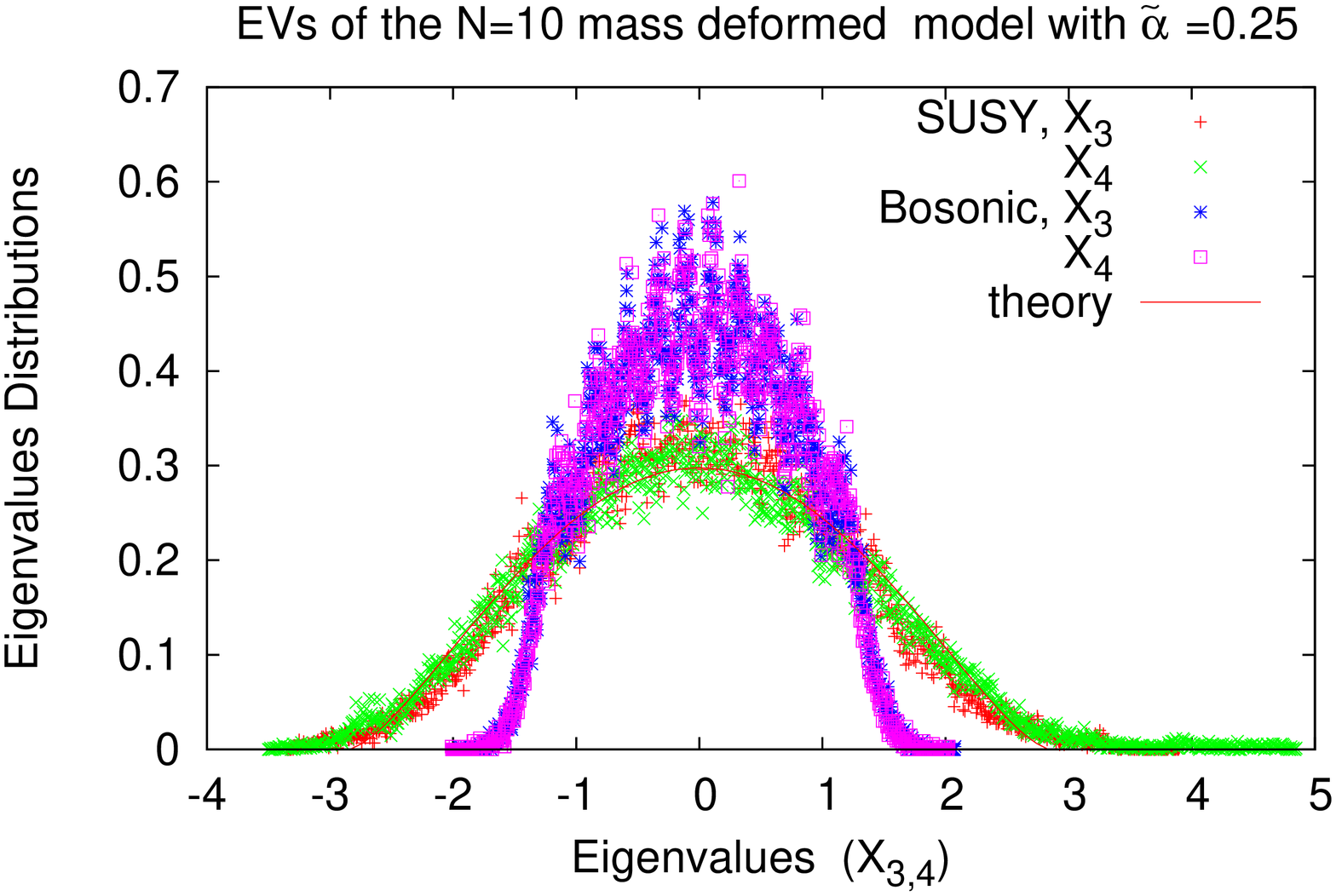}
\caption{The eigenvalues distributions of the mass deformed model. }\label{distrF2}
\end{center}
\end{figure}

\begin{figure}[htbp]
\begin{center}
\includegraphics[width=10.0cm,angle=-90]{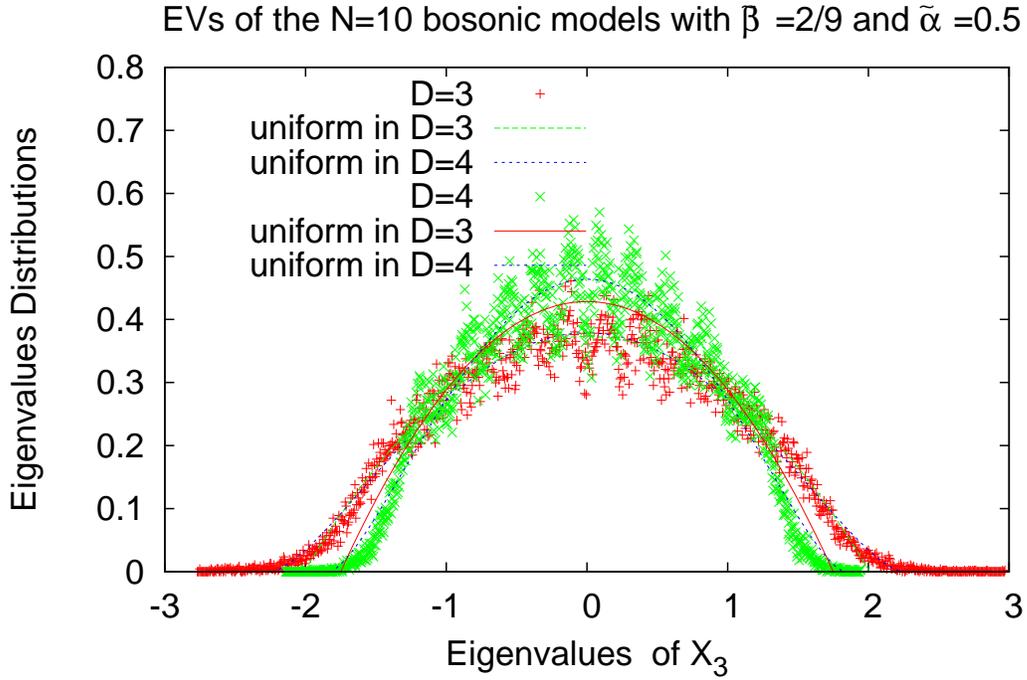}
\includegraphics[width=10.0cm,angle=-90]{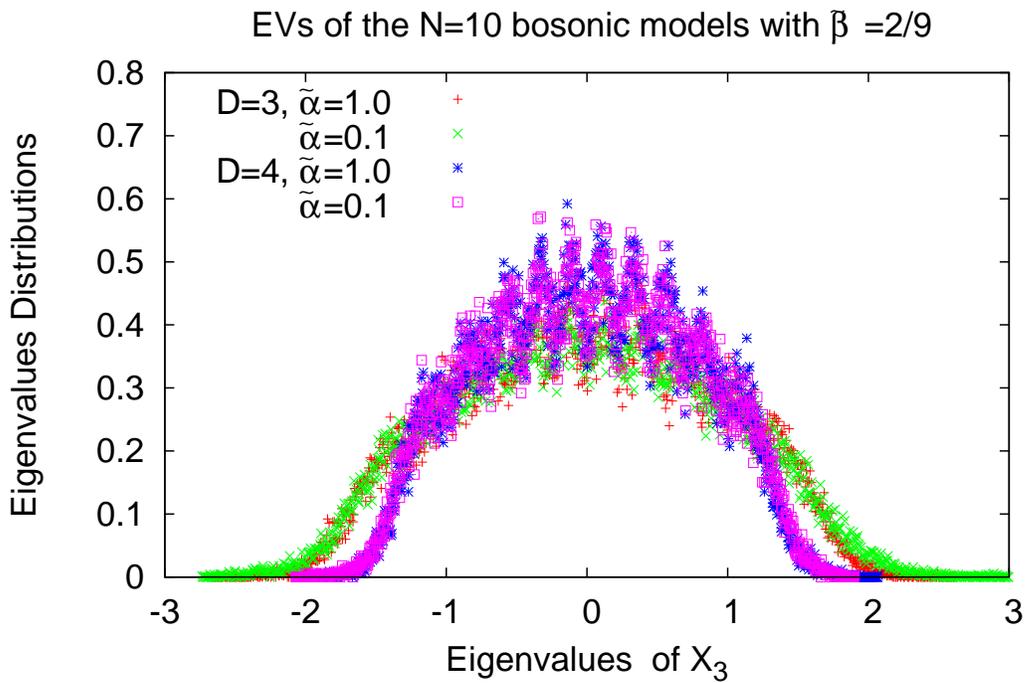}
\caption{The eigenvalues distributions of $X_3$ of the $=3,4$ bosonic models with $\tilde{\beta}=2/9$ in the matrix phase. }\label{D3D4}
\end{center}
\end{figure}

\begin{figure}[htbp]
\begin{center}
\includegraphics[width=10.0cm,angle=-90]{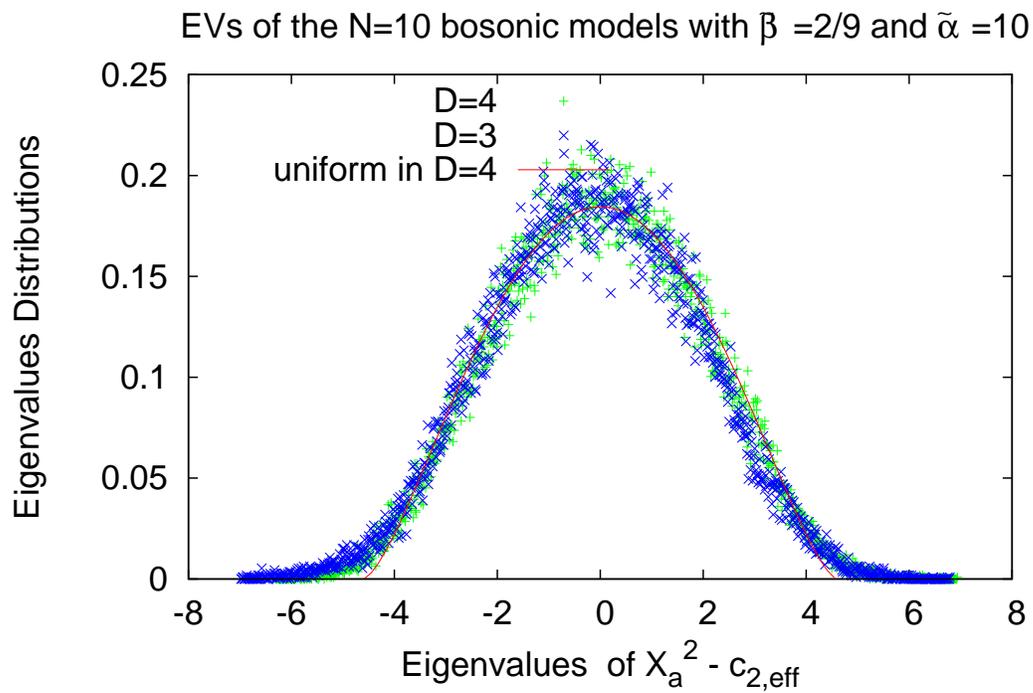}
\caption{The eigenvalues distributions of the normal scalar fluctuation $X_a^2-c_2$ of the $D=3,4$ bosonic models with $\tilde{\beta}=2/9$. }\label{scalar}
\end{center}
\end{figure}

\end{document}